\documentclass[a4paper,twoside]{ociamthesis}

\correctionstrue

\title{Relativistic mean field study of neutron stars and hyperon stars}
\author{Ishfaq Ahmad Rather}
\college{Department of Physics, \\
	Aligarh Muslim University, Aligarh, 202002}

\degree{Doctor of Philosophy}
\degreedate{2021}

\def\k{0.55191496}

\tikzset{
	sphere color/.store in=\spherecolor,
	sphere scale/.store in=\spherescale,
	sphere color=blue,
	sphere scale=1,
	sphere/.style={
		ultra thick,
		line join=round,
		draw=#1!75!black,
		ball color=#1,
	},
	sphere inside/.style={
		shading angle=180,
		sphere=#1!25!gray!75!black
	}
}

\newenvironment{sphere}[1][]
{
	\begin{scope}[x=(0:1cm), y=(90:1cm), z=(260:0.25cm), #1]
		\path [sphere inside=\spherecolor, scale=\spherescale] 
		circle [radius=0.9];
	}
	{
		\path let \n1={cos 10}, \n2={sin 10} in [sphere=\spherecolor, scale=\spherescale, even odd rule, opacity=0.5]
		circle [radius=0.9] 
		[x={(\n1, \n2^2, \n2*\n1)},
		y={(0, \n1, \n2)}, 
		z={(-\n2, -\n1*\n2, \n1^2)}] (0,1,0) 
		.. controls ++( 0, 0,\k) and ++(0,\k, 0) .. (0, 0, 1)
		.. controls ++(\k, 0, 0) and ++(0, 0,\k) .. (1, 0, 0) 
		.. controls ++(0, \k, 0) and ++(\k,0, 0) .. (0, 1, 0);
	\end{scope}
}

\begin{document}

\setlength{\textbaselineskip}{22pt plus2pt}

\setlength{\frontmatterbaselineskip}{17pt plus1pt minus1pt}

\setlength{\baselineskip}{\textbaselineskip}


\setcounter{secnumdepth}{5}
\setcounter{tocdepth}{5}


\begin{romanpages}

\begin{titlepage}
	\begin{center}
		\vspace*{0.15cm}
		
		\Huge
		\textbf{Relativistic Mean Field Study of Neutron Stars and Hyperon Stars}

	
		\vspace{1.3cm}
				\Large
		\textbf{A THESIS}\\
	     Submitted in partial fulfillment of the\\
	     requirements for the award of the degree \\
	     of\\
		\textbf{DOCTOR OF PHILOSOPHY}\\
		in\\
		\textbf{PHYSICS}\\
		\vspace{0.4cm}
		\textit{by}
		\vspace{0.5cm}
		
		{\Large\textbf{Ishfaq Ahmad Rather}\\
		 Enroll. No: GG-1951}
		\vspace{0.9cm}
		
		\includegraphics[width=0.35\textwidth]{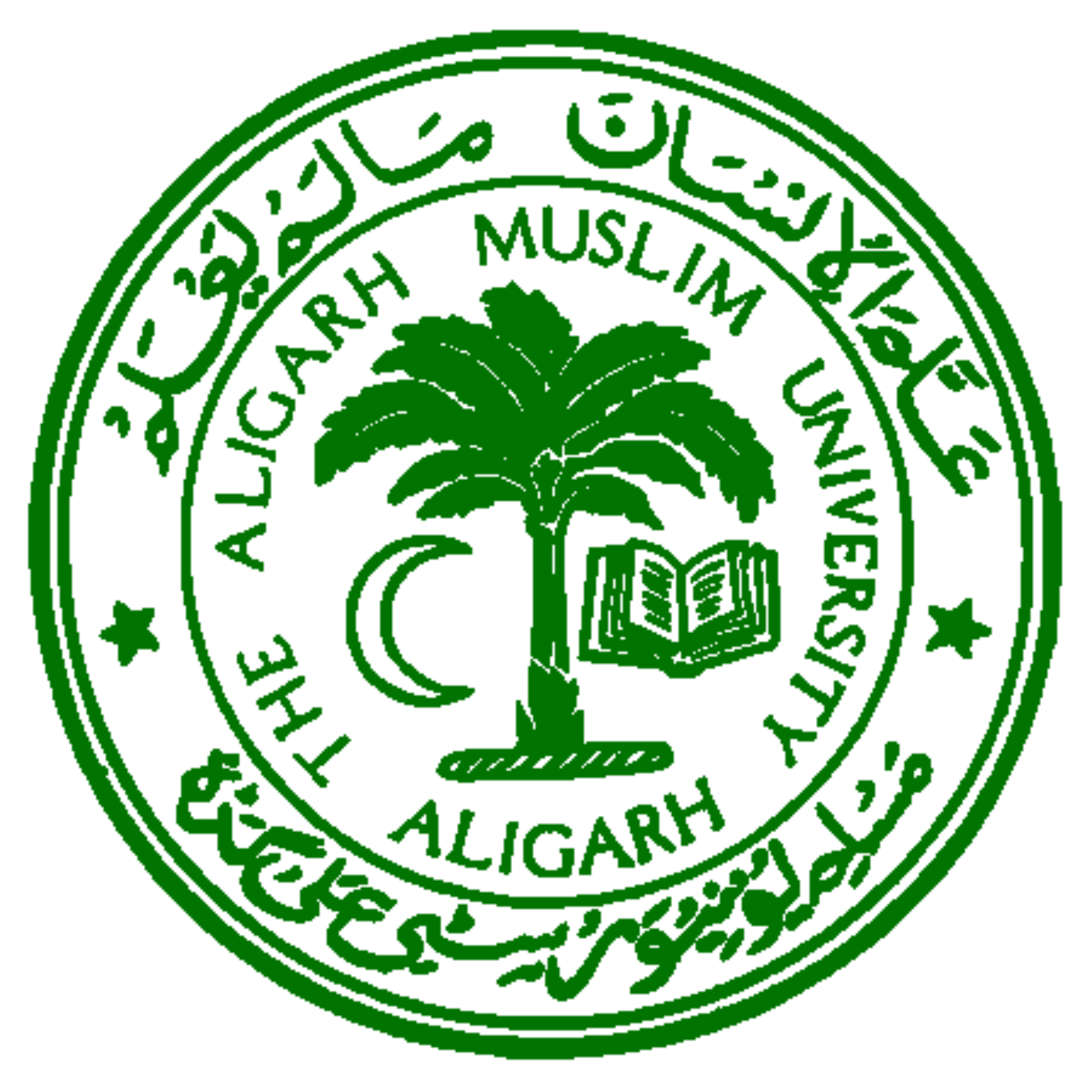}
		
		\vspace{0.4cm}
		Supervisor: {\Large\textbf{Prof. Anisul Ain Usmani}}
		\vspace{0.8cm}
		
		\large
		\textbf{DEPARTMENT OF PHYSICS\\
		ALIGARH MUSLIM UNIVERSITY\\
		ALIGARH-202002 (INDIA)\\
		2021}
		
	\end{center}
\end{titlepage}

\newpage

\vspace{-0.5cm}
\begin{center}
	\includegraphics[width=0.22\textwidth]{figures/amu.pdf}
\end{center}
\vspace{0.1cm}
\begin{center}
	\underline{\textbf{ANNEXURE-I}}\\
	\vspace{0.1cm}
{\large \textbf{CANDIDATE'S DECLARATION}}
\end{center}
\hspace{0.5cm}
I, {\color{blue}\bf Ishfaq Ahmad Rather} (Enroll. No.: GG-1951), Ph.D. student in the Department of Physics certify that 
the work embodied in this Ph.D. thesis entitled {\color{blue} \bf "Relativistic Mean Field Study of Neutron Stars and Hyperon Stars"} is my own bonafide work 
carried out under the supervision of {\bf Prof.~Anisul Ain Usmani} 
at Aligarh Muslim University, Aligarh. 
The content embodied in this Ph.D. thesis has not been submitted 
for the award of any other degree.

\hspace{1.0cm}
I declare that I have faithfully acknowledged, given credit to and referred
to the research workers wherever their works have been cited in the text
and the body of the thesis. I further certify that I have not wilfully
lifted up some other's work, para, text, data, results, etc., reported in the 
journals, books, magazines, reports, dissertations, thesis, etc., or available at websites
and included them in this Ph.D. thesis and cited as my own work. 

\vspace{1.2cm}
	Date: .............. \hspace*{5.5cm}{(\bf Signature of the candidate)}\\
	Place: AMU, Aligarh \hspace*{6.0cm}{Ishfaq Ahmad Rather}
\vspace*{0.2cm} \\
\rule{\textwidth}{0.4pt}
\hspace*{4cm}{\bf CERTIFICATE FROM THE SUPERVISOR}\\
\vspace{0.1cm}
 This is to certify that the above statement made by the candidate is correct to the best of my knowledge.
\vspace{0.8cm}\\
\begin{flushright}
\hspace*{4.5cm}{(\bf Signature of the Supervisor)}
\vspace{0.3cm}\\
\hspace*{4.0cm}{\bf Prof. Anisul Ain Usmani} 
\vspace{0.2cm}\\
\hspace*{5.3cm}Department of Physics,\\
Aligarh Muslim University, \\
Aligarh, Inida.
\end{flushright}
\vspace*{1.0cm}
\hspace*{1.4cm}{\bf(Signature of the chairperson of the Department with seal)}

\newpage
\begin{center}
\includegraphics[width=0.22\textwidth]{figures/amu.pdf}
\end{center}
\vspace{0.2cm}
\begin{center}	
		\underline{\textbf{ANNEXURE-II}}\\
	\vspace{0.1cm}
{\large \bf {COURSE/COMPREHENSIVE EXAMINATION/PRE-
		SUBMISSION SEMINAR COMPLETION CERTIFICATE}}
\end{center}
\vspace{1.5cm}

This is to certify that Mr. Ishfaq Ahmad Rather, Enroll. No: GG-1951, Department of Physics,
 Aligarh Muslim University has satisfactorily completed the course work/comprehensive examination and presubmission seminar requirement {\bf which is the part of his Ph.D. programme}.

\vspace{2.5cm}
\begin{flushleft}
	Date: ...........
\end{flushleft}
\begin{flushright}
{\bf( Signature of the Chairperson)}\\
Department of Physics,\\
A. M. U., Aligarh, India.
\end{flushright}

\newpage
\begin{center}
	\includegraphics[width=0.22\textwidth]{figures/amu.pdf}
\end{center}
\vspace{0.2cm}
\begin{center}
		\underline{\textbf{ANNEXURE-III}}\\
	\vspace{0.1cm}
{\large \bf COPYRIGHT TRANSFER CERTIFICATE}	
\end{center}
\vspace{1cm}
{\bf \large Title of the thesis:} {\color{blue} \bf \large Relativistic Mean Field Study of Neutron \hspace*{4.6cm}Stars and Hyperon Stars}\\

{\bf \large Candidate's Name}: {\color{blue}\bf \large Ishfaq Ahmad Rather}
\vspace{2.2cm}\\
\hspace*{5.0cm}\underline{{\bf COPYRIGHT TRANSFER}}\\
\vspace{0.5cm}

The undersigned hereby assigns to the Aligarh Muslim University, Aligarh copyright that may exist in and for the above thesis submitted
for the award of the Ph.D. degree.\par
\vspace{3.0cm}
\hspace{9.0cm}{\bf (Ishfaq Ahmad Rather)}\par
\vspace{2.0cm}
{\bf Note:} However, the author may reproduce or authorize others to reproduce material extracted verbatim from the thesis or derivative of the
thesis for author's personal use provided the source \& the University's copyright notices are indicated.

\begin{dedication}
\Large{\color{blue} \bf{\textit{Dedicated to\\
My Beloved Family,\\
My Mentors and Friends.}}}\\
\end{dedication}
\newpage
\begin{acknowledgements}


First and foremost, I must acknowledge and thank The Almighty Allah, the most benevolent and compassionate, for bestowing bravery, patience, and strength on me to finish this task.
Looking back on the last five years, I am overwhelmed with a wide range of emotions and recollections. It would be impossible for me to complete my thesis without the assistance of the many lovely people that surround me. I would want to express my gratitude to everyone who contributed to this thesis and helped me during my Ph.D. studies.

I want to thank my supervisor, \textbf{Prof.~Anisul Ain Usmani}, for his direction, paternal care, constructive criticism, kind attitude, continual support, guidance, counsel, and fascinating conversation, as well as for countless opportunities to study and attend conferences and workshops. In several instances, his tolerance and constant and generous monitoring saved me from despair.
I am always grateful for his compassion, belief in me, and tireless efforts as a mentor.
I'd want to convey my heartfelt appreciation once more to my supervisor, without whose academic vigilance, sense of excellence, and essential advice this task would never have been completed.
I am grateful to my supervisor for providing me with the chance to research with a famous theoretical physics group at the Institute of Physics (I.O.P.) in Bhubaneswar, India.

I am grateful to the Chairperson of the Department of Physics at Aligarh Muslim University, Aligarh, who has always been supportive and has supplied me with all of the resources needed to complete this study.

I want to express my heartfelt thanks to \textbf{Prof.~Suresh Kumar Patra} for providing me with the chance to pursue my Ph.D. studies at I.O.P., Bhubaneswar, and be a member of his outstanding group. I learned a lot from our conversations in his office, where he carefully explained the concepts and mechanics to me. Even during the difficult times of Covid-19, he was available for all the discussions and manuscript preparation. His vast knowledge, creative ideas, and zeal for scientific study taught me a lot. Also, I'd want to thank you for his encouragement, which came in handy when I was feeling down and out due to the disappointing results. I also recall how happy I was when I finally had some productive time and finished my first manuscript. I would also like to thank \textbf{Harish Chandra Das}, \textbf{Ankit Kumar}, and \textbf{Manpreet Kaur} for all of their assistance with the smooth running of codes, submitting the initial paper, and during my travels to I.O.P. Harish Chandra Das was constantly accessible to discuss new ideas and reply to referee criticisms, even during Covid-19. I also want to thank \textbf{Dr.~Abdul Quddus} and \textbf{Dr.~Bharat Kumar} for helping with the codes and motivating me to challenge my preconceptions and examine the situations from other angles.

\textbf{Prof.~Veronica Dexheimer} of Kent State University deserves my heartfelt gratitude. It was a wonderful experience and a rewarding partnership with you. Thank you very much for your patience and effort in talking about incorporating new ideas, data analysis, and paper writing. Your outstanding efforts aided me in my job, and I learned a great deal from working with you.

I'd want to thank all of my Ph.D. colleagues, with whom I've shared moments of great worry as well as great delight. Their presence was crucial in a procedure that is frequently perceived as entirely alone. A kind word for my colleague and my friend \textbf{Ishfaq Majeed Bhat}, who always made me feel unique and with whom I had the most pleasing tea breaks of my life.

My heartfelt gratitude goes to the anonymous referees for their thorough examination, constructive criticism, and helpful assistance during the manuscript submission process, which aided me in creating my thesis.

I want to express my heartfelt appreciation to my family for their unwavering love, assistance, and support. I am grateful to \textbf{Irfan Rather}, \textbf{Nadia Nabi} and \textbf{Firdous Rather} for always being a friend to me. I will be eternally grateful to my parents for providing me with the chances and experiences that have shaped who I am. They selflessly pushed me to try new things and follow my path in life. This journey would not have been possible for them, and I want to dedicate this milestone to them.

I'll never forget the friends that stood with me through the tough times, cheered me on, and celebrated each accomplishment:  \textbf{Pir Mohammad Junaid}, \textbf{Ishfaq Hamid Dar}, \textbf{Shafqat Ul Islam}, \textbf{Shariq Hussian Bhat}, \textbf{Junaid Khan}, \textbf{Junaid Rather}.

Special thanks go to my Aligarh friends, who have always been a tremendous source of support when times have been rough. The days would have passed much more slowly if not for my Aligarh friends, who put up with my eccentricities and gave such a rich amount of debate, information, and humor. Special thanks to, \textbf{Dr.~Aadil Rashid}, \textbf{Arshied Manzoor}, \textbf{Asloob Ahmad Rather}, \textbf{Zahoor Ahmad Bhat}, \textbf{Zia ul Haq}, \textbf{Athar Dar}, \textbf{Furqan Bashir}, \textbf{Meer Nasrullah}, \textbf{Irfan Ayoub}, \textbf{Dr.~Sofi Suhail Majid}, \textbf{Shabbir Ahmad Dar}, \textbf{Dr.~Mohsin Rasool}, \textbf{Dr.~Mohd. Shuaib}, \textbf{Prabhat Solanki}.

Friends like \textbf{Mudassir Mohammad} and \textbf{Pomposh Aalam} come along just once in a lifetime. I don't think there are enough words to describe how grateful I am for all the times you've listened to me and been there when I needed someone. It's nice to know you care enough to remain by my side through good and terrible times. I hope I can be as good a friend to you as you have been to me.

I owe a great deal to my lab colleagues {\textbf{Dr.~Mohd. Ikram}}, \textbf{Dr.~Mohd. Imran}, and \textbf{Usuf Rahaman} for their helpful talks, critical reading of the dissertation, and constructive suggestions that have been included to strengthen the thesis content.

Financial assistance provided by U.G.C., Government of India, 
during my Ph.D. program is highly acknowledged.

I am also grateful to all of the staff members of the seminar library at Aligarh Muslim University, Aligarh, for giving me access to the books and periodicals I needed for my study.

When it comes to my alma mater, Aligarh Muslim University, its rich legacy, and magnificent cultural ethos have always served as a source of inspiration for all I've done in my Aligarh life. The lyrical University Tarana and A.M.U. Logo has always expected me to lift my head and achieve in the area of education. Of course, the day I entered this magnificent seat of learning that has Sir Syed's vision and purpose in its air and that I inhaled here day in and day out for ten years will be a memorable period in my life for the rest of my life.
A decade-long stay from B.Sc. to Ph.D. will be the most valuable portion of my life for the rest of my life.
A.M.U. became my second home as a result of this.
\vspace{1.0cm}
\begin{flushright}
	Ishfaq Ahmad Rather
\end{flushright}

\end{acknowledgements}

%
%
%
%
%
{\centering 
		\section*{\huge Abstract}
	}
Nuclear physicists have been attempting to comprehend nuclear systems ranging from finite nuclei to hot and dense nuclear matter in a systematic manner within a single theoretical framework. The nucleus is a many-body system composed of protons and neutrons that are highly interacting and self-bound. As a result, describing the nucleus in a concise manner becomes quite challenging. The relativistic mean-field (RMF) theory, which takes into account relativistic nucleons interacting with one other by exchanging mesons, is one of the most successful and widely employed approaches. In this approach, a nuclear system is composed of relativistic nucleons whose self-energy is governed by meson fields created by nuclear density. It has also been quite successful in explaining a wide range of nuclear events, both in the low-density and high-density regions. The recent detection of gravitational waves from LIGO/Virgo detectors and heavy-ion colliders have greatly enlarged the window through which nuclear matter and neutron stars (NSs) can be investigated in the high-density area. Such investigations contribute to a deeper knowledge of the RMF model as well as new suggestions for how to strengthen it further. Since it is a phenomenological model, its effectiveness may be evaluated by comparing it to the experiment. This thesis focuses on a variety of active research topics, such as nuclear matter, neutron stars, and phase transition within the framework of the RMF model.

We use the previously successful effective field theory-driven Relativistic Mean-Field (RMF) and density-dependent RMF formalisms for analyzing hadron matter to examine the infinite nuclear matter and neutron stars. The presence of exotic phases such as quarks will be investigated using the MIT Bag model and its variants, such as the vBag model, at various bag constants. The other exotic phases, such as hyperons, are also studied under the influence of a strong magnetic field.

After a detailed description of the theoretical models developed in the study of nuclear matter and neutron stars, Chapter 1, Chapter 2 provides a brief introduction to various modern examples of infinite nuclear matter, the formation of NSs, their properties, and possible exotic phases. Chapter 3 concentrates on the mathematical derivations used throughout our study. We begin with the extended RMF Lagrangian density with meson and cross-coupling, which has a large number of terms with various levels of self-and cross-coupling. The DD-RMF model is also used to analyze hadron matter, allowing for a consistent computation of neutron stars and yielding findings equivalent to previous models. It uses microscopic interactions at various densities as input to include the characteristics of the Dirac-Brueckner model. Extrapolation to higher densities is more confined than in phenomenological RMF calculations, which employ only information from the finite nuclei's narrow density range to determine their parameters. The RMF and DD-RMF parameter sets, as well as their nuclear matter properties, are also explored in this thesis study. In addition, the equation of state is computed using the energy-momentum tensor, and various formulae for the characteristics of symmetric nuclear matter, pure neutron matter, and infinite nuclear matter are created in the RMF approximation.

In Chapter 4, we study the hybrid Equation of State (EoS) produced by mixing hadron and quark matter under Gibbs conditions. The E-RMF model for hadron matter with newly published parameter sets, as well as the MIT bag model for quark matter with variable bag constants, are employed. The MIT bag model is a degenerate Fermi gas comprising of quarks (u,d, and s) and electrons stabilized in chemical equilibrium by a number of weak interactions. The quarks are thought to be contained in a colorless region where they are free to move in this scenario. To balance the behavior of the bag and determine its size, a bag constant $B$ is introduced into the system as a constant energy density. For the phase transition from hadron matter to the quark matter, the Gibbs condition, which accounts for the global charge neutrality, is used. Nuclear matter properties including symmetry energy ($J$), slope parameter ($L$), and incompressibility ($K$) are determined for hybrid EoS. It is observed that the values of symmetry energy $J$ and other variables for a hybrid EoS are extremely large and that they grow with the bag constant, but incompressibility decreases with $B$. Star matter characteristics such as mass and radius are calculated for various bag constants. A bag constant in the range $B^{1/4}$ = 130 - 160 MeV is shown to be enough for describing quark matter in neutron stars. The findings obtained with bag values less than 130 MeV predict the presence of quarks below the nuclear saturation density ($\rho_0$), which is unphysical. The bag values greater than 160 MeV contradict previous gravitational wave observables.

The RMF model is used in Chapter 5 to investigate NS properties such as mass, radius, and tidal deformability. To determine the influence of symmetry energy slope parameter on an NS, inner crust EoSs with varying symmetry energy slope parameters are used. For the outer crust, the BPS EoS is used for all sets as the outer crust part doesn't affect the NS maximum mass and radius. For the inner crust part, parameter sets such as NL3, TM1, FSU, NL3$\omega \rho$, DD-ME$\delta$, DD-ME2, and IU-FSU are used whose slope parameter varies from 118.3 - 47.2 MeV. For the core part, NL3, TM1, IU-FSU, IOPB-I, and G3 parameter sets are used. The unified EoSs are constructed using the thermodynamic approach by correctly matching the crust EoS with the outer crust and the core EoS. As the slope parameter is adjusted from low to high values, the radius $R_{1.4}$ increases. The effect of $L_{sym}$ on the NS maximum mass, radius, and radius at 1.4$M_{\odot}$ is calculated. Although the variation in the maximum mass and the corresponding radius of an NS are very small, a difference of about 2 km in radius at the canonical mass is identified. The same combination of EoSs is used to estimate the properties of a maximally rotating star, such as mass, radius, the moment of inertia, and $T/W$ ratio. The maximum mass and corresponding radius for a rotating neutron star, like a static neutron star, do not fluctuate significantly. However, the radius at the canonical mass is affected by the slope parameter, but the radius at the canonical mass is affected by the slope parameter. The moment of inertia and the kinetic to potential energy ratio also vary with the change in the symmetry energy slope parameter of the crust. However, such rotating NS properties are more related to the mass and the radius of the star.

The hadron-quark phase transition is studied in Chapter 6 in the context of the recent discovery of the gravitational wave event GW190814 with a secondary component of mass 2.50 - 2.67$M_{\odot}$. A set of recent DD-RMF parameter sets such as DDV, DDVT, DDVTD, DD-LZ1, and DD-MEX, along with the widely used parameter sets DD-ME1 and DD-ME2, are used to investigate the properties of stellar matter. The Vector-Enhanced Bag (vBag) model for quark matter is used to investigate the phase transition from hadron matter to quark matter. The vBag model is an efficient method that takes into account dynamic chiral symmetry breaking (D$\chi$SB) and repulsive vector interactions. It also considers the phenomenological correction to the quark matter EoS that characterizes the deconfinement, which is dependent on the hadron EoS used to construct the phase transition. The repulsive vector interaction is important because it allows the hybrid stars to reach the 2$M_{\odot}$ maximum mass limit. The inclusion of flavor-dependent chiral bag constants is justified by fits to the pressure of the chirally restored phase. Furthermore, a deconfined bag constant is provided to reduce energy per particle, favoring strange stable matter. The Maxwell and Gibbs methods are both used to build the mixed-phase between hadrons and quarks. The effective bag constant B$^{1/4}_{eff}$ is used, with values of 130 and 160 MeV. 

We see that the hadronic EoSs generated using the most recent DD-RMF parameterizations satisfy the mass constraint from the GW190814 data, allowing us to investigate the idea that the GW190814 secondary component could be a possible massive NS. In order to meet the requirements from the GW170817 data, the phase transition from hadron matter to quark matter reduces NS parameters such as mass, radius, and tidal deformability, imposing further limits on the NS maximum mass and thus on the dense matter EoS. In addition to static NS properties, maximally rotating NS properties such as mass, stellar radius, the moment of inertia, Kerr parameter, and so on are investigated. With the given parameter sets, it is seen that the secondary component of GW190814 is a possible supermassive NS with a strange quark core.

In Chapter 7, we use a DD-RMF model to simulate massive nucleonic and hyperonic stars that meet the constraints imposed by the discovery of the possibly most massive neutron star (NS) ever discovered (GW190814). The hyperonic counterparts of the used DD-RMF parameter sets soften the EoS, hence producing an NS with a maximum mass lower than the pure nucleonic ones with all configurations satisfying the mass-radius limits inferred from the NICER experiment.

Strong magnetic field effects on nucleonic and hypeornic EoSs are also observed. We investigate the EoS and particle populations by creating a realistic chemical potential-dependent magnetic field by solving Einstein-Maxwell equations. Under very strong magnetic fields, the spherically symmetric solutions obtained by solving the TOV equations result in an overestimation of the mass and underestimation of the radius and thus cannot be used to determine stellar parameters. As a result, we use the LORENE library to determine the stellar properties of magnetic NSs. At low values of the magnetic dipole moment, the EoS resembles the non-magnetic one, implying lower magnetic field intensities. The maximum mass and the corresponding radius increase with the increase in the magnetic field. The radius at the canonical mass increases by about 1 km. The variation obtained in the mass radius is larger for hyperonic stars than for the pure nucleonic stars due to the additional effect of de-hyperonization that takes place due to the magnetic field. The dimensionless tidal deformability for NSs with and without hyperons reaches a value of $\Lambda_{1.4}$ $\approx$ 1550, which is much larger than the constraints from GW170817 data. For an NS and a hyperon star with dimensionless tidal deformability well within the limit of GW170817 at 90\% confidence, a magnetic field with a maximum value of $\approx$ 2 $\times$ 10$^{16}$ G is required.

The mass-radius profile for DD-MEX parameter set without magnetic field and with magnetic field considering different magnetic dipole moments by solving general relativity spherically symmetric solutions (TOV) are also shown. As the magnetic field increases by changing magnetic dipole moment, the maximum mass increases by about 0.1$M_{\odot}$ for NS and 0.2$M_{\odot}$ for hyperon star. It is seen that neglecting the deformation effects by solving the spherically symmetric TOV equations leads to an overestimation of the mass and an underestimation of the radius. This happens because the extra magnetic energy that would deform the star is being added to the mass due to the imposed spherical symmetry.  

When different coupling schemes are considered, the maximum mass reproduced satisfies the GW190814 mass limit implying that its secondary component can be a possible hyperonic magnetar. It is seen that for a central magnetic field approaching 10$^{17}$ G, the radius at canonical mass increases by about 1.5 km as compared to the previous couplings, where the radius changes by around 1 km. For an even stronger magnetic field, the different coupling scheme for hyperons increases the radius by 0.2 km in comparison to the previous one, in which case it increases by 1 km. Thus, we see that different hyperon couplings and different hyperon potentials populate the star matter differently and, hence, change the stellar properties significantly.

Chapter 8 comprises an overall summary of the thesis work and few conclusions are presented along with future plans.

	\newpage 
\section*{List of Publications}
\textbf{Published in Journals}
\begin{enumerate}
		\item \qq{Heavy Magnetic Neutron Stars},\\ 
	{\textbf{Ishfaq A. Rather}}, Usuf Rahaman, V. Dexheimer, A. A. Usmani, and S. K. Patra,\\
	\textbf{\textit{\color{blue}The Astrophys. Jour. 917, 46 (2021)}}.
	
	\item \qq{Rotating Neutron stars with Quark cores},\\
	{\textbf{Ishfaq A. Rather}}, Usuf Rahaman, M. Imran, H. C. Das, A. A. Usmani, and S. K. Patra,\\
	\textbf{\textit{\color{blue} Phys. Rev. C 103, 055814 (2021)}}.
	
	\item \qq{Hadron-Quark Phase transition in the context of GW190814},\\
	{\textbf{Ishfaq A. Rather}}, A. A. Usmani, and S. K. Patra,\\ 
	\textbf{\textit{\color{blue} J. Phys. G: Nucl. and Part. Phys. 48, 085201 (2021)}}.

	\item \qq{Effect of Inner crust EoS on Neutron star properties},\\ {\textbf{Ishfaq A. Rather}}, A. A. Usmani, and S. K. Patra,\\ 
\textbf{\textit{\color{blue}Nucl. Phys. A, 122189 (2021)}}.
	
	\item \qq{Constraining Bag constant for Hybrid neutron stars},\\ {\textbf{Ishfaq A. Rather}}, A. Kumar, H.C. Das, M. Imran, A. A. Usmani, and S. K. Patra, \\
	\textbf{\textit{\color{blue} Int. J. Mod. Phys. E 29, 2050044 (2020)}}.
	
	\item \qq{Study of Nuclear matter properties for Hybrid EoS},\\
	{\textbf{Ishfaq A. Rather}}, A. A. Usmani, and S. K. Patra,\\
	\textbf{\textit{\color{blue} J. Phys. G: Nucl. and Part. Phys. 47,105104 (2020)}}.
	
\end{enumerate}
\newpage
\vspace{2.0cm}
{\Large
	\textbf{Conferences/Seminars/Workshops}
}
\vspace{1.0cm}
\begin{enumerate}
	\item \textbf{Talk}: \qq{Magnetic deformation in Neutron stars},\\
	{\textbf{Ishfaq A. Rather}}, Usuf Rahaman, V. Dexheimer, A. A. Usmani, and S. K. Patra, \\
	\textbf{\textit{\color{blue}Proceedings of the 65$^{th}$ DAE Symp. on Nucl. Phys. (2021)}},\\
	BARC, Mumbai.
	
	\item \textbf{Talk}: \qq{GW190814 secondary component as maximally rotating hybrid star},\\
	{M. Imran, \textbf{Ishfaq A. Rather}}, Usuf Rahaman, A. A. Usmani, and S. K. Patra, \\
	\textbf{\textit{\color{blue}Proceedings of the 65$^{th}$ DAE Symp. on Nucl. Phys.  (2021)}},\\
	BARC, Mumbai.
	
		\item \textbf{Talk}: \qq{Magnetic deformation in Neutron and Hyperon stars},\\
	\textbf{Ishfaq A. Rather}\\
	\textbf{\textit{\color{blue}International Conference of International Academy of Physical Sciences (CONIAPS-XXVIII) on Frontiers in Physics (2021) }},\\
	Dept. of Physics, Kashmir University \& Dept. of Physics, Islamic University of Science and Technology, J\&K, India.
	
	\item \textbf{Talk}: \qq{GW190814 secondary component as a Neutron star with Hadron-Quark Phase Transition},\\
	\textbf{Ishfaq A. Rather}\\
	\textbf{\textit{\color{blue}International Workshop on Emerging trends in High Energy and Condensed Matter Physics (2021)}},\\
	GDC, Budgam, Kashmir, J\&K, India.
	
	\item \textbf{Poster}: \qq{Constraining value of Bag constant for Hybrid stars},\\
	{\textbf{Ishfaq A. Rather}}, M. Imran, M. Ikram, A. A. Usmani, and S. K. Patra \\
	\textbf{\textit{\color{blue}Centenary Celebration Conference on Nuclear Structure and Nuclear Reactions (2020)}},\\
	Department of Physics, Aligarh Muslim University, Aligarh.
	
	\item \textbf{Poster}: \qq{Study of nuclear matter properties for Hybrid EoS},\\
	{\textbf{Ishfaq A. Rather}}, M. Ikram, A. A. Usmani, and S. K. Patra \\
	\textbf{\textit{\color{blue}Proceedings of the DAE Symp. on Nucl. Phys. 64 (2019)}},\\
	Department of Physics, Lucknow University.	
	
	\item \textbf{Poster}: \qq{Role of Hyperons in Neutron stars},\\
	{\textbf{Ishfaq A. Rather}}, M. Ikram, and M. Imran, \\
	\textbf{\textit{\color{blue}Proceedings of the DAE Symp. on Nucl. Phys. 63 (2018)}},\\
	BARC, Mumbai.

	\item \textbf{Poster}: \qq{Relativistic mean field study of Neutron and Hyperon stars},\\
	{\textbf{Ishfaq A. Rather}}, Asloob A. Rather, M. Ikram, and A. A. Usmani \\
	\textbf{\textit{\color{blue}Proceedings of the international conference in Nuclear physics with Energetic Heavy Ion Beams 48, 35 (2017)}},\\
	Punjab University, Punjab.
\end{enumerate}
\newpage
\dominitoc 

\flushbottom


\tableofcontents
\newpage
\listoffigures
	\mtcaddchapter
\newpage
\listoftables
	\mtcaddchapter

\begin{mclistof}{List of Abbreviations}{3.2cm}

\item[RMF] Relativistic Mean field

\item[QHD] Quantum Hydrodynamics

\item[DD-RMF] Density-Dependent Relativistic Mean Field
\item[EoS] Equation of State
\item[NM] Nuclear Matter
\item[SNM] Symmetric Nuclear Matter
\item[INM] Infinite Nuclear Matter
\item[PNM] Pure Neutron Matter
\item[NSM] Neutron Star Matter
\item[NS] Neutron Star
\item[HS] Hybrid Star 
\item[D$\chi$SB] Dynamic Chiral Symmetry Breaking
\item[vBag] Vector-Enhanced Bag 
\item[LIGO] Laser Interferometer Gravitational wave Observatory
\item[VIRGO] Variability of solar IRradiance and Gravity Oscillations
\item[NICER]  Neutron star Interior Composition ExploreR
\end{mclistof} 

\newpage
\begin{mclistof}{ List of Constants and Unit Conversions}{3.2cm}

\item[{\color{blue}Newton's constant}] \hspace{1.0cm} $G$ = 6.674 $\times$ 10$^{-11}$ m$^3$ kg$^{-1}$ s$^{-1}$\\

\item[{\color{blue}Speed of light in vacuum}] \hspace{2.0cm} $c$ = 3 $\times$ 10$^8$ m/s\\

\item[{\color{blue}Reduced Planck's constant}] \hspace{2.3cm} $\hbar$ = 1.055 $\times$ 10$^{-34}$ Js\\

\item[{\color{blue}Solar Mass}] \hspace{0.4cm} $M_{\odot}$ = 1.988 $\times$ 10$^{30}$ kg\\

\item[{\color{blue}Neutron Mass}] \hspace{0.4cm} $m_n$ = 939.565 MeV\\

\item[{\color{blue}Nuclear Saturation Density}] \hspace{2.5cm} $\rho_{0}$ = 0.16 fm$^{-3}$\\

\item[{\color{blue}MeV to fm$^{-1}$}] \hspace{0.7cm}: 1/197.33\\

\item[{\color{blue}MeV/fm$^3$ to g/cc}] \hspace{0.7cm}: 1.7827 $\times$ 10$^{12}$ g/cc\\

\item[{\color{blue}MeV/fm$^3$ to 1/km$^2$}] \hspace{0.7cm}: 1.3234 $\times$ 10$^{-6}$\\

\item[{\color{blue}g/cc to 1/km$^2$}] : 7.4261 $\times$ 10$^{-19}$\\

\item[{\color{blue}kg to km}] : 7.4261 $\times$ 10$^{-31}$\\

\item[{\color{blue}MeV to s$^{-1}$}] : 1.52 $\times$ 10$^{21}$\\

\item[{\color{blue}Dyne/cm$^2$ to 1/km$^2$}] \hspace{0.9cm}: 7.4261 $\times$ 10$^{-40}$
\end{mclistof} 

\end{romanpages}

\flushbottom
\begin{savequote}[8cm]
\textlatin{Acquire knowledge and teach people. Learn along with it dignity and tranquility and humility for those who teach you and humility for those whom you teach. Do not be tyrannical scholars and thus base your knowledge upon your ignorance.}

  \qauthor{--- \textit{Umar Ibn Al-Khattab}}
\end{savequote}

\chapter{\label{cha-lit}Introduction} 

\section{Introduction}
Finite nuclei and Neutron Stars (NSs) are many-body nuclear systems dominated by strong force. Despite the fact that Quantum Chromodynamics (QCD) \cite{Fritzsch:1973pi} is the basic theory of the strong interaction, solving the theory in the non-perturbative domain, applicable to nuclear systems, has been challenging. Until recently, these systems could only be examined within the context of an effective theory with adequate degrees of freedom. The approach based on the density functional theory (DFT) \cite{PhysRev.136.B864,PhysRev.140.A1133} is an effective approach, which may be applied to the whole nuclear landscape as well as the study of neutron stars. Several energy density functionals (EDFs), which may be classified into two categories: non-relativistic \cite{RevModPhys.75.121} and relativistic \cite{edf}, have been developed over the past few decades. Within the non-relativistic realm, where nucleons interact through density-dependent effective potentials, Skyrme-type functionals are the most prevalent \cite{CHABANAT1997710,CHABANAT1998231}. Relativistic mean-field (RMF) models, on the other hand, have been utilized effectively which offer a covariant description of finite nuclei as well as extended nucleonic matter \cite{WALECKA1974491,Serot:1984ey,HOROWITZ1981503,PhysRevLett.86.5647}. They are based on a quantum field theory in which nucleons interact via the interchange of different mesons.

Density functional theory offers a distinct framework for estimating the ground-state properties and collective excitations of medium and heavy nuclei. Based on the foundational work of \citet{PhysRev.136.B864,PhysRev.140.A1133}, DFT focuses on the considerably simpler one-body density rather than the complicated many-body wave function. The substantial complexity of calculating the correct ground-state energy and one-body density from the many-body wave function is therefore reduced to minimizing an appropriate density functional. 

\citet{PhysRev.140.A1133} provided the necessary formalism to derive the ground-state energy and corresponding one-body density from a variational problem that reduces to a self-consistent solution of a sequence of mean-field-like (“Kohn-Sham”) equations. Despite having the same form as the self-consistent Hartree (or Hartree-Fock) problem in the presence of nucleon-nucleon (NN) interaction, the constants parametrizing the Kohn-Sham potential were fitted directly to many-body properties such as binding energies and charge radii, rather than two-body data. The parameters implicitly capture the complex many-body dynamics in this manner. In theory, a good formulation of DFT and the Kohn-Sham equations integrate many-body effects into parameters that are functionals of the one-body density, such as the ground-state energy. 


Nuclear experimental data obtained under normal laboratory circumstances, particularly at or near nuclear saturation density and with small-to-moderate isospin asymmetries, are used to calibrate both non-relativistic and relativistic EDFs. The scarcity of experimental data at higher densities and/or with substantial isospin asymmetries produces a broad range of model predictions, even when all models are calibrated to the identical experimental data. As a consequence, important nuclear characteristics such as neutron density in medium-to-heavy nuclei \cite{PhysRevLett.102.122502}, neutron and proton drip lines \cite{Erler2012} and a vast range of neutron star properties \cite{doi:10.1126/science.1090720} remain unknown.

The non-relativistic, as well as relativistic theories have been developed and used to explore the infinite-body nuclear and neutron-matter properties. We briefly summarize them in the following sections. 

\section{Non-Relativistic Models}

A quantitative microscopic description of actual nuclei has only been feasible on a phenomenological level thus far. In order to account for ground-state characteristics, effective interactions based on density have been built. These effective forces are calculated in certain studies using a local density approximation (LDA) to BHF calculations \cite{PhysRev.140.A1133,PhysRevC.74.047304,Baldo:2016jhp}, with extra density dependence to imitate the influence of the missing higher-order corrections \cite{PhysRevC.1.1260}.
Purely phenomenological models with density-dependent interactions, such as the (zero range) Skyrme force \cite{Skyrme:1956zz} or the (limited range) Gogny force \cite{Gogny1975}, are very effective approximations.
These forces can be considered as effective parametrizations of the G-matrix but with only a few free adjustable parameters. To reproduce the nuclear matter densities and binding energies, these forces have shown the capability of not only yielding excellent nuclear ground-state properties of both spherical and deformed nuclei but also of quantitatively describing nuclear dynamical phenomena such as fission and heavy-ion collisions at low temperatures. The density-dependent Hartree-Fock (DDHF) technique is reviewed and explained in Refs.~\cite{doi:10.1146/annurev.ns.28.120178.002515,RevModPhys.54.913}.

We focus primarily on neutrons and their interactions since they are the fundamental constituents of matter within neutron stars.
The experimental data on nucleon-nucleon scattering and known deuteron properties do not uniquely define the NN potential. As a result, in order to build up the equation of state, it must be consistent with the known features of the nuclear matter properties such as binding energy per nucleon, symmetry energy, compressibility, and so on at saturation density. In this section, we will briefly cover some of the most important work done on high-density equations of state in the last century. 

\citet{1968AnPhy..50..411R} developed a phenomenological NN potential model that matches the scattering data notably and this model has been widely used to compute the NS structure \cite{doi:10.1146/annurev.ns.25.120175.000331}.
The lowest order governed variational technique appears to be precise enough for calculating the equation of state P($\rho$) for the central section of the Reid potential \cite{PhysRevC.7.1312}. However, when \citet{PANDHARIPANDE1976269,RevModPhys.51.821} used the Reid potential to compute the nuclear matter characteristics, they discovered that both the saturation density and binding energy were very large. As a result, the Reid potential model was considered as impractical.

With the nucleon-nucleon scattering data not able to determine the nucleon-nucleon short-range interaction uniquely, \citet{BETHE19741} presented a phenomenological potential model, in which they provided various potential models for nucleon-nucleon interaction.
By fitting the scattering data, they hypothesized several reasonable intensities for short-range repulsion.
Using this equation of state, \citet{1975ApJ...199..741M} estimated the maximum mass of neutron stars. Because we expect nucleons (N) and hyperons (Y) to be present at large densities, the hyperon-nucleon interactions are not completely described in that model. These two models are static potential models with separate phenomenological densities. According to their estimates, at large densities, the nuclear matter energy rises linearly.

When several tensor interaction models were proposed \cite{GREEN1974429},
\citet{PANDHARIPANDE1975507} generalized various tensor interaction models and suggested that the attraction between nucleons is generated by the pion exchange tensor interaction with contribution from higher orders. These interactions, however, only matched the s-wave scattering data and could not explain the anticipated strength specific form of the short-range repulsion.
Smith and Pandharipande demonstrated that all the nucleon attraction may be explained by attributing all nucleon scattering data to tensor interaction \cite{SMITH1976327}.
The tensor interaction model was unable to adequately represent the nuclear matter features since calculations utilizing the tensor interaction employing the lowest order variational and Brueckner approaches only fulfilled half of the nuclear matter binding energy at saturation density.
Then Pandharipande and Smith developed a model in which nucleon attraction was generated by the exchange of an effective scalar meson \cite{PANDHARIPANDE197515}. The generic name for this approach is the mean-field model.

A comprehensive evaluation of the attractive interaction owing to all conceivable tensor potentials \cite{SMITH1976327} suggests that it is virtually independent of the spin as well as isospin of the interacting nucleons and therefore its contribution to matter may be analogous to that due to nucleon coupling to a scalar field. In the mean-field approximation (MFA), Walecka used the scalar field \cite{WALECKA1974491} where the nucleons interact through a central potential produced by $\sigma$, $\omega$ and $\pi$ meson exchanges.
While the center regions of the pion exchange potentials have a minimally small influence, the potential approximation may be appropriate for $\omega$ and $\rho$ vector fields. Because the $\omega$-$\rho$ exchange potential has a range of $\sim$ 0.2 fm, significantly less than the mean interparticle spacing of $\sim$ 1.2 fm, the MFA is inappropriate for the vector field.
The variational approach with a hypernetted chain formalism is used to address the short-range corrections produced by the $\omega$-$\rho$ exchange potentials \cite{PhysRevC.7.1312}.
The coupling constants $\sigma$, $\omega$, and $\rho$ are calculated using nuclear matter binding energy, symmetry energy, and saturation density. According to this theory, the incompressibility parameter is $\sim$ 310 MeV. The interactions used in the mean-field model could not explain nucleon-nucleon scattering results. This model, on the other hand, fulfills all experimentally observed nuclear matter characteristics. 
In tensor and mean-field models, the attraction between nucleons decreases as density increases. This is, however, a typical property of microscopic models based on mean-field theoretical calculations. The shortcoming of these models is that energy is proportional to the density at low densities, but tends to saturate at large concentrations.

 \citet{FRIEDMAN1981502} developed a model to compute the equation of state for a broad variety of densities of dense neutron and nuclear matter, using the variational technique described previously by \citet{PANDHARIPANDE1971641}. The phenomenological nucleon-nucleon interactions were used \cite{PANDHARIPANDE1976269} with the contribution from short and intermediate-range components as well as the pion-exchange. A complicated three-nucleon interaction (which is a function of strength parameters, inter-particle distance and alignment angles) also contributed and the parameters of the three-nucleon interactions were determined by procreating the nuclear matter saturation density, binding energy, and incompressibility, using variational calculations. The results from nucleon-nucleon scattering cross-sections and nuclear matter characteristics were accurately explained by this model.

 \citet{PhysRevC.38.1010} proposed a model based on available nuclear data that improved the previous work by \citet{FRIEDMAN1981502}.
The two-nucleon potential is assumed to be the Argonne v14 (AV14) \cite{PhysRevC.29.1207} or Urbana (UV14) potential \cite{LAGARIS1981331} in this model. The structure of the Argonne v14 (AV14) and Urbana (UV14) potentials is the same, but the magnitude of the short-range tensor force varies. They are referred to as v14 models since they are composed of v14 operator components (such as $\sigma_i$, $\sigma_j$, $\pi_i$, $\pi_j$ and so on).
Each v14 model component contains three radial pieces: a long-range one-pion exchange process, an intermediate-range two-pion exchange process, and a short-range portion resulting from heavier meson exchange or composite quark system overlap. The data from nucleon-nucleon scattering and deuteron characteristics are used to fit all of the free parameters.
The Urbana VII potential \cite{PhysRevC.59.682} is used for the three-nucleon interaction, which includes a two-pion exchange component and an intermediate range repulsive contribution. Calculations using the Lagaris and Pandharipande's Urbana v14 plus three-nucleon interaction (TNI) model have been carried out \cite{LAGARIS1981349}. The variational principle is utilized in the many-body computations and the Fermi Hypernetted Chain–Single Operator Chain (FHNC–SOC) integral equations are used \cite{PANDHARIPANDE1976269,Fantoni:2001dfk}. For the UV14 plus TNI model, the nuclear matter binding energy and incompressibility value at saturation density of 0.157 fm$^{-3}$ are -16.6 MeV and 261.0 MeV, respectively. This method outperforms all other non-relativistic strategies. Using this non-relativistic method, the equation of state for beta-stable matter violates causality above $\rho$ = 1 fm$^{-3}$ ($dp/d\epsilon \le c^2$) which is an unfavorable feature of this technique at high concentrations. They predicted an NS with a maximum mass of 2.2$M_{\odot}$ and for a 1.4$M_{\odot}$ neutron star, the central density is found to be considerably less than 1 fm$^{-3}$, which is very feasible. It should be noted that current potentials, when coupled with actual three-body interactions, provide remarkably comparable models of NS structural parameters. The 1.4$M_{\odot}$ models predict a radius in the range 10.4-11.2 km with a core density of roughly 6$\rho_0$.

With the substantial development of high-precision NN potentials and computational techniques, different advanced nuclear many-body methods with realistic NN potentials, such as the Brueckner-Hartree-Fock method \cite{PhysRevC.74.047304,Baldo:2016jhp}, quantum Monte Carlo methods \cite{RevModPhys.87.1067,RevModPhys.65.231}, self-consistent Green's function method \cite{DICKHOFF2004377}, coupled-cluster method \cite{Hagen_2014}, and many-body perturbation theorem \cite{PhysRevC.90.054322,Schmidt:1999lik} were developed in a non-relativistic framework.

These methods can approximate the saturation behaviors for symmetric nuclear matter using modern high-precision potentials like Nijmegen \cite{PhysRevC.49.2950}, Reid93, AV18 \cite{PhysRevC.51.38}, CD-Bonn \cite{PhysRevC.63.024001} and chiral N$^3$LO \cite{PhysRevC.68.041001} and N$^4$LO potentials \cite{PhysRevLett.115.122301}. The saturation properties obtained from these calculations with only two-body force are not able to reproduce the empirical data such as $E/A$ = -16 $\pm$ 1 MeV at $\rho_0$ = 0.16 $\pm$ 0.01 fm$^{-3}$. The three-body nuclear force must be added to these non-relativistic frames to add repulsion contributions to reproduce the acceptable saturation properties \cite{PhysRevC.96.034307}.

\section{Relativistic Models}
Despite the amazing effectiveness of these advanced non-relativistic calculations, inconsistencies with experimental data remain, suggesting that such a conventional approach has reached its limitations. To begin with, the degenerate neutrons Fermi momentum is high in the interior of the neutron star. Secondly, the form of the baryonic potential is yet unknown at extremely small inter-particle separations ($\le$ 0.5 fm) and it is uncertain if the potential description will still be valid at such short distances. As a consequence, numerous authors have gravitated towards the relativistic viewpoint. Typically, the relativistic method starts with a local, renormalizable field theory containing degrees of freedom for baryons and explicit mesons. 
Although these models have the advantage of being relativistic, they cannot be connected to data on nucleon-nucleon scattering.
The theory is designed to be renormalizable such that the empirical nuclear matter at saturation properties such as binding energy, saturation density, compression modulus, effective mass, and symmetry energy may be used to generate coupling constants and mass parameters.

 \citet{ANASTASIO1983327} proposed the relativistic version of the Brueckner-Hartree-Fock approach in the 1980s, which was later refined by \citet{HOROWITZ1987613} and \citet{PhysRevC.42.1965}. The relativistic effect produces a repulsive contribution in the relativistic Brueckner-Hartree-Fock (RBHF) model, which may accurately explain nuclear saturation properties with two-body realistic NN potential.

It was also shown that the Z diagram and the three-body force contributed to the nucleon-antinucleon excitation from the relativistic effect. The contributions of the three-body force, Z diagram and the relativistic effect are partly in accordance with Ref.~\cite{PhysRevC.77.034316}, as the nucleon/antinucleon excitations affect the nuclear matter energy in RBHF models via the second order term of scalar meson. This can be regarded as a component of the microscopic three-body force generated through the two-meson exchange between nucleon excitation states. The RBHF was also used to investigate superfluidity, properties of neutron stars \cite{Schulze:2010zz} and to fit the free parameters nuclear density functional theory \cite{PhysRevC.74.025808}. \citet{PhysRevC.96.014316,SHEN2019103713} achieved a completely consistent calculation of RBHF model for finite nuclei and expanded this framework on neutron drops \cite{SHEN2019103713}. \citet{PhysRevC.98.054302} also proposed a method to calculate the angular integral of center-of-mass momentum for asymmetric nuclear material within the RBHF-model.

In order to use the RBHF Model, the nuclear media effect must be considered in the NN Potential. Bonn potentials \cite{MACHLEIDT19871} is one example of NN interaction that can be considered. With a large amount of two nucleon scattering information, many high precision NN potentials were suggested based on charge-dependent partial wave analyses from 1990s \cite{PhysRevC.48.792,Naghdi2014}. The chiral NN potentials derived in chiral perturbation theories were also rapidly developed. These state-of-the-art chiral possibilities have been widely used to explain the structures of finite, and infinite nuclei, as well as the saturation properties of infinite nuclear matter. The properties of light nuclei and nuclear matter were perfectly reproduced when the four-body (and three-body) forces obtained from chiral perturbation theory are included. This controlled hierarchy allows for easy estimation of uncertainties due to the few-body forces.
The breakdown scale of these chiral potentials was discovered to be approximately 600 MeV and the uncertainty from high-order potentials increase with density. With such studies, the characteristics of nuclear matter below 0.4 fm$^{-3}$ should be conceivable given the current chiral potentials. The study of compact stars, on the other hand, necessitates the equation of state of nuclear matter at 0.8 fm$^{-3}$. As a result, it is critical to use a readily available many-body technique and high-precision NN potentials for a more accurate description of nuclear matter, particularly in the high-density area.

The mean-field approximation (MFA) is employed as a starting point, which should be adequate at extremely high densities (a few times normal nuclear density) \cite{Glendenning:1982nn}. 
The second stage involves the inclusion of one-loop vacuum fluctuations, which results in the Relativistic Hartree Approximation (RHA) \cite{Serot:1984ey}. This method has been utilized as a viable method of parameterizing the equation of state.

A fine description of nuclei and the nuclear matter was introduced by J. D. Walecka in 1974 \cite{WALECKA1974491}. This description based on the interaction between baryons and mesons is referred to as the Quantum Hydrodynamics (QHD). In nuclear matter, nucleons interact through the exchange of mesons and hence the relativistic effects are incorporated naturally.

With nuclei and nuclear matter being complex systems, various models exist among which the QHD is one. For all the models, some experimental inputs are necessary to constraint them and in the case of QHD, these are the coupling constants between nucleons and different mesons. These coupling constants are determined by fitting the calculated properties of nuclei with the experimentally observed values. By fitting various observed parameters, different QHD parameter sets have been developed which differ from each other in terms of the meson fields considered and different couplings between the fields.

Quantum Hydrodynamics I (QHD-I), also known as the $\sigma-\omega$ model, is the original and simplest QHD parameter set \cite{Serot:1984ey,WALECKA1975109}. This model involves the exchange of isoscalar sigma  $\sigma$ and isoscalar vector mesons $\omega$ with the baryons (neutron and proton) which are found to be important in the description of the nuclear matter.
The correct values of nuclear matter binding energy, $E/A$ = -15.75 MeV and saturation density, $k_f$ = 1.42 fm$^{-1}$ are used to fix the model parameters. Within this model, a phase transition similar to liquid-gas transition is obtained, which at asymptotically high densities approaches the causal limit $P=\mathcal{E}$, breaking causality. The NS with a maximum mass of 2.57$M_{\odot}$ as a function of central density is obtained.
To improve the models' realism, the interplay of charged vector mesons such as $\rho$-meson was included (QHD-II) which stiffened the EoS in low-density regime \cite{WALECKA1974491}. This model, however, produced a very large value of nuclear matter incompressibility at saturation density, although it was not mentioned that the isospin triplet vector meson is essential in the NS interior. The NS maximum mass increases slightly to 2.6$M_{\odot}$.
 \citet{gled} developed a theoretical relativistic field model at densities approaching to and beyond nuclear matter density, including isospin-asymmetric baryon matter. Nucleons (neutron, proton), mesons ($\sigma$, $\omega$ and $\rho$) and leptons (e$^-$ and $\mu^-$) were included along with the scalar meson self-interaction.

To reproduce the properties of neutron-rich nuclei more accurately, the NL-SH parameter set was developed by fitting the coupling constants to reproduce the observed values of neutron-rich nuclei \cite{SHARMA1993377}. In late 1994, the TM1 and TM2 parameter sets were developed by Sughara and Toki, by introducing a non-linear self-coupling $\omega$-meson field in the Lagrangian density \cite{1994NuPhA.579..557S}. To improve the value of nuclear incompressibility, \citet{PhysRevC.55.540,PhysRevC.76.064310} developed an NL3 parameter set based on the QHD-I Lagrangian density with the addition of $\rho$ meson and scalar field self-coupling. Applying these parameter sets to the study of high dense matter predicted heavy NSs with a maximum mass > 2.5$M_{\odot}$. Several other parameter sets were developed over time to explain the nuclear matter properties more accurately \cite{PhysRevC.69.034319,PhysRevLett.95.122501}. These parameter sets provided a soft EoS to decrease the NS maximum mass to a value $\approx$ 2$M_{\odot}$ which satisfied the mass constraints from various measurements.

\citet{PhysRev.140.B1452} considered pions ($\pi^-$) as free particles, as they are light mesons and could replace electrons to condensate and showed that they could be present at densities $\rho$ $\ge$ $\rho_0$. It was shown that although the pion condensation softens the EoS, the short-range correlations in the dense matter make them less likely to appear in the core of an NS. \citet{KAPLAN198657} introduced the negative kaons ($K^-$) in NSs, which are strange mesons, using a simplified dense-matter model, and showed that $K^-$ condensates at densities 2-3 times the normal nuclear density. The effect on the nucleonic component of dense matter, particularly, neutrino emissivity, was observed due to Kaon condensation. It was seen that the presence of kaons initiates fast cooling of NSs. \citet{PhysRevC.63.035802,PhysRevD.82.123010,PhysRevC.86.045803,Ramos:2000mv} considered the condensation of both negative kaon ($K^-$) and negative antikaon ($\bar{K}^0$) within the RMF formalism and showed that the antikaon condensate depends upon the optical potential of antikaon and hence is quite sensitive to the EoS. Furthermore, with hyperons taken into account, the EoS becomes softer, thereby delaying the onset of antikaon condensation \cite{PhysRevC.64.055805,2000NuPhA.674..553P}.

An earlier study of Fermi gases by \citet{1960SvA.....4..187A} presented a very reasonable argument for the presence of a hyperon charge on neutron stars. Because of the weak equilibrium, the core of NSs is supposed to be dense enough to allow the appearance of new particles with strangeness content in addition to the usual nucleons and leptons. At approximately 2-3$\rho_0$, hyperons appear as the first odd baryons in an NS.
To construct an EoS of hyperonic matter, Glendenning expanded the preceding formalism of RMF by adding the $K$ and $K^*$ meson exchanges, as well as the self-interaction form, while applying the same constraints as previously to fix the different parameters \cite{1985ApJ...293..470G,GLENDENNING1982392}. The coupling of hyperons was determined by symmetry relations and hypernuclear observables.
This allowed us to assume that hyperons dominate the cores of heavier neutron stars and the overall hyperon population varies between 15\% and 20\% for such stars, depending on whether pions condense or not. The presence of hyperons softened the EoS as the high energy neutrons are replaced by massive low energy hyperons, which reduced the maximum mass of an NS. The hyperonic contribution to the core of NS has been studied with both non-relativistic and relativistic formalisms  \cite{PhysRevC.73.058801,Vidana:2002rg}. 

\citet{1985ApJ...291..308D} developed a model where the nucleons interact with one other via scalar ($\sigma$) mesons, vector ($\omega$ and $\rho$ ) mesons as well as pions ($\pi$). The renormalized Hartree approximation was employed, which generated two sets of EoSs, I \& II, with I determined by fitting the characteristics of symmetric nuclear matter at nuclear saturation density adequately. This is compatible with Baym, Bethe and Pethick's equation of state for neutron matter in the area above the neutron drip line \cite{1971NuPhA.175..225B}. 
The nuclear matter incompressibility reported in this model is far too high ($\sim$ 460 MeV), which may be attributable to the Lagrangian's omission of a $\sigma$-self interaction component. The II EoS is built using the chirally invariant $\sigma$-model Lagrangian, coupled to $\omega$ and $\rho$ mesons, along with an explicit symmetry breaking component.
This EoS accurately describes all of the known symmetry energy properties at nuclear density. At saturation density, the nuclear incompressibility is predicted to be 225 MeV.

Chiral symmetry is an excellent hadron symmetry, second only to isotope spin symmetry \cite{1975PhR....16..219P}. As a consequence, in any theory of dense hadronic matter, chiral symmetry is desired. Simultaneously, the theory should be competent to describe the bulk nuclear matter properties. However, there is no hypothesis that satisfies both of these criteria. Glendenning developed an EoS based on the mean-field approximation
where the $\omega$ meson is analyzed in terms of its dynamical masses \cite{gled}.
The gauge field $\omega_{\mu}$ of a massless vector meson is incorporated into the chiral model via covariant derivatives. Furthermore, the symmetry breaking term that gives the pion a definite mass is represented by the linear term from the $\sigma$ field present. The theory also contains the scalar meson and the pseudoscalar pion, in addition to $\omega$-mesons. The nuclear saturation density and the binding energy per nucleon in the typical symmetric matter are used to determine the parameters in this theory.
However, the incompressibility predicts a high value, $K$ = 650 MeV. Based on the mean-field approximation, Glendenning expanded the chiral sigma model, with no dynamical mass for the $\omega$-meson, even when the vacuum renormalization corrections are considered \cite{Glendenning:1987gk}. Glendenning only included the typical non-pion condensed state of matter, with hyperons incorporated in beta-equilibrium with nucleons and leptons. By generating accurate nuclear matter, he fitted the theory's parameters to obtain two equations of state, a \textit{"stiff"} ($K$ = 300 MeV ) and a \textit{"soft"} ($K$ = 200 MeV ) one, corresponding to two nuclear incompressibility values. \citet{PhysRevC.36.346} presented a chiral sigma model-based equation of state. They investigated the relevance of chiral sigma-models many-body effects in the symmetric nuclear matter and neutron-rich matter EoS. They incorporated the $\sigma$-meson one-loop contributions, but since the isoscalar vector field is not generated dynamically, its contribution is reduced to that of an empirical one.
The empirical nuclear matter saturation density, binding energy, and symmetry energy are all fitted by a set of equations of state. They determined the incompressibility of nuclear matter at saturation density for the symmetric matter that differs from the estimated experimental result. As a result, they allowed for arbitrarily varying the values of coupling parameters in order for the theory to produce the appropriate value of nuclear incompressibility. In this method, the vector field has no influence on the value of the nucleon's effective mass.

For the equation of state of neutron matter, \citet{PhysRevLett.55.126} proposed a phenomenological model with nuclear pressure given in terms of baryon density compression factor $\mu^{\gamma}$.
This model was further modified to fit well with the pure hadron as well as mixed EoSs \cite{BETHE19741,FRIEDMAN1981502,Ramos:2021drk}.

Similar to the Walecka model, \citet{PhysRevC.42.1416} presented a model, but with the scalar field linked through the derivative scalar coupling (DSC). The intriguing characteristic of this model is that the scalar field equation of motion becomes non-linear without the inclusion of any additional parameters. This model predicts a good effective nucleon mass and a suitable value of nuclear matter incompressibility at saturation density. \citet{PhysRevC.45.844} generalized the model to include hyperonic matter and utilized it to determine star matter properties. Instead of solely coupling the scalar field to the vector meson fields and baryons as presented in the preceding (DSC) model, they linked it here with both Yukawa point and derivative coupling to both vector fields and baryons. When compared to the previous calculation, this improved the value of the nuclear incompressibility ($K$ = 224.9 MeV) and effective nucleon mass ($M^*$ = 797.64 MeV) at saturation density. They also included the contribution from $\rho$-meson to account for the asymmetry impact.

The DSC models were shown to be related to SU(6) model for the meson-baryon couplings and hence applied to the study of nuclear matter properties, finite nuclei, and neutron star models (including hyperons) \cite{PhysRevC.45.844,1996PhRvC..53..522B}. 

Since quarks are the fundamental constituents of hadrons, a basic description of dense matter should necessitate quark degrees of freedom. \citet{Ivanenko1965,Ivanenko1969} first proposed the presence of quarks in NS cores, suggesting that the baryons should convert to the quark matter at relatively high densities. A free degenerate gas quark model was developed for superdense low mass stars \cite{Itoh:1970uw}. After the foundation of Quantum Chromodynamics (QCD), perturbative calculations to study the quark matter EoS were established, but it was restricted to very high densities. 
 \citet{PhysRevD.9.3471} developed the MIT Bag model to study the quark matter, using non-perturbative effects of confinement via the bag constant. This model was widely used to calculate the quark matter EoS by varying the bag constant and strange quark mass \cite{PhysRevD.69.074001}. \citet{PhysRevD.30.272} used the simple MIT Bag model with non-interacting quarks to show that the $uds$ quark matter could be absolutely stable for reasonable values of the bag constant . \citet{PhysRevD.30.2379} explored the properties of $uds$ matter including the finite mass of $s$ quark and low-order QCD interactions. This simple version of MIT Bag model is referred to as \textit{"Strange Quark Matter"} (SQM). With the discovery of new pulsars and heavier NSs, models with massive quarks, such as Nambu-Jona-Lasinio and its extensions \cite{PhysRev.122.345,PhysRev.124.246}, which restores the chiral symmetry, were frequently used to study such heavy NSs and to describe the conversion of NSs into strange stars via phase transition at higher densities \cite{Dey:2001ny,Nandi:2017rhy,Banik:2004ju}. 

The density-dependent quark mass phenomenological models via a scalar density-dependent potential were also developed in order to show that the quark masses tend to current quark masses as $\rho_b$ $\rightarrow$ $\infty$ \cite{CHAKRABARTY1989112,PhysRevD.43.627}. These models predicted stars, completely made up of quarks, with a maximum mass $\sim$ 2$M_{\odot}$ and thus allowing the existence of Bare strange stars \cite{Dey:1998rz}. Several other models include  Color-flavor-locked (CFL) \cite{PhysRevLett.86.3492,PhysRevD.64.074017}, Quark mean-field (QMF) \cite{PhysRevC.61.045205}, and Quark meson coupling (QMC) \cite{SAITO20071}.
The phase transition from pure hadronic matter to the quark matter leads to a strong softening of the EoS, thereby decreasing the NS maximum mass \cite{Dey:1988dq}.  

\citet{1991ApJ...383..745L} studied the effect of a strong magnetic field on the NSs by employing the scalar virial theorem. A magnetic field of the order of 10$^{18}$ Gauss was found in the inner surface. Calculations by various groups limited the range between 10$^{17}$ - 10$^{20}$ G, depending upon the models used and studied the properties of static as well as accreting NSs \cite{PhysRevLett.79.2176,PhysRevD.86.125032,Gomes:2017zkc,Konar:1998cg}.

One of the fundamental findings in the success of RMF models is that nonlinear self-interactions for the scalar meson must be added to give adequate flexibility \cite{Serot:1992ti,Boguta:1977xi,GAMBHIR1990132}. Because these models were meant to be renormalizable, scalar self-interactions are restricted to a quartic polynomial, and scalar–vector and vector–vector interactions are not permitted \cite{Boulware:1970zc}. As mentioned by Walecka, one of the incentives for renormalizability is that once the model parameters are calibrated to observable nuclear characteristics, one may extrapolate into regimes of high density or temperature without the introduction of additional, unknown parameters.

Effective field theories, such as chiral perturbation theory \cite{Weinberg:1968de,Weinberg:1978kz}, successfully describe the low-energy phenomenology of hadronic Goldstone bosons \cite{Gasser:1983yg}, inspire an alternate approach. Although a lagrangian is generally used as the starting point, the meson and baryon fields are no longer termed elementary and the renormalizability constraint is eliminated.

Mean-field models of nuclear structure and the EoS must be evaluated in a new context within the methodology of effective field theory. Near normal nuclear density, the mean scalar and vector fields, denoted as $\Phi$ and $W$, are large on nuclear energy scales but small in comparison to the nucleon mass $M$ and vary slowly in finite nuclei. This implies that the $\Phi/M$ and $W/M$ ratios, as well as their gradients $|\nabla \Phi|/M^2$ and $|\nabla W|/M^2$, are useful expansion parameters. The implication of "naturalness" is also important in effective field theory. Naturalness signifies that one should include all the possible terms via a given order and that the coefficients of the various terms in the lagrangian should all be of order unity when interpreted in appropriate dimensionless form. 

From this perspective, it is necessary to stabilize nuclear mean-field models that only contain scalar self-interactions \cite{Boguta:1977xi,GAMBHIR1990132} and extensions that include quartic self-interactions for the neutral vector meson \cite{Bodmer:1991hz}. In addition, a comprehensive study including all meson self-interactions through fourth-order in the isoscalar-scalar and vector fields has been carried out \cite{FURNSTAHL1996539}. These additions result in new model parameters i.e, coupling constants, that must be constrained by comparing to observed nuclear properties. The parameters obtained should be natural for the truncation at fourth-order to be reasonable.

Over the last decade, there has been significant progress in linking laboratory observables to astronomical data. Indeed, significant advancements in theory, experiment, and observation have taken us precariously closer to determining the EoS of neutron-rich matter. In a consistent approach, appropriately optimized energy density functionals are now commonly utilized to compute the ground-state properties of finite nuclei, their collective excitations, and the structure of neutron stars.
\section{Plan of Thesis}
The primary objective of this thesis work is to investigate NSs using the RMF model and its extended variants, which are successful in recreating nuclear matter properties at saturation density and describing the structure and properties of NSs while fulfilling current constraints. The existence of exotic phases, namely quarks and hyperons, is investigated, as is its impact on NS properties. Through astronomical measurements and gravitational wave detections, we investigate whether the occurrence of such exotic particles is supported by NSs.

After an extensive description of the nuclear models (both non-relativistic and relativistic) used in the study of dense matter objects in Chapter \ref{cha-lit}, Chapter 2 provides information about the formation and structure of NSs with extension to the exotic matter present in their cores.
In Chapter 3, a detailed description of the relativistic mean-field (RMF) model is presented. We begin with the fundamental concept of mean-field theory, followed by the RMF method influenced by effective field theory and its EoS. Following a discussion of the density-dependent RMF (DD-RMF) models, the nuclear matter properties and parameter sets utilized in our calculations for both RMF and DD-RMF models are explored in depth. Finally, the infinite nuclear matter (symmetric and asymmetric), as well as the beta-equilibrium and charge neutrality criteria for NSs are explored.

The role of quark matter in the core of NSs is investigated in Chapter 4. The phase transition characteristics of hadron matter to quark matter are investigated using the basic MIT bag model. The nuclear matter properties such as symmetry energy and slope parameter are computed for hybrid EoS for different values of the bag constant $B^{1/4}$ using the unified EoS. The mass-radius profile for various hybrid EoSs created is derived by solving TOV equations and the value of bag constant is constrained in the context of gravitational wave data GW170817.

The influence of inner crust EoS with varied symmetry energy slope parameters on the properties of NS is investigated in Chapter 5. The inner crust and core EoSs from several RMF models are combined to form a single EoS. The characteristics of both static and rotating NSs are investigated, including mass, radius, tidal deformability, and the moment of inertia.

The hadron-quark phase transition is studied in Chapter 6 in the context of the recently detected gravitational wave GW190814. The DD-RMF model is used to investigate the hadronic matter, whereas the vector enhanced bag (vBag) model, an extended version of the MIT bag model, is used to explore quark matter. The characteristics of NS are investigated utilizing various phase transition fabrication approaches. Properties such as the moment of inertia, Kerr parameter, mass, star radius, and redshift are computed for maximally rotating NS to investigate if the secondary component of GW190814 may be a potential supermassive NS.

Chapter 7 examines the influence of a strong magnetic field on the characteristics of neutron and hyperon stars using various hyperonic parameters. On the microscopic level, a chemical potential-dependent magnetic field is utilized to describe matter. The deformation caused by the magnetic field and how its absence results in an overestimation of the mass and an underestimation of the stellar radius of NS and hyperon stars have been explored.

Finally, Chapter 8 combines our research of NS and the findings obtained to make some significant conclusions about NS and the occurrence of unusual phases. The potential expansion of this thesis work is also explored. 
\begin{savequote}[8cm]
\textlatin{The neutron stars/black holes of nature are the most perfect macroscopic objects there are in the universe: the only elements in their construction are our concepts of space and time. }

  \qauthor{--- \textit{Subrahmanyan Chandrasekhar}}
\end{savequote}

\chapter{\label{ch:1-intro}Neutron Stars} 

\minitoc
\section{Introduction}
Neutron stars (NSs) are dense, compact objects that have a significant appeal as probes for studying many aspects of physics. They are the ideal astrophysical laboratories with the greatest magnetic fields and gravity known (except black holes). 
Within a radius of 10 km, NSs contain over a solar mass of stuff at densities of the order of 10$^{15}$ g/cc  and hence give chances to investigate the characteristics of matter at extremely high densities. They have also proved to be excellent test bodies for theories incorporating general relativity. In a larger sense, NS gives users access to the phase diagram of matter at high densities and temperatures, which serves as the foundation for comprehending a variety of astrophysical phenomena.

NSs are great observatories for testing our current knowledge of matter's fundamental characteristics under the impact of strong gravitational and magnetic fields at high density, isospin asymmetry, and temperature conditions. They provide a fascinating interaction between nuclear processes and astrophysical observables. Their research is one of the fascinating areas of study, necessitating skills from several disciplines such as general relativity (GR), high-energy physics, nuclear and hadronic physics, and quantum chromodynamics (QCD). Massive theoretical gains have been made in comprehending the extraordinary and one-of-a-kind features of these exceedingly dense objects.
NSs are good gravitational wave emitters. Exotic materials are attracted when the matter is squeezed beyond nuclear densities.
Many physical processes violently transfer significant quantities of mass at relativistic speeds, altering spacetime and hence emitting huge amounts of gravitational radiation.

The complementary efforts in theory and experiment have resulted in new fields of nuclear physics, such as the extension of the nuclear chart and access to the nuclear matter at various densities \cite{Thoennessen2011, CAPRIO2013179}. Testing/developing nuclear models while describing these additional aspects is a challenge and also aids in the validation of ideas and underlying assumptions. Nuclear matter in NS, which has extreme isospin and density, might be an extreme testing ground in this regard. In this thesis, we have chosen a class of nuclear models, particularly relativistic mean-field models, from among the many available. We investigate how the most current and successful versions of this model may explain some aspects of the NS while also expanding the model with the introduction of quarks and hyperons. The following text provides the bare minimum of information on the NS.

After Chadwick discovered the neutron in 1932 \cite{Chadwick1932}, scientists predicted the presence of neutron stars. In 1934, Baade and Zwicky theorized that neutron stars might originate in supernovae, which occur when the iron core of a large star surpasses the Chandrasekhar limit and collapses \cite{1934PNAS...20..254B}. The immense amount of energy released during the collapse vaporizes the rest of the star and the collapsing core may form a neutron star. For this mechanism to create neutron stars efficiently, the maximum mass of neutron stars should be more than 1.4$M_{\odot}$.

In 1939, \citet{PhysRev.55.364,PhysRev.55.374} performed the first theoretical calculations of neutron stars, assuming that they are gravitationally confined states of neutron Fermi gas. Maximum masses were calculated to be 0.7$M_{\odot}$, with densities of up to $\sim$ 6 $\times$ 10$^{15}$ g/cm$^3$ and radii of 10 km. In comparison, the density of nuclear matter within a large nucleus such as $^{208}$Pb is $\sim$ 0.16 nucleons/fm$^3$, or $\simeq$ 2.7 $\times$ 10$^{14}$ g/cm$^3$ \cite{doi:10.1142/3530}. Their expected maximum mass was less than the Chandrasekhar mass limit of $\sim$ 1.4$M_{\odot}$ for white dwarfs with iron-group nuclei and densities as high as $\sim$ 10$^9$ g/cm$^3$ \cite{1983bhwd.book.....S}. The pressure required to counteract the gravitational attraction of white dwarfs and Oppenheimer-Volkoff NSs is provided by degenerate electron and neutron Fermi gases, respectively.
Tsuruta and Cameron proved in the 1960s that using schematic nuclear force models, they could increase neutron star masses over 1.4$M_{\odot}$ \cite{1974asgr.proc..221C}.

In 1967, Bell and Hewish discovered radio pulsars, which Gold soon identified as rotating neutron stars \cite{GOLD1969}. The subsequent finding of the Crab pulsar in the remnants of the Crab supernova in China in 1054 confirmed the connection to supernovae and inspired ongoing efforts to better understand neutron stars. Hulse and Taylor's 1974 discovery of the first binary pulsar, PSR 1913+16 \cite{1975ApJ...195L..51H} (PSR stands for pulsar and 1913+16 marks the pulsar's position in the sky), ushered in a scientific revolution. 

A basic introduction about the NS, its formation, and properties are discussed below.

\subsection{Formation}
\begin{figure}[h]
	\centering
	\begin{tikzpicture}
	\node[] at (-3,6) {\large\textbf{\color{blue}Core Radius $\sim$ 1 $R_{e}$}};
	\node[] at (-3.3,-2.1) {\large\textbf{\color{blue}Envelope Radius $\sim$ 5 AU}};
	\filldraw[fill=amber(sae/ece), draw=black] (2,2) circle (4.1cm);	
	\filldraw[fill=antiquebrass, draw=black] (2,2) circle (3.5cm);	
	\filldraw[fill=apricot, draw=black] (2,2) circle (2.9cm);	
	\filldraw[fill=yellow, draw=black] (2,2) circle (2.2cm);	
	\filldraw[fill=cadmiumorange, draw=black] (2,2) circle (1.6cm);
	\filldraw[fill=green!80!white, draw=black] (2,2) circle (1.1cm);
	\filldraw[fill=blue!70!white, draw=black] (2,2) circle (0.7cm);
	\node[] at (2,2) {\large\textbf{Fe}};	
	\node[] at (2,2.8) {\normalsize\textbf{Si}};
	\node[] at (2,3.4) {\normalsize\textbf{O}};		
	\node[] at (2,4.0) {\normalsize\textbf{Ne}};	
	\node[] at (2,4.6) {\normalsize\textbf{C}};
	\node[] at (2,5.2) {\normalsize\textbf{He}};	
	\node[] at (2,5.8) {\normalsize\textbf{H}};		
	\coordinate (a) at (3,5.5);
	\coordinate (b) at (1,5.0);
	\coordinate (c) at (3,4.3);
	\coordinate (d) at (1,3.6);
	\coordinate (e) at (3,3.0);
	\coordinate (f) at (1,2.3);
	\draw[->,black] (a) -- ( $ (a)!.12!(0,0) $ ) node [midway, sloped, above] {};
	\draw[->,black] (b) -- ( $ (b)!.12!(2,0) $ ) node [midway, sloped, above] {};
	\draw[->,black] (c) -- ( $ (c)!.12!(0,0) $ ) node [midway, sloped, above] {};
	\draw[->,black] (d) -- ( $ (d)!.12!(2,0) $ ) node [midway, sloped, above] {};
	\draw[->,black] (e) -- ( $ (e)!.12!(0,0) $ ) node [midway, sloped, above] {};
	\draw[->,black] (f) -- ( $ (f)!.15!(4,0) $ ) node [midway, sloped, above] {};
	
	\end{tikzpicture}
	\caption{ Schematic representation of a star burning through a succession of nuclear fusion fuels. R$_e$ represents the radius of the earth $\approx$ 6400 km and AU represents the astronomical units (1.5$\times$ 10$^8$ km). }\label{fig1.2}
\end{figure}
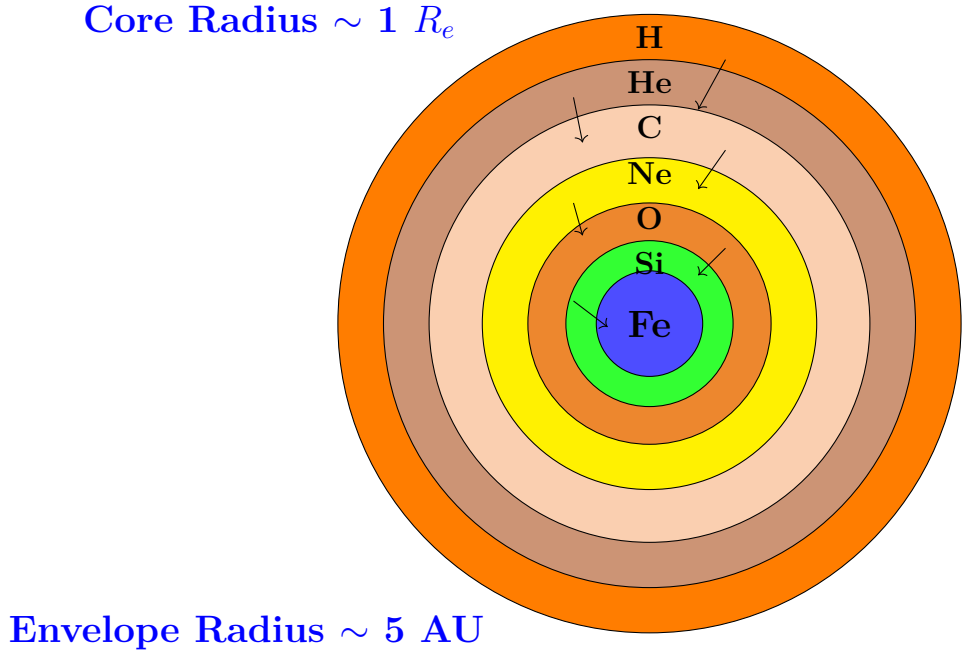

The stuff in the very early cosmos was diffuse gases of light elements.
These dispersed gases are drawn together and form stars as a result of gravity's attraction. Main sequence stars spend the majority of their lives fusing Hydrogen to produce Helium \cite{Seeds}. When all of the Hydrogen has been turned into Helium, the star begins to burn Helium, followed by Carbon, Oxygen, Silicon, and so on, until they seek to fuse iron, as seen in Fig.~\ref{fig1.2}. The approximate time scales of burnings are: Hydrogen burning $\approx$ 10$^7$ years, Helium burning $\approx$ 10$^6$ years, Carbon burning $\approx$ 500 years, Neon burning $\approx$ 10 years, Oxygen burning $\approx$ 1 year, and Silicon burning  $\approx$ 1 day \cite{2002nsps.conf..273P}.

 For a star with mass greater than 8$M_{\odot}$, the nuclei in the outer portions of the star tug against each other, forming elements heavier than Iron, which eventually leads to core-collapse supernovae. As the core
runs out of fuel, it contracts and the outer layers of the star expand and the stars become less bright, which further become either a red giant or a red super giant star, depending upon the initial mass of the star. Finally, these giant stars will collapse and explode, will either become a white dwarf, an NS, or a  black hole. 

Fig.~\ref{fig1.3} is a schematic representation of the life cycle of a star. In the heart of the core-collapse supernova, an extreme new state of matter (NS) develops for $M$ $\approx$ 8 - 20$M_{\odot}$. The elements of the star melt into the uniform nuclear matter due to the star's high density. Finally, nuclear interactions and nucleon degeneracy balance the gravitational pull and a neutron star emerges. Because the typical densities of NSs are equal to that of nuclei, it is thought that NS is composed of baryonic matter (such as protons and neutrons) and may thus be considered as gigantic nuclei, albeit with a mass number of about 10$^{57}$ \cite{gled}. More discussion on the discovery of NSs and other key contributions to their understanding may be found in the work of \citet{Novak2008, walter} and references therein.

\begin{figure}
	\centering
	\begin{tikzpicture}[>={LaTeX[width=8mm,length=8mm]},->]
	\shade[ball color=amber(sae/ece)] (-10.0,2) circle (.8cm);
	\shade[ball color=violet!60!white] (-8.5,4.0) circle (.6cm);
	\shade[ball color=violet!80!white] (-8.5,0.0) circle (1.0cm); 
	\shade[ball color=blue!60!white] (-6.2,4.0) circle (.5cm);
	\shade[ball color=blue!90!white] (-6.0,0.0) circle (0.8cm); 
	\shade[ball color=red!60!white] (-3.7,4.0) circle (.6cm);
	\shade[ball color=red] (-3.5,0.0) circle (1.0cm);
	\shade[ball color=yellow!80!white] (-1.2,4.0) circle (.6cm);
	\shade[ball color=white] (1.0,4.0) circle (.4cm);
	\coordinate (y) at (2.0,2.25);
	\coordinate (z) at (2.4,0.57);
	\draw[line width=2mm, gray] (y) -- ( $ (y)!.08!(1.5,5.0) $ ) node [midway, sloped, above] {};
	\draw[line width=2mm, gray] (z) -- ( $ (z)!.08!(5.50,-6.5) $ ) node [midway, sloped, above] {};
	\shade[ball color=red!40!blue] (2.2,1.2) circle (0.6cm);  
	\shade[ball color=black] (2.8,4.0) circle (.4cm);
	\node[] at (-8,2.2) {\textbf{Nebula}};	
	\node[align=left] at (-9.0,5.1) {Low/Medium \\	Mass Protostar};
	\node[align=left] at (-9.0,-2.0) {High-Mass\\ Protostar};	
	\node[align=left] at (-6.0,5.1) {Low/Medium \\	Mass Star};
	\node[align=left] at (-6.2,-2.0) {High-Mass\\ Star};
	\node[align=left] at (-3.5,5.1) {Red Giant};
	\node[align=left] at (-3.5,-2.0) {Super Giant};
	\node[align=left] at (-1.0,5.1) {Planetary\\ Nebula};
	\node[align=left] at (-0.5,-2.0) {Supernova};
	\node[align=left] at (1.0,5.1) {White\\ Dwarf};
	\node[align=left] at (2.0,-3.0) {Black Hole};
	\node[align=left] at (2.9,5.1) {Black \\Dwarf};
	\node[align=left] at (2.0,2.6) {Neutron Star};
	\coordinate (a) at (-9.5,2.55);
	\coordinate (b) at (-9.5,1.35);
	\draw[line width=2mm] (a) -- ( $ (a)!.08!(0,15.0) $ ) node [midway, sloped, above] {};
	\draw[line width=2mm] (b) -- ( $ (b)!.08!(0,-4.5) $ ) node [midway, sloped, above] {};
	\draw[line width=4mm] (-7.85,4.0) -- (-6.6,4.0);
	\draw[line width=4mm] (-5.6,4.0) -- (-4.2,4.0);
	\draw[line width=4mm] (-3.15,4.0) -- (-1.7,4.0);
	\draw[line width=4mm] (-0.5,4.0) -- (0.6,4.0);
	\draw[line width=4mm] (1.5,4.0) -- (2.4,4.0);
	\draw[line width=4mm] (-7.5,0.0) -- (-6.6,0.0);
	\draw[line width=4mm] (-5.3,0.0) -- (-4.4,0.0);
	\draw[line width=4mm] (-2.55,0.0) -- (-1.5,0.0);
	\coordinate (c) at (0.6,0.7);
	\coordinate (d) at (0.7,-0.2);
	\draw[line width=2mm] (c) -- ( $ (c)!.12!(12.0,3.0) $ ) node [midway, sloped, above] {};
	\draw[line width=2mm] (d) -- ( $ (d)!.12!(9.5,-6.0) $ ) node [midway, sloped, above] {};
	{
		\path
		[decoration={
			text effects along path, text={},
			text effects/.cd,
			character count=\i,
			character total=\n,
			characters={
				text along path,
				font=\sffamily\Huge\bfseries,
				text=red,
				scale=\i/\n*1.5+.75,
				anchor=center,
			}
		},
		decorate,
		local bounding box=boom
		] (-0.8,0) coordinate [left] (a) -- (0.0,0.1);
	}
	\scoped[on background layer]
	\node[starburst, starburst point height=5mm, draw=red, ball color=yellow, line width=1pt, double distance=2.5pt, double=yellow, decorate, decoration={random steps, segment length=2mm, amplitude=1pt}, fit=(boom)] {};
	;
	\end{tikzpicture}
	\caption{Schematic representation of life cycle of stars.} \label{fig1.3}
\end{figure}
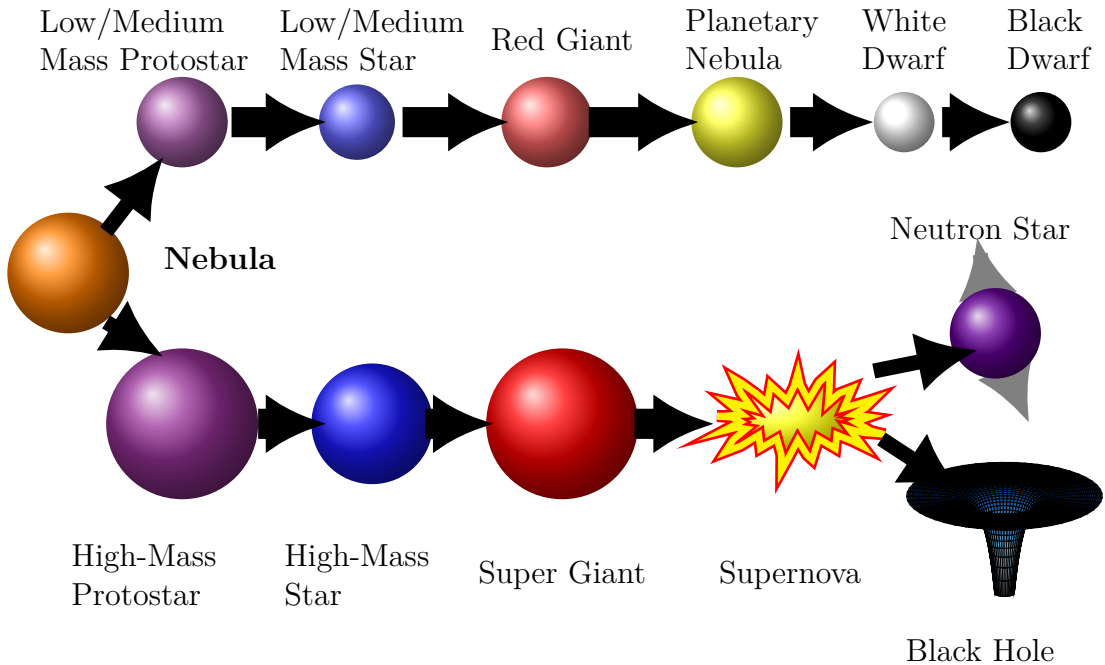
\begin{tikzpicture}[remember picture,overlay]
\tikzset{shift={(11,11.5)},yshift=-3.0cm}
\begin{axis}[
axis line style={draw=none},
tick style={draw=none},
colormap/Blues,
data cs=polar,
samples=50,
domain=0:360,
y domain=1:10,
declare function={darkhole(\r)={-exp(-\r)};
	pol2cartX(\angle,\radius) = \radius * cos(\angle);
	pol2cartY(\angle,\radius) = \radius * sin(\angle);
},
xtick={\empty},
ytick={\empty},
ztick={\empty},
]
\addplot3 [surf,shader=flat,draw=black,z buffer=sort] {darkhole(y)} ;
\end{axis}

\end{tikzpicture}
\subsection{Mass and radius}
\begin{figure}[hbt!]
	\centering
	\includegraphics[width=12cm,height=16cm]{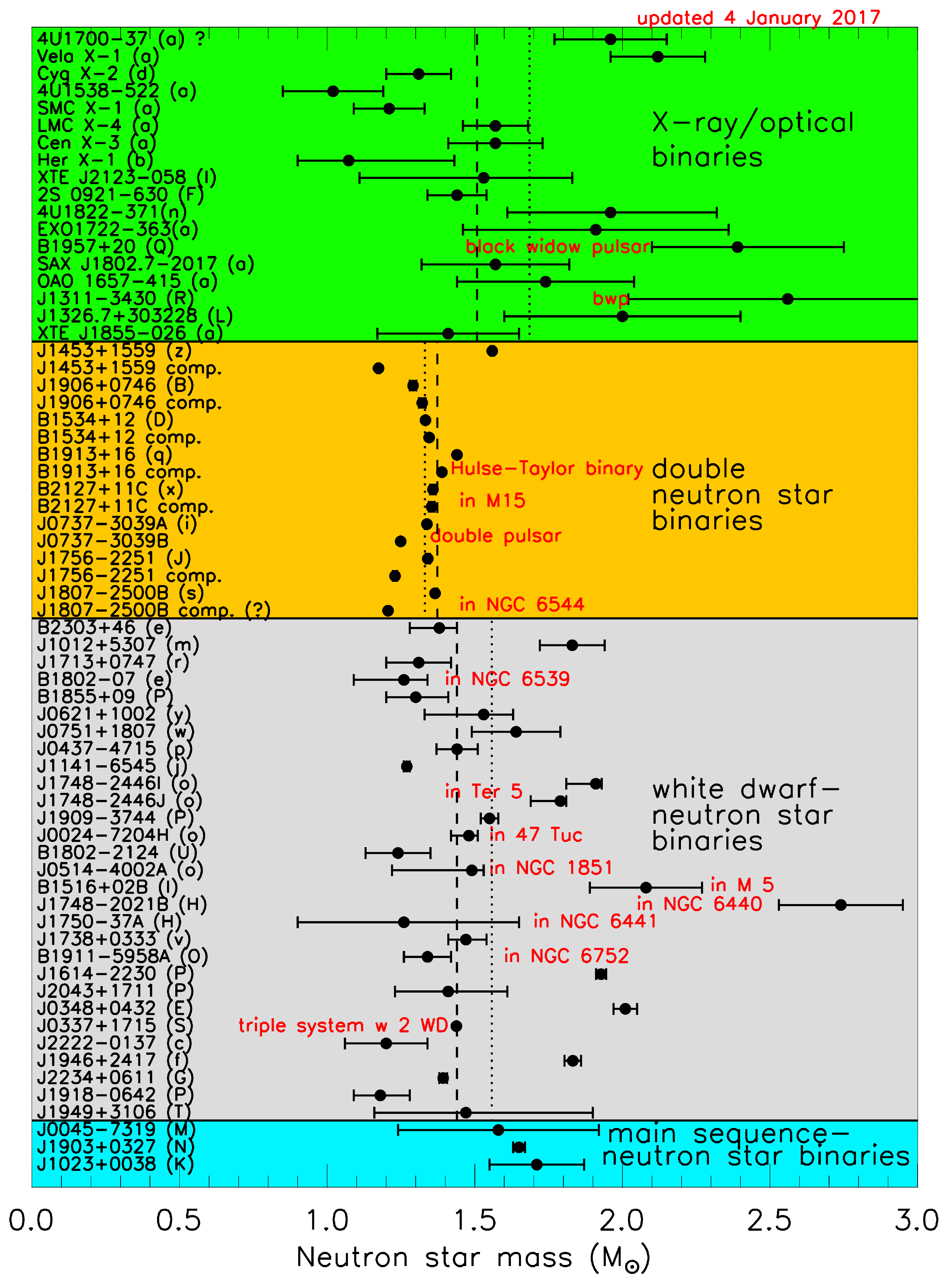}
	\caption{NS masses in binary pulsars (yellow, grey, and blue areas) and X-ray binaries (green region) are depicted graphically. The dashed and dotted vertical lines indicate the masses' simple and weighted averages, respectively. Figure taken from Ref. \cite{doi:10.1146/annurev-nucl-102711-095018}.}
	\label{fig1.4}
\end{figure}

The most visible quantity is the mass of an NS. The most precise mass determinations of NS are obtained by timing observations of pulsars in binary systems, i.e., with a white dwarf or a second NS. The Doppler effect typically allows us to estimate the orbital size as well as the overall mass of the binary system; subsequently, in many circumstances, the discovery of relativistic effects such as Shapiro delay or orbit contraction owing to gravitational wave emission provides a measure of the two masses. In certain situations, the masses are determined with remarkable precision (for example, the Hulse-Taylor binary PSR 1913+16 with masses $m_1$ = 1.4398 $\pm$ 0.0002$M_{\odot}$ and $m_2$ = 1.3886 $\pm$ 0.0002$M_{\odot}$), but the mass measurement of an accreting NS in X-ray binaries yields less precise findings. Fig.~\ref{fig1.4} displays the measured masses of neutron stars in binary systems.

The precise measurements of pulsar masses PSR J1614-2230 (1.928 $\pm$ 0.017)$M_{\odot}$ \cite{Demorest2010},PSR J0348+0432(2.01 $\pm$ 0.04)$M_{\odot}$ \cite{Antoniadis1233232}, and PSR J0740+6620 (2.04$^{+0.10}_{-0.09}$)$M_{\odot}$ \cite{Cromartie2020} show that the maximum mass of an NS should be at least 2$M_{\odot}$.  
The Laser Interferometer Gravitational-wave Observatory (LIGO)-Variability of solar IRradiance and Gravity Oscillations (VIRGO) detector network detected a gravitational-wave signal, GW170817, from the inspiral of two low-mass compact objects on August 17, 2017, compatible with a binary neutron star (BNS) merger. The total mass of the GW170817 BNS merger was found to be around 2.7$M_{\odot}$ with the heavier component of 1.16 - 1.60$M_{\odot}$ for low spin priors, and the maximum mass approached 1.9$M_{\odot}$ for high spin priors \cite{PhysRevX.9.011001}. The gravitational wave event GW170817 is interpreted as the possibility of an upper limit on the NS maximum mass which is around 2.3$M_{\odot}$ \cite{PhysRevD.100.023015}. A recent gravitational wave detection, GW190814, with a black hole merger of mass 22.2 - 24.3$M_{\odot}$ and a secondary component with mass 2.50 - 2.67$M_{\odot}$ \cite{Abbott_2020a} gained a lot of attention about the nature of its secondary component, whether it's a light black hole, supermassive NS, or some other exotic object. 

The LIGO Scientific and Virgo Collaborations (LVC) \cite{Abbott_2021} announced two gravitational wave occurrences, GW200105, and GW200115, that are consistent with neutron star-black hole (NSBH) binaries based on the second part of the third observation run. At 90\% confidence, the major components were discovered to be black holes with masses of 8.9$^{+1.2}_{-1.5} M_{\odot}$ and 5.7$^{+1.8}_{-2.1}$$M_{\odot}$, respectively and secondary ones with masses of 1.9$^{+0.3}_{-0.2}$$M_{\odot}$ and 1.5$^{+0.7}_{-0.3}$$M_{\odot}$, respectively.
When compared to the maximum mass of NSs, these secondary components are compatible with NSs with a probability of $\approx$ 90\%.
The radius measurement with mass is critical for understanding the structure of the NS. There is only one type of neutron star, known as an X-ray burster \cite{doi:10.1146/annurev.astro.34.1.607}, that, despite losing all of its magnetic field, continues to rotate around a partner and permits radius measurements.
The measurement of the thermal spectra of NS offers information on the gravitational redshift, which relies on the NSs mass and radius \cite{PhysRevLett.3.439}. As a result, this finding is valuable in determining the mass and radius of the NS. However, the mass and radius cannot be calculated precisely with a single observation and substantial work is being done to make such exact and simultaneous measurements \cite{2010ApJ...722...33S,Miller_2015}.

The Neutron Star Interior Composition ExploreR (NICER) is an International Space Station (ISS) experiment that uses soft X-ray timing to study neutron stars. Through rotation-resolved X-ray spectroscopy, NICER challenges nuclear physics theory by investigating unusual states of matter within neutron stars. Riley and Miller estimated the size and mass of pulsar J0030+0451 using NICER data. Riley obtained the inferred mass $M$ and equatorial radius $R_{eq}$ as ${1.34}_{-0.16}^{+0.15}\,{M}_{\odot}$ and ${12.71}_{-1.19}^{+1.14}\ \mathrm{km}$ \cite{Riley_2019}, while the radius and mass estimates obtained by Miller are 
${R}_{e}={13.02}_{-1.06}^{+1.24}$ km and $M={1.44}_{-0.14}^{+0.15}\,{M}_{\odot }$ (68\%) \cite{Miller_2019}. Very recently, data from the NICER X-ray Timing Instrument (NICER XTI) event constrained the equatorial radius
and mass of PSR J0740+6620 in the limit ${12.39}^{+1.30}_{-0.98} \ \mathrm{km}$ and ${2.072}^{+0.067}_{-0.066}{M}_{\odot}$respectively \cite{riley2021nicer}. The equatorial circumferential radius
of PSR J0740+6620 is found to be ${13.7}^{+2.6}_{-1.5}\,\mathrm{km}$ (68\%) from NICER and X-ray Multi-Mirror
(XMM-Newton) X-ray observations \cite{miller2021radius}. The inferred radius range for 1.4${M}_{\odot}$ neutron stars has been considerably narrowed by new measurements. Before the new NICER measurements, the radius of PSR J0030+0451 had been constrained to 11.2-13.3 km at 90\% credibility and 11.9-13.0 km at 68\% credibility using a Gaussian process EoS model, nuclear data, information about neutron star tidal deformability from the gravitational wave event GW170817 and previous mass and radius measurements.
With the addition of the new data from the PSR J0740+6620 measurement, these ranges are reduced to 11.8-13.4 km (90\%) and 12.2-13.1 km (68\%) \cite{miller2021radius}.
Based on the recent measurements reported by the Lead Radius EXperiment (PREX-2) exploiting the strong correlation between $R_{skin}^{208}$ and the symmetry energy slope parameter $L$, the radius at 1.4$M_{\odot}$ is constrained in the region 13.25 $\le R_{1.4}$ (km) $\le$ 14.26 \cite{PhysRevLett.126.172503}.
\subsection{Tidal Deformability}
The gravitational waves heralded the start of a spectacular light show. Because black holes are the gravitational fields left behind when extremely massive stars collide, they contain nothing that might generate light when an isolated pair of them merges. Neutron stars, on the other hand, are the dead cores left behind after relatively smaller stars explode in supernovae and they are composed of nearly pure neutrons in the densest materials known. When such orbs collide, they should release debris that emits light at all wavelengths. The discovery of the first binary pulsar, PSR 1913+16, by Hulse and Taylor in 1974 ushered in a scientific revolution \cite{1975ApJ...195L..51H}. In the BNS system, we have two astronomical entities with a radius of approximately ten kilometers, but a mass equivalent to that of the sun, and they are just a few times the moon's distance from the Earth. This finding confirmed the existence of gravitational waves, which Einstein's general theory of relativity anticipated.
\vspace{0.8cm}
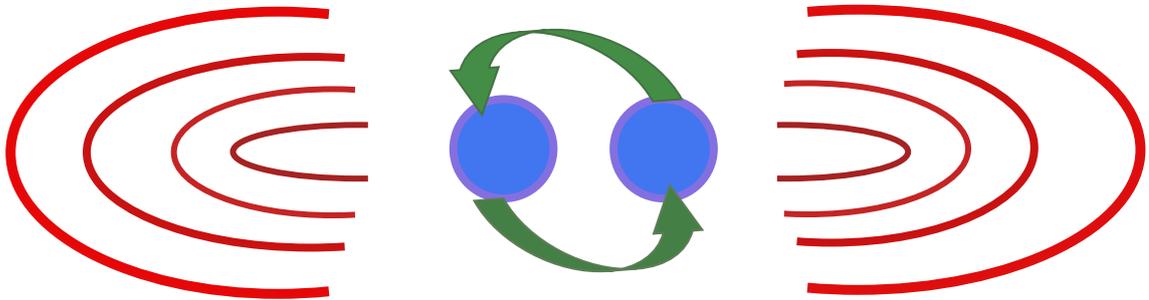
\begin{figure}[htb!]
	\centering
	
	\tikzset{every picture/.style={line width=0.75pt}} 
	
	\begin{tikzpicture}[x=0.75pt,y=0.75pt,yscale=-1,xscale=1]
	
	\draw  [draw opacity=0][line width=2.25]  (205.4,161.96) .. controls (203.78,161.99) and (202.15,162) .. (200.5,162) .. controls (165.98,162) and (138,155.96) .. (138,148.5) .. controls (138,141.04) and (165.98,135) .. (200.5,135) .. controls (202.15,135) and (203.78,135.01) .. (205.4,135.04) -- (200.5,148.5) -- cycle ; \draw  [color={rgb, 255:red, 165; green, 34; blue, 34 }  ,draw opacity=1 ][line width=2.25]  (205.4,161.96) .. controls (203.78,161.99) and (202.15,162) .. (200.5,162) .. controls (165.98,162) and (138,155.96) .. (138,148.5) .. controls (138,141.04) and (165.98,135) .. (200.5,135) .. controls (202.15,135) and (203.78,135.01) .. (205.4,135.04) ;
	\draw  [draw opacity=0][line width=2.25]  (198.94,180.17) .. controls (195.2,180.39) and (191.38,180.5) .. (187.5,180.5) .. controls (143.87,180.5) and (108.5,166.29) .. (108.5,148.75) .. controls (108.5,131.21) and (143.87,117) .. (187.5,117) .. controls (191.38,117) and (195.2,117.11) .. (198.94,117.33) -- (187.5,148.75) -- cycle ; \draw  [color={rgb, 255:red, 197; green, 36; blue, 36 }  ,draw opacity=1 ][line width=2.25]  (198.94,180.17) .. controls (195.2,180.39) and (191.38,180.5) .. (187.5,180.5) .. controls (143.87,180.5) and (108.5,166.29) .. (108.5,148.75) .. controls (108.5,131.21) and (143.87,117) .. (187.5,117) .. controls (191.38,117) and (195.2,117.11) .. (198.94,117.33) ;
	\draw  [draw opacity=0][line width=2.25]  (409.6,161.96) .. controls (411.22,161.99) and (412.85,162) .. (414.5,162) .. controls (447.91,162) and (475,155.96) .. (475,148.5) .. controls (475,141.04) and (447.91,135) .. (414.5,135) .. controls (412.85,135) and (411.22,135.01) .. (409.6,135.04) -- (414.5,148.5) -- cycle ; \draw  [color={rgb, 255:red, 165; green, 34; blue, 34 }  ,draw opacity=1 ][line width=2.25]  (409.6,161.96) .. controls (411.22,161.99) and (412.85,162) .. (414.5,162) .. controls (447.91,162) and (475,155.96) .. (475,148.5) .. controls (475,141.04) and (447.91,135) .. (414.5,135) .. controls (412.85,135) and (411.22,135.01) .. (409.6,135.04) ;
	\draw  [draw opacity=0][line width=2.25]  (413.12,179.64) .. controls (417,179.88) and (420.96,180) .. (425,180) .. controls (469.18,180) and (505,165.23) .. (505,147) .. controls (505,128.77) and (469.18,114) .. (425,114) .. controls (420.96,114) and (417,114.12) .. (413.12,114.36) -- (425,147) -- cycle ; \draw  [color={rgb, 255:red, 197; green, 36; blue, 36 }  ,draw opacity=1 ][line width=2.25]  (413.12,179.64) .. controls (417,179.88) and (420.96,180) .. (425,180) .. controls (469.18,180) and (505,165.23) .. (505,147) .. controls (505,128.77) and (469.18,114) .. (425,114) .. controls (420.96,114) and (417,114.12) .. (413.12,114.36) ;
	\draw  [draw opacity=0][line width=3]  (193.76,196.18) .. controls (188.14,196.55) and (182.37,196.75) .. (176.5,196.75) .. controls (114.92,196.75) and (65,175.26) .. (65,148.75) .. controls (65,122.24) and (114.92,100.75) .. (176.5,100.75) .. controls (182.37,100.75) and (188.14,100.95) .. (193.76,101.32) -- (176.5,148.75) -- cycle ; \draw  [color={rgb, 255:red, 201; green, 18; blue, 18 }  ,draw opacity=1 ][line width=3]  (193.76,196.18) .. controls (188.14,196.55) and (182.37,196.75) .. (176.5,196.75) .. controls (114.92,196.75) and (65,175.26) .. (65,148.75) .. controls (65,122.24) and (114.92,100.75) .. (176.5,100.75) .. controls (182.37,100.75) and (188.14,100.95) .. (193.76,101.32) ;
	\draw  [draw opacity=0][line width=3]  (419.49,193.57) .. controls (425.06,194.02) and (430.79,194.25) .. (436.63,194.25) .. controls (492.61,194.25) and (538,172.87) .. (538,146.5) .. controls (538,120.13) and (492.61,98.75) .. (436.63,98.75) .. controls (430.79,98.75) and (425.06,98.98) .. (419.49,99.43) -- (436.63,146.5) -- cycle ; \draw  [color={rgb, 255:red, 201; green, 18; blue, 18 }  ,draw opacity=1 ][line width=3]  (419.49,193.57) .. controls (425.06,194.02) and (430.79,194.25) .. (436.63,194.25) .. controls (492.61,194.25) and (538,172.87) .. (538,146.5) .. controls (538,120.13) and (492.61,98.75) .. (436.63,98.75) .. controls (430.79,98.75) and (425.06,98.98) .. (419.49,99.43) ;
	\draw  [draw opacity=0][line width=3.75]  (185.88,218.72) .. controls (177.66,219.56) and (169.18,220) .. (160.5,220) .. controls (86.77,220) and (27,188.21) .. (27,149) .. controls (27,109.79) and (86.77,78) .. (160.5,78) .. controls (169.18,78) and (177.66,78.44) .. (185.88,79.28) -- (160.5,149) -- cycle ; \draw  [color={rgb, 255:red, 230; green, 8; blue, 8 }  ,draw opacity=1 ][line width=3.75]  (185.88,218.72) .. controls (177.66,219.56) and (169.18,220) .. (160.5,220) .. controls (86.77,220) and (27,188.21) .. (27,149) .. controls (27,109.79) and (86.77,78) .. (160.5,78) .. controls (169.18,78) and (177.66,78.44) .. (185.88,79.28) ;
	\draw  [draw opacity=0][line width=3.75]  (424.75,216.87) .. controls (432.94,217.61) and (441.38,218) .. (450,218) .. controls (527.87,218) and (591,186.44) .. (591,147.5) .. controls (591,108.56) and (527.87,77) .. (450,77) .. controls (441.38,77) and (432.94,77.39) .. (424.75,78.13) -- (450,147.5) -- cycle ; \draw  [color={rgb, 255:red, 223; green, 14; blue, 14 }  ,draw opacity=1 ][line width=3.75]  (424.75,216.87) .. controls (432.94,217.61) and (441.38,218) .. (450,218) .. controls (527.87,218) and (591,186.44) .. (591,147.5) .. controls (591,108.56) and (527.87,77) .. (450,77) .. controls (441.38,77) and (432.94,77.39) .. (424.75,78.13) ;
	\draw  [color={rgb, 255:red, 132; green, 111; blue, 224 }  ,draw opacity=1 ][fill={rgb, 255:red, 65; green, 118; blue, 240 }  ,fill opacity=1 ][line width=3]  (248,147) .. controls (248,133.19) and (259.19,122) .. (273,122) .. controls (286.81,122) and (298,133.19) .. (298,147) .. controls (298,160.81) and (286.81,172) .. (273,172) .. controls (259.19,172) and (248,160.81) .. (248,147) -- cycle ;
	\draw  [color={rgb, 255:red, 132; green, 111; blue, 224 }  ,draw opacity=1 ][fill={rgb, 255:red, 65; green, 118; blue, 240 }  ,fill opacity=1 ][line width=3]  (328,147) .. controls (328,133.19) and (339.19,122) .. (353,122) .. controls (366.81,122) and (378,133.19) .. (378,147) .. controls (378,160.81) and (366.81,172) .. (353,172) .. controls (339.19,172) and (328,160.81) .. (328,147) -- cycle ;
	\draw  [color={rgb, 255:red, 68; green, 114; blue, 68 }  ,draw opacity=1 ][fill={rgb, 255:red, 68; green, 141; blue, 70 }  ,fill opacity=1 ] (294.56,87.27) .. controls (320.1,85.24) and (350.3,100.79) .. (362,122) -- (347.6,123.14) .. controls (335.89,101.93) and (305.7,86.38) .. (280.15,88.41) ;\draw  [color={rgb, 255:red, 68; green, 114; blue, 68 }  ,draw opacity=1 ][fill={rgb, 255:red, 68; green, 141; blue, 70 }  ,fill opacity=1 ] (280.15,88.41) .. controls (264.67,89.64) and (254.45,97.05) .. (251.41,107.41) -- (246.61,107.8) -- (262.3,129.92) -- (270.61,105.89) -- (265.81,106.27) .. controls (268.85,95.9) and (279.07,88.5) .. (294.56,87.27)(280.15,88.41) -- (294.56,87.27) ;
	\draw  [color={rgb, 255:red, 93; green, 124; blue, 86 }  ,draw opacity=1 ][fill={rgb, 255:red, 68; green, 127; blue, 69 }  ,fill opacity=1 ] (325.07,208.51) .. controls (300.11,210.49) and (270.26,194.67) .. (258.4,173.16) -- (273,172) .. controls (284.87,193.51) and (314.72,209.34) .. (339.67,207.35) ;\draw  [color={rgb, 255:red, 93; green, 124; blue, 86 }  ,draw opacity=1 ][fill={rgb, 255:red, 68; green, 127; blue, 69 }  ,fill opacity=1 ] (339.67,207.35) .. controls (354.8,206.15) and (364.65,198.68) .. (367.42,188.2) -- (372.29,187.81) -- (356.06,165.4) -- (347.95,189.74) -- (352.82,189.36) .. controls (350.05,199.84) and (340.19,207.31) .. (325.07,208.51)(339.67,207.35) -- (325.07,208.51) ;
	\end{tikzpicture}
	\caption{The schematic representation of the gravitational waves produced by the inspiral of compact bodies in a binary neutron star system.}
	\label{fig1.5}
\end{figure}

Two stars spin around a shared center of mass in this mechanism. As they revolve, they emit gravitational waves, causing the orbits to lose energy and get more close, a process known as inspiralling. As they grow closer, they emit additional gravitational waves and get even closer, finally crashing. Just before the merger, the star is tidally disrupted by the other companion stars' exterior tidal field, causing a slight adjustment in the phase of the gravitational waves. During the inspiral phase of an NS-NS merger, an extraordinarily powerful tidal gravitational field is created, which deforms the stars' multipolar structure (Fig.~\ref{fig1.5}). This impact may be expressed in terms of the stars tidal deformability or tidal Love number, which provides information on the internal structure of the NS \cite{PhysRevD.81.123016,Hinderer_2008}.

The Love numbers have a direct effect on the amount of tidal bulge on bodies caused by a non-uniform external gravitational field. To demonstrate this, imagine the Sun and Earth, where the Sun is seen as a point mass. It has been discovered that the Sun's gravitational field is strongest on the side of the Earth that is closest to the Sun. As a result, there is relative acceleration, resulting in quadrupole deformation as seen from the Earth's center-of-mass frame. As a result, the production of two high tides every day at a particular place on Earth is the overall result.

A spherical star  placed in a static external quadrupolar tidal field $\mathcal{E}_{ij}$ causes star deformation as well as quadrupole deformation, resulting in leading order disruption. This type of deformation is assessed by \cite{PhysRevD.81.123016,Hinderer_2008}
\begin{align}\label{eq1.1}
\lambda&=-\frac{Q_{ij}}{\mathcal{E}_{ij}} = \frac{2}{3}k_2R^5,
\end{align}
\begin{align}\label{eq1.2}
\Lambda&=\frac{2k_2}{3C^5},
\end{align}
where $Q_{ij}$ is a binary star's induced quadrupole moment and $\mathcal{E}_{ij}$ is the companion star's static external quadrupole tidal force. $\lambda$ is the tidal deformability parameter, which is determined by the EoS through the NS radius and a dimensionless quantity $k_2$, also known as the second Love number \cite{Hinderer_2008}. $C=M/R$ is the compactness parameter. $\Lambda$ is the dimensionless equivalent of $\lambda$. In general relativity, we must differentiate $k_2$ gravitational fields produced by masses (electric type) from those produced by mass motion, i.e., mass currents (magnetic type), which have no counterpart in Newtonian gravity \cite{PhysRevD.81.084016,PhysRevD.89.124011}.
According to Eq.~(\ref{eq1.1}), $\lambda$ is strongly influenced by the radius of the NS and the value of $k_2$. Furthermore, $k_2$ is influenced by the internal structure of the component body and enters the gravitational wave phase of the inspiraling BNS directly, conveying information about the EoS. Because the gravitational gradient grows with the radius of the NS, so does the deformation produced by the external field. In other words, stiff (soft) EoS generates significant (little) deformation in the BNS system.

Recently, upgraded LIGO and Virgo detectors reported for the first time the direct detection of gravitational waves from a spinning NS-NS binary, dubbed GW170817 \cite{PhysRevLett.119.161101}. The binary chirp mass is found to be ${1.188}^{+0.004}_{-0.002}\,{M}_{\odot}$ at the 90\% credible intervals, as determined by data analysis of GW170817. The dimensionless tidal deformability $\Lambda_1$ and $\Lambda_2$ with 90\% and 50\% confidence limits derived for two stars in the BNS merger seen by GW170817 are presented in Fig.~5 of Ref.~\cite{PhysRevLett.119.161101}. The measurement is stated as a limit for average dimensionless tidal deformability $\Lambda$ $\le$ 800 for low-spin prior. In their subsequent analysis \cite{PhysRevLett.121.161101}, the LVC suggests a much smaller upper limit of 580 on dimesionless tidal deformability, ruling out the stiffer EoSs \cite{PhysRevD.99.121301}.
\section{Static and Rotating Neutron stars}
\subsection{Static neutron star}
\label{tovall}
For a spherically symmetric, static NS (SNS), the metric element has the Schwarzschild form ($G$ = $c$ = 1)
\begin{equation}
ds^2=-e^{2\phi(r)}dt^2+e^{2\Lambda(r)}dr^2+r^2(d\theta^2+sin^2\theta d\phi^2),
\end{equation}
where the metric functions $e^{-2\phi(r)}$ and $e^{2\Lambda(r)}$ are defined as
\begin{equation}
e^{-2\phi(r)} =(1-\gamma(r))^{-1},
\end{equation}
\begin{equation}
e^{2\Lambda(r)}=(1-\gamma(r)),
\end{equation}
with
\begin{equation}
\gamma(r)=2M(r)/r.
\end{equation}
The energy-momentum tensor reduces the Einstein field equations to well-known 
Tolman-Oppenheimer-Volkoff coupled differential equations given by \cite{PhysRev.55.364,PhysRev.55.374}
\begin{equation}\label{tov1}
\frac{dP(r)}{dr}= -\frac{[\mathcal{E}(r) +P(r)][M(r)+4\pi r^3 P(r)]}{r^2(1-2M(r)/r) },
\end{equation}
and
\begin{equation}\label{tov2}
\frac{dM(r)}{dr}= 4\pi r^2 \mathcal{E}(r),
\end{equation}
where $M(r)$, $\mathcal{E}(r)$ and $P(r)$ represent the gravitational mass at radius $r$ with fixed central density, energy density, and pressure, respectively. The boundary conditions  $P(0)=P_c$, $M(0)=0$ allow one to solve the above differential equations and determine the properties of a NS. \par
The tidal deformability $\lambda$ is defined as the ratio of the induced quadrupole mass $Q_{ij}$ to the external tidal field $\mathcal{E}_{ij}$ as defined by Eqs.~(\ref{eq1.1}) and (\ref{eq1.2}) \cite{PhysRevD.81.123016,PhysRevC.95.015801}.
The expression for the Love number is written as \cite{PhysRevD.81.123016}
\begin{equation}\label{l3}
\begin{split}
k_2=\frac{8}{5}(1-2C)^2 [2C(y-1)]\Bigl\{2C(4(y+1)C^4
+(6y-4)C^3\\
+(26-22y)C^2+3(5y-8)C-3y+6)\\
-3(1-2C)^2(2C(y-1)-y+2)log\Big(\frac{1}{1-2C}\Big)\Bigr\}^{-1}.
\end{split}
\end{equation}
The function $y=y(R)$ can be computed by solving the differential equation \cite{PhysRevC.95.015801,Hinderer_2008}
\begin{equation}\label{l4}
r\frac{dy(r)}{dr}+y(r)^2+y(r)F(r)+r^2 Q(r)=0,
\end{equation}
where
\begin{equation}\label{l5}
F(r)=\frac{r-4\pi r^3 [\mathcal{E}(r)-P(r)]}{r-2M(r)},
\end{equation}
\begin{equation}\label{l6}
\begin{split}
Q(r)=\frac{4\pi r\Big(5\mathcal{E}(r)+9P(r)+\frac{\mathcal{E}(r)+P(r)}{\partial P(r)/\partial\mathcal{E}(r)}-\frac{6}{4\pi r^2}\Big)}{r-2M(r)}\\
-4\Bigg[\frac{M(r)+4\pi r^3 P(r)}{r^2 (1-2M(r)/r)}\Bigg]^2.
\end{split}
\end{equation}
The above equations are solved for spherically symmetric and static NS to determine the properties such as mass, radii, and tidal deformability.
\subsection{Rotating neutron star}
For a rapidly rotating NS (RNS) with a nonaxisymmetric configuration, they would emit gravitational waves until they achieve axisymmetric configuration. The rotation deforms the NS. Here we study the rapidly rotating NS assuming a stationary, axisymmetric space-time. The energy-momentum tensor for such a perfect fluid describing the matter is given by
\begin{equation}
T^{\mu \nu} = (\mathcal{E}+P)u^{\mu}u^{\nu}+Pg^{\mu \nu},
\end{equation}
where the first term represents the contribution from matter. $u^{\mu}$ denotes the fluid-four-velocity, $\mathcal{E}$ is the energy density, and $P$ is the pressure. For RNS, the metric tensor is given by \cite{PhysRevLett.62.3015}
\begin{equation}
\begin{gathered}
ds^2=-e^{2\nu(r,\theta)}dt^2+e^{2\psi(r,\theta)}(d\phi-\omega(r)dt)^2\\ +e^{2\mu(r,\theta)}d\theta^2 +e^{2\lambda(r,\theta)}dr^2,\\
\end{gathered}
\end{equation}
where the gravitational potentials $\nu$, $\mu$, $\psi$, and $\lambda$ are the functions of $r$ and $\theta$ only. The Einstein's field equations are solved for the given potential to determine the physical properties that govern the structure of the RNS. Global properties such as gravitational mass, equitorial radius, moment of inertia, angular momentum, etc. are calculated.\par 
For a RNS, the angular momentum $J$ is easy to calculate. By defining the angular velocity of the fluid relative to a local inertial frame, $\bar{\omega}(r) =\Omega-\omega(r)$, $\bar{\omega}$ satisfies the following differential equation
\begin{equation}
\frac{1}{r^4}\frac{d}{dr}\Bigg(r^4 j \frac{d\bar{\omega}}{dr}\Bigg)+\frac{4}{r}\frac{dj}{r}\bar{\omega}=0,
\end{equation}
where $j=j(r)=e^{-(\nu+\lambda)/2}$.\par 
The angular momentum of the star is then given by the relation
\begin{equation}
J=\frac{1}{6}R^4 \Bigg(\frac{d\bar{\omega}}{dr}\Bigg)_{r=R},
\end{equation}
which relates the angular velocity as
\begin{equation}
\Omega=\bar{\omega}(R)+\frac{2J}{R^3}.
\end{equation}
The properties of a rotating neutron star studied in this thesis are calculated by using the RNS code \cite{Stergioulas2003,Stergioulas_1995,rnscode}.
\section{Neutron star structure}
\begin{figure}[hbt!]
	\centering
	\begin{tikzpicture}[>={LaTeX[width=5mm,length=2mm]},->]
	\draw [fill=blue!40!white,opacity=1] (-3.5,4.98) -- (3.5,4.98) -- (0,-2.0) -- cycle;
	\draw [fill=blue!20!white,opacity=1,] (0,5) circle (3.5cm and 0.4cm);
	\draw [fill=blue!60!red,opacity=1] (-3.1,4.18) -- (3.1,4.18) -- (0,-2.0) -- cycle;
	\draw [fill=blue!30!red,opacity=1,] (0,4.2) circle (3.1cm and 0.3cm);
	\draw [fill=green!40!white,opacity=1] (-2.65,3.28) -- (2.65,3.28) -- (0,-2.0) -- cycle;
	\draw [fill=green!20!white,opacity=1,] (0,3.3) circle (2.65cm and 0.4cm);
	\draw [fill=yellow!40!white,opacity=1] (-1.4,0.78) -- (1.4,0.78) -- (0,-2.0) -- cycle;
	\draw [fill=yellow!20!white,opacity=1,] (0,0.8) circle (1.4cm and 0.3cm);
	
	\node[align=left] at (-5.0,4.7) {Outer Crust\\ 0.3-0.5 km};
	\node[align=left] at (-4.2,3.5) {Inner Crust\\ 1-2 km};
	\node[align=left] at (-3.4,1.9) {Outer Core\\ $\approx$ 9 km};
	\node[align=left] at (-2.7,-0.2) {Inner Core\\ 0-3 km};
	
	\node[align=left] at (-0.15,3.3) {0.3-0.5 $\rho_{0}$};
	\node[align=left] at (-0.15,1.9) {0.5-2.0 $\rho_{0}$};
	\node[align=left] at (-0.12,-0.2) {2-15 $\rho_{0}$};
	\end{tikzpicture}
	\caption{Schematic representation of the NS structure in terms of its radial distance. $\rho_{0}$ is the nuclear saturation density ($\approx$ 0.15 fm$^{-3}$).}\label{fig1.7}
\end{figure}
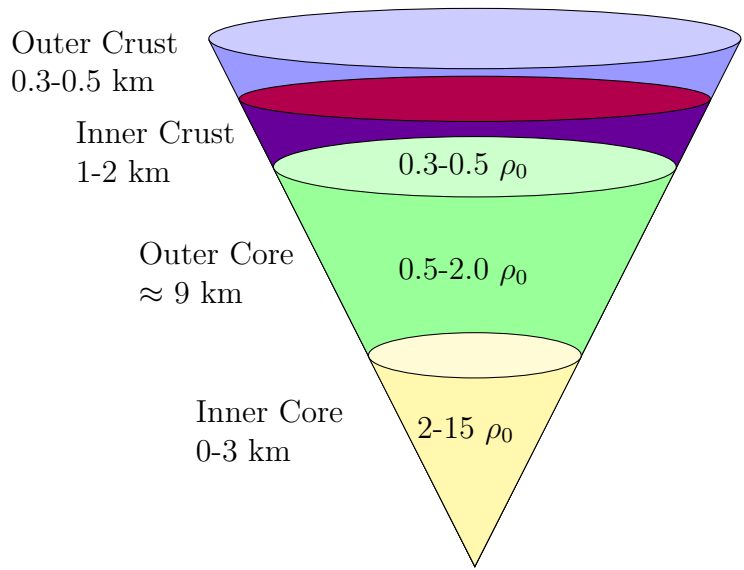

The many possibilities of the NS structure are discussed below, albeit as will be demonstrated, the contents are not always neutrons. In light of this, further definitions of NS include hyperon stars, quark stars, and hybrid stars to characterize compact stellar objects \cite{gled, 2001PhR...342..393G}. Beyond basic interactions, we need physics to comprehend these stars. The physics of neutron (and nucleon) matter has developed from very basic explanations consistent with observation to numerous more complicated descriptions incorporating many species such as baryons, mesons, leptons, and quarks. The current study contributes to a deeper understanding of matter at both the microscopic and macroscopic sizes, as well as the theory and formalism that connects these two extremes.
With increasing density, the radial structure of NS may be separated into the atmosphere, crust, and core of NS (see Fig.~\ref{fig1.7}) \cite{haensel}.

The NSs atmosphere or surface is a thin plasma layer that creates the thermal electromagnetic radiation spectrum. This radiation contains valuable information about NS, such as effective surface temperature, surface gravity, chemical composition, surface magnetic field strength and shape. The atmosphere's thickness ranges from 10 cm to a few millimeters.
\vspace{0.5cm}

\subsection{Crust}
The crust of a neutron star accounts for just a small fraction of the star's mass, but it has a significant influence on phenomena such as cooling rate and the production of spectacular gamma-ray bursts. The microscopic structure of the outer crust is a lattice of neutron-rich nuclei surrounded by a homogenous cloud of electrons. The outer crust (the outer envelope) stretches from the bottom of the atmosphere to a layer with a density of $\rho$ $\approx$ 4 $\times$ 10$^{11}$ g cm$^{-3}$ and a thickness of about 0.3-0.5 km. The higher pressure fuses more electrons and protons into neutrons as we get closer to the star's core, increasing the neutron density in the nucleus. When the inner crust's nuclei can no longer accept any more neutrons, the released neutrons form a superfluid that penetrates the lattice.

The inner crust is approximately 1-2 km thick. The inner crust's density ranges from $\rho$ $\approx$ 4 $\times$ 10$^{11}$ g cm$^{-3}$ (at the upper border) to $\sim$ 0.5 $\rho_0$ (at the base). Electrons, free neutrons, and neutron-rich atomic nuclei make up the inner crust. The fraction of free neutrons increases as density increases. The inner crust of NS is made up of several components known together as the pasta structure. The outer layer of the NS crust, with a density less than the nuclear saturation density, presents special challenges. Nucleons are correlated over short distances by attractive strong interactions, but anti-correlated over long distances by Coulomb repulsion at this density. Complex and unique nuclear structures such as spheres, bubbles, rods, slabs, and tubes develop as a result of the rivalry between these short and long-range interactions. The term "pasta phases" has been used to describe these complex structures. Various approaches have been used to investigate the pasta phases \cite{PhysRevC.103.055807}. 

Deformations and fissures in neutron star crusts have been related to phenomena such as gravitational waves, bursts of gamma rays, and "glitches"-events in which a star's spin rapidly accelerates \cite{2018ApJ...866...94B}. The structure of the inner crust is critical to understanding these occurrences. The inner crust's structure influences its strength and rigidity, which can have serious consequences on a star's behavior. A neutron star's crust, for example, can sustain mountain-like formations on its surface if it is sufficiently strong. These mountains rotate more than 600 times per second, causing disturbances in spacetime known as gravitational waves.

\subsection{Core} 
The core of an NS is roughly 10 km thick and comprises the majority of the NSs mass. The core of the NS, like the crust, is split into two parts: the outer core and the inner core.
The outer core has a density range of 0.5$\rho_{0} \le \rho \le 0.20 \rho_{0}$ and a thickness of several kilometers. It is composed of neutrons with a significant admixture (5 to 15\%) of protons, electrons, and potentially muons called $npe\mu$ composition. The state of this matter is controlled by electric neutrality and beta equilibrium, which are complemented with a microscopic model of many-body nucleon interaction.

The center area of NS is occupied by the inner core, where $\rho \ge 2\rho_{0}$. It has a radius of many kilometers with a core density of $\sim$ (10-15)$\rho_{0}$. The inner core's composition and EoS are poorly understood and heavily model-dependent. Several hypotheses have been proposed, each of which predicts the arrival of new fermions and/or bosons.
\vspace{0.2cm}
\begin{figure}[hbt!]
	\centering
	\begin{tikzpicture}[>={LaTeX[width=5mm,length=5mm]},->]
	\begin{sphere}[sphere scale=4.3, sphere color=blue]
	\begin{sphere}[sphere scale=3.6, sphere color=green]
	\begin{sphere}[sphere scale=2.7, sphere color=red]
	\begin{sphere}[sphere scale=1.8, sphere color=orange]
	\begin{sphere}[sphere scale=0.8, sphere color=yellow]
	
	\end{sphere}
	\end{sphere}
	\end{sphere}
	\end{sphere}
	\end{sphere}
	\draw[line width=2mm, blue] (-3.95,4.5) -- (-1.1,3.0);
	\node[align=left] at (-4.0,5.0) {{\large\textbf{Surface}}\\ Hydrogen/Helium plasma};
	\draw[line width=2mm, green] (2.0,4.5) -- (1.2,2.7);
	\node[align=left] at (2.4,5.5) {{\large\textbf{Outer Crust}}\\ Ions\\ Electron gas };
	\draw[line width=2mm, red] (4.7,3.9) -- (1.1,2.1);
	\node[align=left] at (6.0,5.0) {{\large\textbf{Inner Crust}}\\ Heavy ions\\ Relativistic $e^-$ gas\\Superfluid neutrons };
	\draw[line width=2mm, yellow!50!red] (4.5,2.5) -- (0.7,1.3);
	\node[align=left] at (6.2,2.5) {{\large\textbf{Outer Core}}\\ $n$, $p$\\ $e^-$, $\mu^-$};
	\draw[line width=2mm, yellow!80!green] (4.0,-0.6) -- (0.6,0.2);
	\node[align=left] at (6.4,-2.0) {{\large\textbf{Inner Core}}\\ $n$,$p$\\ $e^-$,$\mu^-$\\Hyperons ($\Lambda, \Sigma$)\\Deltas ($\Delta$)\\Boson ($\pi, K$)\\Deconf. quarks (u,d,s)};
	\end{tikzpicture}
	\caption{Hypothetical view of the NS structure depicting many conceivable phases in various types of NS represented by different sectors such as hyperon stars, strange stars, and so on \cite{Weber:2004kj}.}\label{fig1.8}
\end{figure}
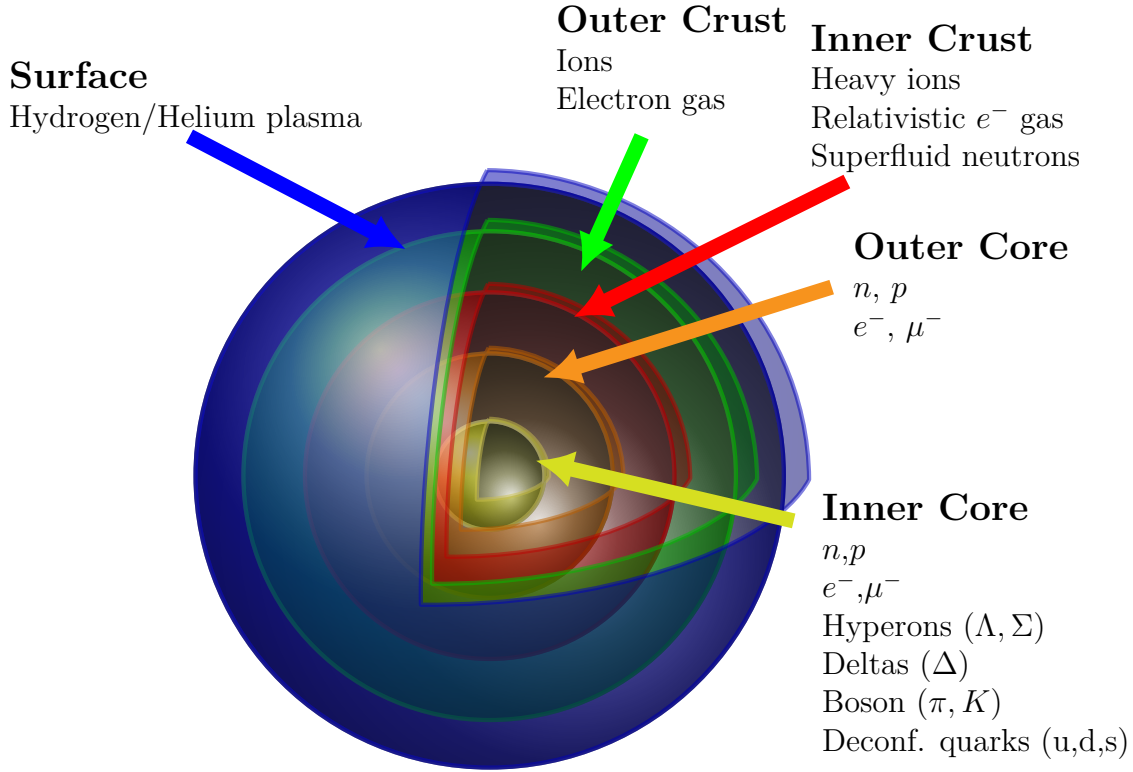

Atoms are so closely packed together at the high densities seen inside neutron stars that new states of matter can arise. While neutron stars are extreme settings in and of themselves, the matter may be transformed into something much more unusual by increasing density. The most obvious example is quark deconfinement, in which basic particles (for example, neutrons) are broken down into their constituent quarks. Quarks do not typically exist as free particles, but this can happen at the high temperatures and densities that occur during quark deconfinement. A quark star might form if quark matter is more stable than conventional matter.

The major possible exotic phases such as Kaons, Hyperons, and deconfined quark matter in the NS core are displayed in Fig.~\ref{fig1.8}. The modelling of these exotic phases in the NS core and their effect on the NS properties are discussed in Chapter 1 of this thesis.

Because we examine quark matter in depth in this thesis, it's important to describe it adequately.

\section{Quark matter}
\subsection{MIT Bag Model}
The Bag Model provides a reasonable phenomenological explanation of quarks in hadrons by confining quarks inside a hadron. While there are many other variants of the model, the MIT Bag Model incorporates the fundamental properties of quark confinement phenomenology. In 1974, the Massachusetts Institute of Technology proposed the MIT bag model to account for hadronic masses in terms of their quark components \cite{PhysRevD.9.3471}. Quarks are represented as massless particles inside a finite-dimensional bag and infinitely enormous outside the bag in the MIT bag model. Confinement in the model is caused by the balance between the inwardly directed bag pressure $B$ and the stress caused by the kinetic energy of the quarks. Bag pressure $B$ is a phenomenological variable established in this case to account for the nonperturbative effects of QCD. It has a value in the range (100-200)$^4$ MeV. If the quarks are confined in the bag, the gluons should be confined as well. Gauss's law requires that the total color charge of the stuff inside the bag be colorless.
\vspace{0.3cm}
\begin{figure}[hbt!]
	\centering

	\tikzset{every picture/.style={line width=0.75pt}} 
	
	\begin{tikzpicture}[x=0.75pt,y=0.75pt,yscale=-1,xscale=1]
	
	\draw  [color={rgb, 255:red, 0; green, 0; blue, 0 }  ,draw opacity=0.04 ][line width=3] [line join = round][line cap = round] (223,67) .. controls (235.16,73.08) and (257.76,84) .. (271,84) .. controls (277.41,84) and (259.89,77.53) .. (254,75) .. controls (250.57,73.53) and (247.73,70) .. (244,70) .. controls (241.02,70) and (240.71,76.59) .. (240,78) .. controls (239.46,79.07) and (238,79.8) .. (238,81) ;
	\draw   (31,128) -- (263,128) -- (263,262) -- (31,262) -- cycle ;
	\draw   (336,128) -- (568,128) -- (568,262) -- (336,262) -- cycle ;
	\draw    (31,101) -- (31,155) ;
	\draw    (568,101) -- (568,155) ;
	\draw    (336,101) -- (336,155) ;
	\draw    (263,101) -- (263,155) ;
	\draw  [fill={rgb, 255:red, 189; green, 183; blue, 183 }  ,fill opacity=1 ] (336,120) -- (568,120) -- (568,128) -- (336,128) -- cycle ;
	\draw  [fill={rgb, 255:red, 194; green, 190; blue, 190 }  ,fill opacity=1 ] (31,120) -- (263,120) -- (263,128) -- (31,128) -- cycle ;
	\draw  [fill={rgb, 255:red, 0; green, 0; blue, 0 }  ,fill opacity=0.98 ] (59,109) -- (65,109) -- (65,91) -- (77,91) -- (77,109) -- (83,109) -- (71,121) -- cycle ;
	\draw  [fill={rgb, 255:red, 14; green, 2; blue, 2 }  ,fill opacity=1 ] (129,108) -- (135,108) -- (135,90) -- (147,90) -- (147,108) -- (153,108) -- (141,120) -- cycle ;
	\draw  [fill={rgb, 255:red, 19; green, 3; blue, 3 }  ,fill opacity=1 ] (500,109) -- (506,109) -- (506,91) -- (518,91) -- (518,109) -- (524,109) -- (512,121) -- cycle ;
	\draw  [fill={rgb, 255:red, 14; green, 2; blue, 2 }  ,fill opacity=1 ] (431,108) -- (437,108) -- (437,90) -- (449,90) -- (449,108) -- (455,108) -- (443,120) -- cycle ;
	\draw  [fill={rgb, 255:red, 8; green, 1; blue, 1 }  ,fill opacity=1 ] (367,109) -- (373,109) -- (373,91) -- (385,91) -- (385,109) -- (391,109) -- (379,121) -- cycle ;
	\draw  [fill={rgb, 255:red, 15; green, 3; blue, 3 }  ,fill opacity=1 ] (195,108) -- (201,108) -- (201,90) -- (213,90) -- (213,108) -- (219,108) -- (207,120) -- cycle ;
	\draw   (31,155) .. controls (31,141.19) and (42.19,130) .. (56,130) .. controls (69.81,130) and (81,141.19) .. (81,155) .. controls (81,168.81) and (69.81,180) .. (56,180) .. controls (42.19,180) and (31,168.81) .. (31,155) -- cycle ;
	\draw   (33,236) .. controls (33,222.19) and (44.19,211) .. (58,211) .. controls (71.81,211) and (83,222.19) .. (83,236) .. controls (83,249.81) and (71.81,261) .. (58,261) .. controls (44.19,261) and (33,249.81) .. (33,236) -- cycle ;
	\draw   (77,236) .. controls (77,222.19) and (88.19,211) .. (102,211) .. controls (115.81,211) and (127,222.19) .. (127,236) .. controls (127,249.81) and (115.81,261) .. (102,261) .. controls (88.19,261) and (77,249.81) .. (77,236) -- cycle ;
	\draw   (118,236) .. controls (118,222.19) and (129.19,211) .. (143,211) .. controls (156.81,211) and (168,222.19) .. (168,236) .. controls (168,249.81) and (156.81,261) .. (143,261) .. controls (129.19,261) and (118,249.81) .. (118,236) -- cycle ;
	\draw   (160,235) .. controls (160,221.19) and (171.19,210) .. (185,210) .. controls (198.81,210) and (210,221.19) .. (210,235) .. controls (210,248.81) and (198.81,260) .. (185,260) .. controls (171.19,260) and (160,248.81) .. (160,235) -- cycle ;
	\draw   (200,234) .. controls (200,220.19) and (211.19,209) .. (225,209) .. controls (238.81,209) and (250,220.19) .. (250,234) .. controls (250,247.81) and (238.81,259) .. (225,259) .. controls (211.19,259) and (200,247.81) .. (200,234) -- cycle ;
	\draw   (195,153) .. controls (195,139.19) and (206.19,128) .. (220,128) .. controls (233.81,128) and (245,139.19) .. (245,153) .. controls (245,166.81) and (233.81,178) .. (220,178) .. controls (206.19,178) and (195,166.81) .. (195,153) -- cycle ;
	\draw   (153,155) .. controls (153,141.19) and (164.19,130) .. (178,130) .. controls (191.81,130) and (203,141.19) .. (203,155) .. controls (203,168.81) and (191.81,180) .. (178,180) .. controls (164.19,180) and (153,168.81) .. (153,155) -- cycle ;
	\draw   (113,154) .. controls (113,140.19) and (124.19,129) .. (138,129) .. controls (151.81,129) and (163,140.19) .. (163,154) .. controls (163,167.81) and (151.81,179) .. (138,179) .. controls (124.19,179) and (113,167.81) .. (113,154) -- cycle ;
	\draw   (72,154) .. controls (72,140.19) and (83.19,129) .. (97,129) .. controls (110.81,129) and (122,140.19) .. (122,154) .. controls (122,167.81) and (110.81,179) .. (97,179) .. controls (83.19,179) and (72,167.81) .. (72,154) -- cycle ;
	\draw   (139,192) .. controls (139,178.19) and (150.19,167) .. (164,167) .. controls (177.81,167) and (189,178.19) .. (189,192) .. controls (189,205.81) and (177.81,217) .. (164,217) .. controls (150.19,217) and (139,205.81) .. (139,192) -- cycle ;
	\draw   (100,190) .. controls (100,176.19) and (111.19,165) .. (125,165) .. controls (138.81,165) and (150,176.19) .. (150,190) .. controls (150,203.81) and (138.81,215) .. (125,215) .. controls (111.19,215) and (100,203.81) .. (100,190) -- cycle ;
	\draw   (55,194) .. controls (55,180.19) and (66.19,169) .. (80,169) .. controls (93.81,169) and (105,180.19) .. (105,194) .. controls (105,207.81) and (93.81,219) .. (80,219) .. controls (66.19,219) and (55,207.81) .. (55,194) -- cycle ;
	\draw   (180,191) .. controls (180,177.19) and (191.19,166) .. (205,166) .. controls (218.81,166) and (230,177.19) .. (230,191) .. controls (230,204.81) and (218.81,216) .. (205,216) .. controls (191.19,216) and (180,204.81) .. (180,191) -- cycle ;
	\draw  [fill={rgb, 255:red, 50; green, 83; blue, 124 }  ,fill opacity=1 ] (265,179) -- (307,179) -- (307,169) -- (335,189) -- (307,209) -- (307,199) -- (265,199) -- cycle ;
	\draw  [fill={rgb, 255:red, 33; green, 71; blue, 223 }  ,fill opacity=1 ] (352,141) .. controls (352,138.79) and (353.79,137) .. (356,137) .. controls (358.21,137) and (360,138.79) .. (360,141) .. controls (360,143.21) and (358.21,145) .. (356,145) .. controls (353.79,145) and (352,143.21) .. (352,141) -- cycle ;
	\draw  [fill={rgb, 255:red, 86; green, 197; blue, 23 }  ,fill opacity=1 ] (351,240) .. controls (351,237.79) and (352.79,236) .. (355,236) .. controls (357.21,236) and (359,237.79) .. (359,240) .. controls (359,242.21) and (357.21,244) .. (355,244) .. controls (352.79,244) and (351,242.21) .. (351,240) -- cycle ;
	\draw  [fill={rgb, 255:red, 206; green, 43; blue, 16 }  ,fill opacity=1 ] (233,241) .. controls (233,238.79) and (234.79,237) .. (237,237) .. controls (239.21,237) and (241,238.79) .. (241,241) .. controls (241,243.21) and (239.21,245) .. (237,245) .. controls (234.79,245) and (233,243.21) .. (233,241) -- cycle ;
	\draw  [fill={rgb, 255:red, 33; green, 71; blue, 223 }  ,fill opacity=1 ] (419,147) .. controls (419,144.79) and (420.79,143) .. (423,143) .. controls (425.21,143) and (427,144.79) .. (427,147) .. controls (427,149.21) and (425.21,151) .. (423,151) .. controls (420.79,151) and (419,149.21) .. (419,147) -- cycle ;
	\draw  [fill={rgb, 255:red, 33; green, 71; blue, 223 }  ,fill opacity=1 ] (225,225) .. controls (225,222.79) and (226.79,221) .. (229,221) .. controls (231.21,221) and (233,222.79) .. (233,225) .. controls (233,227.21) and (231.21,229) .. (229,229) .. controls (226.79,229) and (225,227.21) .. (225,225) -- cycle ;
	\draw  [fill={rgb, 255:red, 33; green, 71; blue, 223 }  ,fill opacity=1 ] (174,226) .. controls (174,223.79) and (175.79,222) .. (178,222) .. controls (180.21,222) and (182,223.79) .. (182,226) .. controls (182,228.21) and (180.21,230) .. (178,230) .. controls (175.79,230) and (174,228.21) .. (174,226) -- cycle ;
	\draw  [fill={rgb, 255:red, 33; green, 71; blue, 223 }  ,fill opacity=1 ] (149,225) .. controls (149,222.79) and (150.79,221) .. (153,221) .. controls (155.21,221) and (157,222.79) .. (157,225) .. controls (157,227.21) and (155.21,229) .. (153,229) .. controls (150.79,229) and (149,227.21) .. (149,225) -- cycle ;
	\draw  [fill={rgb, 255:red, 33; green, 71; blue, 223 }  ,fill opacity=1 ] (98,222) .. controls (98,219.79) and (99.79,218) .. (102,218) .. controls (104.21,218) and (106,219.79) .. (106,222) .. controls (106,224.21) and (104.21,226) .. (102,226) .. controls (99.79,226) and (98,224.21) .. (98,222) -- cycle ;
	\draw  [fill={rgb, 255:red, 33; green, 71; blue, 223 }  ,fill opacity=1 ] (55,252) .. controls (55,249.79) and (56.79,248) .. (59,248) .. controls (61.21,248) and (63,249.79) .. (63,252) .. controls (63,254.21) and (61.21,256) .. (59,256) .. controls (56.79,256) and (55,254.21) .. (55,252) -- cycle ;
	\draw  [fill={rgb, 255:red, 33; green, 71; blue, 223 }  ,fill opacity=1 ] (215,203) .. controls (215,200.79) and (216.79,199) .. (219,199) .. controls (221.21,199) and (223,200.79) .. (223,203) .. controls (223,205.21) and (221.21,207) .. (219,207) .. controls (216.79,207) and (215,205.21) .. (215,203) -- cycle ;
	\draw  [fill={rgb, 255:red, 33; green, 71; blue, 223 }  ,fill opacity=1 ] (157,183) .. controls (157,180.79) and (158.79,179) .. (161,179) .. controls (163.21,179) and (165,180.79) .. (165,183) .. controls (165,185.21) and (163.21,187) .. (161,187) .. controls (158.79,187) and (157,185.21) .. (157,183) -- cycle ;
	\draw  [fill={rgb, 255:red, 33; green, 71; blue, 223 }  ,fill opacity=1 ] (116,180) .. controls (116,177.79) and (117.79,176) .. (120,176) .. controls (122.21,176) and (124,177.79) .. (124,180) .. controls (124,182.21) and (122.21,184) .. (120,184) .. controls (117.79,184) and (116,182.21) .. (116,180) -- cycle ;
	\draw  [fill={rgb, 255:red, 33; green, 71; blue, 223 }  ,fill opacity=1 ] (76,190) .. controls (76,187.79) and (77.79,186) .. (80,186) .. controls (82.21,186) and (84,187.79) .. (84,190) .. controls (84,192.21) and (82.21,194) .. (80,194) .. controls (77.79,194) and (76,192.21) .. (76,190) -- cycle ;
	\draw  [fill={rgb, 255:red, 33; green, 71; blue, 223 }  ,fill opacity=1 ] (228,149) .. controls (228,146.79) and (229.79,145) .. (232,145) .. controls (234.21,145) and (236,146.79) .. (236,149) .. controls (236,151.21) and (234.21,153) .. (232,153) .. controls (229.79,153) and (228,151.21) .. (228,149) -- cycle ;
	\draw  [fill={rgb, 255:red, 33; green, 71; blue, 223 }  ,fill opacity=1 ] (171,139) .. controls (171,136.79) and (172.79,135) .. (175,135) .. controls (177.21,135) and (179,136.79) .. (179,139) .. controls (179,141.21) and (177.21,143) .. (175,143) .. controls (172.79,143) and (171,141.21) .. (171,139) -- cycle ;
	\draw  [fill={rgb, 255:red, 33; green, 71; blue, 223 }  ,fill opacity=1 ] (144,162) .. controls (144,159.79) and (145.79,158) .. (148,158) .. controls (150.21,158) and (152,159.79) .. (152,162) .. controls (152,164.21) and (150.21,166) .. (148,166) .. controls (145.79,166) and (144,164.21) .. (144,162) -- cycle ;
	\draw  [fill={rgb, 255:red, 33; green, 71; blue, 223 }  ,fill opacity=1 ] (90,138) .. controls (90,135.79) and (91.79,134) .. (94,134) .. controls (96.21,134) and (98,135.79) .. (98,138) .. controls (98,140.21) and (96.21,142) .. (94,142) .. controls (91.79,142) and (90,140.21) .. (90,138) -- cycle ;
	\draw  [fill={rgb, 255:red, 33; green, 71; blue, 223 }  ,fill opacity=1 ] (42,141) .. controls (42,138.79) and (43.79,137) .. (46,137) .. controls (48.21,137) and (50,138.79) .. (50,141) .. controls (50,143.21) and (48.21,145) .. (46,145) .. controls (43.79,145) and (42,143.21) .. (42,141) -- cycle ;
	\draw  [fill={rgb, 255:red, 86; green, 197; blue, 23 }  ,fill opacity=1 ] (485,143) .. controls (485,140.79) and (486.79,139) .. (489,139) .. controls (491.21,139) and (493,140.79) .. (493,143) .. controls (493,145.21) and (491.21,147) .. (489,147) .. controls (486.79,147) and (485,145.21) .. (485,143) -- cycle ;
	\draw  [fill={rgb, 255:red, 86; green, 197; blue, 23 }  ,fill opacity=1 ] (548,200) .. controls (548,197.79) and (549.79,196) .. (552,196) .. controls (554.21,196) and (556,197.79) .. (556,200) .. controls (556,202.21) and (554.21,204) .. (552,204) .. controls (549.79,204) and (548,202.21) .. (548,200) -- cycle ;
	\draw  [fill={rgb, 255:red, 86; green, 197; blue, 23 }  ,fill opacity=1 ] (539,147) .. controls (539,144.79) and (540.79,143) .. (543,143) .. controls (545.21,143) and (547,144.79) .. (547,147) .. controls (547,149.21) and (545.21,151) .. (543,151) .. controls (540.79,151) and (539,149.21) .. (539,147) -- cycle ;
	\draw  [fill={rgb, 255:red, 86; green, 197; blue, 23 }  ,fill opacity=1 ] (354,189) .. controls (354,186.79) and (355.79,185) .. (358,185) .. controls (360.21,185) and (362,186.79) .. (362,189) .. controls (362,191.21) and (360.21,193) .. (358,193) .. controls (355.79,193) and (354,191.21) .. (354,189) -- cycle ;
	\draw  [fill={rgb, 255:red, 86; green, 197; blue, 23 }  ,fill opacity=1 ] (393,146) .. controls (393,143.79) and (394.79,142) .. (397,142) .. controls (399.21,142) and (401,143.79) .. (401,146) .. controls (401,148.21) and (399.21,150) .. (397,150) .. controls (394.79,150) and (393,148.21) .. (393,146) -- cycle ;
	\draw  [fill={rgb, 255:red, 86; green, 197; blue, 23 }  ,fill opacity=1 ] (525,246) .. controls (525,243.79) and (526.79,242) .. (529,242) .. controls (531.21,242) and (533,243.79) .. (533,246) .. controls (533,248.21) and (531.21,250) .. (529,250) .. controls (526.79,250) and (525,248.21) .. (525,246) -- cycle ;
	\draw  [fill={rgb, 255:red, 86; green, 197; blue, 23 }  ,fill opacity=1 ] (501,189) .. controls (501,186.79) and (502.79,185) .. (505,185) .. controls (507.21,185) and (509,186.79) .. (509,189) .. controls (509,191.21) and (507.21,193) .. (505,193) .. controls (502.79,193) and (501,191.21) .. (501,189) -- cycle ;
	\draw  [fill={rgb, 255:red, 86; green, 197; blue, 23 }  ,fill opacity=1 ] (403,229) .. controls (403,226.79) and (404.79,225) .. (407,225) .. controls (409.21,225) and (411,226.79) .. (411,229) .. controls (411,231.21) and (409.21,233) .. (407,233) .. controls (404.79,233) and (403,231.21) .. (403,229) -- cycle ;
	\draw  [fill={rgb, 255:red, 86; green, 197; blue, 23 }  ,fill opacity=1 ] (433,164) .. controls (433,161.79) and (434.79,160) .. (437,160) .. controls (439.21,160) and (441,161.79) .. (441,164) .. controls (441,166.21) and (439.21,168) .. (437,168) .. controls (434.79,168) and (433,166.21) .. (433,164) -- cycle ;
	\draw  [fill={rgb, 255:red, 86; green, 197; blue, 23 }  ,fill opacity=1 ] (214,240) .. controls (214,237.79) and (215.79,236) .. (218,236) .. controls (220.21,236) and (222,237.79) .. (222,240) .. controls (222,242.21) and (220.21,244) .. (218,244) .. controls (215.79,244) and (214,242.21) .. (214,240) -- cycle ;
	\draw  [fill={rgb, 255:red, 86; green, 197; blue, 23 }  ,fill opacity=1 ] (185,235) .. controls (185,232.79) and (186.79,231) .. (189,231) .. controls (191.21,231) and (193,232.79) .. (193,235) .. controls (193,237.21) and (191.21,239) .. (189,239) .. controls (186.79,239) and (185,237.21) .. (185,235) -- cycle ;
	\draw  [fill={rgb, 255:red, 86; green, 197; blue, 23 }  ,fill opacity=1 ] (143,236) .. controls (143,233.79) and (144.79,232) .. (147,232) .. controls (149.21,232) and (151,233.79) .. (151,236) .. controls (151,238.21) and (149.21,240) .. (147,240) .. controls (144.79,240) and (143,238.21) .. (143,236) -- cycle ;
	\draw  [fill={rgb, 255:red, 86; green, 197; blue, 23 }  ,fill opacity=1 ] (94,236) .. controls (94,233.79) and (95.79,232) .. (98,232) .. controls (100.21,232) and (102,233.79) .. (102,236) .. controls (102,238.21) and (100.21,240) .. (98,240) .. controls (95.79,240) and (94,238.21) .. (94,236) -- cycle ;
	\draw  [fill={rgb, 255:red, 86; green, 197; blue, 23 }  ,fill opacity=1 ] (41,228) .. controls (41,225.79) and (42.79,224) .. (45,224) .. controls (47.21,224) and (49,225.79) .. (49,228) .. controls (49,230.21) and (47.21,232) .. (45,232) .. controls (42.79,232) and (41,230.21) .. (41,228) -- cycle ;
	\draw  [fill={rgb, 255:red, 86; green, 197; blue, 23 }  ,fill opacity=1 ] (195,181) .. controls (195,178.79) and (196.79,177) .. (199,177) .. controls (201.21,177) and (203,178.79) .. (203,181) .. controls (203,183.21) and (201.21,185) .. (199,185) .. controls (196.79,185) and (195,183.21) .. (195,181) -- cycle ;
	\draw  [fill={rgb, 255:red, 86; green, 197; blue, 23 }  ,fill opacity=1 ] (154,199) .. controls (154,196.79) and (155.79,195) .. (158,195) .. controls (160.21,195) and (162,196.79) .. (162,199) .. controls (162,201.21) and (160.21,203) .. (158,203) .. controls (155.79,203) and (154,201.21) .. (154,199) -- cycle ;
	\draw  [fill={rgb, 255:red, 86; green, 197; blue, 23 }  ,fill opacity=1 ] (125,190) .. controls (125,187.79) and (126.79,186) .. (129,186) .. controls (131.21,186) and (133,187.79) .. (133,190) .. controls (133,192.21) and (131.21,194) .. (129,194) .. controls (126.79,194) and (125,192.21) .. (125,190) -- cycle ;
	\draw  [fill={rgb, 255:red, 86; green, 197; blue, 23 }  ,fill opacity=1 ] (62,196) .. controls (62,193.79) and (63.79,192) .. (66,192) .. controls (68.21,192) and (70,193.79) .. (70,196) .. controls (70,198.21) and (68.21,200) .. (66,200) .. controls (63.79,200) and (62,198.21) .. (62,196) -- cycle ;
	\draw  [fill={rgb, 255:red, 86; green, 197; blue, 23 }  ,fill opacity=1 ] (210,141) .. controls (210,138.79) and (211.79,137) .. (214,137) .. controls (216.21,137) and (218,138.79) .. (218,141) .. controls (218,143.21) and (216.21,145) .. (214,145) .. controls (211.79,145) and (210,143.21) .. (210,141) -- cycle ;
	\draw  [fill={rgb, 255:red, 86; green, 197; blue, 23 }  ,fill opacity=1 ] (174,151) .. controls (174,148.79) and (175.79,147) .. (178,147) .. controls (180.21,147) and (182,148.79) .. (182,151) .. controls (182,153.21) and (180.21,155) .. (178,155) .. controls (175.79,155) and (174,153.21) .. (174,151) -- cycle ;
	\draw  [fill={rgb, 255:red, 86; green, 197; blue, 23 }  ,fill opacity=1 ] (128,145) .. controls (128,142.79) and (129.79,141) .. (132,141) .. controls (134.21,141) and (136,142.79) .. (136,145) .. controls (136,147.21) and (134.21,149) .. (132,149) .. controls (129.79,149) and (128,147.21) .. (128,145) -- cycle ;
	\draw  [fill={rgb, 255:red, 86; green, 197; blue, 23 }  ,fill opacity=1 ] (89,154) .. controls (89,151.79) and (90.79,150) .. (93,150) .. controls (95.21,150) and (97,151.79) .. (97,154) .. controls (97,156.21) and (95.21,158) .. (93,158) .. controls (90.79,158) and (89,156.21) .. (89,154) -- cycle ;
	\draw  [fill={rgb, 255:red, 86; green, 197; blue, 23 }  ,fill opacity=1 ] (60,150) .. controls (60,147.79) and (61.79,146) .. (64,146) .. controls (66.21,146) and (68,147.79) .. (68,150) .. controls (68,152.21) and (66.21,154) .. (64,154) .. controls (61.79,154) and (60,152.21) .. (60,150) -- cycle ;
	\draw  [fill={rgb, 255:red, 206; green, 43; blue, 16 }  ,fill opacity=1 ] (180,246) .. controls (180,243.79) and (181.79,242) .. (184,242) .. controls (186.21,242) and (188,243.79) .. (188,246) .. controls (188,248.21) and (186.21,250) .. (184,250) .. controls (181.79,250) and (180,248.21) .. (180,246) -- cycle ;
	\draw  [fill={rgb, 255:red, 206; green, 43; blue, 16 }  ,fill opacity=1 ] (132,226) .. controls (132,223.79) and (133.79,222) .. (136,222) .. controls (138.21,222) and (140,223.79) .. (140,226) .. controls (140,228.21) and (138.21,230) .. (136,230) .. controls (133.79,230) and (132,228.21) .. (132,226) -- cycle ;
	\draw  [fill={rgb, 255:red, 206; green, 43; blue, 16 }  ,fill opacity=1 ] (109,245) .. controls (109,242.79) and (110.79,241) .. (113,241) .. controls (115.21,241) and (117,242.79) .. (117,245) .. controls (117,247.21) and (115.21,249) .. (113,249) .. controls (110.79,249) and (109,247.21) .. (109,245) -- cycle ;
	\draw  [fill={rgb, 255:red, 206; green, 43; blue, 16 }  ,fill opacity=1 ] (66,227) .. controls (66,224.79) and (67.79,223) .. (70,223) .. controls (72.21,223) and (74,224.79) .. (74,227) .. controls (74,229.21) and (72.21,231) .. (70,231) .. controls (67.79,231) and (66,229.21) .. (66,227) -- cycle ;
	\draw  [fill={rgb, 255:red, 206; green, 43; blue, 16 }  ,fill opacity=1 ] (81,205) .. controls (81,202.79) and (82.79,201) .. (85,201) .. controls (87.21,201) and (89,202.79) .. (89,205) .. controls (89,207.21) and (87.21,209) .. (85,209) .. controls (82.79,209) and (81,207.21) .. (81,205) -- cycle ;
	\draw  [fill={rgb, 255:red, 206; green, 43; blue, 16 }  ,fill opacity=1 ] (114,199) .. controls (114,196.79) and (115.79,195) .. (118,195) .. controls (120.21,195) and (122,196.79) .. (122,199) .. controls (122,201.21) and (120.21,203) .. (118,203) .. controls (115.79,203) and (114,201.21) .. (114,199) -- cycle ;
	\draw  [fill={rgb, 255:red, 206; green, 43; blue, 16 }  ,fill opacity=1 ] (171,196) .. controls (171,193.79) and (172.79,192) .. (175,192) .. controls (177.21,192) and (179,193.79) .. (179,196) .. controls (179,198.21) and (177.21,200) .. (175,200) .. controls (172.79,200) and (171,198.21) .. (171,196) -- cycle ;
	\draw  [fill={rgb, 255:red, 206; green, 43; blue, 16 }  ,fill opacity=1 ] (201,195) .. controls (201,192.79) and (202.79,191) .. (205,191) .. controls (207.21,191) and (209,192.79) .. (209,195) .. controls (209,197.21) and (207.21,199) .. (205,199) .. controls (202.79,199) and (201,197.21) .. (201,195) -- cycle ;
	\draw  [fill={rgb, 255:red, 206; green, 43; blue, 16 }  ,fill opacity=1 ] (216,157) .. controls (216,154.79) and (217.79,153) .. (220,153) .. controls (222.21,153) and (224,154.79) .. (224,157) .. controls (224,159.21) and (222.21,161) .. (220,161) .. controls (217.79,161) and (216,159.21) .. (216,157) -- cycle ;
	\draw  [fill={rgb, 255:red, 206; green, 43; blue, 16 }  ,fill opacity=1 ] (180,164) .. controls (180,161.79) and (181.79,160) .. (184,160) .. controls (186.21,160) and (188,161.79) .. (188,164) .. controls (188,166.21) and (186.21,168) .. (184,168) .. controls (181.79,168) and (180,166.21) .. (180,164) -- cycle ;
	\draw  [fill={rgb, 255:red, 206; green, 43; blue, 16 }  ,fill opacity=1 ] (144,141) .. controls (144,138.79) and (145.79,137) .. (148,137) .. controls (150.21,137) and (152,138.79) .. (152,141) .. controls (152,143.21) and (150.21,145) .. (148,145) .. controls (145.79,145) and (144,143.21) .. (144,141) -- cycle ;
	\draw  [fill={rgb, 255:red, 206; green, 43; blue, 16 }  ,fill opacity=1 ] (101,160) .. controls (101,157.79) and (102.79,156) .. (105,156) .. controls (107.21,156) and (109,157.79) .. (109,160) .. controls (109,162.21) and (107.21,164) .. (105,164) .. controls (102.79,164) and (101,162.21) .. (101,160) -- cycle ;
	\draw  [fill={rgb, 255:red, 206; green, 43; blue, 16 }  ,fill opacity=1 ] (49,165) .. controls (49,162.79) and (50.79,161) .. (53,161) .. controls (55.21,161) and (57,162.79) .. (57,165) .. controls (57,167.21) and (55.21,169) .. (53,169) .. controls (50.79,169) and (49,167.21) .. (49,165) -- cycle ;
	\draw  [fill={rgb, 255:red, 206; green, 43; blue, 16 }  ,fill opacity=1 ] (491,248) .. controls (491,245.79) and (492.79,244) .. (495,244) .. controls (497.21,244) and (499,245.79) .. (499,248) .. controls (499,250.21) and (497.21,252) .. (495,252) .. controls (492.79,252) and (491,250.21) .. (491,248) -- cycle ;
	\draw  [fill={rgb, 255:red, 206; green, 43; blue, 16 }  ,fill opacity=1 ] (378,213) .. controls (378,210.79) and (379.79,209) .. (382,209) .. controls (384.21,209) and (386,210.79) .. (386,213) .. controls (386,215.21) and (384.21,217) .. (382,217) .. controls (379.79,217) and (378,215.21) .. (378,213) -- cycle ;
	\draw  [fill={rgb, 255:red, 206; green, 43; blue, 16 }  ,fill opacity=1 ] (377,245) .. controls (377,242.79) and (378.79,241) .. (381,241) .. controls (383.21,241) and (385,242.79) .. (385,245) .. controls (385,247.21) and (383.21,249) .. (381,249) .. controls (378.79,249) and (377,247.21) .. (377,245) -- cycle ;
	\draw  [fill={rgb, 255:red, 206; green, 43; blue, 16 }  ,fill opacity=1 ] (474,168) .. controls (474,165.79) and (475.79,164) .. (478,164) .. controls (480.21,164) and (482,165.79) .. (482,168) .. controls (482,170.21) and (480.21,172) .. (478,172) .. controls (475.79,172) and (474,170.21) .. (474,168) -- cycle ;
	\draw  [fill={rgb, 255:red, 206; green, 43; blue, 16 }  ,fill opacity=1 ] (512,166) .. controls (512,163.79) and (513.79,162) .. (516,162) .. controls (518.21,162) and (520,163.79) .. (520,166) .. controls (520,168.21) and (518.21,170) .. (516,170) .. controls (513.79,170) and (512,168.21) .. (512,166) -- cycle ;
	\draw  [fill={rgb, 255:red, 206; green, 43; blue, 16 }  ,fill opacity=1 ] (550,232) .. controls (550,229.79) and (551.79,228) .. (554,228) .. controls (556.21,228) and (558,229.79) .. (558,232) .. controls (558,234.21) and (556.21,236) .. (554,236) .. controls (551.79,236) and (550,234.21) .. (550,232) -- cycle ;
	\draw  [fill={rgb, 255:red, 206; green, 43; blue, 16 }  ,fill opacity=1 ] (547,171) .. controls (547,168.79) and (548.79,167) .. (551,167) .. controls (553.21,167) and (555,168.79) .. (555,171) .. controls (555,173.21) and (553.21,175) .. (551,175) .. controls (548.79,175) and (547,173.21) .. (547,171) -- cycle ;
	\draw  [fill={rgb, 255:red, 33; green, 71; blue, 223 }  ,fill opacity=1 ] (348,167) .. controls (348,164.79) and (349.79,163) .. (352,163) .. controls (354.21,163) and (356,164.79) .. (356,167) .. controls (356,169.21) and (354.21,171) .. (352,171) .. controls (349.79,171) and (348,169.21) .. (348,167) -- cycle ;
	\draw  [fill={rgb, 255:red, 33; green, 71; blue, 223 }  ,fill opacity=1 ] (461,145) .. controls (461,142.79) and (462.79,141) .. (465,141) .. controls (467.21,141) and (469,142.79) .. (469,145) .. controls (469,147.21) and (467.21,149) .. (465,149) .. controls (462.79,149) and (461,147.21) .. (461,145) -- cycle ;
	\draw  [fill={rgb, 255:red, 33; green, 71; blue, 223 }  ,fill opacity=1 ] (529,217) .. controls (529,214.79) and (530.79,213) .. (533,213) .. controls (535.21,213) and (537,214.79) .. (537,217) .. controls (537,219.21) and (535.21,221) .. (533,221) .. controls (530.79,221) and (529,219.21) .. (529,217) -- cycle ;
	\draw  [fill={rgb, 255:red, 33; green, 71; blue, 223 }  ,fill opacity=1 ] (522,188) .. controls (522,185.79) and (523.79,184) .. (526,184) .. controls (528.21,184) and (530,185.79) .. (530,188) .. controls (530,190.21) and (528.21,192) .. (526,192) .. controls (523.79,192) and (522,190.21) .. (522,188) -- cycle ;
	\draw  [fill={rgb, 255:red, 33; green, 71; blue, 223 }  ,fill opacity=1 ] (517,147) .. controls (517,144.79) and (518.79,143) .. (521,143) .. controls (523.21,143) and (525,144.79) .. (525,147) .. controls (525,149.21) and (523.21,151) .. (521,151) .. controls (518.79,151) and (517,149.21) .. (517,147) -- cycle ;
	\draw  [fill={rgb, 255:red, 33; green, 71; blue, 223 }  ,fill opacity=1 ] (372,161) .. controls (372,158.79) and (373.79,157) .. (376,157) .. controls (378.21,157) and (380,158.79) .. (380,161) .. controls (380,163.21) and (378.21,165) .. (376,165) .. controls (373.79,165) and (372,163.21) .. (372,161) -- cycle ;
	\draw  [fill={rgb, 255:red, 33; green, 71; blue, 223 }  ,fill opacity=1 ] (392,182) .. controls (392,179.79) and (393.79,178) .. (396,178) .. controls (398.21,178) and (400,179.79) .. (400,182) .. controls (400,184.21) and (398.21,186) .. (396,186) .. controls (393.79,186) and (392,184.21) .. (392,182) -- cycle ;
	\draw  [fill={rgb, 255:red, 33; green, 71; blue, 223 }  ,fill opacity=1 ] (434,212) .. controls (434,209.79) and (435.79,208) .. (438,208) .. controls (440.21,208) and (442,209.79) .. (442,212) .. controls (442,214.21) and (440.21,216) .. (438,216) .. controls (435.79,216) and (434,214.21) .. (434,212) -- cycle ;
	\draw  [fill={rgb, 255:red, 33; green, 71; blue, 223 }  ,fill opacity=1 ] (442,248) .. controls (442,245.79) and (443.79,244) .. (446,244) .. controls (448.21,244) and (450,245.79) .. (450,248) .. controls (450,250.21) and (448.21,252) .. (446,252) .. controls (443.79,252) and (442,250.21) .. (442,248) -- cycle ;
	\draw  [fill={rgb, 255:red, 33; green, 71; blue, 223 }  ,fill opacity=1 ] (433,190) .. controls (433,187.79) and (434.79,186) .. (437,186) .. controls (439.21,186) and (441,187.79) .. (441,190) .. controls (441,192.21) and (439.21,194) .. (437,194) .. controls (434.79,194) and (433,192.21) .. (433,190) -- cycle ;
	\draw  [fill={rgb, 255:red, 33; green, 71; blue, 223 }  ,fill opacity=1 ] (484,217) .. controls (484,214.79) and (485.79,213) .. (488,213) .. controls (490.21,213) and (492,214.79) .. (492,217) .. controls (492,219.21) and (490.21,221) .. (488,221) .. controls (485.79,221) and (484,219.21) .. (484,217) -- cycle ;
	\draw  [fill={rgb, 255:red, 206; green, 43; blue, 16 }  ,fill opacity=1 ] (455,230) .. controls (455,227.79) and (456.79,226) .. (459,226) .. controls (461.21,226) and (463,227.79) .. (463,230) .. controls (463,232.21) and (461.21,234) .. (459,234) .. controls (456.79,234) and (455,232.21) .. (455,230) -- cycle ;
	\draw  [fill={rgb, 255:red, 206; green, 43; blue, 16 }  ,fill opacity=1 ] (455,188) .. controls (455,185.79) and (456.79,184) .. (459,184) .. controls (461.21,184) and (463,185.79) .. (463,188) .. controls (463,190.21) and (461.21,192) .. (459,192) .. controls (456.79,192) and (455,190.21) .. (455,188) -- cycle ;
	\draw  [fill={rgb, 255:red, 206; green, 43; blue, 16 }  ,fill opacity=1 ] (508,222) .. controls (508,219.79) and (509.79,218) .. (512,218) .. controls (514.21,218) and (516,219.79) .. (516,222) .. controls (516,224.21) and (514.21,226) .. (512,226) .. controls (509.79,226) and (508,224.21) .. (508,222) -- cycle ;
	\draw  [fill={rgb, 255:red, 206; green, 43; blue, 16 }  ,fill opacity=1 ] (437,145) .. controls (437,142.79) and (438.79,141) .. (441,141) .. controls (443.21,141) and (445,142.79) .. (445,145) .. controls (445,147.21) and (443.21,149) .. (441,149) .. controls (438.79,149) and (437,147.21) .. (437,145) -- cycle ;
	\draw  [fill={rgb, 255:red, 206; green, 43; blue, 16 }  ,fill opacity=1 ] (346,215) .. controls (346,212.79) and (347.79,211) .. (350,211) .. controls (352.21,211) and (354,212.79) .. (354,215) .. controls (354,217.21) and (352.21,219) .. (350,219) .. controls (347.79,219) and (346,217.21) .. (346,215) -- cycle ;
	\draw  [fill={rgb, 255:red, 206; green, 43; blue, 16 }  ,fill opacity=1 ] (403,206) .. controls (403,203.79) and (404.79,202) .. (407,202) .. controls (409.21,202) and (411,203.79) .. (411,206) .. controls (411,208.21) and (409.21,210) .. (407,210) .. controls (404.79,210) and (403,208.21) .. (403,206) -- cycle ;
	\end{tikzpicture}	
	\caption{Representation of quarks in a bag. At high densities, the quark bags overlap (left panel). As the density increases, the quarks exist in a deconfined phase (right panel).}
	\label{fig1.9}
\end{figure}
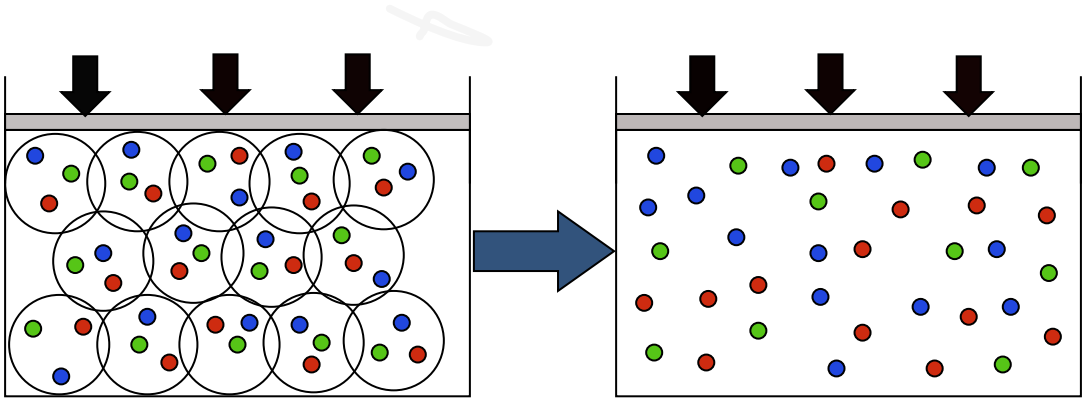

Quarks in the bag model inhabit single-particle orbitals. If all of the quarks are in the ground state, the bag has a spherical form (Fig.~\ref{fig1.9}). A suitable boundary condition at the bag surface assures that no quark may escape. This indicates that no quarks exist outside of the bag. The “vacuum” of QCD is supposed to have the property of excluding quarks.
The vacuum must be ejected inside the hadronic region, or more broadly within any volume containing quarks. This requires a lot of energy. The bag constant, indicated by $B$, is the energy per unit volume. The energy thus associated with the simple existence of quarks in volume $V$ is denoted by the symbol $BV$. Because of the energy associated with the kinetic motion of quarks in the volume, the QM energy density, and pressure will have two components: the contribution of the confining bag and the kinetic motion of the quarks \cite{gled}.

The total energy of the bag system is thus
\begin{equation}
E=E_q +BV,
\end{equation}
where $E_q$ is the energy associated with the kinetic motion of quarks and $BV$ is the energy due to the mere presence of quarks.

\begin{figure}[hbt!]
	\centering
	\includegraphics[width=0.75\textwidth]{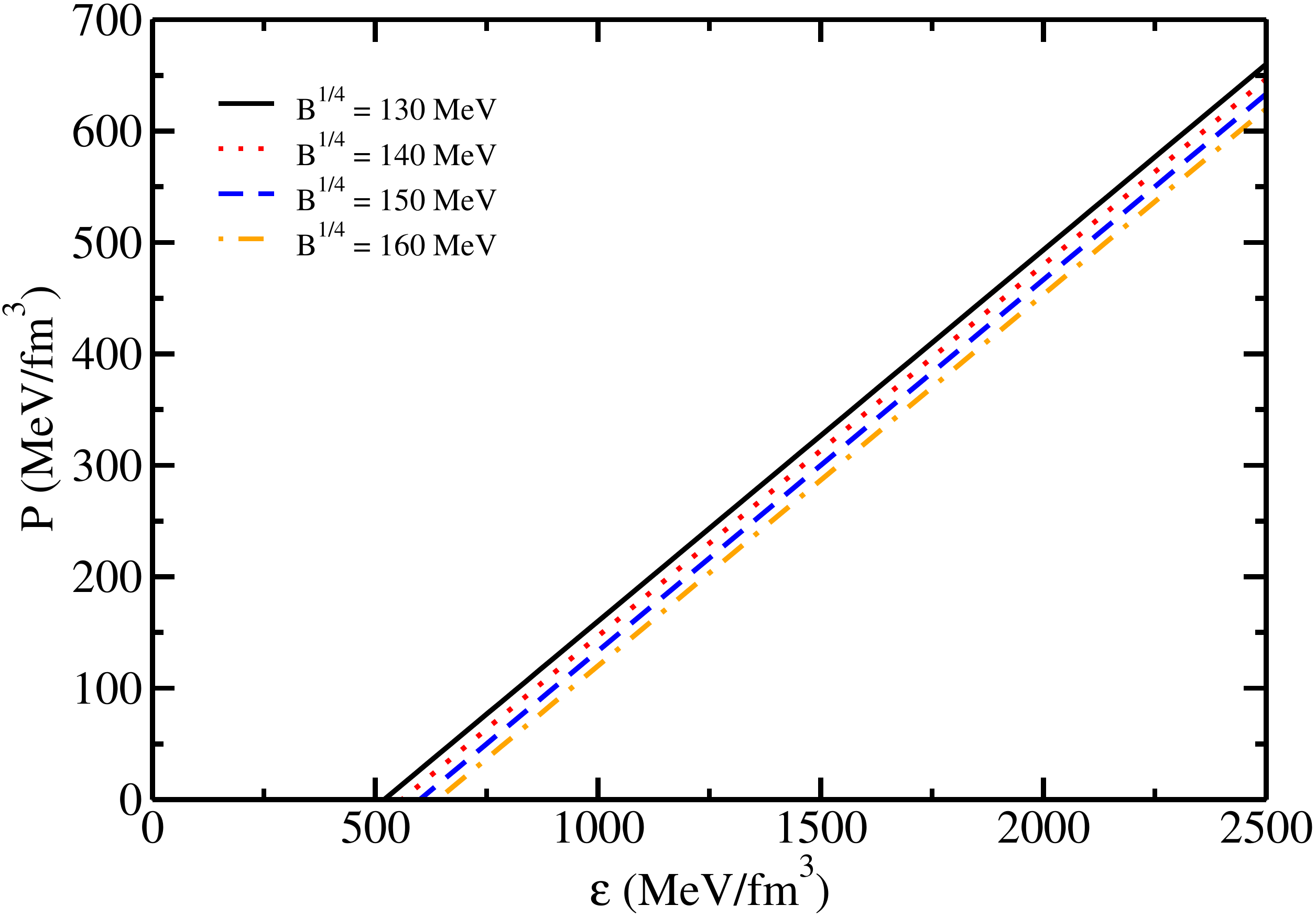}
	\caption{Energy density vs pressure for simple bag model at different values of bag constant.}
	\label{fig1.10}
\end{figure}

In the simplest of bag model, only light quarks ($u$ and $d$) are considered with vanishing mass. The baryon density is then given as
\begin{equation}
\rho_B = \frac{\gamma_{q}\mu^3}{18\pi^2},
\end{equation}
where $\mu$ is the baryonic chemical potential and $\gamma_{q}$ is the degeneracy factor =24 (2 flavors $\times$ 3 colors $\times$ 2 spin $\times$ 2 anti-quarks).  The energy density can be written as 
\begin{equation}
\mathcal{E}=\frac{\gamma_{q} \mu^4}{8\pi^2} +B.
\end{equation}
The corresponding pressure is
\begin{equation}
P=\frac{\gamma_{q} \mu^4}{24\pi^2} -B.
\end{equation}
From the above equations, a linear relationship between energy density and pressure is obtained.
\begin{equation}
\mathcal{E}=3P+4B.
\end{equation}
Fig.~\ref{fig1.10} displays the variation of pressure with energy density for simple MIT bag model at different values of bag constant $B^{1/4}$ = 130, 140, 150, and 160 MeV. As seen, the pressure decreases with increase in the value of $B$. Refs. \cite{Rather_2020,Rather2020_1,BAYM1976241,Klahn:2015mfa,PhysRevD.30.2379,doi:10.1142/S0218271819410062} discuss the MIT bag model, its extensions, and bag constant constraints. 

%
\subsection{Color-Flavored Locked Phase}

Quark matter is generally color-superconducting at very high densities and low temperatures. Quarks create a degenerate Fermi liquid at sufficiently high concentrations and low temperatures. At these densities, QM is weakly interactive, with one gluon exchange providing the main interaction between quarks, resulting in quarks forming condensates of cooper pairs of equal and opposite momenta with distinct flavors around the Fermi surface. Because the cores of NSs are not dense enough to contain any charm or heavier quarks, we always use the high-density limit with up, down, and strange quarks when considering the matter at high densities \cite{RevModPhys.80.1455}.

The thermodynamic potential or free energy at zero temperature is $\Omega$ = $E-\mu N$, where $E$ represents the overall energy of the system, $\mu$ represents the chemical potential and $N$ represents the number of fermions. If no interactions occurred, then adding a particle to the system would require the energy equal to the Fermi energy $E_F = \mu$, meaning that adding or removing particles or holes at the Fermi surface would cost no free energy. When we add a pair of particles or holes with the attractive channel's quantum numbers to a weak attractive interaction in any channel, the potential energy of their attraction reduces the free energy. As a result, the creation of cooper pairs on the Fermi surface is encouraged.

As a result of the attractive color-antisymmetric channel in the interaction between quarks, a pairing gap is formed in the system's free energy. As a result of colour superconductivity in dense quark matter, one of the primary consequences is the emergence of a non-zero energy gap $\Delta$ in the one-particle spectrum. In a standard BCS method, this is stated as 
\begin{equation}
Ek = \sqrt{(\epsilon_k -\mu)^2 +\Delta^2},
\end{equation}
where $E_k$ is the modified (quasi-particle) energy of the $kth$ level and $\epsilon_k$ is the equivalent single-particle energy. Because the quark pairs cannot be color singlets, the Cooper pair condensate in quark matter will violate the local color symmetry $SU(3)_c$, thereby giving rise to the phrase \textit{color superconductivity}.

The Color-Flavor Locked (CFL) phase, in which quarks of all three colors and flavors form conventional zero momentum spinless Cooper pairs, is the most symmetric and appealing option in quark matter at sufficiently high densities, where the up, down, and strange quarks can be treated on an equal footing and the disruptive effects of the strange quark mass can be ignored. As a result, the CFL phase with an equal amount of \textit{u}, \textit{d}, and \textit{s} quarks is most likely the ground state of QCD. The symmetry of state enforces the same number of flavors and therefore electrons are missing since the combination is inherently neutral \cite{PhysRevD.64.074017}.

The thermodynamic potential for CFL phase is given as \cite{PhysRevD.67.074024}
\begin{equation}
\Omega_{CFL}= \Omega_{f}-\frac{3}{\pi^2} \Delta^2 \mu^2 +B.
\end{equation}
The pressure and energy density for the CFL phase arise from thermodynamic potential and are given as
\begin{equation}
P=\sum_{i=1}^3 \frac{1}{4\pi^2}\Big[\mu_i \nu\Big(\mu_i^2-\frac{5}{2}m_i^2\Big)+\frac{3}{2}m_i^2 \log\Big(\frac{\mu_i +\nu}{m_i}\Big)\Big]+\frac{3}{\pi^2} \Delta^2 \mu^2 -B,
\end{equation}
\begin{equation}
\mathcal{E}=\sum_{i=1}^3 \mu_i \rho_i -P =3\mu \rho_B -P,
\end{equation}
where, $3\mu=\mu_u +\mu_d+\mu_s$. $\rho_{B}$ is the baryon density written as
\begin{equation}
\rho_B= \rho_u=\rho_d=\rho_s = \frac{\nu^3+2\Delta^2\mu}{\pi^2},
\end{equation}
and  $\nu$ is the common fermi momentum defined as
\begin{equation}
\nu=2\mu-\sqrt{\mu^2+\frac{m_s^2}{3}}.
\end{equation}

\begin{figure}
	\centering
	\includegraphics[width=0.80\textwidth]{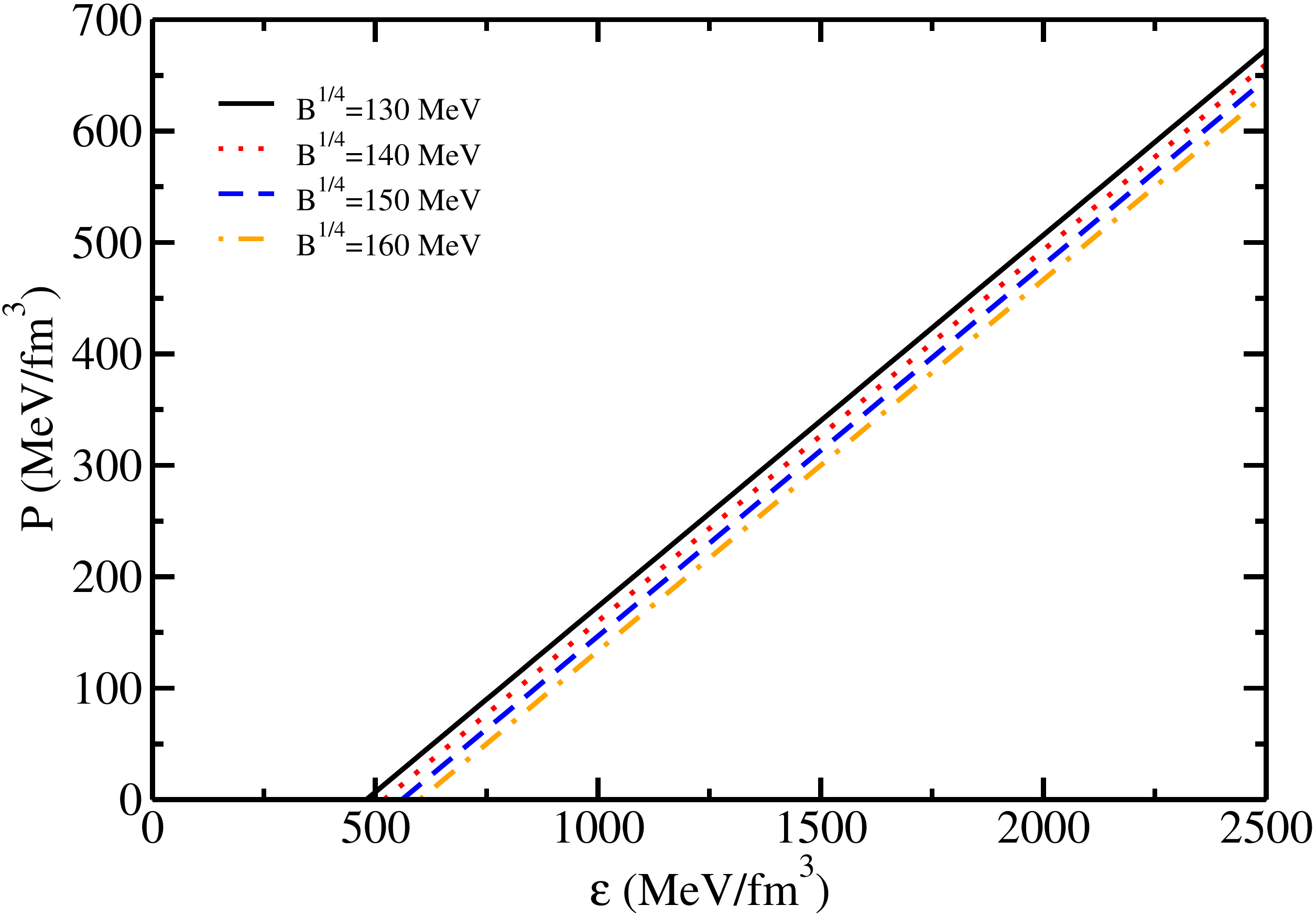}
	\caption{Energy density vs pressure for CFL quark matter at $\Delta$ = 100 MeV and bag values of $B^{1/4}$ = 130, 140, 150, and 160 MeV.}
	\label{fig1.11}
\end{figure}

Fig.~\ref{fig1.11} shows the variation of pressure with energy density for CFL quark phase at $\Delta$ = 100 MeV and bag values of $B^{1/4}$ = 130, 140, 150, and 160 MeV. Comparing this with the simple bag model (Fig.~\ref{fig1.10}), we see that the CFL phase yields more pressure at a given value of bag constant than the simple bag model. This shows that the CFL phase is most confined state with the lowest energy density (at a given density). This finding also explains why the presence of CFL phases is more common in high-density matter.

Other approaches, such as the Nambu-Jona-Lasinio model and its extensions, can also be used to study the quark matter EoS \cite{PhysRev.122.345,PhysRev.124.246}, although the parameters involved in most quark matter models are poorly constrained. However, the addition of vector-isoscalar and vector-isovector interactions stiffen the EoS and hence produce a neutron star with maximum mass $\approx$ 2$M_{\odot}$, satisfying constraints from various measurements. The confining quark matter (CQM) model with extension to isospin-dependency (ICQM), can describe pure quark stars with mass around 2$M_{\odot}$ \cite{PhysRevD.96.083019}.

\begin{savequote}[8cm]

\textlatin{All intelligent thoughts have already been thought;\\
what is necessary is only to try to think them again.}
  \qauthor{--- \textit{Johann Wolfgang von Goethe}}
\end{savequote}

\chapter{\label{ch:2-litreview}Relativistic mean field models}

\minitoc

\section{Quantum Hydrodynamics}
\label{qhd}
A fine description of nuclei and nuclear matter was introduced by J. D. Walecka in 1974. This description based on the interaction between baryons and mesons is referred to as the Quantum Hydrodynamics (QHD). In nuclear matter, nucleons interact through the exchange of mesons and hence the relativistic effects are incorporated naturally.

With nuclei and nuclear matter being complex systems, various models exist among which the QHD is one. For all the models, some experimental inputs are necessary to constraint them and in the case of QHD, these are the coupling constants between nucleons and different mesons. These coupling constants are determined by fitting the calculated properties of nuclei with the experimentally observed values. By fitting various observed parameters, different QHD parameter sets have been developed which differ from each other in terms of the meson fields considered and different couplings between the fields.

\begin{figure}
	\centering
	\includegraphics[width=0.75\textwidth]{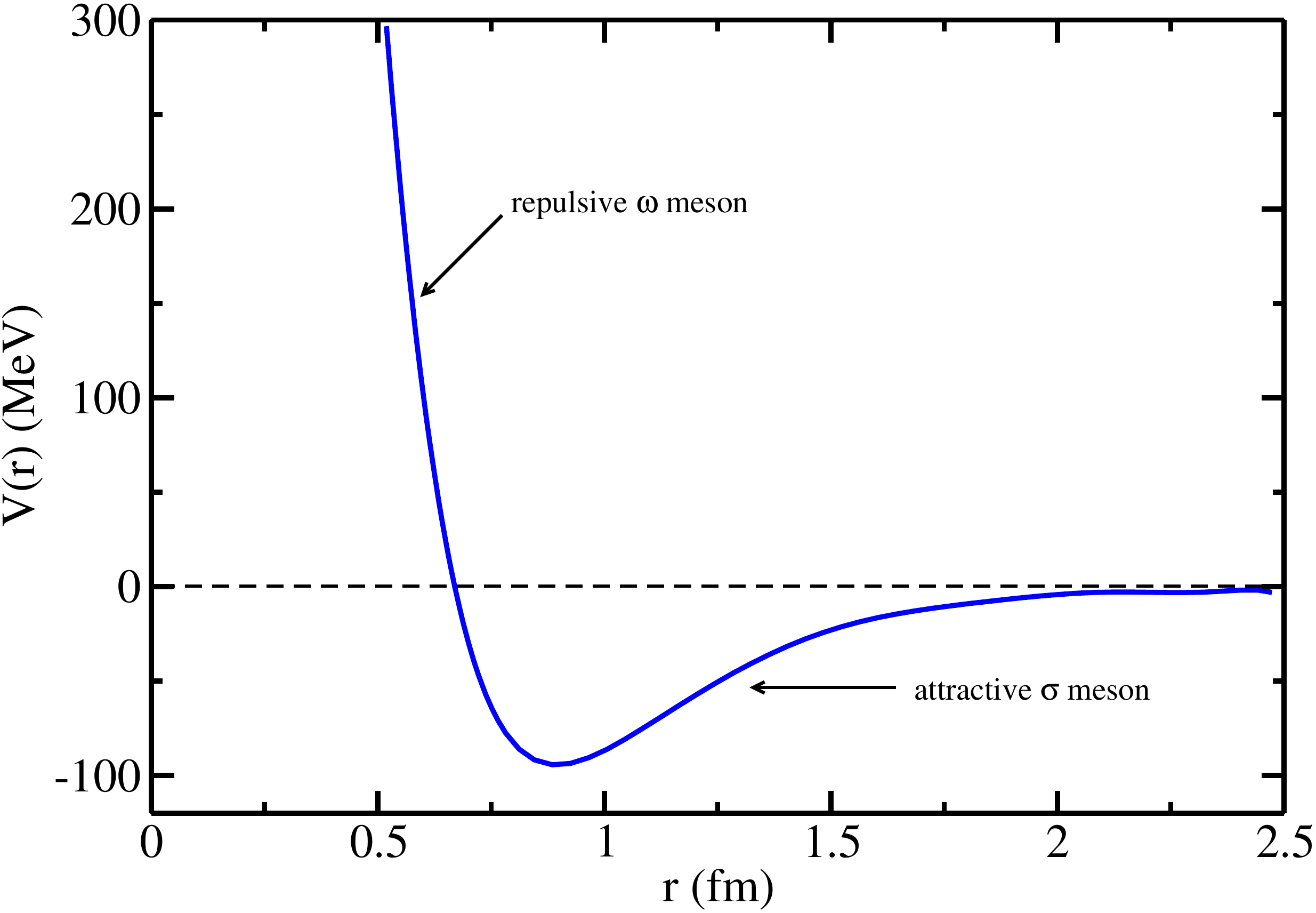}
	\caption{Nucleon-nucleon potential.}
	\label{fig2.1}
\end{figure}

Quantum Hydrodynamics I (QHD-I), also known as the $\sigma-\omega$ model, is the original and simplest QHD parameter set \cite{Serot:1984ey,WALECKA1975109}. This model involves the exchange of isoscalar sigma  $\sigma$ and isoscalar vector mesons $\omega$ with the baryons (neutron and proton) which are found to be important in the description of the nuclear matter. The effective nucleon-nucloen potential of QHD-I is defined as
\begin{equation}
V_{eff}=\frac{g_{\omega}^2}{4\pi}\frac{e^{-m_{\omega}r}}{r}
	-\frac{g_{\sigma}^2 }{4\pi}\frac{e^{-m_{\sigma}r}}{r}.
\end{equation} 
The $\sigma$ meson represents the strong attractive central force while as the $\omega$ meson represents the strong replusive central force as shown in Fig.~\ref{fig2.1}.
The Lagrangian density of QHD-I is written as \cite{Serot:1984ey}
\begin{align}\label{2.2}
\mathcal{L}&= \bar{\psi}[\gamma_{\mu}(i\partial^{\mu}-g_{\omega}V^{\mu})-(M-g_{\sigma}\sigma)]\psi+\frac{1}{2}(\partial_{\mu}\sigma\partial^{\mu}\sigma-m_{\sigma}^2 \sigma^2)\nonumber \\
& -\frac{1}{4}V_{\mu \nu}V^{\mu \nu}+\frac{1}{2}m_{\omega}^2 V_{\mu}V^{\mu},
\end{align}
where $\psi$ represents the baryonic field, $\sigma$ and $\omega$ denote the scalar and vector meson fiels, $m_{\sigma}$, $m_{\omega}$, and $M$ represent the sigma, omega meson, and nucleon mass, respectively. $g_{\sigma}$ and $g_{\omega}$ are the scalar and vector coupling constants and $V_{\mu \nu}=\partial_{\mu}V_{\nu}-\partial_{\nu}V_{\mu}$.  The first term of Eq.~(\ref{2.2}) represents the non-interactive dynamics of baryonic field with the meson fields, the second, and third terms represent the dynamics of $\sigma$ and $\omega$ fields, repsectively.

Using the Euler-Lagrange equation
\begin{equation}\label{2.3}
\frac{\partial}{\partial x^{\mu}}
\Bigg[\frac{\partial \mathcal{L}}
{\partial(\partial q_i/\partial x^{\mu})}\Bigg]-\frac{\partial \mathcal{L}}{\partial q_i}=0,
\end{equation}
the different field equations are obtained as
\begin{equation}\label{2.4}
(\partial_{\mu}\partial^{\mu}+m_{\sigma}^2)\sigma =g_{\sigma}\bar{\psi}\psi,
\end{equation} 
\begin{equation}\label{2.5}
(\partial_{\mu}V^{\mu \nu}+m_{\omega}^2)V^{\mu} =g_{\omega}\bar{\psi}\gamma^{\mu}\psi,
\end{equation}
and
\begin{equation}\label{2.6}
[\gamma^{\mu}(i\partial_{\mu}-m_{\omega}V_{\mu}) -(M-g_{\sigma}\sigma)]\psi=0.
\end{equation}
The energy momentum tensor is given by the expression \cite{Serot:1984ey}
\begin{equation}\label{2.7}
T_{\mu \nu} = \sum_{i} \partial_{\nu}\phi_i \frac{\partial \mathcal {L}}{\partial (\partial ^{\mu} \phi_i)} - g_{\mu \nu} \mathcal{L},
\end{equation}
where, $g_{\mu \nu}$ is the Lorentz transformation matrix given by
\begin{equation}
g_{\mu \nu} = \begin{pmatrix}
1&0&0&0\\
0&-1&0&0\\
0&0&-1&0\\
0&0&0&-1\\
\end{pmatrix} 
\end{equation}
The energy density and pressure are obtained from the above tensor expression as the zeroth annd third component respectively.
\begin{equation}
\mathcal{E}=\frac{g_{\omega}^2}{2m_{\omega}^2}\rho_b^2 +\frac{m_{\sigma}^2}{2g_{\sigma}^2} (M-M^*)^2 +\frac{\gamma}{(2\pi)^3}\int_{0}^{k_f} (k^2 +M^{*2})^{1/2} d^3k,
\end{equation}
\begin{equation}
P=\frac{g_{\omega}^2}{2m_{\omega}^2}\rho_b^2 -\frac{m_{\sigma}^2}{2g_{\sigma}^2} (M-M^*)^2 +\frac{1}{3}\frac{\gamma}{(2\pi)^3}\int_{0}^{k_f} \frac{k^2}{(k^2 +M^{*2})^{1/2}} d^3k,
\end{equation}
where, $k_f$ represents the fermi momentum of nucleons, $\gamma$ denotes the spin-isospin degeneracy which is 4 for nuclear matter and 2 for neutron matter. $\rho_b$ represents the baryon density given by the expression
\begin{equation}
\rho_b=\frac{\gamma}{(2\pi)^3}\int_{0}^{k_f} (k^2 +M^{*2})^{1/2} d^3k = \frac{\gamma}{6\pi^2}k_f^3.
\end{equation}
The effective mass of nucleons $M^*$ is obtained by minimizing the energy density concerning the effective mass which leads to the relation
\begin{align}
M^* &=M-\frac{g_{\sigma}^2}{m_{\sigma}^2} \frac{\gamma}{(2\pi)^3}\int_{0}^{k_f} \frac{k^2}{(k^2 +M^{*2})^{1/2}} d^3k\nonumber \\
&=M-\frac{g_{\sigma}^2}{m_{\sigma}^2}\frac{\gamma M^*}{4\pi^2}\Bigg[k_fE_f^*-M^{*2}
\ln\Bigg(\frac{k_f+E_f^*}{M^*}\Bigg)\Bigg],
\end{align}
with $E_f^* = \sqrt{k_f^2+M^{*2}}$.

The equations of motion (Eqs. \ref{2.4}-\ref{2.6}) are non-linear, coupled equations that are difficult to solve and hence are approximated. This approximation is provided by the \textit{relativistic mean-field theory}.
In the RMF theory, the meson fields are replaced by the classical fields i.e., their ground state expectation values, to simplify the solution of the field equations \cite{Serot:1984ey}. The addition of several other mesons along with the self-and cross-coupling terms lead to the development of new parameter sets which have been discussed in chapter \ref{cha-lit}.
\section{Effective field theory motivated RMF model}
\label{rmf}
Quantum Hadrodynamics (QHD), the Effective Field Theory (EFT) for strong interaction \cite{WALECKA1974491,Reinhard:1989zi,Serot:1992ti} at low energies has been studied extensively to describe the properties of both finite nuclei \cite{HOROWITZ1981503,Boguta:1977xi,GAMBHIR1990132,RING1996193} and infinite nuclear matter (INM) \cite{WALECKA1974491}. In this theory, the interaction of nucleons occurs with the exchange of mesons like $\sigma$, $\omega$, $\rho$, and $\delta$.\par 
  A systematic formalism based on naturalness and Naive Dimensional Analysis (NDA), the effective field theory motivated relativistic-mean-field (E-RMF) lagrangian is constructed. The E-RMF is an extension to the basic RMF theory in which all possible self- and cross-couplings between the mesons are included \cite{FURNSTAHL1996539,FURNSTAHL1997441,bharat}. The E-RMF Lagrangian contains the contribution from $\sigma$-, $\omega$- mesons upto 4th order expansion along with the $\rho$- and $\delta$- mesons upto 2nd order and is given by  \cite{FURNSTAHL1996539,bharat,PhysRevC.97.045806}
\begin{align}\label{eq1}
\mathcal{E} (r) &=\sum_{\alpha} \phi^{\dagger}_{\alpha} (r) \Biggl\{-i \bm{\alpha} \cdot \bm{\nabla} +\beta [M-\Phi(r) -\tau_3 D(r)] 
+W(r) + \frac{1+\tau_3}{2} A(r)\nonumber \\ 
&+\frac{1}{2}\tau_3 R(r)  
-\frac{i\beta \bm{\alpha}}{2M}.\Bigg(f_{\omega}\bm{\nabla} W(r)
+\frac{1}{2}f_{\rho}\tau_3 \bm{\nabla} R(r)\Bigg)\Biggr\} \phi_{\alpha}(r) -\frac{\zeta_0}{4!}\frac{1}{g_{\omega}^2}W^4(r) \nonumber \\
&+\Bigg(\frac{1}{2}+\frac{k_3}{3!}\frac{\Phi(r)}{M}+\frac{k_4}{4!}\frac{\Phi^2(r)}{M^2}\Bigg) \frac{m_s^2}{g_s^2}\Phi^2 (r)
+\frac{1}{2 g_s^2}\Bigg(1+\alpha_1 \frac{\Phi(r)}{M}\Bigg)(\bm{\nabla} \Phi(r))^2 \nonumber \\ 
&-\frac{1}{2 g_{\omega}^2}\Bigg(1+\alpha_2 \frac{\Phi(r)}{M}\Bigg) \times (\bm{\nabla} W(r))^2-\frac{1}{2}\Bigg(1+\eta_{\rho}\frac{\Phi(r)}{M}\Bigg)\frac{m_{\rho}^2}{g_{\rho}^2} R^2 (r)\nonumber \\
&-\frac{1}{2} \Bigg(1+\eta_1\frac{\Phi(r)}{M}+\frac{\eta_2}{2}\frac{\Phi^2(r)}{M^2}\Bigg)\frac{m_{\omega}^2}{g_{\omega}^2} W^2(r)-\frac{1}{2 e^2}(\bm{\nabla} A(r))^2 -\frac{1}{g_{\rho}^2}(\bm{\nabla} R(r))^2 \nonumber \\
&  -\Lambda_{\omega}(R^2 (r) W^2(r))+\frac{1}{2 g_{\delta}^2}(\bm{\nabla} D(r))^2
+ \frac{1}{2}\frac{m_{\delta}^2}{g_{\delta}^2} (D(r)^2),
\end{align}
where $\Phi$, $W$, $R$, $D$, and $A$ are $\sigma$, $\omega$, $\rho$, $\delta$, and photon fields respectively, $g_{\sigma}$, $g_{\omega}$, $g_{\rho}$, $g_{\delta}$, and $\frac{e^2}{4\pi}$ are the corresponding coupling constants and $m_{\sigma}$, $m_{\omega}$, $m_{\rho}$, and $m_{\delta}$ are the masses for  $\sigma$, $\omega$, $\rho$, and $\delta$ mesons respectively. $\phi_{\alpha}$ is the nucleonic field. The addition of parameters like $\eta_1$, $\eta_2$, $\eta_{\rho}$, $\alpha_1$, $\alpha_2$ in G3 set have their own importance in explaining various properties of finite as well as INM. For example, the non linear interaction of $\eta_1$ and $\eta_2$ parameters analyze the surface properties of finite nuclei \cite{PhysRevC.63.024314}.\par
Using the equation $\Big(\frac{\partial \mathcal{E}}{\partial \phi_i}\Big)_{\rho = const} =0$, we obtain equation of motion for mesons.
The energy eigenvalue of the Dirac equation constraining the normalisation condition $\sum_{\alpha}\phi_{\alpha}^{\dagger}(r)\phi_{\alpha}(r)$=1 is used to calculate the single particle energy for nucleons using the Lagrange multiplier $\mathcal{E}_{\alpha}$. For the wave function $\phi_{\alpha}(r)$, the Dirac equation becomes
\begin{equation}
\frac{\partial}{\partial \phi_{\alpha}^{\dagger}(r)}\Bigg[\mathcal{E}(r)-\sum_{\alpha}\phi_{\alpha}^{\dagger}(r)\phi_{\alpha}(r)\Bigg]=0,
\end{equation} 
which when solved together with the Eq.~(\ref{eq1}) gives
\begin{eqnarray}
\Biggl\{-i\bm{\alpha} \cdot\bm{\nabla}+\beta[M-\Phi(r)-\tau_3 D(r)]+W(r)+\frac{1}{2}\tau_3R(r)+\frac{1+\tau_3}{2}A(r) \nonumber \\
-\frac{i\beta \bm{\alpha}}{2M}.\Bigg[f_{\omega}\bm{\nabla} W(r)+\frac{1}{2}f_{\rho}\tau_3\bm{\nabla}R(r)\Bigg] \Biggr\}\phi_{\alpha}(r)=\mathcal{E}_{\alpha}\phi_{\alpha}(r).
\end{eqnarray}
 The mean-field equation of motion for $\Phi$, $W$, $R$, $D$, and $A$ field are then given as
 \begin{align}\label{eq4}
 -\Delta \Phi(r)+ m_s^2\Phi(r)&=g_s^2 \rho_s(r)-\frac{m_s^2}{M}\Phi^2(r)\Bigg(\frac{k_3}{2}+\frac{k_4}{3!}\frac{\Phi(r)}{M}\Bigg)\nonumber \\
 &+\frac{g_s^2}{2M}\Bigg(\eta_1+\eta_2\frac{\Phi(r)}{M}\Bigg)\frac{m_{\omega}^2}{g_{\omega}^2}W^2(r)+\frac{\eta_{\rho}}{2M}\frac{g_s^2}{g_{\rho}^2}m_{\rho}^2R^2(r)\nonumber \\
 &+\frac{\alpha_1}{2M}[(\bm{\nabla}\Phi(r))^2
 +2\Phi(r)\Delta\Phi(r)]
 +\frac{\alpha_2}{2M}\frac{g_s^2}{g_{\omega}^2}(\bm{\nabla}W(r))^2,
 \end{align} 
\begin{align}\label{eq5}
-\Delta W(r)+ m_{\omega}^2W(r)&=g_{\omega}^2 \Bigg(\rho(r)+\frac{f_{\omega}}{2}\rho_T (r)\Bigg) -\frac{\Phi(r)}{M}\Bigg(\eta_1+\eta_2\frac{\Phi(r)}{M}\Bigg)m_{\omega}^2W(r)\nonumber \\
& -\frac{1}{3!}\zeta_0 W^3(r) 
+\frac{\alpha_2}{M}[\bm{\nabla}\Phi(r)\cdot \bm{\nabla}W(r)+\Phi(r)\Delta W(r)]\nonumber \\
&-2\Lambda_{\omega}g_{\rho}^2 R(r)W^2(r),
\end{align} 
\begin{align}\label{eq6}
-\Delta R(r)+m_{\rho}^2 R(r)&=\frac{1}{2}g_{\rho}^2\Bigg(\rho_3(r)
+\frac{1}{2}f_{\rho}\rho_{T,3}(r)\Bigg)-\eta_{\rho}\frac{\Phi(r)}{M}m_{\rho}^2 R(r) \nonumber \\ &-2\Lambda_{\omega}g_{\rho}^2 R(r)W^2(r),
\end{align}
\begin{align}\label{eq7}
-\Delta A(r)=e^2 \rho_p(r),
\end{align}
\begin{align}\label{eq8}
-\Delta D(r)+m_{\delta}^2 D(r)=g_{\delta}^2 \rho_{s3}(r),
\end{align}
where $\rho(r)$, $\rho_s(r)$, $\rho_3(r)$, and $\rho_{s3}(r)$, $\rho_p(r)$,  $\rho_T(r)$, and $\rho_{T,3}(r)$ are the corresponding baryon, scalar, isovector, proton, and tensor densities, respectively, given as
\begin{align}
\rho(r)&=\sum_{\alpha}\phi_{\alpha}^{\dagger}(r)\phi_{\alpha}(r)=\rho_p(r)+\rho_n(r)\nonumber \\ &=\frac{2}{(2\pi)^3}\Bigg(\int_{0}^{k_p}d^3k+\int_{0}^{k_n}d^3k\Bigg),
\end{align}
\begin{align}
\rho_s(r)&=\sum_{\alpha}\phi_{\alpha}^{\dagger}(r)\beta\phi_{\alpha}(r)=\rho_{sp}(r)+\rho_{sn}(r)\nonumber \\ &=\frac{2}{(2\pi)^3}\sum_{\alpha}\int_{0}^{k_{\alpha}}d^3k
\frac{M_{\alpha}^*}{(k_{\alpha}^2 +M_{\alpha}^{*2})^{1/2}},
\end{align}
\begin{align}
\rho_3(r)&=\sum_{\alpha}\phi_{\alpha}^{\dagger}(r)\tau_3\phi_{\alpha}(r)=\rho_p(r)-\rho_n(r),
\end{align}
\begin{align}
\rho_{s3}(r)&=\sum_{\alpha}\phi_{\alpha}^{\dagger}(r)\tau_3\beta\phi_{\alpha}(r)=\rho_{sp}(r)-\rho_{sn}(r),
\end{align}
\begin{align}
\rho_p(r)&=\sum_{\alpha}\phi_{\alpha}^{\dagger}(r)\Big(\frac{1+\tau_3}{2}\Big)\phi_{\alpha}(r),
\end{align}
\begin{align}
\rho_T(r)&=\sum_{\alpha}\frac{i}{M} \bm{\nabla}\cdot [\phi_{\alpha}^{\dagger}(r)\beta\bm{\alpha}\phi_{\alpha}(r)],
\end{align}
and
\begin{align}
\rho_{T3}(r)&=\sum_{\alpha}\frac{i}{M} \bm{\nabla}\cdot [\phi_{\alpha}^{\dagger}(r)\beta\bm{\alpha}\tau_3\phi_{\alpha}(r)],
\end{align}
where $k_{\alpha}$ is the nucleonic Fermi momentum with the summation over all occupied states. 
During the fitting process, the coupling constants of the effective Lagrangian are determined from a set of experimental data while accounting for vacuum polarisation effects in the \textit{no-sea approximation}, which is important in determining the stationary solutions of the RMF equations that describe the nuclear ground-state properties.
 The effective masses of proton and neutron $M_p^*$ and $M_n^*$  which are splitted due to $\sigma$ and $\delta$ meson are written as
\begin{equation}
M_p^* = M-\Phi(r)-D(r),
\end{equation}
and
\begin{equation}
M_n^*=M-\Phi(r)+D(r).
\end{equation}
\section{Nuclear Equation of State}
The nuclear equation of state (EoS) plays a vital role in explaining the properties of NM and NS. The EoS provides a very good explanation of the matter at low and high density, especially in NSs. The energy per nucleon is described by the nuclear EoS as a function of the nucleonic density (neutron $\rho_n$ and proton $\rho_p$ densities) of a uniform and infinite system at zero temperature that interacts via the residual strong interaction, or nuclear force \cite{doi:10.1146/annurev-nucl-102711-095018}.
For the uniform, infinite, and isotropic nuclear matter, the field gradient in Eqs. (\ref{eq4}-\ref{eq8}) vanish. The electromagnetic interaction $A(r)$ is also neglected in the context of infinite nuclear matter. The expression for the EoS (energy density and pressure) for such a system is obtained using the energy-momentum tensor given by the Eq.~(\ref{2.7}).

The zeroth component of the energy-momentum tensor corresponds to the  energy density while the third component calculates the pressure for the given system
\begin{align} \label{eq9}
\mathcal{E}_H=<0|T_{00}|0>&=\sum_{i=n,p} \frac{2}{(2\pi)^3}\int_0^{k_i} d^3k E^*_i (k)+\rho W+\frac{m_s^2 \Phi^2}{g_s^2}\Bigg(\frac{1}{2}+\frac{k_3}{3!}\frac{\Phi}{M}+\frac{k_4}{4!}\frac{\Phi^2}{M^2}\Bigg)\nonumber \\
&-\frac{1}{4!}\frac{\zeta_0 W^4}{g_{\omega}^2}+\frac{1}{2}\rho_3 R -\frac{1}{2}m_{\omega^2}\frac{W^2}{g_{\omega}^2}\Bigg(1+\eta_1\frac{\Phi}{M}+\frac{\eta_2}{2}\frac{\Phi^2}{M^2}\Bigg)\nonumber \\
&-\frac{1}{2}\Bigg(1+\frac{\eta_{\rho}\Phi}{M}\Bigg)\frac{m_{\rho}^2}{g_{\rho}^2}R^2 -\Lambda_{\omega}(R^2 W^2)
+\frac{1}{2}\frac{m_{\delta}^2}{g_{\delta}^2}(D^2),
\end{align}
and
\begin{align}\label{eq10}
P_H=\frac{1}{3}\sum_{i=1}^{3}<0|T_{ii}|0> &= \sum_{i=n,p}\frac{2}{3(2\pi)^3}\int_0^{k_i} d^3k E^*_i (k)-\frac{m_s^2 \Phi^2}{g_s^2}\Bigg(\frac{1}{2}+\frac{k_3}{3!}\frac{\Phi}{M}+\frac{k_4}{4!}\frac{\Phi^2}{M^2}\Bigg) \nonumber \\ 
&+\frac{1}{4!}\frac{\zeta_0 W^4}{g_{\omega}^2}
+\frac{1}{2}m_{\omega^2}\frac{W^2}{g_{\omega}^2}\Bigg(1+\eta_1\frac{\Phi}{M}+\frac{\eta_2}{2}\frac{\Phi^2}{M^2}\Bigg)\nonumber \\
&+\frac{1}{2}\Bigg(1+\frac{\eta_{\rho}\Phi}{M}\Bigg)\frac{m_{\rho}^2}{g_{\rho}^2}R^2
+\Lambda_{\omega}(R^2 W^2)
-\frac{1}{2}\frac{m_{\delta}^2}{g_{\delta}^2}(D^2),
\end{align}
where

$E^*_i(k) =\sqrt{k^2+ M^{*2}_i}$ is the effective energy of nucleons. 

\section{Density dependent RMF model}
The self- and cross-coupling of various mesons in the RMF model can be replaced by the density-dependent nucleon-meson coupling constants in the density-dependent RMF (DD-RMF) models \cite{PhysRevLett.68.3408}. The density-dependent coupling constants enable a consistent computation of NSs and strange matter, with findings comparable to previous models. It uses microscopic interactions at various densities as input to include the characteristics of the Dirac-Brueckner model. Extrapolation to higher densities is more confined than in phenomenological RMF calculations, which employ only information from the finite nuclei's narrow density range to determine their parameters \cite{TYPEL1999331}.

The contribution of the rearrangement term self-energy to DD-RMF field equations is the most significant variation from the RMF model. The rearrangement term physically accounts for the effects of static polarisation in the nuclear medium. The contribution of rearrangement term to pressure implies that by not considering its contribution, it violates thermodynamic consistency because the mechanical pressure obtained from the energy-momentum tensor must coincide with the thermodynamic derivation.

 The coupling constants can be either dependent on the scalar density $\rho_s$ or the vector density $\rho_B$, but usually the vector density parameterizations are considered which influences only the self-energy instead of the total energy.\par
The DD-RMF Lagrangian is given as:
\begin{align}\label{ddeq}
\mathcal{L} & =\sum_{\alpha=n,p} \bar{\psi}_{\alpha} \Biggl\{\gamma^{\mu}\Bigg(i\partial_{\mu}-g_{\omega}(\rho_B)\omega_{\mu}
-\frac{1}{2}g_{\rho}(\rho_B)\gamma^{\mu}\rho_{\mu}\tau\Bigg) \nonumber \\
&-\Bigg(M-g_{\sigma}(\rho_B)\sigma-g_{\delta}(\rho_B)\delta\tau\Bigg)\Biggr\} \psi_{\alpha} 
+\frac{1}{2}\Bigg(\partial^{\mu}\sigma \partial_{\mu}\sigma-m_{\sigma}^2 \sigma^2\Bigg) \nonumber \\
&+\frac{1}{2}\Bigg(\partial^{\mu}\delta \partial_{\mu}\delta-m_{\delta}^2 \delta^2\Bigg)
-\frac{1}{4}W^{\mu \nu}W_{\mu \nu}
+\frac{1}{2}m_{\omega}^2 \omega_{\mu} \omega^{\mu}\nonumber \\
&-\frac{1}{4}R^{\mu \nu} R_{\mu \nu}
+\frac{1}{2}m_{\rho}^2 \rho_{\mu} \rho^{\mu},
\end{align}
where $\psi$ denotes the nucleonic wave-function. $\sigma$, $\omega_{\mu}$, $\rho_{\mu}$, and $\delta$ represent the sigma, omega, rho, and delta meson fields, respectively. $g_{\sigma}$, $g_{\omega}$, $g_{\rho}$, and $g_{\delta}$ are the meson coupling constants which are density-dependent and $m_{\sigma}$, $m_{\omega}$, $m_{\rho}$, and $m_{\delta}$ are the masses for  $\sigma$, $\omega$, $\rho$, and $\delta$ mesons respectively. The anti-symmetric tensor fields $W^{\mu \nu}$ and $R^{\mu \nu}$ are given by 
	\begin{align}
	W^{\mu \nu}&=\partial^{\mu}W^{\nu}-\partial^{\nu}W^{\mu},\\
	R^{\mu \nu}&=\partial^{\mu}R^{\nu}-\partial^{\nu}R^{\mu}.
	\end{align}
The density-dependent coupling constants are represented as:
\begin{equation}
g_i(\rho_B) = g_i(\rho_0) f_i(x),
\end{equation}
where
\begin{equation}\label{eq2.36}
f_i(x) = a_i \frac{1+b_i (x+d_i)^2}{1+c_i(x+d_i)^2},~~~i=\sigma,\omega
\end{equation}
is a function of $ x=\rho_B/\rho_{0}$ with $\rho_0$ as the NM saturation density.\par 
For the function $f_i(x)$, one has five constraint conditions $f_i(1)=1$, $f^{''}_{\sigma}(1)=f^{''}_{\omega}(1)$, $f^{''}_i(0)=0$ which reduce the number of free parameters from eight to three in Eq.~(\ref{eq2.36}). The first two constraints lead to
\begin{equation}
a_i=\frac{1+c_i(1+d_i)^2}{1+b_i(1+d_i)^2},~~~ 3c_id_i^2=1.
\end{equation} 
For $\rho$ and $\delta$ mesons, the coupling constants are given by an exponential dependence as
\begin{equation}
g_i(\rho_B)=g_i(\rho_0)exp[-a_i(x-1)].
\end{equation}
Following the Euler-Lagrange equation, we obtain equation of motion for nucleons and mesons as
\begin{equation}
\sum_{\alpha=n,p}\Bigg[i\gamma^{\mu}\partial_{\mu}-\gamma^0 \Bigg(g_{\omega}(\rho_B)\omega 	+\frac{1}{2}g_{\rho}(\rho_B)\rho \tau_3\\ +\sum_R (\rho_B)\Bigg)
-M_{\alpha}^*\Bigg]\psi_{i}=0,\\
\end{equation} 
	\begin{align}
	m_{\sigma}^2 \sigma &= g_{\sigma}(\rho_B)\rho_s,\\
	m_{\omega}^2 \omega &= g_{\omega}(\rho_B)\rho_B,\\
	m_{\rho}^2 \rho &= \frac{g_{\rho}(\rho_B)}{2}\rho_3,\\
	m_{\delta}^2 \delta &= g_{\delta}(\rho_B)\rho_{s3}.
	\end{align} 
$\sum\limits_{R}$ is the rearrangment term introduced in the equation of motion of mesons due to the density dependent coupling constants.
\begin{equation}
\sum_R(\rho_B) = -\frac{\partial g_{\sigma}}{\partial \rho_B}\sigma \rho_s +\frac{\partial g_{\omega}}{\partial \rho_B}\omega \rho_B+\frac{1}{2}\frac{\partial g_{\rho}}{\partial \rho_B}\rho \rho_3-\frac{\partial g_{\delta}}{\partial \rho_B}\delta \rho_{s3},
\end{equation}
where $\rho_s$, $\rho_B$, $\rho_{s3}$, and $\rho_3$ are the scalar, baryon and isovector densities, respectively, given by
\begin{align}
\rho_s &= \sum_{\alpha=n,p}\bar{\psi}\psi =\rho_{sp} +\rho_{sn}=\sum_{\alpha}\frac{2}{(2\pi)^3}\int_{0}^{k_{\alpha}}d^3k \frac{M_{\alpha}^*}{E_{\alpha}^*}, \\
\rho_B &= \sum_{\alpha=n,p}\psi^{\dagger}\psi =\rho_{p} +\rho_{n}=\sum_{\alpha}\frac{2}{(2\pi)^3}\int_{0}^{k_{\alpha}}d^3k,\\ 
\rho_{s3} &= \sum_{\alpha}\bar{\psi}\tau_3\psi =\rho_{sp} -\rho_{sn},
\end{align}
and 
\begin{align}
\rho_3 &= \sum_{\alpha}\psi^{\dagger}\tau_3\psi =\rho_p -\rho_n.
\end{align}
The effective masses of nucleons are given as:
\begin{align}
M_p^* &=M-g_{\sigma}(\rho_B)\sigma -g_{\delta}(\rho_B)\delta,
\end{align}
and
\begin{align}
M_n^* &=M-g_{\sigma}(\rho_B)\sigma +g_{\delta}(\rho_B)\delta.
\end{align}
Also,
\begin{equation}
E_{\alpha}^*=\sqrt{k_{\alpha}^2+M_{\alpha}^{*2}}
\end{equation}
is the effective mass of nucleons with $k_{\alpha}$ as the nucleon momentum.
The energy-momentum tensor determines the energy density and pressure for the NM as
\begin{align} \label{eq18}
\mathcal{E}_H &= \frac{1}{2}m_{\sigma}^2 \sigma^2-\frac{1}{2}m_{\omega}^2 \omega^2-\frac{1}{2}m_{\rho}^2 \rho^2+\frac{1}{2}m_{\delta}^2 \delta^2 \nonumber \\
&+g_{\omega}(\rho_B)\omega \rho_B+\frac{g_{\rho}(\rho_B)}{2}\rho \rho_3 +\mathcal{E}_{kin},
\end{align}
\begin{align}\label{eq19}
P_H &= -\frac{1}{2}m_{\sigma}^2 \sigma^2+\frac{1}{2}m_{\omega}^2 \omega^2+\frac{1}{2}m_{\rho}^2 \rho^2-\frac{1}{2}m_{\delta}^2 \delta^2 \nonumber \\
&-\rho_B \sum_R (\rho_B)+P_{kin},
\end{align}
where
$\mathcal{E}_{kin}$ and $P_{kin}$ are the contributions to the energy density and pressure from the kinetic part,
\begin{align} \label{eq20}
\mathcal{E}_{kin} &=\frac{1}{\pi^2}\int_{0}^{k_{\alpha}}k^2 \sqrt{k^2 +M_{\alpha}^{*2}} dk,\nonumber \\
P_{kin} &= \frac{1}{3\pi^2}\int_{0}^{k_{\alpha}}\frac{k^4 dk}{\sqrt{k^2 +M_{\alpha}^{*2}}}.
\end{align}
\section{Nuclear matter properties}
The characteristics of nuclear matter are deduced from the experimentally observed properties of finite nuclei. The semi-empirical mass formula developed by Bethe-Weizsacker in 1936, based on the liquid drop model, is quite successful in describing the properties of finite nuclei. In this model, the binding energy per particle of a nucleus is defined as
\begin{equation}
\frac{E(Z,N)}{A}=M-a_{vol}+a_{surf}\frac{1}{A^{1/3}}+a_{coul}\frac{Z^2}{A^{2/3}}+a_{asym}\frac{(N-Z)^2}{A^2}+...,
\end{equation}
where, $A=Z+N$  is the total number of nucleons. $a_{vol}$, $a_{surf}$, $a_{coul}$, and $a_{asym}$ are the strength parameters corresponding to volume, surface, coulomb, and asymmetry term, respectively. For INM, the liquid drop is extended by switching off the coulomb term $a_{coul}$ = 0 and neglecting the contribution from surface part $a_{surf}$.  The binding energy per nucleon of the system thus becomes
\begin{align}
\frac{E(Z,N)}{A}-M &=-a_{vol}+a_{asym}\frac{(N-Z)^2}{A^2},\nonumber\\
e(\rho,\alpha)&=\frac{\mathcal{E}}{\rho_B}-M=-a_{vol}+a_{asym}\alpha^2,
\end{align}
where $\alpha=\frac{(N-Z)}{A}=\frac{\rho_n -\rho_p}{\rho_n +\rho_p}$ is termed as the neutron excess of INM. For compressible matter around $\alpha=0$, the binding energy per particle can  be approximated by the parabolic law as \cite{PhysRevC.44.1892}
\begin{equation}
e(\rho, \alpha)= e(\rho,\alpha=0) + S(\rho) \alpha^2 + \mathcal{O}(\alpha^4),
\end{equation}
where $e(\rho)$ is the binding energy per particle of symmetric nuclear matter ($\alpha=0$). $\alpha=1$ corresponds to the pure neutron matter. Due to the charge symmetry of nuclear force, the linear terms in $\alpha$ vanish. The second term in above equation, $S(\rho)$, is the symmetry energy defined as
\begin{equation}
S(\rho) = \frac{1}{2}\Big[\frac{\partial^2 e(\rho,\alpha)}{\partial \alpha^2}\Big]_{\alpha=0}.
\end{equation}
The symmetry energy $S$ for a nuclear system with mass number $A$ is defined as $S =\frac{E}{A}(A,N=A)-\frac{E}{A}(A,N= Z)$. Huge literature is devoted to the calculation of the symmetry energy $S$ and its slope parameter $L$. Different phenomenological approaches like Hartree-Fock \cite{FARINE1978317} and Thomas-Fermi \cite{PEARSON19911} have been used to study the symmetry energy which predicts the value in the range 27-38 MeV at saturation. Such studies have also shown the correlation between the slope parameter of symmetry energy and the neutron skin thickness.\par 
This isospin asymmetry arise as a result of difference in the masses and densities of proton and neutron. The isovector-vector meson $\rho$ takes care of asymmetry density while the isovector-scalar meson $\delta$ takes care of mass asymmetry. The combined expression of the $\rho$ and $\delta$ meson symmetry energies gives the overall symmetry energy of the system  \cite{PhysRevC.63.024314,PhysRevC.84.054309}
\begin{equation}
S(\rho) = S^{kin}(\rho)+ S^{\rho}(\rho) +S^{\delta}(\rho),
\end{equation}
where
\begin{equation}
S^{kin}(\rho) =\frac{k_F^2}{6 E_F^*},
\end{equation}
and
\begin{equation}
S^{\rho}(\rho) =\frac{g_{\rho}^2 \rho}{8 m_{\rho}^{*2}}.
\end{equation}
Due to the cross-coupling between the $\rho$-$\omega$ fields, the mass of the $\rho$ meson is modified as
\begin{equation}
m_{\rho}^{*2} = \Big(1+\eta_{\rho} \frac{\Phi}{M}\Big)m_{\rho}^2 +2 g_{\rho}^2 (\Lambda_{\omega} W^2).
\end{equation}
The contribution to the symmetry energy due to the $\delta$ meson is
\begin{equation}
S^{\delta}(\rho) =-\frac{1}{2}\rho \frac{g_{\delta}^2}{m_{\delta}^2} \Bigg(\frac{M^*}{E_F}\Bigg)^2 u_{\delta}(\rho, M^*).
\end{equation}
The discreteness of the Fermi momentum leads to the function $u_{\delta}$. This momentum is relatively big in nuclear matter, thus the system may be regarded as continuous, implying that the function $u_{\delta}$$\approx$1. So the final expression for the symmetry energy becomes
\begin{equation}
S(\rho) = \frac{k_F^2}{6 E_F^*}+\frac{g_{\rho}^2 \rho}{8 m_{\rho}^{*2}} -\frac{1}{2}\rho \frac{g_{\delta}^2}{m_{\delta}^2} \Bigg(\frac{M^*}{E_F}\Bigg)^2.
\end{equation}
Numerically, the symmetry energy $S(\rho)$ is calculated as the difference in the energy of the Symmetric Nuclear Matter (SNM) and Pure Neutron Matter (PNM). The symmetry energy around the saturation density $\rho_0$ can be expanded by Taylor series as:
\begin{equation}
S(\rho) = J+L \mathcal{Y}+ \frac{1}{2} K_{sym} \mathcal{Y}^2 + \frac{1}{6} Q_{sym} \mathcal{Y}^3 +\mathcal{O}[\mathcal{Y}^4],
\end{equation}
where \par 
$J$=$S(\rho_0)$ corresponds to the symmetry energy at the saturation density $\rho_0$ and $\mathcal{Y} = (\rho- \rho_0)/(3\rho_0)$. The derivatives of $S(\rho)$ are $L$, $K_{sym}$, and $Q_{sym}$ are defined as:
\begin{equation}
L = 3 \rho_0 \frac{\partial S(\rho)}{\partial \rho}\Bigg|_{\rho = \rho_0},
\end{equation}
\begin{equation}
K_{sym} = 9 \rho_0^2 \frac{\partial^2 S(\rho)}{\partial \rho^2}\Bigg|_{\rho = \rho_0},
\end{equation}
and
\begin{equation}
Q_{sym} = 27 \rho_0^3 \frac{\partial^3 S(\rho)}{\partial \rho^3}\Bigg|_{\rho = \rho_0}.
\end{equation}
Here $L$ is the slope parameter and $K_{sym}$ represents the symmetry energy curvature at saturation density. $Q_{sym}$ is the skewness of $S(\rho)$ at $\rho_0$. To fix the values of all these quantities, a large number of attempts have been made \cite{Newton_2012,PhysRevLett.108.081102,PhysRevC.86.025804}. The density dependence of symmetry energy is an important quantity for understanding the characteristics of both finite and infinite matter \cite{PhysRevLett.102.122502}. The presently recognised symmetry energy and slope values are $J$ = 31.6 $\pm$ 2.66 MeV and $L$ = 58.9 $\pm$ 16 MeV, as determined by different astrophysical observations \cite{LI2013276}. \citet{PhysRevC.96.054311} calculated the value of nuclear matter fourth-order symmetry energy. The following limits for symmetry energy and slope parameter have been determined by combining the original PREX finding with the recently released PREX-2 measurement: $J$ = (38.1 $\pm$ 4.7) MeV and $L$ = (106 $\pm$ 37) MeV \cite{PhysRevLett.126.172502,PhysRevLett.126.172503}.
 
 The incompressibility of nuclear matter $K$ at saturation is the amount of nuclear matter that can be compressed and is defined as
\begin{equation}
K = 9\rho_0 \frac{\partial^2 \mathcal{E}}{\partial \rho^2}\Bigg|_{\rho = \rho_0}.
\end{equation}
The current accepted value of $K$ = 240 $\pm$ 20 MeV is determined from the isoscalar giant monopole resonance (ISGMR) for $^{90}Zr$ and $^{208}Pb$ \cite{Colo:2013yta,Piekarewicz:2013bea}.

The symmetry energy and its density dependence have a strong correlation between the pressure (at $\rho \approx \rho_0$) inside a neutron star and its radius \cite{Lattimer_2001}. Studies have also shown that the slope parameter $L$ is related to the neutron skin thickness \cite{PhysRevC.72.064309}. A large value of $L$ corresponds to a higher  neutron matter pressure, and a thicker neutron skin \cite{PhysRevLett.106.252501,PhysRevLett.86.5647}. It is found that the value of the parameters $L$, $K_{sym}$, and $Q_{sym}$ have a huge impact on the radius-mass relation of a neutron star. The more accurate values of these parameters may come from future experiments or better knowledge of neutron star MR relation.
\section{Parameter sets}
\subsection{RMF model}
The nucleon coupling constants for a given parameter sets are denoted by the symbols $g_{\sigma}$, $g_{\omega}$, $g_{\rho}$, $k_3$, $k_4$, $\zeta_0$, $\eta_1$, $\eta_2$, $\eta_{\rho}$, and $\Lambda_{\omega}$. All these coupling constants display different values for different parameter sets. The coupling constants for QHD-I were fitted to reproduce the nuclear matter saturation at Fermi momentum of 1.30 fm$^{-1}$ and binding energy of -15.75 MeV. This basic relativistic Lagrangian has the contribution from $\sigma$ and $\omega$ mesons without any self-coupling terms which is the original Walecka model \cite{HOROWITZ1981503} as discussed in Sec. \ref{qhd}. The prediction of the nuclear incompressibility $K$ by this model is very large ($\approx$ 550 MeV) \cite{WALECKA1974491} and hence the self-coupling terms were added by Boguta and Bodmer in $\sigma$ meson to minimize the value of $K$. With the added coupling terms, several parameter sets like NL1 \cite{Reinhard:1989zi}, NL2 \cite{Reinhard:1989zi}, NL3 \cite{PhysRevC.55.540} were produced, which provided the results well within the range \cite{GAMBHIR1990132}. With this, the problem of incompressibility and finite nuclei was solved, but the equation of states at high-density region was quite stiff and the mass-radius of neutron stars were quite high. The addition of vector meson self-coupling allowed the formation of new parameter sets \cite{Bodmer:1991hz,PhysRevC.68.054318}, which explained both finite nuclei and infinite nuclear matter properties with greater accuracy. \par 
The contribution of isoscalar and isovector cross-couplings with new parameter sets such as FSUGold \cite{PhysRevLett.95.122501} and IU-FSU \cite{PhysRevC.82.055803} have a huge effect on neutron star radius without compromising the predictive power of finite nuclei. The introduction of $\delta$  meson \cite{KUBIS1997191,PhysRevC.89.044001} influences various quantities like symmetry energy, neutron skin thickness, neutron-proton effective masses. While the effect of $\delta$ meson on the properties of finite nuclei is minimal due to low isospin asymmetry, its contribution to the strong isospin asymmetry matter at high densities like neutron stars is large and hence the contribution of $\delta$ meson should be considered.  The inclusion of cross-couplings has a huge impact on neutron-skin thickness, symmetry energy, and radius of NSs. The different coupling constants, nucleon masses, and meson masses for different parameter sets are shown in Table \ref{tab2.1}.

To calculate the symmetry energy and all other parameters for a hadron EoS, different parameter sets NL3 \cite{PhysRevC.55.540}, FSUGarnet \cite{CHEN2015284}, G3 \cite{KUMAR2017197} and IOPB-I \cite{PhysRevC.97.045806} have been used. The NM  properties for the hadron EoS at saturation density $J$, $L$, $K_{sym}$, and $Q_{sym}$ for all parameter sets are listed in Table \ref{tab2.2}. For NL3 set, the symmetry energy $J$ = 37.43 MeV and slope parameter $L$ = 118.65 MeV are little higher than the empirical value $J$ = 31.6 $\pm$ 2.66 MeV, and $L$ = 58.9 $\pm$ 16 MeV  \cite{LI2013276}. The $J$ and $L$ for other parameter sets lie well within the given range. The incompressibility of the given parameter sets lies within the range 240 $\pm$ 20 MeV with NL3 set producing a little higher value than the rest. The G3 set predicts a more accurate value of $K$ = 243.96 MeV, which shows that the contribution of $\delta$ mesons is necessary for the high dense matter. The value of incompressibility for different parameter sets is compatible with the observational data from various experiments. The value of the incompressibility parameter at saturation density is an important feature of nuclear matter. It appears as a parameter in the calculations of mass spectrum and properties of neutron stars, which are important in understanding nuclear matter at high densities.
\vspace{0.2cm}

\begin{table}[hbt!]
	\centering    
	\caption{\label{tab2.1} Nucleon mass, meson mass, and coupling constants of various RMF parameter sets for hadron matter. For all the sets, the nucleon mass is $M$= 939.0 MeV. All the coupling constants are dimensionless.}
		\begin{tabular}{ ccccc }
			\hline
			&NL3&FSUGarnet&G3&IOPB-I \\
			\hline
			$m_s/M$ & 0.541&0.529&0.559&0.533\\
			$m_{\omega}/M$ &0.833&0.833&0.832&0.833\\
			$m_{\rho}/M$&0.812&0.812&0.820&0.812\\
			$m_{\delta}/M$&0.0&0.0&1.043&0.0\\
			$g_s/{4\pi}$&0.813&0.837&0.782&0.827\\
			$g_{\omega}/{4\pi}$&1.024&1.091&0.923&1.062\\
			$g_{\rho}/{4\pi}$&0.712&1.105&0.962&0.885\\
			$g_{\delta}/{4\pi}$&0.0&0.0&0.160&0.0\\
			$k_3$&1.465&1.368&2.606&1.496\\
			$k_4$&-5.688&-1.397&1.694&-2.932\\
			$\zeta_0$&0.0&4.410&1.010&3.103\\
			$\eta_1$&0.0&0.0&0.424&0.0\\
			$\eta_2$&0.0&0.0&0.114&0.0\\
			$\eta_{\rho}$&0.0&0.0&0.645&0.0\\
			$\Lambda_{\omega}$&0.0&0.043&0.038&0.024\\
			$\alpha_1$&0.0&0.0&2.000&0.0\\
			$\alpha_2$&0.0&0.0&-1.468&0.0\\
			$f_{\omega}/4$&0.0&0.0&0.220&0.0\\
			$f_{\rho}/4$&0.0&0.0&1.239&0.0\\
			$\beta_{\sigma}$&0.0&0.0&-0.087&0.0\\
			$\beta_{\omega}$&0.0&0.0&-0.484&0.0\\
			\hline
		\end{tabular}
\end{table}

\begin{table}[hbt!]
	\centering    
	\caption{\label{tab2.2} Nuclear matter properties like symmetry energy $J$, slope parameter $L$, and other higher order derivatives of various RMF parameter sets for hadron matter. The empirical/experimental values of nuclear matter properties are also shown. }
	\begin{tabular}{ cccccc }
		\hline
		&NL3&FSUGarnet&G3&IOPB-I &Emp./Exp.\\
		\hline
		$\rho_0$ (fm$^{-3}$) & 0.148 &0.153&0.148&0.149&0.148 - 0.185 \cite{doi:10.1146/annurev.ns.21.120171.000521}\\
		$\epsilon_0 $(MeV) & -16.29&-16.23&-16.02&-16.10&-(15.00 - 17.00) \cite{doi:10.1146/annurev.ns.21.120171.000521} \\
		M*/M&0.595&0.578&0.699&0.593&---\\
		$J $(MeV)  & 37.43&30.95&31.84&33.30&30.20 - 33.70 \cite{DANIELEWICZ20141} \\
		$L $(MeV) &118.65&51.04&49.31&63.58 &35.00 - 70.00 \cite{DANIELEWICZ20141}\\
		$K_{sym}$ (MeV) &101.34&59.36&-106.07&-37.09 &-(174 - 31) \cite{zimmerman2020measuring}\\
		$Q_{sym}$ (MeV)& 177.90&130.93&915.47&862.70 &---\\
		$K$ (MeV)&271.38&229.5&243.96&222.65&220 - 260
		 \cite{Colo2014}\\
		\hline
	\end{tabular}
\end{table}

\subsection{DD-RMF model}
In this thesis work, several DD-RMF parameterizations are also used. Recently proposed DD-RMF parameters like DD-MEX \cite{TANINAH2020135065}, DD-LZ1 \cite{ddmex}, and DDV, DDVT, DDVTD \cite{typel} have been used. All these parameter sets were obtained by different groups by fitting the ground state properties of finite nuclei. These parameter sets include the necessary tensor couplings of the vector mesons to nucleons apart from the basic couplings. Apart from the above, we also used DD-ME1 \cite{PhysRevC.66.024306} and DD-ME2 \cite{PhysRevC.71.024312} parameter sets.

The nucleon and meson masses and the coupling constants between nucleon and mesons for DD-LZ1, DD-ME1, DD-ME2, DD-MEX, DDV, DDVT, and DDVTD  parameter sets are shown in Table \ref{tab2.3}. The independent parameters $a$, $b$, $c$, and $d$ for $\sigma$, $\omega$, and $\rho$ mesons are also shown. None of the mentioned parameter sets in the Table \ref{tab2.3} includes the contribution from delta meson and hence its mass and coupling constants are not shown here.

\begin{table}[hbt!]
	\centering
	\caption{Nucleon and meson masses and different coupling constants for various DD-RMF parameter sets. }
	\begin{tabular}{ cccccccc }
		\hline
		\hline
		&DD-LZ1&DD-ME1&DD-ME2&DD-MEX&DDV&DDVT&DDVTD \\
		\hline
		$m_n$ & 938.900&939.000&939.000&939.000&939.565&939.565&939.565\\
		$m_p$&938.900&939.000&939.000&939.000&938.272&938.272&938.272\\
		$m_{\sigma}$&538.619&549.525&550.124&547.333&537.600&502.599&502.620\\
		$m_{\omega}$&783.000&783.000&783.000&783.000&783.000&793.000&783.000\\
		$m_{\rho}$&769.000&763.000&763.000&763.000&763.000&763.000&763.000\\
		$g_{\sigma}(\rho_0)$&12.001&10.443&10.539&10.707&10.137&8.383&8.379\\
		$g_{\omega}(\rho_0)$&14.292&12.894&13.019&13.339&12.770&10.987&10.980\\
		$g_{\rho}(\rho_0)$&15.151&7.611&7.367&7.238&7.848&7.697&80.060\\
		\hline
		$a_{\sigma}$&1.063&1.385&1.388&1.397&1.210&1.204&1.196\\
		$b_{\sigma}$&1.764&0.978&1.094&1.335&0.213&0.192&0.192\\
		$c_{\sigma}$&2.309&1.534&1.706&2.067&0.308&0.278&0.274\\
		$d_{\sigma}$&0.380&0.466&0.442&0.402&1.040&1.095&1.103\\
		$a_{\omega}$&1.059&1.388&1.389&1.393&1.237&1.161&1.169\\
		$b_{\omega}$&0.418&0.852&0.924&1.019&0.039&0.0456&0.026\\
		$c_{\omega}$&0.539&1.357&1.462&1.606&0.072&0.067&0.042\\
		$d_{\omega}$&0.787&0.496&0.477&0.456&2.146&2.227&2.806\\
		$a_{\rho}$&0.776&0.501&0.565&0.620&0.333&0.549&0.558\\
		\hline
		\hline
	\end{tabular}
	\label{tab2.3}
\end{table}

It is necessary to mention that the coefficients of meson coupling constants $g_i$, $i$ = $\sigma$, $\omega$, $\rho$ in DD-LZ1 parameter set are the values at zero density, while for other parameter sets, the values obtained are at nuclear saturation density ($\rho_0$). 

The symmetry energy parameter $J$ for the listed parameter sets are compatible with the $J$ = (31.6 $\pm$ 2.66) and (38.1 $\pm$ 4.7) MeV obtained from various astrophysical observations \cite{LI2013276}. The symmetry energy slope parameter $L$ also satisfies the recent constraints $L$ = (59.57 $\pm$ 10.06) MeV \cite{DANIELEWICZ20141}. The $K_0$ value for all the given parameter sets satisfies the range $K_0$ = 240 $\pm$ 20 MeV except for the DD-MEX which predicts a little higher value.
\begin{table}[hbt!]
	\centering
	\caption{NM properties Binding energy ($E/A$), incompressibility ($K_0$), symmetry energy ($J$), slope parameter ($L$) in units of MeV at saturation density $\rho_0$ (fm$^{-3}$) for various DD-RMF parameter sets. }
	\begin{tabular}{ cccccccc }
		\hline
		\hline
		&DD-LZ1&DD-ME1&DD-ME2&DD-MEX&DDV&DDVT&DDVTD \\
		\hline
		$\rho_0$ &0.158&0.152&0.152&0.152&0.151&0.154&0.154\\
		$E/A$&-16.126&-16.668&-16.233&-16.140&-16.097&-16.924&-16.915\\
		$K_0$&231.237&243.881&251.306&267.059&239.499&239.999&239.914\\
		$J$&32.016&33.060&32.310&32.269&33.589&31.558&31.817\\
		$L$&42.467&55.428&51.265&49.692&69.646&42.348&42.583\\
		$M_n^*/M$&0.558&0.578&0.572&0.556&0.586&0.667&0.667\\
		$M_p^*/M$&0.558&0.578&0.572&0.556&0.585&0.666&0.666\\
		\hline
		\hline
	\end{tabular}
	\label{tab2.4}
\end{table}

\begin{figure}[hbt!]
	\centering
	\includegraphics[width=0.75\textwidth]{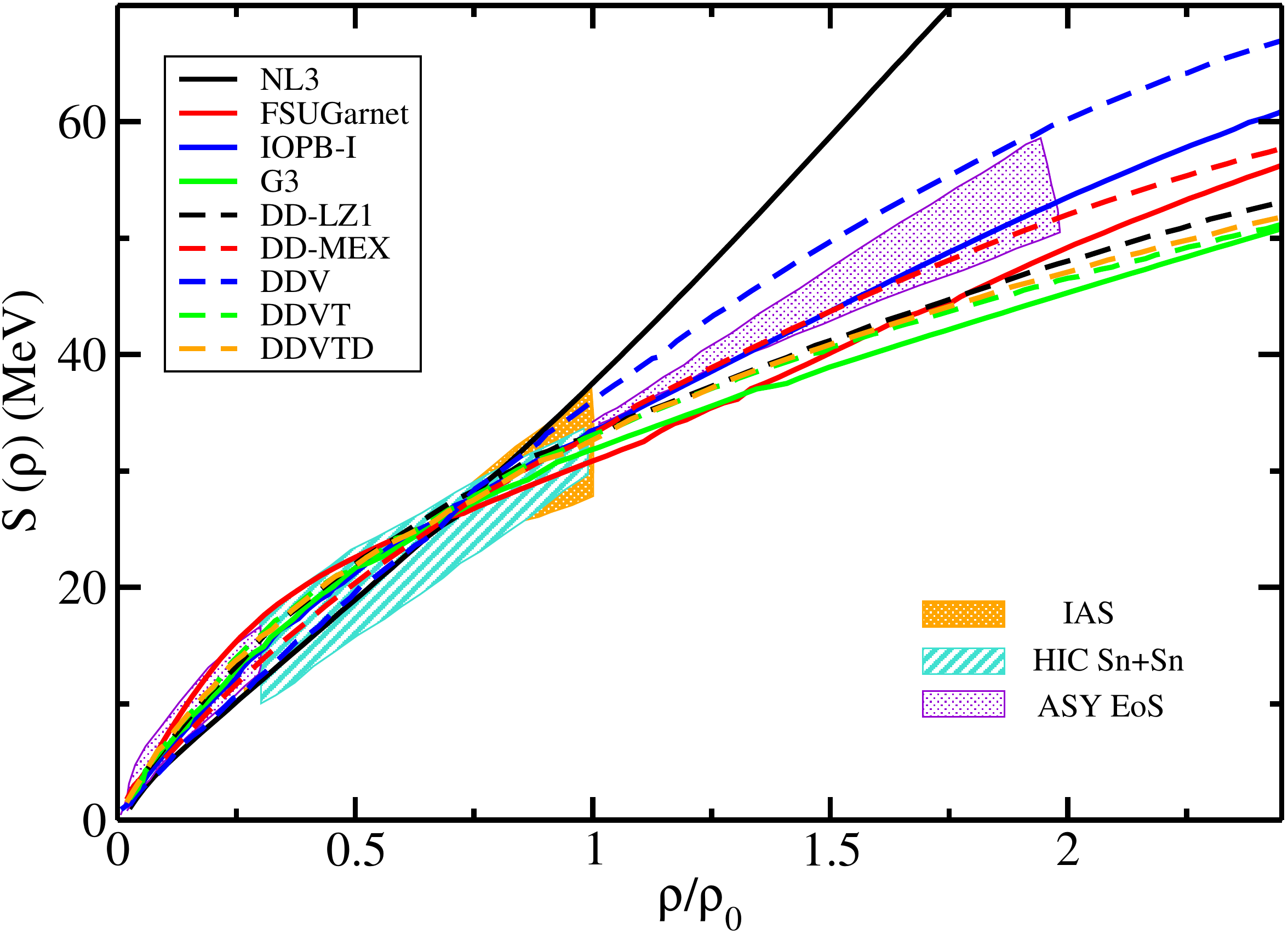}
	\caption{Density dependence of Symmetry energy for the given RMF (solid lines) and DD-RMF (dashed lines) parameter sets. The shaded regions represent the constraint on the symmetry energy from IAS \cite{DANIELEWICZ20141}, HIC Sn+Sn \cite{PhysRevLett.102.122701}, and ASY-EoS experimental data \cite{PhysRevC.94.034608}. }
	\label{figsym}
\end{figure}

Fig.~\ref{figsym} shows the density-dependent symmetry energy for the RMF parameter sets NL3, FSUGarnet, IOPB-I, and G3 as well as DD-RMF parameter sets DD-LZ1, DD-MEX, DDV, DDVT, and DDVTD. The constraints from various experimental measurements are also shown. The symmetry energy for NL3 set satisfies the IAS and HIC data at low-density region, but predicts a stiff $S(\rho)$ at high-density. The DDV parameter set satisfies the symmetry energy constraint in the low-density region, but stiffens as the density increases. DDVT and DDVTD parameter sets provide a soft $S(\rho)$ with the increasing density similar to FSUGarnet and G3 parameter sets. All the RMF and DD-RMF parameter sets satisfy all the low-density symmetry energy constraints from various experimental data. 

\section{Infinite Nuclear Matter}
\subsection{Symmetric nuclear matter}
An infinite system with equal number of protons and neutrons is termed as symmetric nuclear matter (SNM). Since there is no asymmetry present between neutrons and protons, the asymmetry parameter $\alpha$ = $(\rho_n-\rho_p)/(\rho_n+\rho_p)$ becomes zero. This implies that 
\begin{equation}
\rho_n=\rho_p 
\end{equation}
Furthermore, the SNM is electrically charge neutral which implies that $\rho_e$ = 0. For a chosen baryon density $\rho$ = $\rho_n+\rho_p$, the field equations are solved to calculate meson fields and fermi momentum of nucleons which is related to the density as

\begin{equation}
\rho_n=\frac{k_{fn}^3}{3\pi^2}, \hspace{0.2cm} \rho_p=\frac{k_{fp}^3}{3\pi^2}.
\end{equation} 

\begin{figure}[hbt!]
	\centering
	\includegraphics[width=0.77\textwidth]{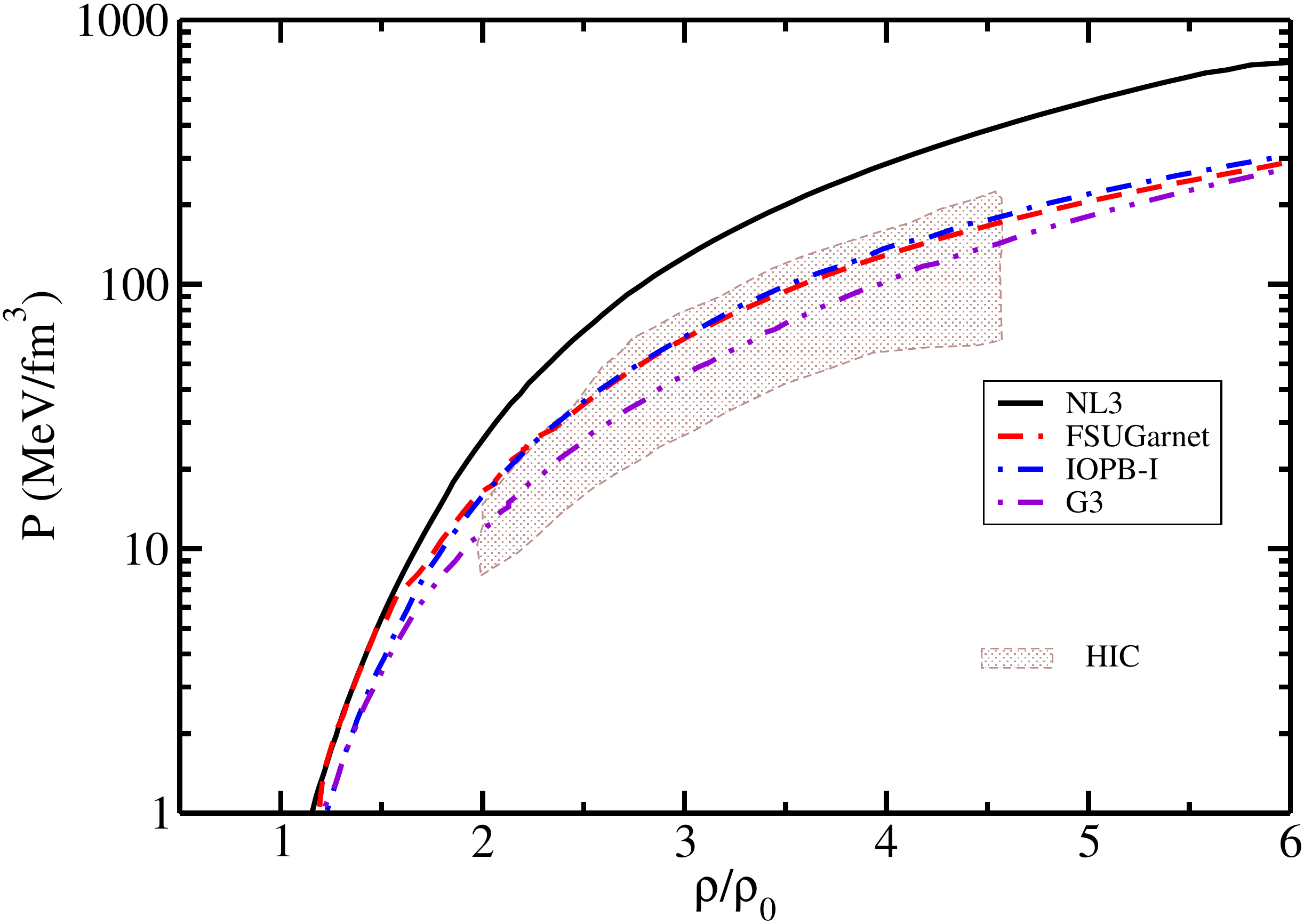}
	\caption{Pressure variation with baryon density for symmetric nuclear matter for the given parameter sets. The shaded area represents the experimental data from HIC \cite{Danielewicz1592}.}
	\label{fig2.2}
\end{figure}

Fig.~\ref{fig2.2} shows the variation of SNM pressure with baryon density for NL3, FSUGarnet, IOPB-I, and G3 parameter sets. The calculated pressure for the G3 set shows an excellent agreement with the heavy-ion collision (HIC) data for the whole density range. The parameter sets like IOPB-I and FSUGarnet, although produce stiff EoSs as compared to G3, also agree with the  HIC data. The NL3 parameter set produces very stiff EoS and disagrees with the HIC data. The addition of higher-order couplings soften the EoS \cite{PhysRevC.82.055803}. The $\zeta_0$ self-coupling of the $\omega$ meson (Eq.~(\ref{eq1})) is efficient in softening the EoS at supra-normal densities and the isoscalar-isovector coupling $\Lambda_{\omega}$ softens the symmetry energy considerably as seen in case of IOPB-I and G3 parameter sets \cite{PhysRevLett.86.5647,PhysRevC.64.062802}.

\subsection{Pure neutron matter}
For the case of pure neutron matter (PNM), the total baryon density is equal to the neutron density, while the proton and electron densities are equal to zero. For such case
\begin{equation}
\alpha=\frac{\rho_n-\rho_p}{\rho_n+\rho_p}=1
\end{equation}
which implies that $\rho_p$ = 0. For charge neutral condition, $\rho_e$ = 0.

\begin{figure}[hbt!]
	\centering
	\includegraphics[width=0.77\textwidth]{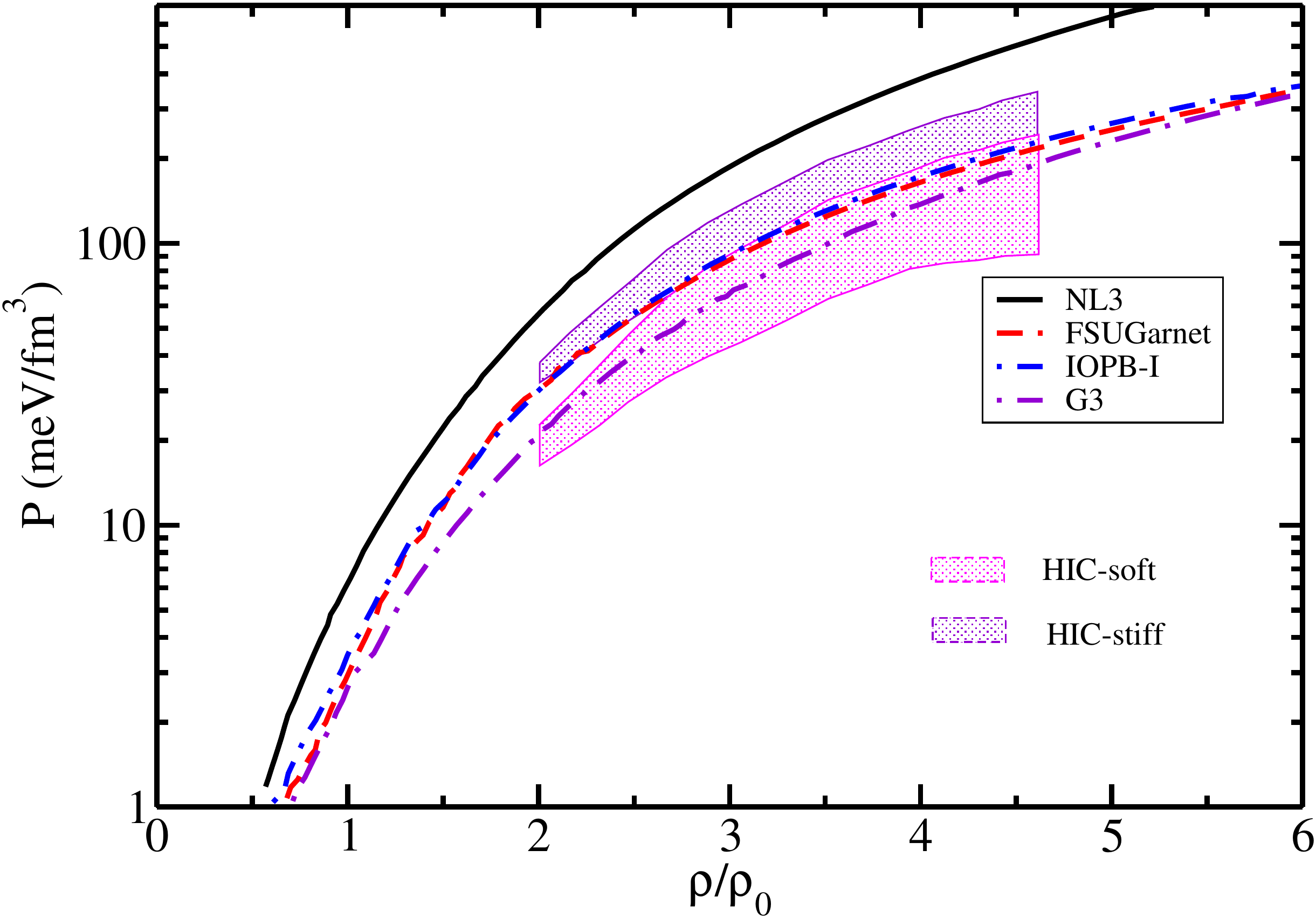}
	\caption{Pressure variation with baryon density for Pure neutron matter for the given parameter sets. The shaded area represents the experimental data from HIC \cite{Danielewicz1592}.}
	\label{fig2.3}
\end{figure}

Fig.~\ref{fig2.3} shows the variation of pressure for PNM with baryon density, which in this case is neutron density, for NL3, FSUGarnet, IOPB-I, and G3 parameter sets. The shaded regions represent the stiff and soft HIC data \cite{Danielewicz1592}. The results produced are similar to SNM, but the PNM shows a soft EoS at low density. FSUGarnet produces more soft EoS than IOPB-I at higher densities. All parameter sets agree with the HIC-soft data except the NL3 parameter set, which produces very stiff EoS at both low and high density. 

\subsection{Neutron star matter}
For neutron star matter, where the baryons are strongly interacting particles, the $\beta$-equilibrium, and charge neutrality are two important conditions to be satisfied to determine the composition of the system. For any baryon $B$, the relation $\mu_B = b_B \mu_n - q_B \mu_e$, where $\mu_B$ is the chemical potential with charge $q_B$ and baryon number $b_B$, represents the beta-equilibrium condition. For the present case with $n,p$, $e$, and $\mu$ only, the $\beta$-equilibrium condition is given by the chemical potential of proton $\mu_p$, neutron $\mu_n$, and electron $\mu_e$ as \cite{gled}
\begin{align}\label{c1}
\mu_p &= \mu_n - \mu_e,\nonumber \\
\mu_e&=\mu_u. 
\end{align}
The chemical potential of a baryon can thus be obtained from these two independent chemcial potentials $\mu_n$ and $\mu_e$. The charge neutrality condition is given by

\begin{equation}
q_{total} = \sum_{i=n,p} q_i k_i^3/(3\pi^2)+\sum_{l=e,\mu} q_l k_l^3/(3\pi^2)=0,
\end{equation}

which implies, $n_p$ = $n_e$ +$n_{\mu}$, where $n_p$, $n_e$, and $n_{\mu}$ are the number densities of proton, electron, and muon, respectively. 
The total energy density and pressure of neutron star matter is then given as

\begin{align}
\mathcal{E}& = \mathcal{E}_H +\mathcal{E}_l,\nonumber \\
P&=P_H+P_l.
\end{align}

Here, $\mathcal{E}_l$ and $P_l$ are the lepton energy density and pressure. $\mathcal{E}_H$ and $P_H$ are the hadronic energy density and pressure which follow from Eqs. (\ref{eq9}) and (\ref{eq10}) for RMF and Eqs. (\ref{eq18}) and (\ref{eq19}) for DD-RMF model. 
\begin{equation}
\mathcal{E}_l= \sum_{l=e,\mu} \frac{2}{(2\pi^3)}\int_0^{k_l} d^3k \sqrt{k^2 +m_l^2},
\end{equation}
and
\begin{equation}
P_l= \sum_{l=e,\mu} \frac{2}{3(2\pi^3)}\int_0^{k_l} d^3k k^2 /(\sqrt{k^2 +m_l^2})
\end{equation}

The total baryon density in the NSM case is given as
\begin{equation}
\rho=\rho_n+\rho_p .
\end{equation}

\begin{figure}[hbt!]
	\centering
	\includegraphics[width=0.77\textwidth]{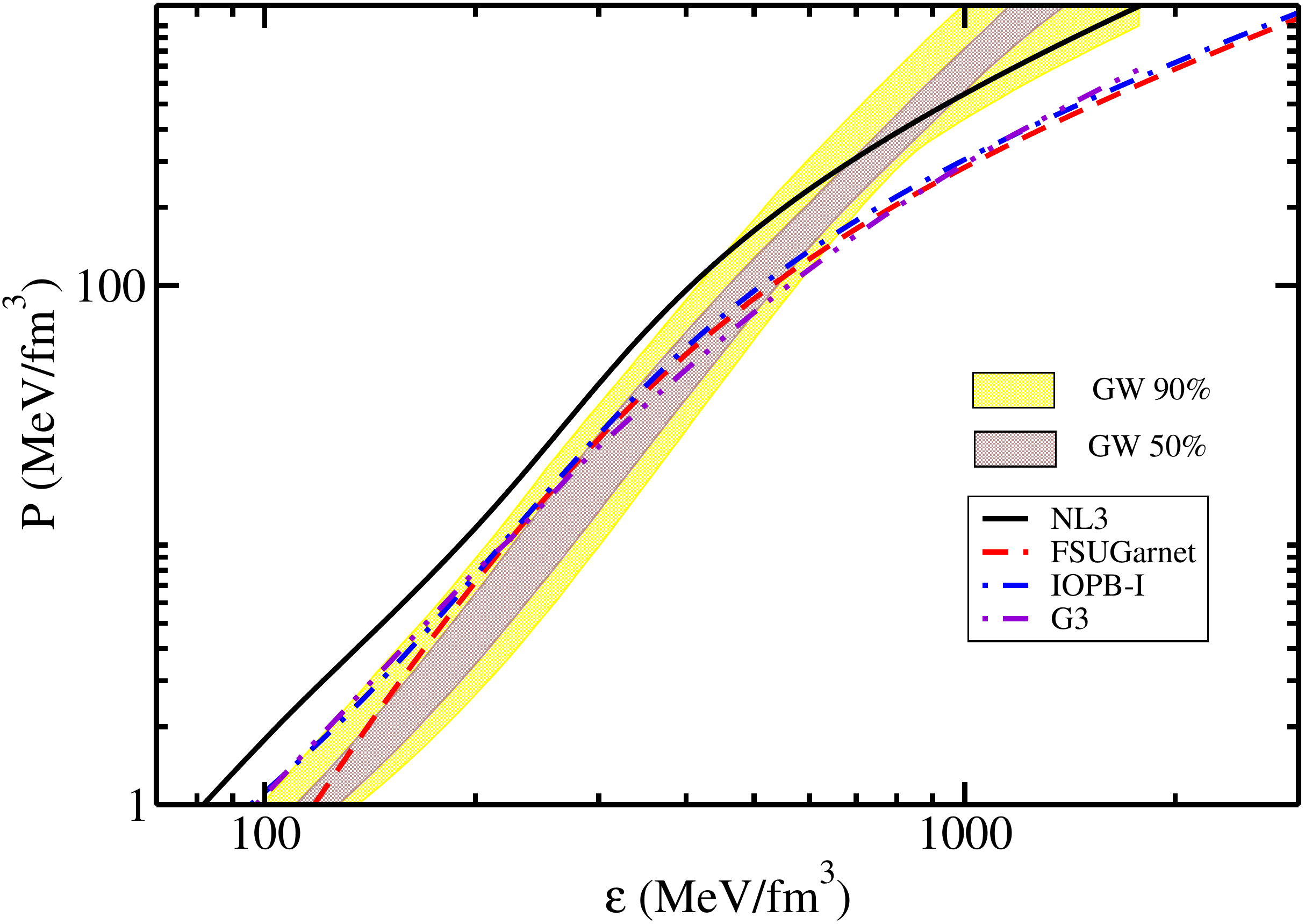}
	\caption{Equation of state of $\beta$-equilibrated  and charge neutral matter for NL3, FSUGarnet, IOPB-I, and G3 parameter sets. The shaded regions represent 50\% (brown) and 90\% (yellow) posterior credible limit set by GW170817 data \cite{PhysRevLett.121.161101}.}
	\label{fig2.4}
\end{figure}

Fig.~\ref{fig2.4} shows the EoS for neutron star matter in $\beta$-equilibrium and charge-neutral conditions for NL3, FSUGarnet, IOPB-I, and G3 parameter sets. The shaded regions represent the 50\% (brown) and 90\% (yellow) posterior credible limits by GW170817 data \cite{PhysRevLett.121.161101}. The NL3 EoS satisfies the posterior credible limits at higher densities while producing stiff EoS at low energy density. FSUGarnet, IOPB-I, and G3 EoSs satisfy the GW170817 constraint at low density and produce a soft EoS at higher densities.  With these EoSs, the TOV equations can be used to determine the properties of neutron star \cite{PhysRev.55.364,PhysRev.55.374}.
\begin{savequote}[8cm]
\textlatin{No research is ever quite complete. It is the glory of a good bit of work that it opens the way for something still better, and this repeatedly leads to its own eclipse.}

  \qauthor{---\textit{Mervin Gordon}}
\end{savequote}

\chapter{\label{ch:3-hybridtstar}Quark Matter in Neutron stars}

\minitoc


\section{Introduction}
Since the Quark matter is by assumption completely stable, it may be the true ground state of the hadronic matter \cite{PhysRevD.30.272,PhysRevD.30.2379}. So the quark matter, the deconfined quark phase, is quite likely to occur in the inner regions of compact objects like neutron stars. It may exist both as a pure phase in the central regions and as a mixed phase with hadronic matter \cite{PhysRevD.46.1274}. The neutron stars with a hadronic crust and a quark core (pure or mixed) are termed as \textit{"Hybrid stars"}.

In nuclear physics and nuclear astrophysics, the EoS plays a very crucial role in understanding the nature of the matter in finite and infinite nuclear matter \cite{Danielewicz1592,Lattimer:2015nhk,RevModPhys.88.021001}. The binding energy per nucleon $e(\rho,\alpha$) = $\mathcal{E}/A$ and the isospin asymmetry $\alpha$ = $(\rho_n - \rho_p)/\rho$ are one of the basic inputs for calculating the pressure and the energy density (EoS) of NSM. The symmetry energy $S(\rho)$ and other quantities have a huge impact on the EoS. However, the $S(\rho)$ cannot be measured directly, so fully depends on the theoretical models. Unfortunately, these models predict a wide range of symmetry energy \cite{PhysRevC.90.055203}. At saturation density $(\rho_0)$, all these quantities are known more or less to a good extent, but the results are very much uncertain for the densities above $\rho_0$. While many theoretical models predict the symmetry energy $S(\rho)$ to be increasing with the density, several other models predict that the $S(\rho)$ increases with the density up to $\rho_0$ and thereafter decreases \cite{PhysRevLett.106.252501}.  At densities around 2-8 $\rho_0$, the symmetry energy and the higher derivatives such as the slope parameter $L=3\rho_0 S'(\rho_0)$, curvature of symmetry energy $K_{sym}=9\rho_0^2 S''(\rho_0)$, $Q_{sym}=27\rho_0^3 S'''(\rho_0)$ and also the incompressibility plays a key role in determining the structure and properties of neutron stars \cite{PhysRevC.98.065801} and the possibility of the exotic phases \cite{THORSSON1994693}. \par 
The properties of a NS such as its composition, mass, radius, etc. depend upon the EoS. The outer part of the neutron star where the density is low ($\approx \rho_0$) is mainly described by the hadronic matter. As the density increases (> 3-4 $\rho_0$), a phase transition from hadronic matter to quark matter is possible, where a mixed hadron-quark phase is formed for a certain density range followed by a pure quark phase.\par 
In the present work, we combine the two phases to build a single hybrid EoS. We calculate the nuclear matter properties for hybrid EoS and the effect of bag constant on these properties. For hadronic matter, the E-RMF model is employed by using recently proposed different parameter sets as discussed in Sec. \ref{rmf}. The MIT Bag Model is used to describe the Unpaired Quark Matter (UQM) \cite{PhysRevD.9.3471,PhysRevD.30.2379}. 

The theoretical approach employed to study the EoS of the quark matter and the phase transition to hadron matter are discussed in Sec. \ref{quarkmatter} and \ref{phase}. The NM properties like symmetry energy and other quantities for hybrid EoS are obtained and discussed in Sec. \ref{c3results}, which is followed by the summary along with the conclusion in Sec. \ref{c3conclusion}. The calculations discussed in this chapter are based on the work from Ref.~\cite{Rather_2020,Rather2020_1}. 

\section{Formalism}
\label{sec:headings}

\subsection{ Quark Matter}
\label{quarkmatter}
In the central part of the NS, the density is presumed to be high enough for the hadron matter (HM) to undergo a phase transition to quark matter (QM). This transition leads to the formation of a mixed-phase at the density that varies from saturation density $\rho_0$ to few times $\rho_0$ depending upon the properties of NS and the models used. For the quark phase, we employ the simple MIT Bag model for the unpaired quark matter \cite{PhysRevD.9.3471,PhysRevD.30.2379}. This model is a degenerate Fermi gas of quarks ($u$, $d$, and $s$) and electrons with chemical equilibrium being maintained by several weak interactions. In this model, the quarks are assumed to be confined in a colorless region where the quarks are free to move. The quark masses considered are $m_u$ = $m_d$ = 5.0 MeV and $m_s$ = 150 MeV. For the present work, we ignore the one gluon exchange inside the gas. The equilibrium condition satisfied by the quark matter is
\begin{equation}\label{c2}
\mu_d=\mu_s=\mu_u + \mu_e.
\end{equation}
The chemical potential of the individual quark follows from the neutron and electron chemical potentials $\mu_n$ and $\mu_e$ respectively as:
\begin{equation}
\mu_u =\frac{1}{3}\mu_n -\frac{2}{3}\mu_e,
\end{equation}
\begin{equation}
\mu_d =\frac{1}{3}\mu_n +\frac{1}{3}\mu_e,
\end{equation}
and
\begin{equation}
\mu_s =\frac{1}{3}\mu_n +\frac{1}{3}\mu_e.
\end{equation}
The charge neutrality condition obtained is
\begin{equation}
\frac{2}{3}n_u-\frac{1}{3}n_d -\frac{1}{3}n_s -n_e=0,
\end{equation}
where $n_q$ $(q=u,d,s)$, the total quark matter density is given as
\begin{equation}
n_q = \frac{1}{3}(n_u +n_d +n_s).
\end{equation}

The pressure of the quarks $(q=u,d,s)$ is given by \cite{book}
\begin{equation}
P_Q = \frac{1}{4\pi^2}\sum_q \Biggl\{\mu_q k_q \Bigg(\mu_q^2 -\frac{5}{2}m_q^2\Bigg)+\frac{3}{2}m_q^4 ln \Bigg(\frac{\mu_q +k_q}{m_q}\Bigg)\Biggr\}.
\end{equation}
The total pressure due to quarks and leptons is given by
\begin{equation}\label{eq16}
P=P_Q +P_l -B,
\end{equation}
The expression for the quark energy density is
\begin{equation}\label{eq17}
\mathcal{E}_Q = \frac{3}{4\pi^2}\sum_q \Biggl\{\mu_q k_q \Bigg(\mu_q^2 -\frac{1}{2}m_q^2\Bigg)-\frac{1}{2}m_q^4 ln \Bigg(\frac{\mu_q +k_q}{m_q}\Bigg)\Biggr\}+ B.
\end{equation} 
where $B$ is the Bag constant. The bag constant is defined as the difference between the energy densities of the perturbative and non-perturbative vacuums (true ground state of QCD). The pressure exerted by the freely moving quarks at the surface of the bag can make the bag unstable. To prevent this an external pressure defined as the Bag pressure $B$ is applied to compensate the internal pressure of the system. The quarks are assumed to have a very low mass inside the bag as compared to that outside, where the mass is very high. To balance the behavior of the bag and to find its size, a bag constant $B$ is introduced as a constant energy density in the system. At the surface of the bag, the outward pressure produced by the quarks is balanced by the inward pressure $B$. Thus the quark pressure decreases with the increasing value of $B$ thereby influencing the structure of the star. With a very low mass of $u$ and $d$ quarks, the value of $B$ depends on the mass of the strange quark. The value of $B$ varies from $B^{1/4} \approx$ 145-160 MeV for massless strange quark \cite{Stergioulas2003}. This range narrows down with the increase in the mass of strange quark. However, different bag values have been used in the literature. The bag value  $B^{1/4} \approx$ 200 MeV is used in the QCD calculations by \citet{Satz}. Also, following the  CERN-SPS and RHIC data, the bag constants are allowed to have a wider range \cite{PhysRevC.66.025802, refId0}. So, we can consider the bag constant as an effective free parameter. 

A range of bag constants have been used in the literature \cite{Baym:2017whm,STEINER2000239,Kalam2013}. In the bag model, the standard value of $B$ is taken as $B^{1/4}$= 140 MeV \cite{PhysRevD.22.1198}. But the definite range of bag values for hybrid stars is yet to be obtained. It is important to obtain a definite range of bag values for hybrid stars that will correspond to their stable configuration. The proper choice of bag constant can explain the hybrid stars with constraints imposed from recent gravitational wave observation using the simple MIT bag model. \par 

\subsection{ Phase Transition}
\label{phase}
The deconfined phase transition from hadron matter to quark matter is assumed to be of the first order, so the transition should produce a mixed-phase between the pure hadron phase and pure quark phase. The mixed-phase region between the pure HM and QM is not well-defined \cite{PhysRevD.46.1274}. Beta-equilibrium and charge neutrality conditions determine the density range over which the mixed-phase can exist. The quark-hadron phase transition in neutron stars has been widely studied using different techniques \cite{PhysRevC.75.035808,PhysRevC.66.025802}. Usually, the technique involved in constructing the mixed-phase depends upon the surface tension. Beyond a critical value of the surface tension, the Maxwell construction (MC) \cite{PhysRevD.88.063001} is used. With no specific value of surface tension being known, the Gibbs construction (GC) \cite{PhysRevD.46.1274} is found to be more relevant.
The MC is appropriate to obtain the liquid-vapor phase transition EoS. However, Glendenning \cite{PhysRevD.46.1274,GLENDENNING2001393} pointed out that MC is not appropriate for the hadron-quark phase transition. Glendenning further pointed out that the standard Maxwell formulation is only valid for systems with one particle species and one chemical potential, but in neutron stars, two variables are relevant: the charge and baryon number chemical potentials. The global charge neutrality constraint is applied in GC, which indicates that both the hadron phase and the quark phase are permitted to be charge neutral independently, whereas the local charge neutrality criterion is utilized in Maxwell construction. Also, in GC, the pressure increases with the density in the mixed-phase contrary to Maxwell construction, where the pressure remains constant throughout the phase transition.\par 
The Gibbs conditions for the mixed-phase are given by:
\begin{equation}\label{d1}
P_{HP}(\mu_{HP}) = P_{QP}(\mu_{QP}) = P_{MP},
\end{equation}
and
\begin{equation}\label{d2}
\mu_{HP,i} = \mu_{QP,i} = \mu_i, ~i=n,e.
\end{equation}
In case of two independent chemical potentials which follow from Eqs.~(\ref{c1}) and (\ref{c2}), the gibbs conditions (Eqs.~(\ref{d1}) and (\ref{d2})) can be fulfilled if the coexisting phases have opposite electric charges with global charge neutrality imposed, the baryon density for the mixed-phases then follows from the equation:
\begin{equation}\label{c3e2}
\rho_{MP} = \chi \rho_{QP} +(1-\chi)\rho_{HP}.
\end{equation}
where $\chi$ = $V_Q/V$ represents the quark volume fraction obtained using the global charge neutrality of the mixed-phase within the volume $V$ which implies that the charge density integral $Q=4\pi \int_V dr r^2 q(r)$, must vanish rather than $q(r)$ itself.
\begin{equation}
0=\frac{Q}{V} = (1-\chi)q_H (\mu^n, \mu^e)+\chi q_Q (\mu^n, \mu^e) +q_L,
\end{equation}
where $q_L$ is the lepton charge density. The energy density in the mixed-phase then reads: 
\begin{equation}\label{c3e1}
\varepsilon_{MP} = \chi \varepsilon_{QP} +(1-\chi)\varepsilon_{HP} +\varepsilon_l,
\end{equation}
By definition, the $\chi$ ranges between 0 and 1 depending on how much hadronic matter has been converted to quark matter.

Once the mixed phase is obtained, the Eqs.~(\ref{c3e2}) and (\ref{c3e1}) can be solved to determine the properties of the mixed-phase.

\section{Results and Discussions}
\label{c3results}
To calculate the symmetry energy and all other parameters for a hybrid EoS, we used NL3, FSUGarnet, G3, and IOPB-I parameter sets for hadron matter. The NM  properties for the hadron EoS at saturation density $J$ , $L$, $K_{sym}$ and $Q_{sym}$ for all parameter sets are listed in Table \ref{tab2.1}. 
To obtain energy density and the pressure for NS matter in $\beta$-equilibrium and charge neutrality condition, we solve Eqs.~(\ref{eq9}) and (\ref{eq10}) for different parameter sets.

\begin{figure}[h]
	\centering
	\includegraphics[width=0.70\textwidth]{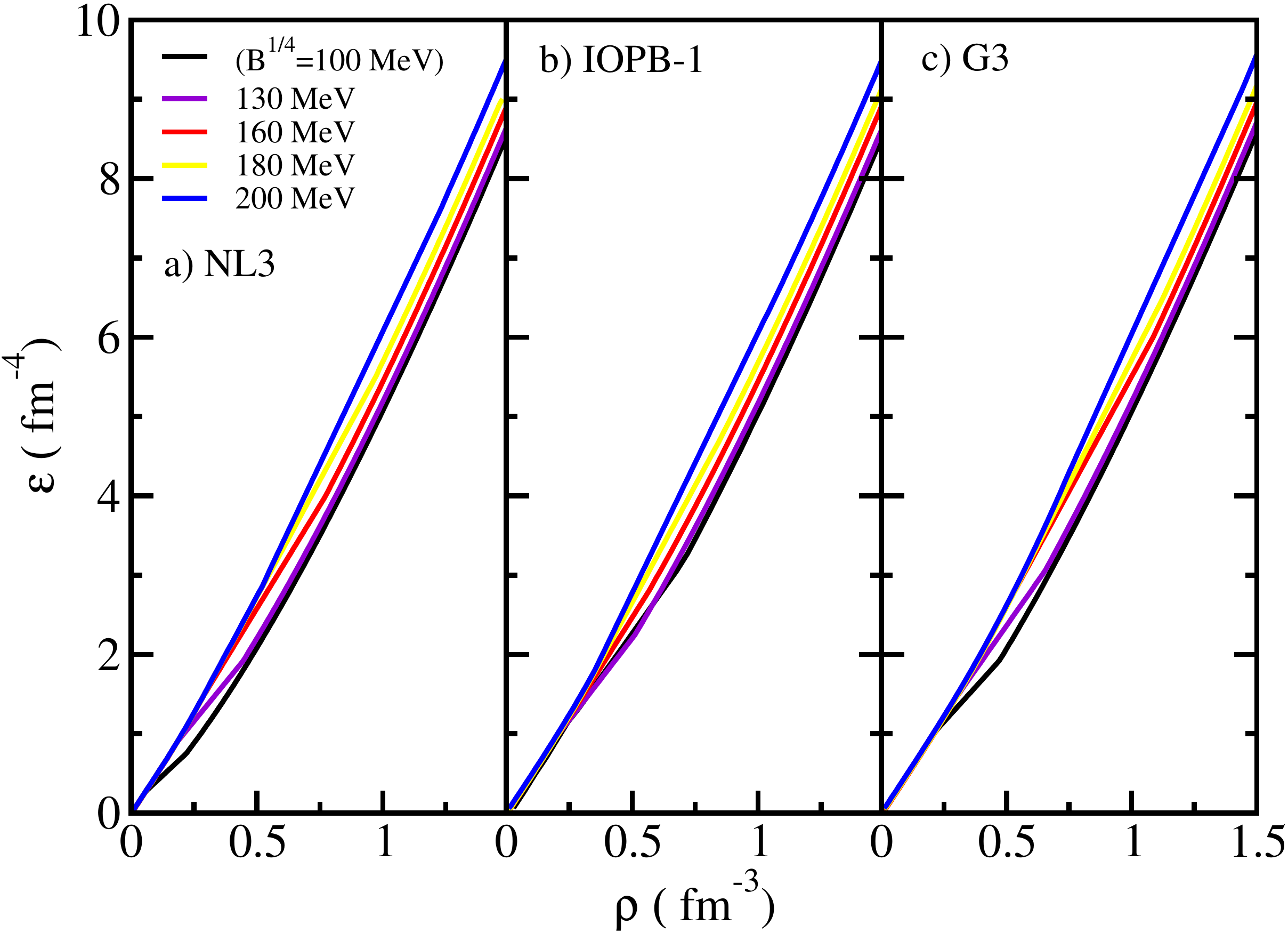}
	\caption{Hybrid EoS for  different bag constants for a) NL3, b) IOPB-I and c) G3 parameter sets.}
	\label{fig3.2}
\end{figure}

To obtain hybrid EoS, we solve Eqs.~({\ref{c3e2}}) and (\ref{c3e1}) together with the hadronic and quark EoS.
All the hybrid EoS for different hadronic matter parameter sets (NL3, IOPB-I, and G3) and different quark matter bag values ($B^{1/4}$ = 100, 130, 160, 180 and 200 MeV) are shown in Fig.~\ref{fig3.2}. The energy density increases with the bag constant and hence the pressure will correspondingly decrease with the bag constant. This implies that the hybrid EoS becomes softer as we increase the bag value. It is to be mentioned that the phase transition density of mixed-phase changes with the bag constant. For small values of $B$, the phase transition takes place below the nuclear saturation density, which is unphysical \cite{Ghosh1995}. As the bag value increases, the phase transition density shifts to higher values. The importance of hybrid EoS lies in the formation of mixed-phase. The transition from HM to QM using Gibbs condition determines the stiffness or softness of the EoS. Due to the stiffness/softness of hybrid EoS by the mixed phase, the nuclear matter properties of hybrid EoS change with the bag constant. 
\begin{table}[ht]
	\centering
	\caption{\label{tab3.1} Transition densities for the mixed phase. $\rho^{MP}_{start}$ and $\rho^{MP}_{end}$ denote the formation and the end of the mixed phase respectively. }
		 \begin{tabular}{ p{2.0cm}p{1.5cm}p{1.5cm}p{1.5cm}p{1.5cm}p{1.5cm} }
			\hline
			\centering
			$B^{1/4}$ (MeV)&100&130&160&180&200 \\
			\hline
			&&NL3\\
			\hline	
			$\rho^{MP}_{start}$($\rho_0$)  & 0.98&1.12&1.81&3.05&4.63 \\
			$\rho^{MP}_{end}$($\rho_0$) &1.43&2.44&3.22&6.12&8.12\\
			
			\hline
			&& IOPB-I\\
			\hline	
			$\rho^{MP}_{start}$($\rho_0$) & 0.96&1.04&1.63&2.96&4.42 \\
			$\rho^{MP}_{end}$($\rho_0$) &1.16&2.14&3.02&5.81&7.93\\
			
			\hline
			&& G3\\
			\hline	
			$\rho^{MP}_{start}$ ($\rho_0$)  & 0.96&1.02&1.58&2.84&4.16\\
			$\rho^{MP}_{end}$ ($\rho_0$) &1.20&2.06&2.92&5.34&7.15\\
			
			\hline
		\end{tabular}
\end{table}

Table \ref{tab3.1} shows the transition densities for the mixed-phase. $\rho^{MP}_{start}$ represents the end of the pure hadron-phase and beginning of hadron-quark mixed-phase, while $\rho^{MP}_{end}$ represents the beginning of pure quark phase. For $B^{1/4}$ = 100 MeV, the transition density from pure hadron phase to mixed phase occurs at around nuclear saturation density. With increasing bag constant, the phase transition density also increases. For $B^{1/4}$ = 200 MeV, the mixed phase region extends from $\approx$ (4-8)$\rho_0$. The mixed phase region broadens with the bag constant.

From the EoSs obtained, quantities like energy density, pressure, and density are now known for the hybrid EoS. The nuclear matter properties like symmetry energy and other quantities for the hybrid EoS at saturation are calculated as shown in Table \ref{tab3.2}. 

\begin{table}[ht]
	\centering
	\caption{\label{tab3.2} NM properties such as symmetry energy, slope parameter, and incompressibility of Mixed EoS for different bag constants. }
		\begin{tabular}{ cccccc }
			\hline
			$B^{1/4}$ (MeV)&100&130&160&180&200 \\
			\hline
			&&NL3\\
			\hline	
			$J $(MeV)  & 45.11&41.72&35.76&32.20&36.84 \\
			$L $(MeV) &130.75&128.12&124.59&121.05&131.78\\
			$K$ (MeV)&580.08&566.95&557.83&554.43&522.84\\
			\hline
			&& IOPB-I\\
			\hline	
			$J $(MeV)  & 35.88&37.86&38.45&43.64&54.54 \\
			$L $(MeV) &69.61&62.28&68.64&72.18&89.32\\
			$K$ (MeV) &455.76&432.61&415.29&401.08&400.85\\
			\hline
			&& G3\\
			\hline	
			$J$ (MeV)  & 37.48&37.89&38.71&51.49&56.17\\
			$L$ (MeV) &55.66&55.82&68.73&76.39&81.05\\
			$K$ (MeV) & 557.03&543.93&540.79&539.58&537.24\\
			\hline
		\end{tabular}
	
\end{table} 
The value of symmetry energy $J$ at saturation and other parameters is very large as compared to the hadronic matter. The $J$ value of hadronic EoS for G3 set is 31.84 MeV, while for G3 hybrid EoS the value is 37.48 MeV for $B^{1/4}$ = 100 MeV and increases with the bag constant. The value of slope parameter for hybrid EoS with G3 set lies in the range (50-80) MeV which is compatible with the astrophysical observations \cite{LI2013276}, but for NL3 hybrid EoS, the $L$ value is very large and lies in the range (120-130) MeV. However, with the recent measurement of slope parameter $L$ = (106 $\pm$ 37) MeV from PREX-2 experiment \cite{PhysRevLett.126.172503}, the values obtained for all the parameter sets at all bag constants satisfy this constraint. 

\begin{figure}[h]
	\centering
	\includegraphics[width=0.75\textwidth]{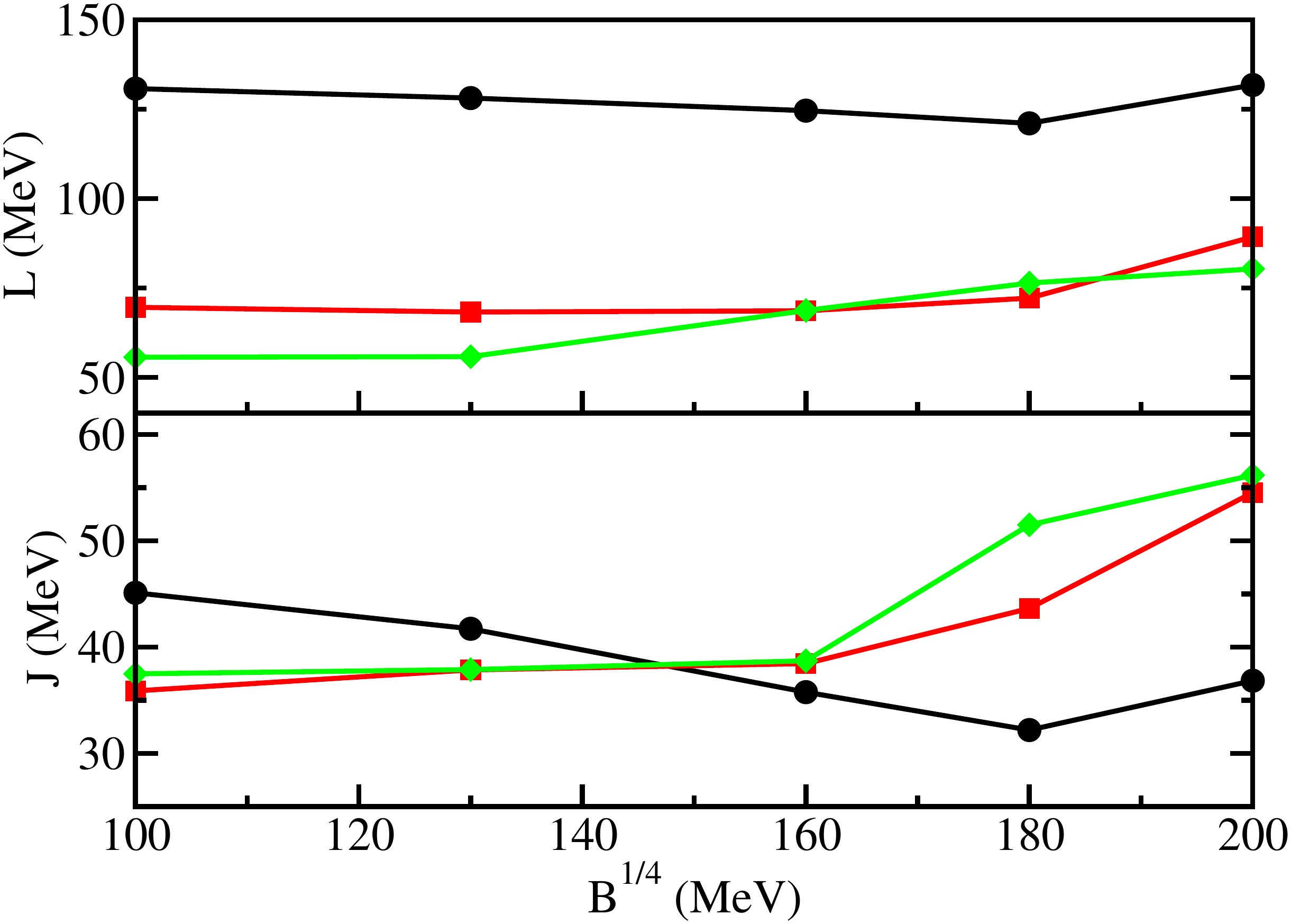}
	\caption{ Symmetry energy $S$ and slope parameter $L$ for Hybrid EoS as a function of bag constant $B^{1/4}$ for NL3, IOPB-I, and G3 parameter sets.}
	\label{fig3.3}
\end{figure}

\begin{figure}[h]
	\centering
	\includegraphics[width=0.75\textwidth]{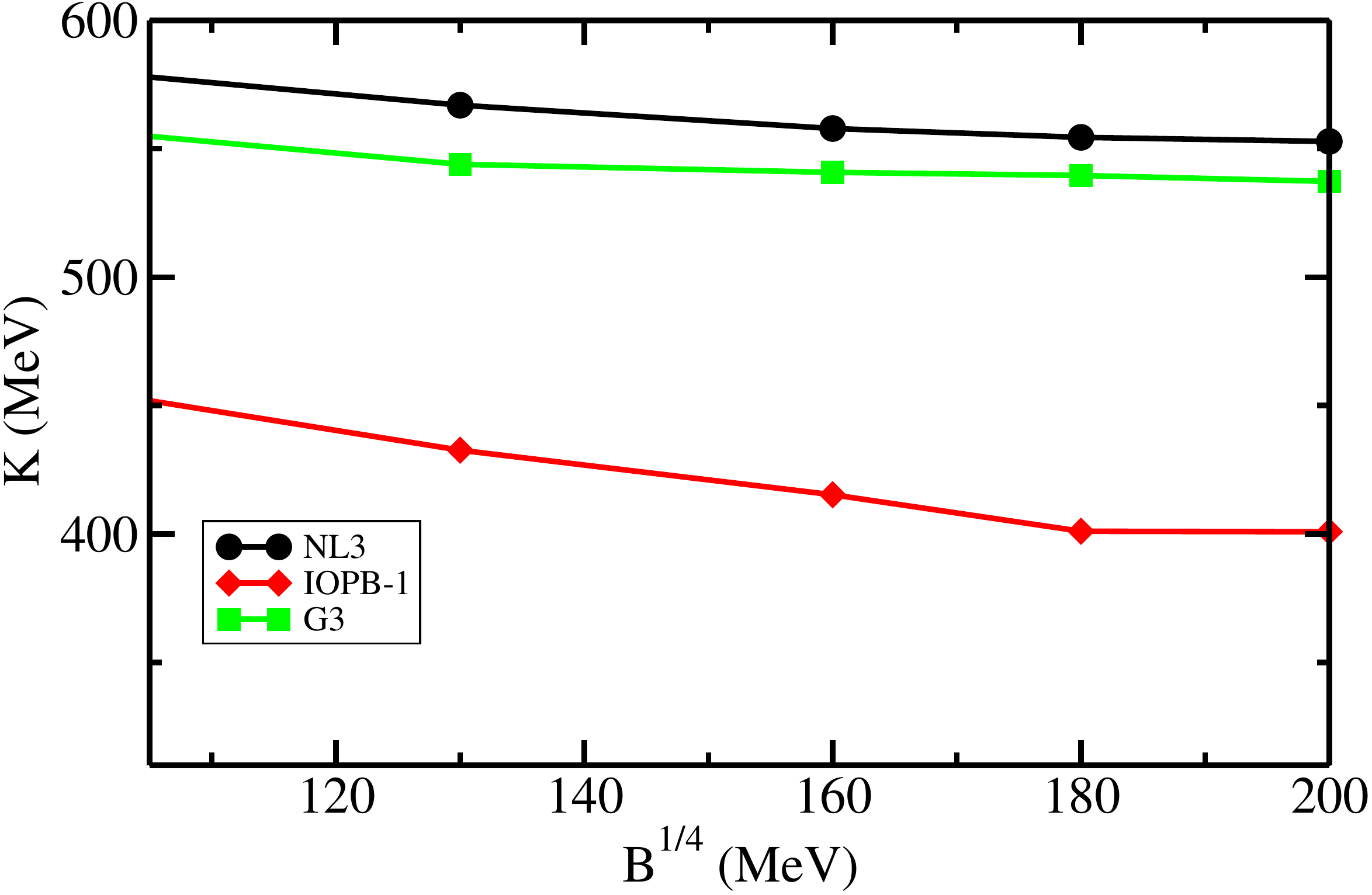}
	\caption{ NM incompressibility for Hybrid EoS as a function of bag constant $B^{1/4}$ for NL3, IOPB-I, and G3 parameter sets.}
	\label{fig3.4}
\end{figure}

The nuclear parameters of hybrid EoS such as symmetry energy $J$ and slope parameter $L$ at saturation are plotted as a function of Bag constant $B^{1/4}$ for different parameter sets as displayed in Fig.~\ref{fig3.3}. The symmetry energy $J$ increases with the bag values for IOPB-I and G3 sets, while for NL3 it decreases initially for $B^{1/4}$ values up to 180 MeV and then increases for 200 MeV, showing a completely different nature than the rest of the parameter sets. The slope parameter $L$ varies almost in a similar fashion for IOPB-I and G3 sets. NL3 has higher value of slope parameter $L$.
  The incompressibility coefficient $K$ for all parameter sets is displayed in Fig.~\ref{fig3.4}. The values lie in the range 400-600 MeV, which is very large compared to the predicted values from ISGMR \cite{Colo:2013yta,Piekarewicz:2013bea}. 

\begin{figure}[h]
	\centering
	\includegraphics[width=0.75\textwidth]{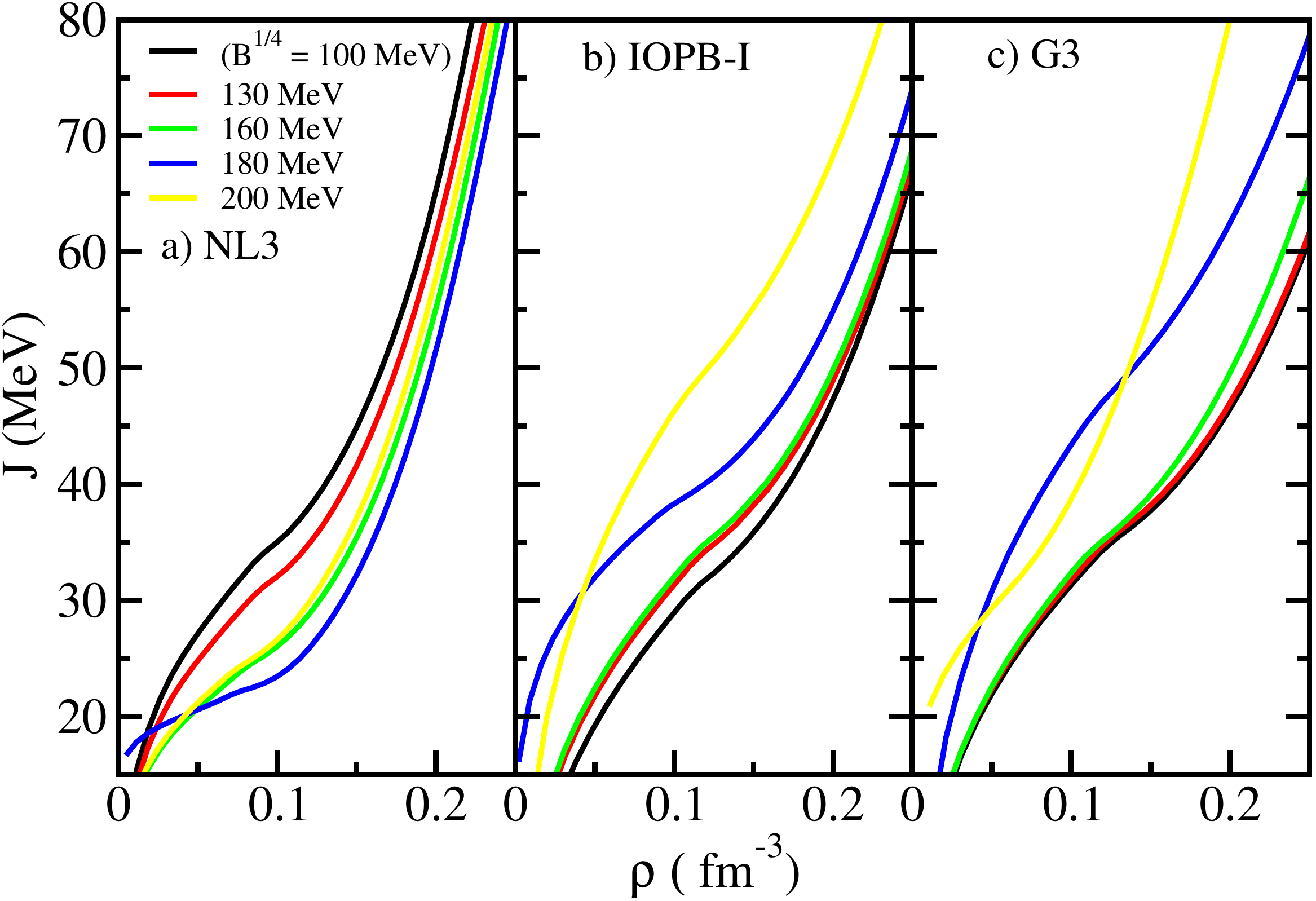}
	\caption{ Symmetry energy $J$ versus density for hybrid EoS with different bag values for a) NL3, b) IOPB-I and c) G3 parameter sets.}
	\label{fig3.5}
\end{figure}

The variation of symmetry energy for hybrid EoS with density for different HM parameter sets and different bag values are shown in Fig.~\ref{fig3.5}.
The symmetry energy for NL3 set increases smoothly with density for all bag constants. However, for IOPB-I and G3 sets, the symmetry energy shows a rapid increase for bag constants $B^{1/4}$ = 180 and 200 MeV. The G3 set produces softer symmetry energy for low bag values in comparison to the IOPB-I and NL3 sets, while as it produces a very stiff value of symmetry energy for bag constants 180 and 200 MeV. The symmetry energy at saturation density $\rho_0$ for NL3 set initially decreases with bag constant up to $B^{1/4}$ = 180 MeV, thereafter it increases for $B^{1/4}$ = 200 MeV. No such variation in the symmetry energy is seen for IOPB-I and G3 sets. The large variation in symmetry energy for 180 MeV and 200 MeV bag values for IOPB-I and G3 sets at higher densities may well contribute to the star matter properties.
\begin{figure}[h]
	\centering
	\includegraphics[width=0.75\textwidth]{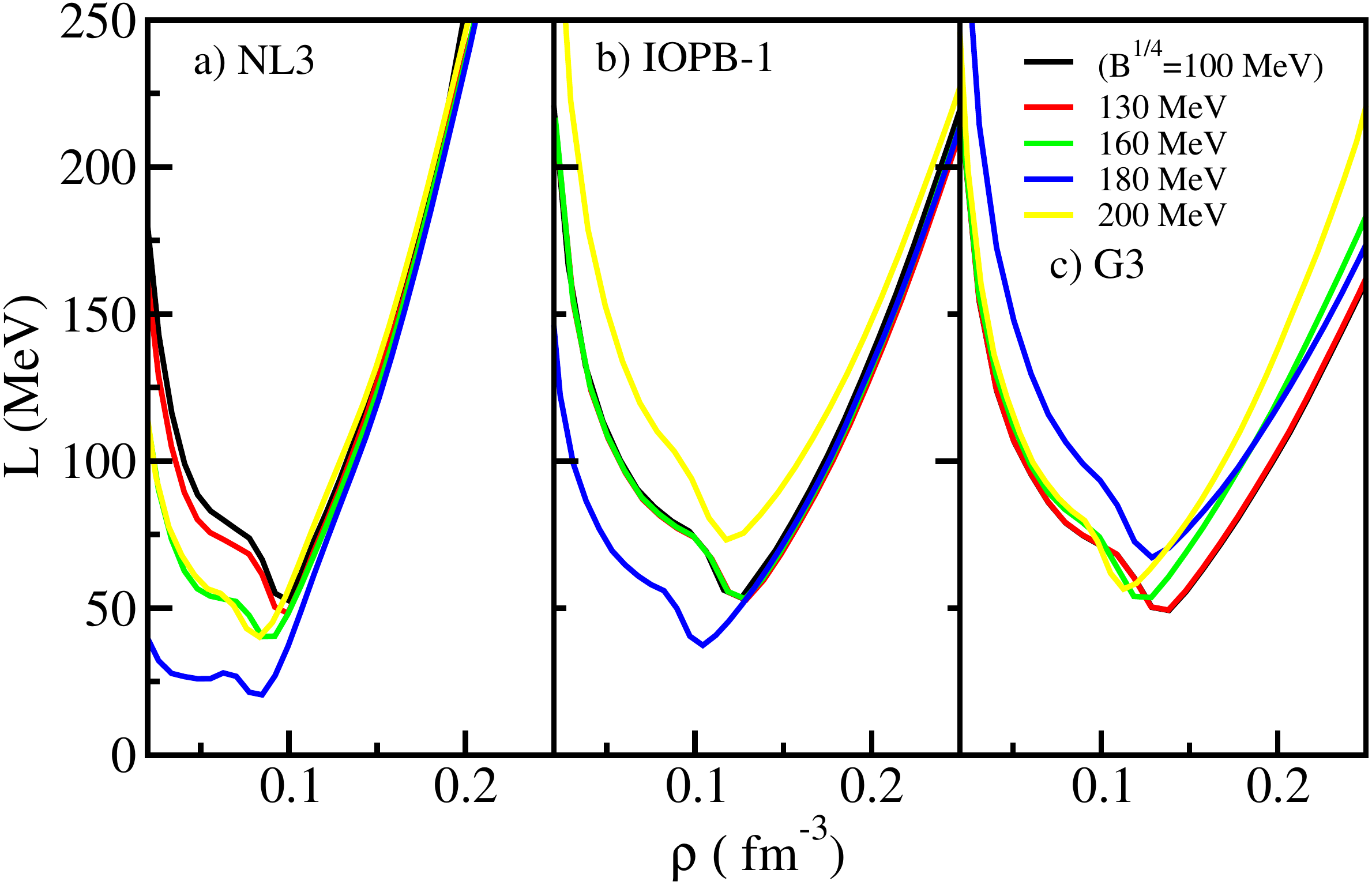}
	\caption{ Slope parameter $L$ as a function of density for hybrid EoS with different bag values for a) NL3, b) IOPB-I and c) G3 parameter sets.}
	\label{fig3.6}
\end{figure}

The slope parameter $L$ versus $\rho$ is displayed in Fig.~\ref{fig3.6}. For NL3 set, $L$ shows similar behavior for all bag constants at density $\rho$ $>$ $\rho_0$. However, for $\rho$ $<$ $\rho_0$, the $L$ value shows more saturation for all bag constants. IOPB-I set follows an almost similar pattern. However, for G3 set, the $L$ value increases with density at $\rho$ $>$ $\rho_0$. The G3 set produces soft slope parameter $L$ as compared to NL3 and IOPB-I parameterizations.

With the EoSs obtained for the hybrid stars, we use the TOV Eqs.~(\ref{tov1}) and (\ref{tov2}) that are used to evaluate the structure of the star. 
\begin{figure}
	\centering
	\includegraphics[width=0.75\textwidth]{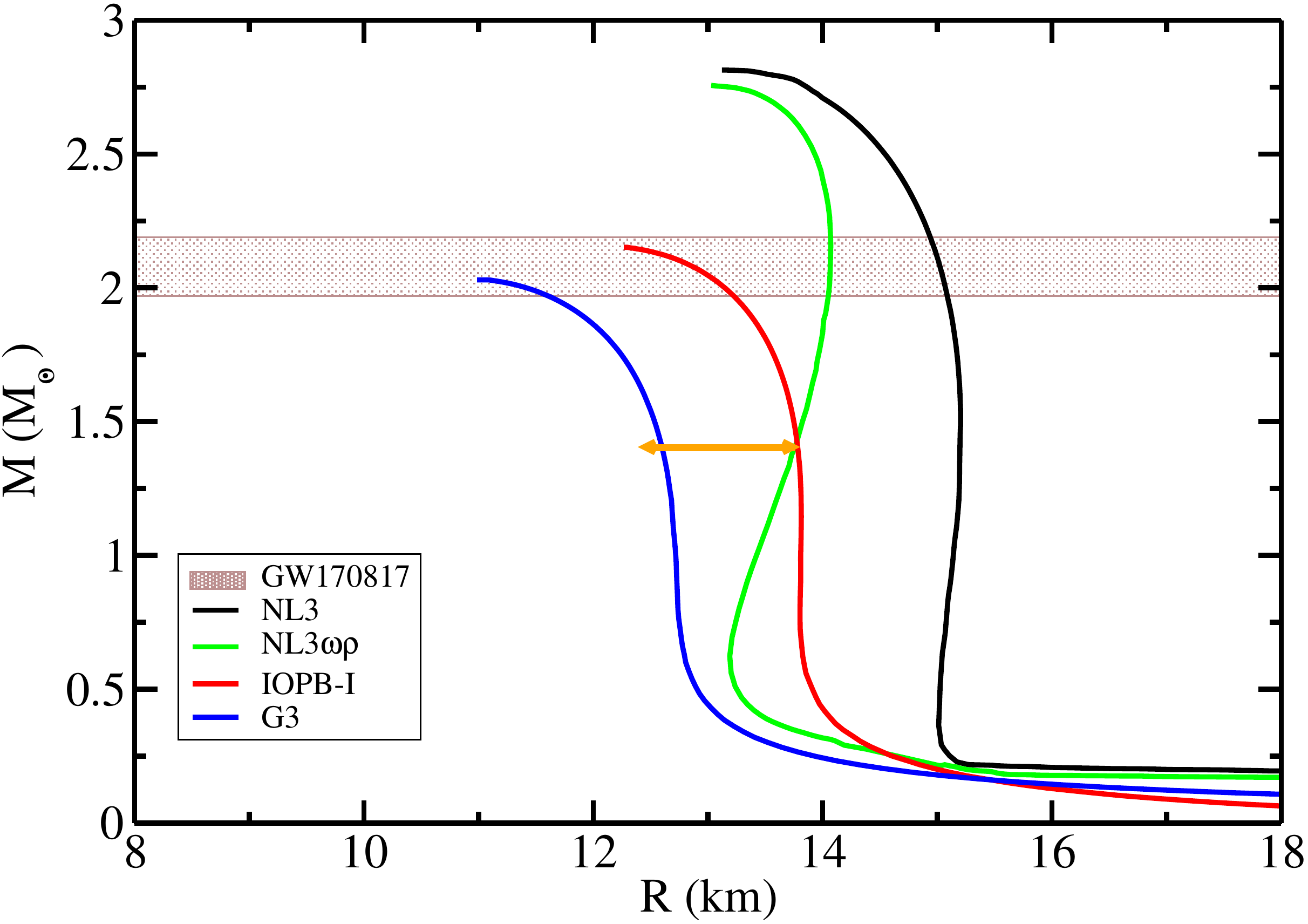}
	\caption{Mass Radius profile of pure hadronic matter for NL3, NL3${\omega \rho}$, IOPB-I, and G3 parameter sets. The recent constraints on mass and radius \cite{PhysRevLett.119.161101,PhysRevLett.120.172702} are also shown.}
	\label{fig3.7}
\end{figure}
 With the higher value of slope parameter for NL3 set, we have used NL3${\omega \rho}$ which includes the non-linear $\omega-\rho$ terms that softens the symmetry energy density dependence \cite{PhysRevLett.86.5647}. NL3${\omega \rho}$ set has symmetry energy and its slope parameter as 31.66 and 55.21 MeV respectively which is much lower than the NL3 set \cite{PhysRevC.90.045803}.
  
Fig.~\ref{fig3.7} shows the mass-radius profile obtained for pure hadronic matter using different parameter sets. The NL3 set predicts a large radius and mass for NS. The NS mass is found to be 2.81$M_{\odot}$ and the corresponding radius is 13.20 km. NL3${\omega \rho}$ produces an NS with a maximum mass of 2.75$M_{\odot}$ which is quite close to NL3 NS maximum mass. This close resemblance is also seen in the NS radius. The NL3${\omega \rho}$ MR curve differs from the usual NL3 MR curve at low mass and low radii. This means that the  NL3${\omega \rho}$ set predicts a smaller radius of NS at the canonical mass. For the IOPB-I parameter set, the NS maximum mass is around 2.15$M_{\odot}$ and the radius is 12.27 km. The G3 set predicts an NS with a maximum mass of 2.03$M_{\odot}$ and the corresponding radius 11.06 km \cite{PhysRevC.97.045806}. The maximum mass of a non-rotating NS is in the range 2.01 $\pm$ 0.04 $\le$ $M(M_{\odot})$ $\le$ 2.16 $\pm$ 0.03. This range was obtained by \citet{Rezzolla_2018} by combining the recent GW observation of a BNS merger (GW170817) \cite{PhysRevLett.119.161101} with the quasi-universal relation between rotating and non-rotating neutron star maximum mass.

All the MR plots shown in Fig.~\ref{fig3.7} represent the complete stellar EoS obtained by properly joining the inner crust EoS with outer crust and core EoS. The inner crust EoS for all the models is taken such that the symmetry energy properties of both the crust and core EoS match with each other. For IOPB-I and G3 families, we do not have a unified EoS for the inner crust. However considering the slope parameter $L$ of symmetry energy for these two families, we have used IU-FSU \cite{PhysRevC.82.055803} and FSU  \cite{PhysRevLett.95.122501} models as the inner crust EoS for G3 and IOPB-I, respectively as they have a close comparison in the slope parameter value. All the inner crust EoSs used can be found in Ref. \cite{PhysRevC.90.045803}. 

\begin{figure}	
	\centering
	\includegraphics[width=13.5cm,height=10.0cm]{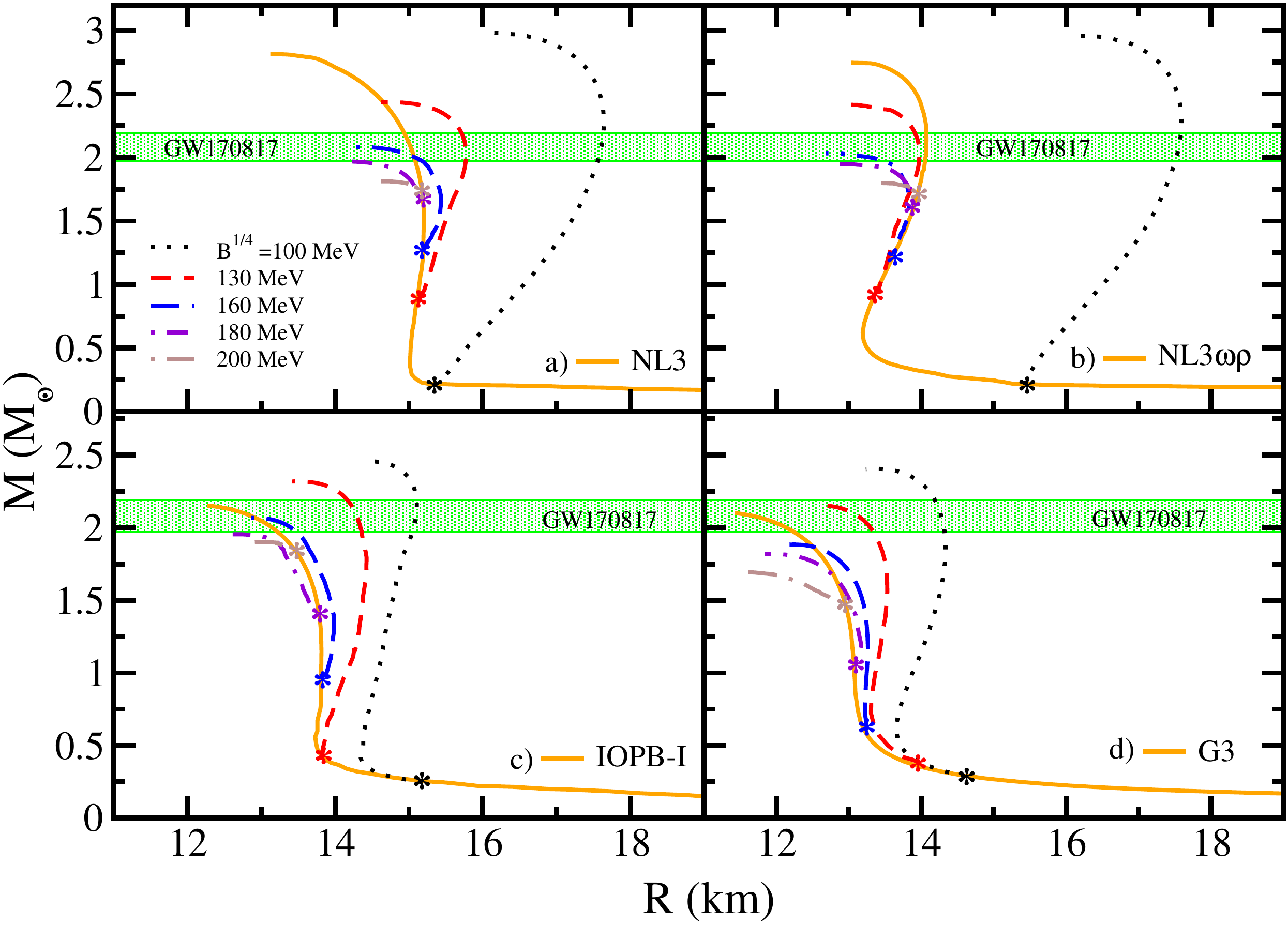}
	\caption{Mass Radius profile of hybrid star for a) NL3, b) NL3${\omega \rho}$, c) IOPB-I and d) G3 parameter sets. The corresponding asterisks denote the transition from pure hadron star to hybrid star.}
	\label{hyb}
\end{figure}
The mass-radius profile for hybrid EoS obtained for different bag constants is shown in Fig.~\ref{hyb}. The green band represents the maximum mass range obtained for a non-rotating star \cite{PhysRevLett.119.161101,Margalit_2017,Antoniadis1233232} . This band also satisfies the precisely measured mass of PSR J0348+0432 and PSR J1614-2230 with mass (2.01 $\pm$ 0.04)$M_{\odot}$ \cite{Antoniadis1233232} and (1.97 $\pm$ 0.04)$M_{\odot}$ \cite{Demorest2010} respectively. These measurements imply that the maximum mass of any NS predicted by any theoretical model should reach the limit of $\approx$ 2.0$M_{\odot}$. The maximum mass of the hybrid star decreases as the bag constant increases. For lower bag values, the maximum mass produced is very high at $\approx$ 2.5$M_{\odot}$ for IOPB-I and G3 sets, and for higher values of the bag, the mass is reduced to $\approx$ 1.8$M_{\odot}$. The mass-radius profile of pure hadronic matter is also shown (solid lines) again for all parameter sets for comparison. The mass of pure hadronic matter for IOPB-I and G3 sets lie well within the bag values $B^{1/4}$ = 130-160 MeV. The MR curves for NL3 and NL3${\omega \rho}$ sets show a similar trend for all bag constants. 
The asterisk symbols denote the transition masses from hadron star to hybrid star. The transition to a hybrid star begins earlier for low bag values. As the bag values increase, the phase transition also increases. The addition of inner crust EoS stabilizes the transition between hadronic matter and quark matter.   
The mass at which the transition from pure hadron matter to hybrid star matter takes place increases with the bag constant, which means that as the bag constant increases, the hybrid star becomes more and more hadronic.

The maximum mass along with the corresponding radius of hybrid star matter for different parameter sets is shown in Table \ref{mixp} for different bag constants. The calculations for the pure hadron matter are also shown. A decrease in the maximum mass is seen with bag constant. The radius also decreases except for NL3 and IOPB-I parameter sets at bag value of 200 MeV. 
\begin{table}[htb!]
	\centering
	\caption{Maximum mass and radius of hybrid stars for different bag constants. The results for pure hadron matter are also shown. }
	\begin{tabular}{ ccccccc }
		\midrule
		\midrule
		$B^{1/4}$ (MeV)&Pure Hadron&100&130&160&180&200 \\
		\midrule
		&&NL3\\
		\midrule	
		$M$ ($M_{\odot}$)  &2.81& 2.98&2.43&2.08&1.97&1.81 \\
		$R$ (km) &13.20&16.15&14.72&14.39&14.13&14.65\\
		
		\midrule
		
		&&NL3${\omega \rho}$\\
		\midrule	
		$M$ ($M_{\odot}$)  &2.75& 2.95&2.41&2.03&1.94&1.79 \\
		$R$ (km) &13.01&16.06&12.82&12.68&12.87&13.40\\
		
		\midrule
		&& IOPB-I\\
		\midrule	
		$M$ ($M_{\odot}$) &2.15 & 2.46&2.32&2.07&1.95&1.90 \\
		$R$ (km) &12.27&14.45&13.50&12.91&12.64&13.00\\
		\midrule
		&& G3\\
		\midrule	
		$M$ ($M_{\odot}$) &2.03 & 2.40&2.15&1.88&1.82&1.69 \\
		$R$ (km) &11.06&13.40&12.59&12.24&11.88&11.57\\
		\midrule
		\midrule
	\end{tabular}
	\label{mixp}
\end{table} 
\begin{figure}[htb!]
	\centering
	\includegraphics[width=0.75\textwidth]{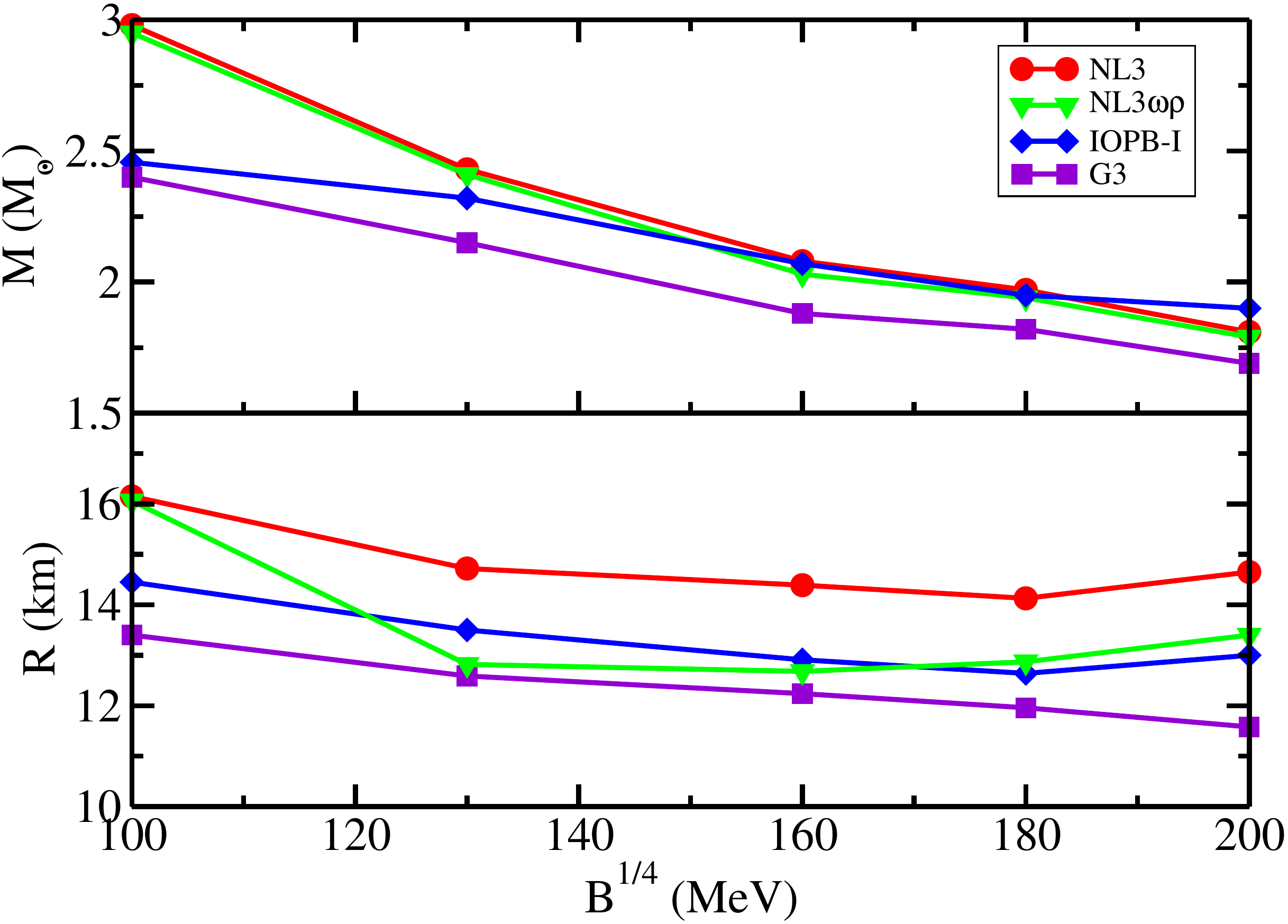}
	\caption{Variation of Maximum Mass and Radius of hybrid stars with different bag constants for NL3, NL3${\omega \rho}$, IOPB-I and G3 parameter sets.}
	\label{fig3.11}
\end{figure}
Fig.~\ref{fig3.11} shows the variation of maximum mass and radius with bag constants. The maximum mass for NL3 set at bag value $B^{1/4}$ = 100 MeV is 2.98$M_{\odot}$ while for pure hadronic matter, it predicts a mass of 2.81$M_{\odot}$. As the bag constant increases, the maximum mass decreases from 2.98 to 1.81$M_{\odot}$. The hybrid NS maximum mass for the NL3${\omega \rho}$ set is almost similar to the NL3 set. However, it predicts a hybrid star with a smaller radius as compared to NL3. So, while the  GW170817 data rules out the pure NL3 and NL3${\omega \rho}$ EoSs, the addition of quarks softens the EoS and hence reduces the maximum mass satisfying the GW170817 mass constraints. Similarly, the maximum mass of IOPB-I and G3 set decreases from 2.46 to 1.90$M_{\odot}$ and 2.40 to 1.69$M_{\odot}$ respectively. From Fig.~\ref{fig3.11}, we see that for NL3 set with B$^{1/4}$ = 100 MeV the maximum mass obtained is 2.98$M_{\odot}$ which doesn't agree with the recent mass measurements from GW data. For IOPB-I and G3 sets, the maximum mass produced at this bag value is 2.46 and 2.40$M_{\odot}$ respectively, which are also larger than the recent mass measurements. Furthermore, for B$^{1/4}$ = 100 MeV, the transition from hadron matter to quark matter takes place well below the normal nuclear density, predicting the presence of quarks below nuclear density, which is unphysical. So we agree that the bag value of 100 MeV produces unphysical results and hence cannot be considered as a proper choice for the bag constant. Similar results follow for bag value of 200 MeV where the maximum mass produced is much less than 2$M_{\odot}$ and thus this bag value isn't considered as a good choice to explain the quark matter in neutron stars.

From the GW observation of the maximum mass of NS in the range 2.01 $\pm$ 0.04 $\le$ $M(M_{\odot})$ $\le$ 2.16 $\pm$ 0.03 \cite{Rezzolla_2018}, one can see that the bag constant $B^{1/4}$ = 130 MeV produces a maximum mass of 2.15$M_{\odot}$ for G3 set, while the same bag constant gives 2.43$M_{\odot}$ for NL3 set and 2.32$M_{\odot}$ for IOPB-I set. For $B^{1/4}$ = 160 MeV, the maximum mass is 2.08, 2.07 and 1.82$M_{\odot}$ for NL3, IOPB-I and G3 respectively. The radius of the canonical mass for the bag constants $B^{1/4}$ = 130-160 MeV is found to be in the range 12.5-13.2 km which is well withing the range of $R_{1.4}$ $\le$ 13.76 km as extracted from the NS tidal deformability of GW170817 event \cite{PhysRevLett.120.172702}. Further, the recently detected GW190425 constrains the NS mass in the range 1.12 to 2.52$M_{\odot}$ \cite{Abbott_2020}. Thus we see that the most probable values of bag constant for the obtained EoSs lies in the range 130 MeV $<$ $B^{1/4}$ $<$ 160 MeV. This range of bag constants determines the properties of hybrid stars that agree with the recent gravitational-wave observations GW170817 and GW190425. \citet{Aziz:2019rgf} have constrained the value of bag constant in the range 150 MeV $\le$ $B^{1/4}$ $\le$ 180 MeV. The precisely measured mass of 1.97 $\pm$ 0.04$M_{\odot}$ for binary millisecond PSR J1614-2230 by \citet{Demorest2010} supports the presence of quarks in the NS \cite{zel_2010,Alford,Lastowiecki2015}.
 
\section{Conclusion}
\label{c3conclusion}
In this chapter, we studied the hybrid EoS by mixing hadron matter and quark matter using Gibbs conditions. The E-RMF model for hadron matter with recently reported parameter sets and MIT bag model for quark matter with different bag constants were employed. The nuclear matter properties such as symmetry energy ($J$), slope parameter ($L$), and incompressibility ($K$) are calculated for hybrid EoS.  It is found that the values of symmetry energy $J$ and other quantities are very high for a hybrid EoS and they increase with the bag constant except for the $J$  and $L$ values (for NL3) and  incompressibility $K$ (for all parameter sets) which decreases with the bag values $B$. The values obtained for symmetry energy and other quantities are very large as compared to their predicted values for the hadronic matter. The predicted values of these quantities from various theoretical models also have large uncertainty.

All these quantities have a huge impact on the neutron star mass-radius relation and other important quantities. The slope parameter influences the properties of both finite and infinite nuclear matter. The phase transition properties of hadron-quark matter and the existence of exotic phases like Kaons, Hyperons, etc., in neutron stars are also dependent on these quantities. Since the symmetry energy cannot be measured directly, it is important to identify the observables that correlate the symmetry energy and its density dependence to impose constraints on the quantities like slope parameter, symmetry energy curvature, etc. The additional information about these quantities can be extracted from the astrophysical observations of high dense matter objects like neutron stars or better knowledge of EoS. The nature of EoS is influenced remarkably with these quantities and since these parameters are controlled by the bag constant $B$, then it will be possible to adjust the mass and radius of a NS by tuning the bag constant $B$.

The star matter properties like mass and radius are calculated for different bag constants. It is found that the value of bag constant in the range 130 MeV $\le$ $B^{1/4}$ $\le$ 160 MeV is suitable for explaining the quark matter in neutron stars. The results obtained with bag values less than 130 MeV and greater than 160 MeV do not agree with the recently measured observables from gravitational wave data. Hence such bag values are not suitable enough to explain the quark matter in neutron stars. 

Since the bag constant has a huge effect on the EoS, a more dominant theoretical approach is required to constrain the value of the bag constant. Furthermore, it may allow us to properly calculate the fraction of quark matter present in neutron stars. Considering color flavor or one gluon exchange in the simple MIT bag model or using other models like NJL for quark matter may further constrain these nuclear matter properties for hybrid EoS. The presence of exotic phases like kaons, hyperons, etc in the neutron stars will further modify the EoS. The presence of quarks will provide new insight into the physics of neutron stars and other high dense objects.
\begin{savequote}[8cm]
\textlatin{Research is seeing what everybody else has seen and thinking what nobody else has thought.}

  \qauthor{---\textit{lbert Szent-Gy$\ddot{o}$rgyi }}
\end{savequote}

\chapter{\label{ch:4-innercrust}Effect of Inner Crust Equation of State}

\minitoc

\section{Introduction}
\label{intro4}
The structure and the properties of NSs have been studied effectively from experimental as well as theoretical models. Such studies reveal the inner structure of NS and the presence of exotic phases. The results obtained from the astrophysical observations require several theoretical inputs for the interpretation. A coextensive effort from theory and experiments has improved and provided new insights into the field. After the discovery of GW170817, more theoretical work has been done to understand the relation between EoS and NS properties through various aspects like a phase transition, symmetry energy \cite{PhysRevLett.120.172703,PhysRevLett.120.172702,PhysRevLett.120.261103}. However, there are still numerous fundamental problems related to the matter under extreme conditions that are yet to be answered. The most important one is the unified model which can describe the overall structure of an NS, from the outer crust to the inner core.

The unified EoS describes the NS from its outer crust to the inner core. However, a unified EoS is generally not available. Hence the complete EoS is divided into three different phases: the outer crust phase, the inner crust, and the core phase. It has been shown \cite{PhysRevC.94.015808} that the NS properties such as mass and radius do not depend upon the outer crust EoS, but a particular choice of inner crust EoS and its matching with the core EoS is critical and the variations larger than 0.5 km have been obtained for a 1.4$M_{\odot}$. For the outer crust which lies in the density range 10$^4$-10$^{11}$ g/cm$^{3}$, the Baym-Pethick-Sutherland (BPS) EoS \cite{Baym:1971pw}, the Haensel-Pichon (HP) EoS \cite{1994A&A...283..313H} are widely used in the literature. Both these EoSs do not affect the mass and radius of an NS. For the matter beyond the neutron drip density (4$\times$10$^{11}$ g/cm$^{3}$) ranging from  10$^{11}$-10$^{14}$ g/cm$^{3}$, the inner crust EoS follows. The Baym-Bethe-Pethick (BBP) EoS \cite{BAYM1971225} is usually used. However to avoid the large uncertainties in mass and radius of an NS, studies have shown that for the complete unified EoS, the inner crust EoS should be either from the same model as core EoS or the symmetry energy slope parameter should match \cite{PhysRevC.94.015808,PhysRevC.90.045803}.  

The observation of the mass and the radius of static as well as rotating NSs may provide some useful constraints on the EoS. The NSs, due to their compactness, can rotate very fast as compared to the other astrophysical observables. Hence the measurements of RNS properties like mass, radius, frequency, the moment of inertia, etc., close to the mass-shielding limit may lead to more constraints on the composition of the nuclear matter at higher densities and the EoS.  Rotating NSs provide more information about the structure and the composition of the star through the measurement of more quantities than the non-rotating ones. Quantities like the moment of inertia, eccentricity provide information about how fast an object can spin with a given angular momentum and the deformation of the mass while spinning. The universal relations between these bulk properties of NSs may help to impose constraints on the radius of an NS. The choice of inner crust EoS on the static and rotating NS will affect the mass and radius and consequently the other star matter properties.
 
This chapter is organized as follows. In Sec. \ref{sec:1}, the Thomas-Fermi approximation within the RMF framework to describe the inner crust part and the transition from crust to the core is discussed. The unified EoS by combining the outer crust, inner crust, and the core EoS will be discussed in Sec. \ref{sec:2}. The NS properties for static as well as rotating NS's with different EoSs are also discussed. Finally, the summary, and conclusions are outlined in Sec. \ref{sec:3}.
The calculations discussed in this chapter is based on the work \cite{rathernpa}.

\section{Formalism}
\label{sec:1}

The inner crust which contains the inhomogenous NM is defined by applying the Thomas-Fermi (TF) approximation \cite{10.1143/PTP.100.1013} within the same RMF model \cite{PhysRevC.82.055807}. The Skyrme type interactions like Lattimer and Swesty EoS \cite{LATTIMER1991331} and the compressible liquid drop model have been widely used in the literature by several authors to describe the nonuniform matter \cite{Abbott_2017,CHABANAT1997710}. \par 
The outer crust of an NS consists of nuclei distributed in a solid body-centered-cubic (bcc) lattice filled by free electron gas. Moving from the outer crust to the inner crust part, the increasing density leads to the complete ionization of all the atoms and beyond a certain density, leads to the formation of a quasiuniform gas. As we move to the greater density regions, more and more electrons are captured by the nuclei, which thus become neutron-rich. At the density around 4.2$\times$10$^{11}$ g cm$^{-3}$, neutron drip sets in. With the help of nuclear models, the neutron drip density determines the boundary between the outer crust and the inner crust. The inner crust part contains the free electrons and the neutron gases, forming different types of pasta structures. Both the inner and the outer crust densities are a fraction of the normal nuclear density.  \par 
The transition density at the crust-core interface is uncertain due to the insufficient knowledge of the neutron-rich NM EoS. Furthermore, the determination of the transition density is very complicated as the inner crust part may contain some internal structures (pasta phases). A well-defined approach is to find the density at which the uniform liquid phase becomes unstable against small density fluctuations, indicating the formation of nuclear clusters. This boundary between the inhomogeneous solid crust and the liquid core is connected to the isospin dependence of nuclear models below the saturation value by the widely used thermodynamic method \cite{PhysRevC.76.025801,PhysRevC.86.015801}. In the present work, we use the thermodynamic method to determine the crust-core transition boundary, which allows the stability of the NS's to be determined in terms of their bulk properties. 
A more detailed description of the crust-core transition is given in Refs. \cite{PhysRevC.86.015801,PhysRevC.76.025801}.
\section{Results and Discussion}
\label{sec:2}
To study the effect of crust EoS on NS properties, we have chosen several parameter sets to construct the core EoS. Since the outer crust EoS does not affect the NS maximum mass and radius, therefore the BPS EoS \cite{Baym:1971pw} has been used for the outer crust part. For the inner crust part, the relativistic mean-field model with constant couplings, non-linear terms \cite{Boguta:1977xi} and density-dependent couplings \cite{Typel:1999yq} have been used. These include NL3 \cite{PhysRevC.55.540} set with non-linear $\sigma$ terms, TM1 \cite{Tm1} with non-linear $\sigma$ and $\omega$ terms, NL3$\omega \rho$ \cite{PhysRevC.64.062802,PhysRevLett.86.5647} which includes the non-linear $\omega \rho$ terms in addition to the previous couplings, FSU \cite{PhysRevLett.95.122501} and IU-FSU \cite{PhysRevC.82.055803}, and the density-dependent DD-ME2 \cite{PhysRevC.71.024312} and DD-ME$\delta$ \cite{PhysRevC.84.054309}.

\begin{figure}[ht]
	\centering
	\includegraphics[width=0.75\textwidth]{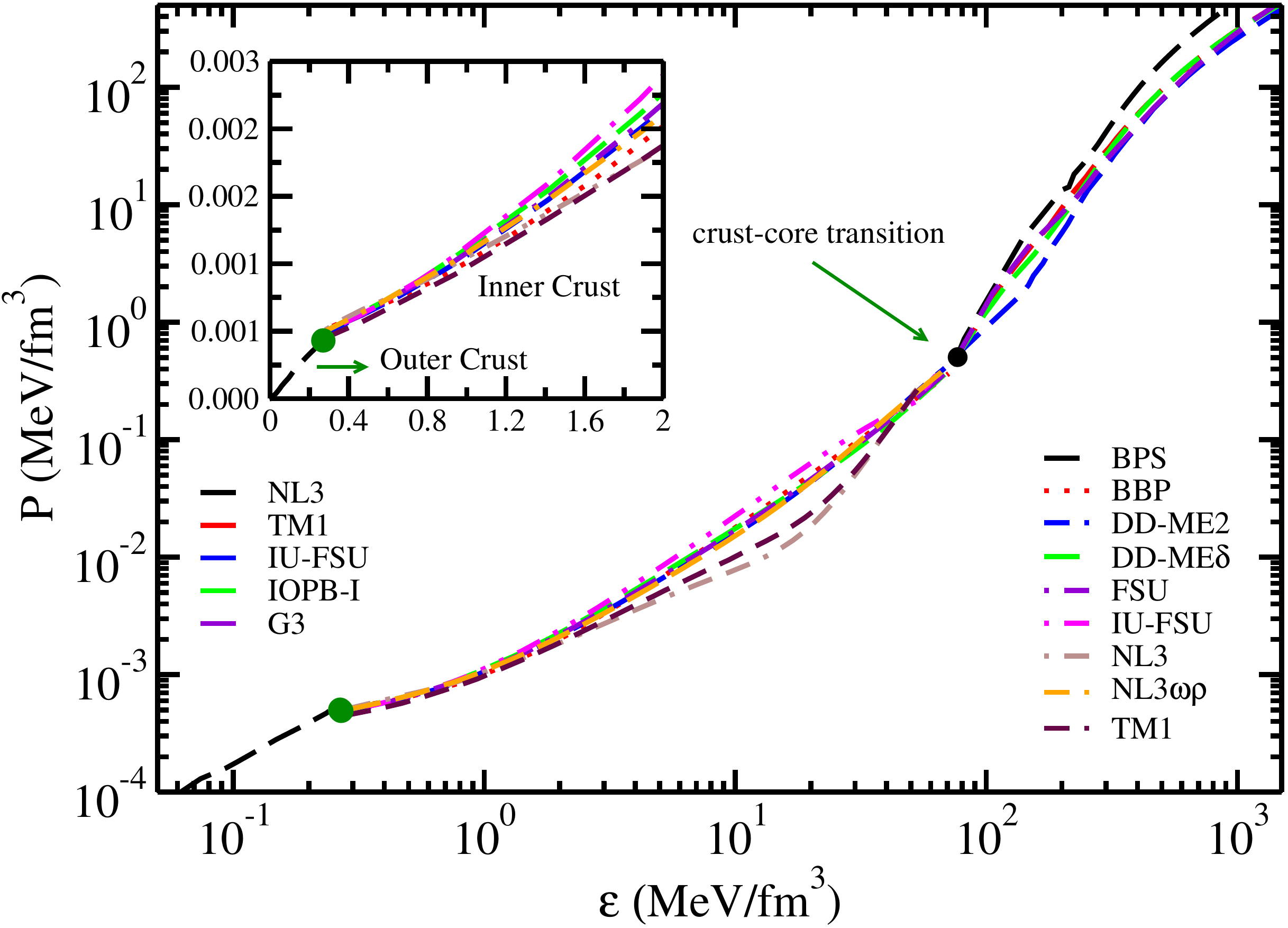}
	\caption{Unified EoS with different inner crust and core EoS. The inset shows the matching of outer crust with the inner crust. The green dot shows the matching point of BPS EoS with the inner crust. The black dot represents the point of crust-core transition.}
	\label{FIG:1}
\end{figure}

The NM properties at saturation density for the above considered crust EoSs are shown in Table \ref{tbl1}. The symmetry energy slope parameter $L_{sym}$ of the given sets lies in the range 47-118 MeV.
\begin{table}[ht]
	\centering
	\caption{NM properties for the inner crust EoS at saturation density ($\rho_0$), energy ($E_0$), symmetry energy ($S$), slope parameter ($L_{sym}$), incompressibility coefficient ($K$), and skewness parameter ($Q_{sym}$). All the values are in MeV except for the ($\rho_0$) which is in fm$^{-3}$.}
{\begin{tabular}{p{1.7cm}p{1.2cm}p{1.2cm}p{1.2cm}p{1.2cm}p{1.2cm}p{1.2cm}}
			\hline 
			Model & $\rho_0$ & $E_0$ & $S$ & $L_{sym}$ & $K$ & $Q_{sym}$\\
			\hline
			NL3 & 0.148 & -16.24 & 37.3 & 118.3 & 270.7 & 203 \\
			TM1 & 0.145 & -16.26 & 36.8 & 110.6 & 280.4 & -295 \\
			FSU & 0.148 & -16.30 & 32.6 & 60.5 & 230.0 & -523 \\
			IU-FSU & 0.155 & -16.40 & 31.3 & 47.2 & 213.2 & -288 \\
			NL3$\omega \rho$ & 0.148 & -16.30 & 31.7 & 55.2 & 272.0 & 203 \\
			DD-ME2 & 0.152 & -16.14 & 32.3 & 51.4 & 250.8 & 478 \\
			DD-ME$\delta$ & 0.152 & -16.12 & 32.4 & 52.9 & 219.1 & -741 \\
			\hline
		\end{tabular}\label{tbl1}}
\end{table}
The E-RMF formalism is used to construct the core EoS.  We covered a wide range of models containing only $\sigma$ self-coupling terms to the models with all types of self-and cross-couplings along with the $\delta$ meson inclusion. NL3, TM1, IU-FSU \cite{PhysRevC.82.055803}, IOPB-I, and G3 parameter sets are used to study the NS core. All the coupling constants and the meson masses for the above parameter sets are given in \cite{Tm1,PhysRevC.82.055803,PhysRevC.97.045806}. The parameter sets for the core EoS used in the present study cover an NS maximum mass range from $\approx$ 2-2.8$M_{\odot}$. This allows us to determine the variation in the properties of the NS for a range of maximum mass with the symmetry energy slope parameter in the range 47-118 MeV.
The complete EoS consisting of the outer crust, the inner crust, and the core are constructed using the above-defined parameter sets. The unified and non-unified EoSs follow as BPS (for the outer crust)+ BBP, NL3, TM1, NL3$\omega \rho$, FSU, IU-FSU, DD-ME2, and DD-ME$\delta$ (for the inner crust)+ NL3, TM1, IU-FSU, IOPB-I, and G3 (for the core). The EoS without inner crust is also constructed to see the impact of inner crust on NS properties. All the inner crust EoSs used are taken from the reference \cite{PhysRevC.90.045803}. The different unified and non-unified EoSs produced are shown in Fig.~\ref{FIG:1}.
The green dot in the inset (as well as in the main plot) of Fig.~\ref{FIG:1} shows the matching of outer crust with inner crust EoS while the black dot in the main plot represents the crust-core transition point.
The NL3 parameter set produces stiff core EoS as compared to other parameter sets. IU-FSU produces soft EoS at low density. G3 set produces soft EoS as compared to NL3, TM1, and IOPB-I at both low and high energy densities. Among the inner crust EoS, NL3 and TM1 set produce soft EoS at very low density and become stiff as the density increases. IU-FSU crust initially produces stiff EoS but becomes soft at higher energy density ($\mathcal{E}$ $\approx$ 45 MeV/fm$^3$). For the outer crust, the BPS EoS is used for all different combinations of inner crust and core EoSs as the outer crust EoS doesn't affect the mass and radius of an NS. 
%
%
The TOV Eqs.~(\ref{tov1}) and (\ref{tov2}) are solved to obtain the NS properties. Apart from obtaining the properties of a static NS, we also see the impact of inner crust on maximally rotating NS (RNS). The properties of rotating NS are obtained using the RNS code  \cite{rnscode}. 

\begin{figure}[hbt!]
	\centering
	\includegraphics[width=0.75\textwidth]{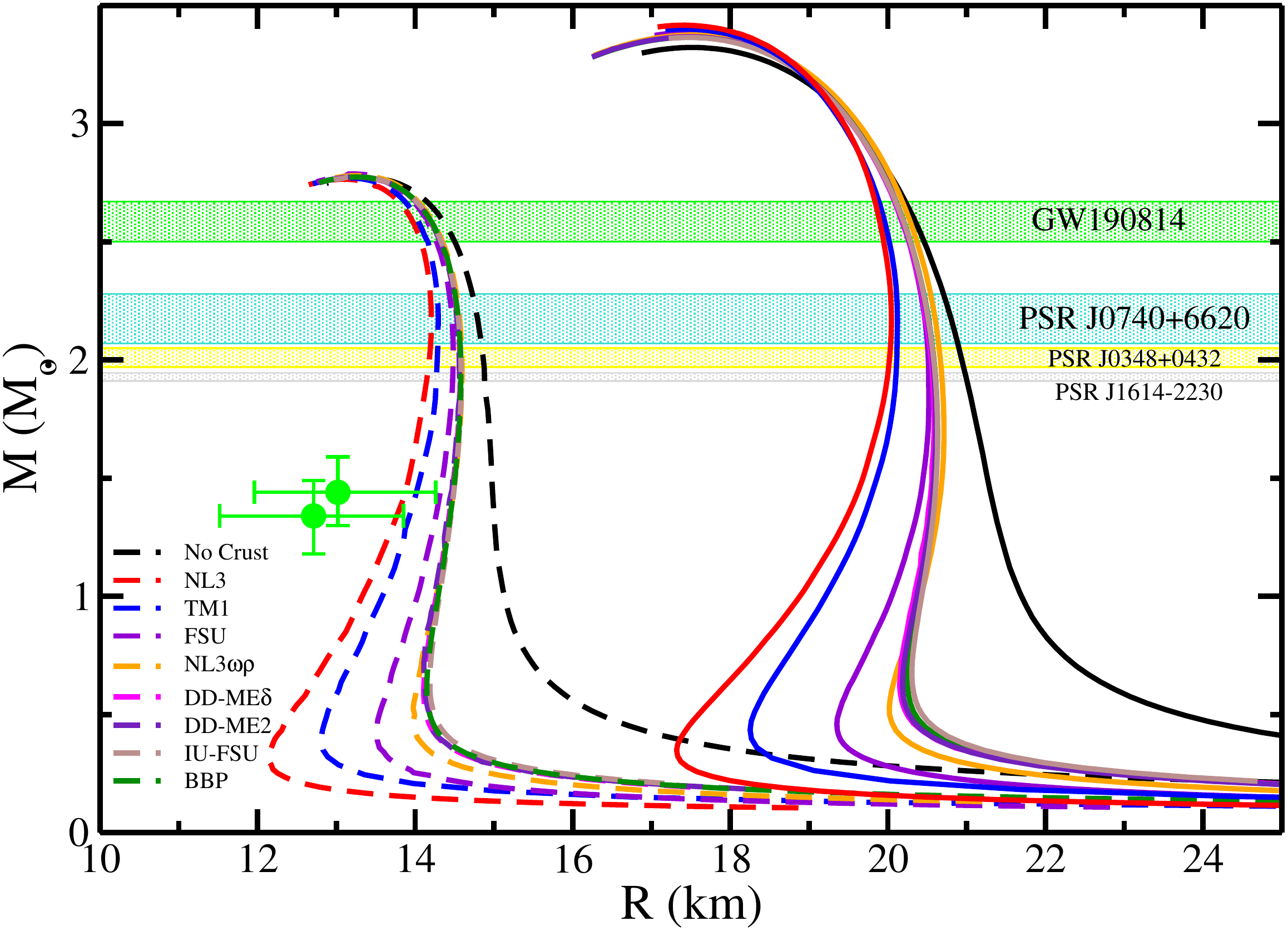}
	\caption{Mass-Radius profile of SNS and RNS for NL3 core with different inner crust EoS. The solid (dashed) lines represent the rotating NS's (static NS's) The recent constraints on the mass \protect\cite{Demorest2010,Antoniadis1233232,Cromartie2020,Abbott_2020a} and radius from NICER measurements \protect\cite{Riley_2019,Miller_2019} of NS are also shown.}
	\label{FIG:2}
\end{figure}

Fig.~\ref{FIG:2} shows the mass-radius relation for a static and maximally rotating NS for NL3 core with different inner crust EoS. The shaded regions represent the constraints on the maximum mass of an NS by pulsars PSR J1614-2230 (1.928 $\pm$ 0.017)$M_{\odot}$ \cite{Demorest2010}, PSR J0348+0432 (2.01 $\pm$ 0.04)$M_{\odot}$ \cite{Antoniadis1233232}, PSR J0740+6620 (2.04$^{+0.10}_{-0.09}$)$M_{\odot}$ \cite{Cromartie2020}, and GW190814 (2.50-2.67$M_{\odot}$) \cite{Abbott_2020a}. The NICER constraints on the stellar radius obtained from the measurement of millisecond pulsar (MSP) PSR J0030+0451 at the inferred mass $M$ = $1.34_{-0.16}^{+0.15}$$M_{\odot}$ and $M$ = $1.44_{-0.14}^{+0.15}$$M_{\odot}$ given by $R$ = $12.71_{-1.19}^{+1.14}$ km and $R$ = $13.04_{-1.06}^{+1.24}$ km are also shown \cite{Riley_2019, Miller_2019}.  
The upper limit on the radius at the canonical mass of an NS is found to be $R_{1.4} \le 13.76$ km \cite{PhysRevLett.120.172702}. The constraints on the maximum mass show that the theoretical prediction of an NS maximum mass should reach the limit $\approx$ 2.0$M_{\odot}$. But the recent observation of gravitational wave data GW190814 has a second component with a maximum mass in the range (2.5-2.67)$M_{\odot}$. This secondary object is considered to be either a light black-hole or a supermassive NS. \par 
From Fig.~\ref{FIG:2}, it is clear that the NS maximum mass produced using different inner crust EoS varies by small margins and lies in the range 2.764-2.787$M_{\odot}$. The corresponding radius varies from 13.027-13.378 km. However, the radius at canonical mass is much more affected than the radius at maximum mass. For an NS without inner crust, the radius at the canonical mass is found out to be $R_{1.4}$ = 14.987 km. The large value for the radius of an NS at 1.4$M_{\odot}$ without inner crust is due to the direct transition from the outer crust to the core part of the star. The neutron drip density determines the outer crust-inner crust boundary and the thermodynamic method provides the transition between the inner crust and the core. With no inner crust considered, the transition density obtained for the outer crust-core boundary affects the core EoS which results in an unexpectedly large value of the radius. With the addition of the inner crust, the proper measurement of the transition density leads to a true value of the NS radius which varies from 14.496-13.853 km. The NS with a small radius at the canonical mass is produced by using NL3 as inner crust EoS which satisfies the constraints by GW170817. The NL3 set has a higher value of symmetry energy slope parameter $L_{sym}$ = 118.3 MeV, but matches completely with the core EoS and hence forms a unified EoS. Thus, a unified EoS produces an NS with a smaller radius at the canonical mass. The other inner crust EoSs have a smaller value of slope parameter than the NL3 set and such low-value slope parameter sets like IU-FSU produces a larger radius at the canonical mass as seen in the figure. Also, the NL3 inner crust EoS satisfies the radius constraints from the NICER measurements. Thus we see that the $R_{1.4}$ has a significant relation with the slope parameter. This is consistent with the work in Refs. \cite{PhysRevLett.120.172702,PhysRevLett.121.062701}.

Fig.~\ref{FIG:2} also shows the MR profile for maximally rotating stars (RNS). Similar to the Static NS (SNS), the RNS maximum mass, and the corresponding radius are not much affected by the inner crust EoS, but the radius at 1.4$M_{\odot}$ varies in a similar fashion as SNS in the range $R_{1.4}$ = 19.5-21.4 km.  
\begin{figure}
	\centering
	\renewcommand{\arraystretch}{0}
	\setlength{\tabcolsep}{0pt}
	
	\begin{tabular}{cc}
		\includegraphics[width=0.70\textwidth]{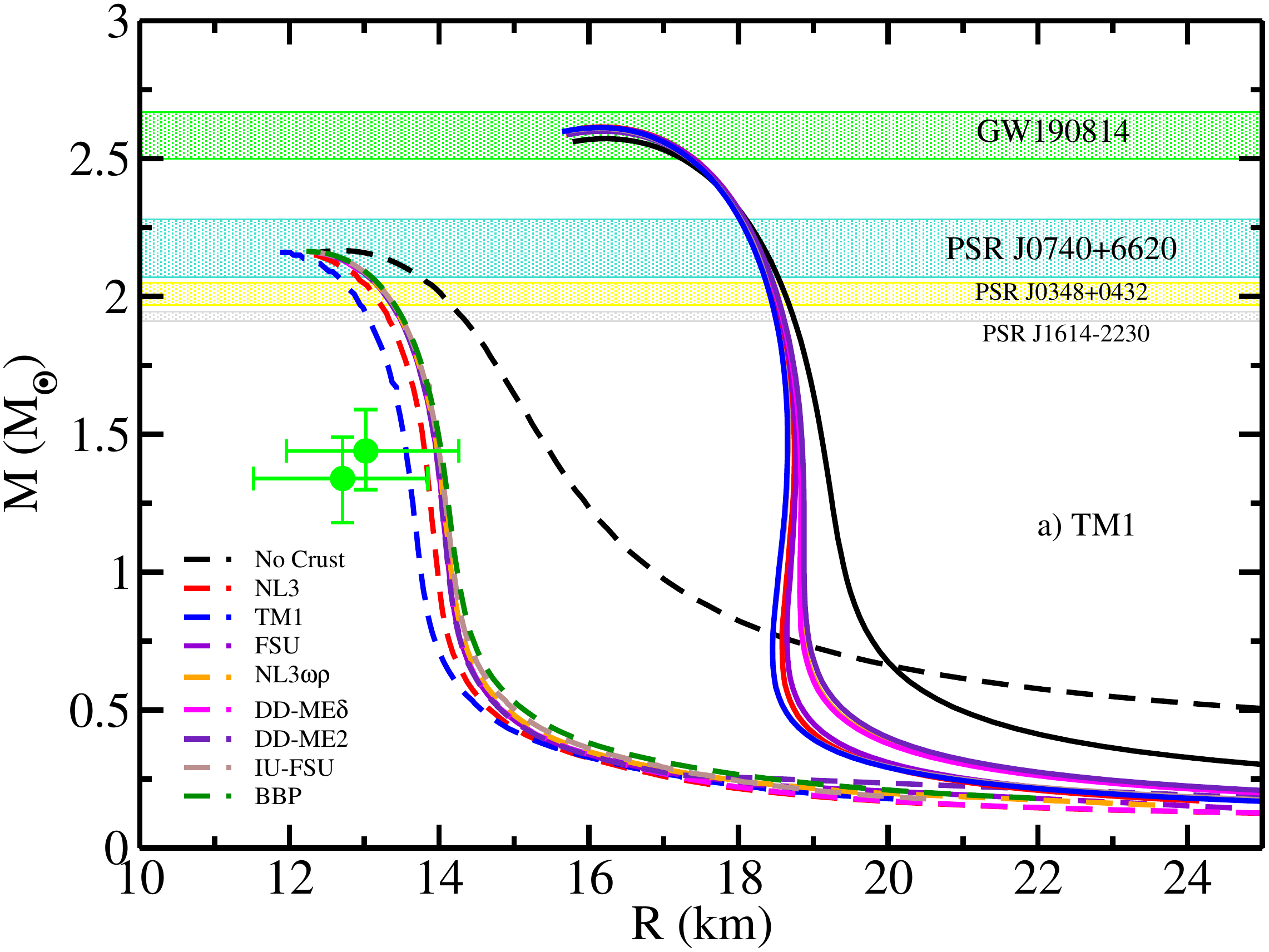} \\
		\includegraphics[width=0.70\linewidth]{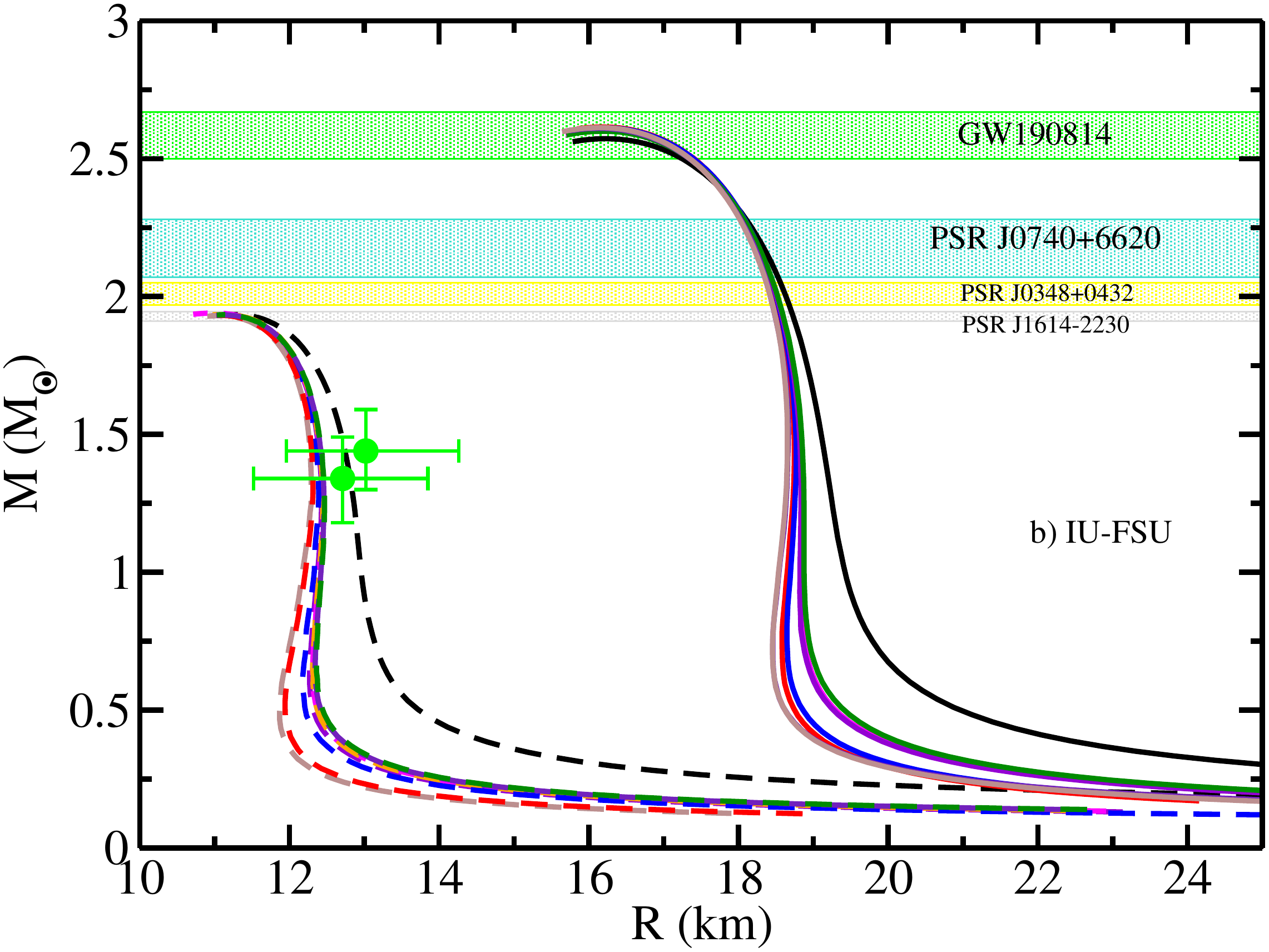} 
	\end{tabular}
	
	\caption{Same as Fig.~\ref{FIG:2}, but for a) TM1 and b) IU-FSU core EoS.}
	\label{tm1}
\end{figure}
The Mass-Radius profile for TM1 and IU-FSU parameter sets are shown in Fig.~\ref{tm1}. For TM1 core EoS with different inner crust EoS, the maximum mass and the corresponding radius lie in the range 2.141-2.177$M_{\odot}$  and 12.234-12.805 km, respectively as shown in figure. The radius at the canonical mass varies from 13.572 km for TM1 crust EoS to 15.549 km produced without using the inner crust. As seen in the figure, although every EoS for TM1 core along with different inner crust satisfies the mass constraint from recently observed GW data, only the unified EoS (TM1 inner crust + TM1 core) satisfies the radius constraint at the canonical mass, $R_{1.4}\le 13.76$ km. For IU-FSU core EoS, as shown in the figure, the maximum mass and radius vary from 1.931-1.940$M_{\odot}$  and 11.030-11.263 km, respectively. The radius at the canonical mass varies from 12.295 km for IU-FSU crust EoS to 12.778 km without inner crust, all satisfying the radius constraint from NICER measurements. The maximum mass and the radius for RNS with TM1 and IU-FSU core are almost identical with mass in the range 2.57-2.63$M_{\odot}$ satisfying the GW190814 mass constraint and radius around 16 km. This also shows the possibility of the secondary component of GW190814 to be a maximally RNS. The unified EoS in TM1 ($L_{crust}$ = $L_{core}$ = 110.6 MeV) and IU-FSU ($L_{crust}$ = $L_{core}$ = 47.2 MeV) produce an NS with smaller radius for SNS as well as RNS. 

\begin{figure}
	\centering
	\renewcommand{\arraystretch}{0}
	\setlength{\tabcolsep}{0pt}
	
	\begin{tabular}{cc}
		\includegraphics[width=0.70\linewidth]{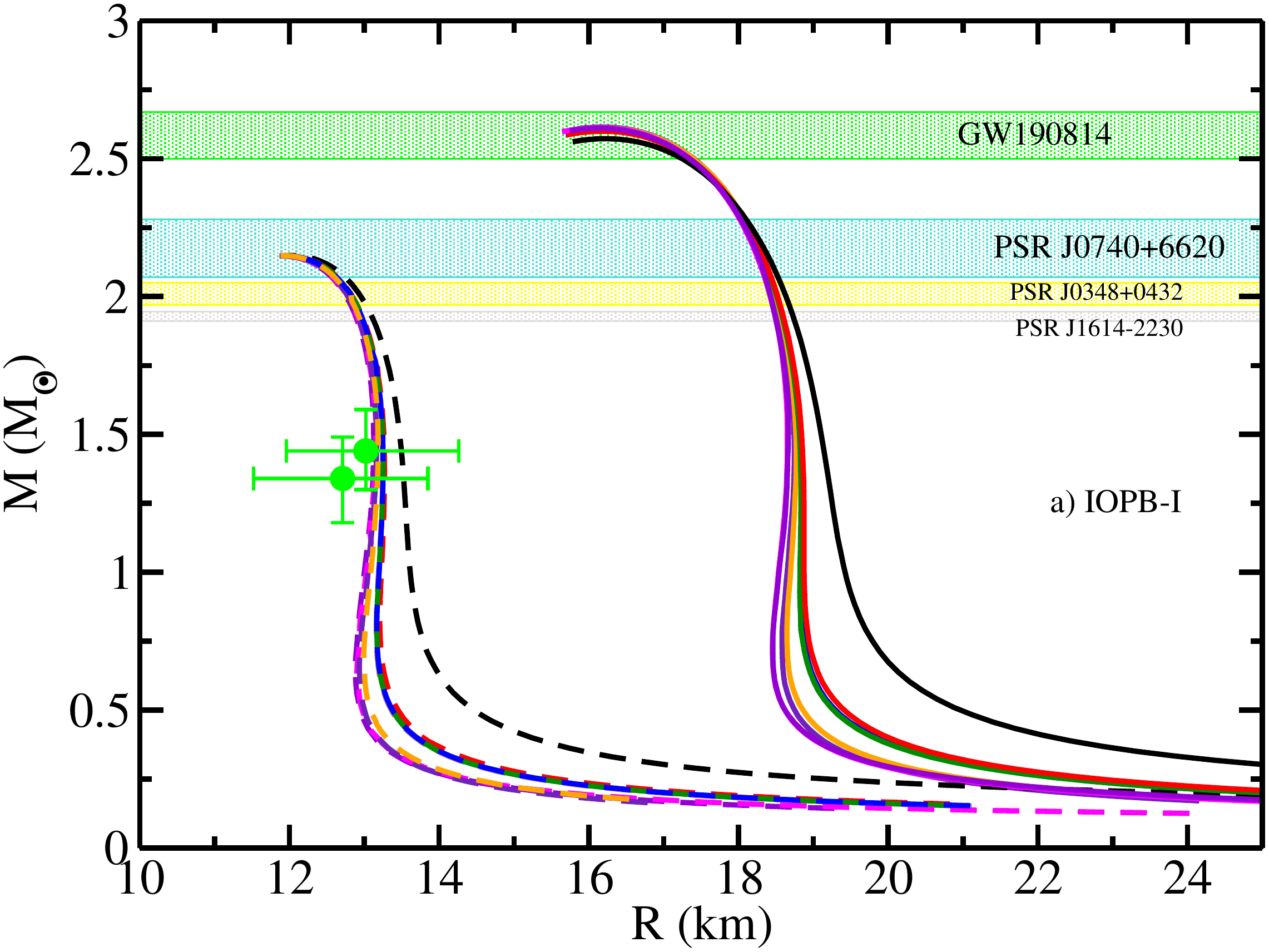} \\
		\includegraphics[width=0.70\linewidth]{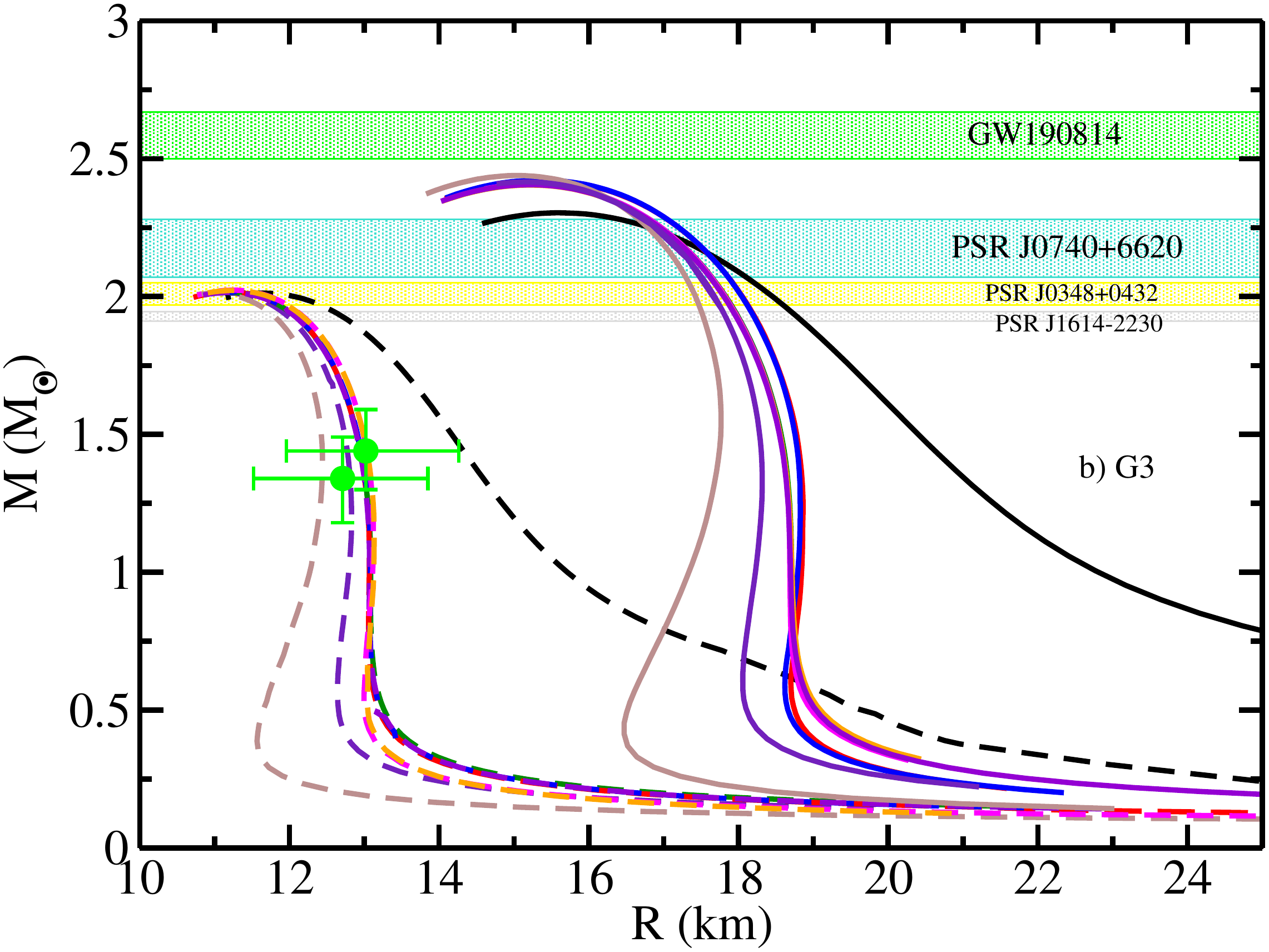} 
	\end{tabular}
	
	\caption{Same as Fig.~\ref{FIG:2}, but for a) IOPB-I and b) G3 core EoS.}
	\label{iopb}
\end{figure}

Fig.~\ref{iopb} shows the MR profile for IOPB-I and G3 core EoS. For IOPB-I set, the maximum mass varies from 2.141-2.156$M_{\odot}$ and the radius 11.872-12.029 km. $R_{1.4}$ varies from 13.118-13.508 km. 
Similar results follow for G3 EoS, where the NS maximum mass and the corresponding radius vary slightly with different inner crust EoS. However, as usual, the radius $R_{1.4}$ varies from 12.436-14.447 km. For RNS, the radius at the canonical mass varies from 18.65 to 19.18 km and 17.72 to 20.86 km for IOPB-I and G3 EoS, respectively. It is to be mentioned here that both IOPB-I and G3 sets do not form a unified EoS i.e, inner crust and core EoS with same symmetry energy slope parameter. However, we see that for IOPB-I core EoS, the FSU inner crust with a similar value of slope parameter as IOPB-I, predicts a smaller radius for NS at 1.4$M_{\odot}$ than any other crust EoS. Similar follows for G3 set ($L_{sym}$ = 49.3 MeV), the IU-FSU inner crust with a slope parameter value of 47.2 MeV gives a smaller radius NS.  
From the above MR relations, we see that the maximum mass and the radius of both static and rotating NS's do not change by a large amount using the inner crust with different slope parameters. The unified EoSs (NL3, TM1, and IU-FSU) produce NS's with a small radius at the canonical mass. However, for non-unified EoSs (IOPB-I and G3), the inner crust EoS from the same or different model with smaller symmetry energy slope parameter $L_{sym}$ predicts a smaller radius at the canonical mass of an NS.  

The effect of the inner crust on the static NS tidal deformability $\lambda$ is also studied for all core EoSs. In addition to this, the variation in the RNS properties like Moment of Inertia (MI) is also discussed. The tidal deformability of an NS is defined by Eqs.~(\ref{eq1.1}) and (\ref{eq1.2}).
\begin{figure}[hbt!]
	\centering
	\includegraphics[width=0.75\textwidth]{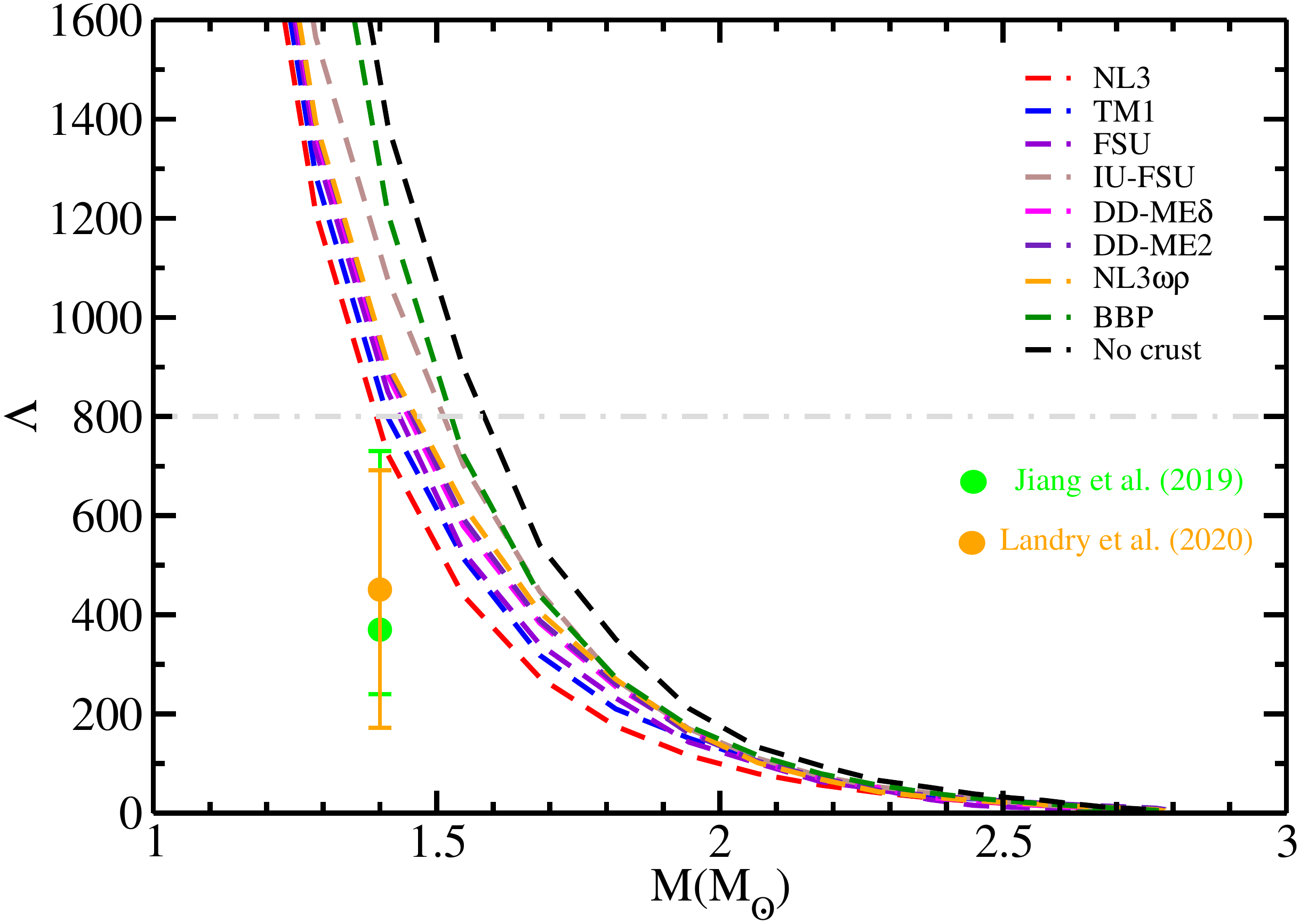}
	\caption{The relation between dimensionless tidal deformability and the mass of an NS for NL3 core EoS with different inner crust EoS. The overlaid arrows (green and orange) represents the combined constraints on the tidal deformability at 1.4$M_{\odot}$ from  PSR J0030+0451 and GW170817 and x ray data \protect\cite{PhysRevLett.121.161101, Jiang_2020,PhysRevD.101.123007} . The grey dashed line represents the upper limit on the $\Lambda_{1.4}$ value \protect\cite{PhysRevLett.119.161101} .}
	\label{FIG:5}
\end{figure}
Fig.~\ref{FIG:5} shows the variation of the dimensionless tidal deformability with the NS mass for NL3 core EoS with different inner crust EoS. The constraint on the $\Lambda$ from the recent GW data is also shown. The green overlaid arrow shows the recent constraints on the $\Lambda_{1.4}$ from the combined data of PSR J0030+0451 and GW170817,  $\Lambda_{1.4}=370_{-130}^{+360}$ \cite{Jiang_2020}, while the orange one shows the non-parametric constraints from PSRs + GWs + X-ray, $\Lambda_{1.4}=451_{-279}^{+241}$  \cite{PhysRevD.101.123007}.The grey dashed line represents the upper limit on the dimensionless tidal deformability at the canonical mass from GW170817 data, $\Lambda_{1.4}$ = 800  \cite{PhysRevLett.119.161101}. The NL3 unified EoS predicts the lowest value of the dimensionless tidal deformability at 1.4$M_{\odot}$, $\Lambda_{1.4}$ = 800, due to the small radius at the canonical mass. The other non-unified NL3 EoSs (NL3 core + other crust EoSs except NL3) predict a value in the range 800-1400 with $\Lambda_{1.4}$ = 1400 for NL3 without the inner crust. This shows that the unified EoS is important in determining the NS properties that support the contraints from recent GW data. 
\begin{figure}[hbt!]
	\centering
	\includegraphics[width=0.75\textwidth]{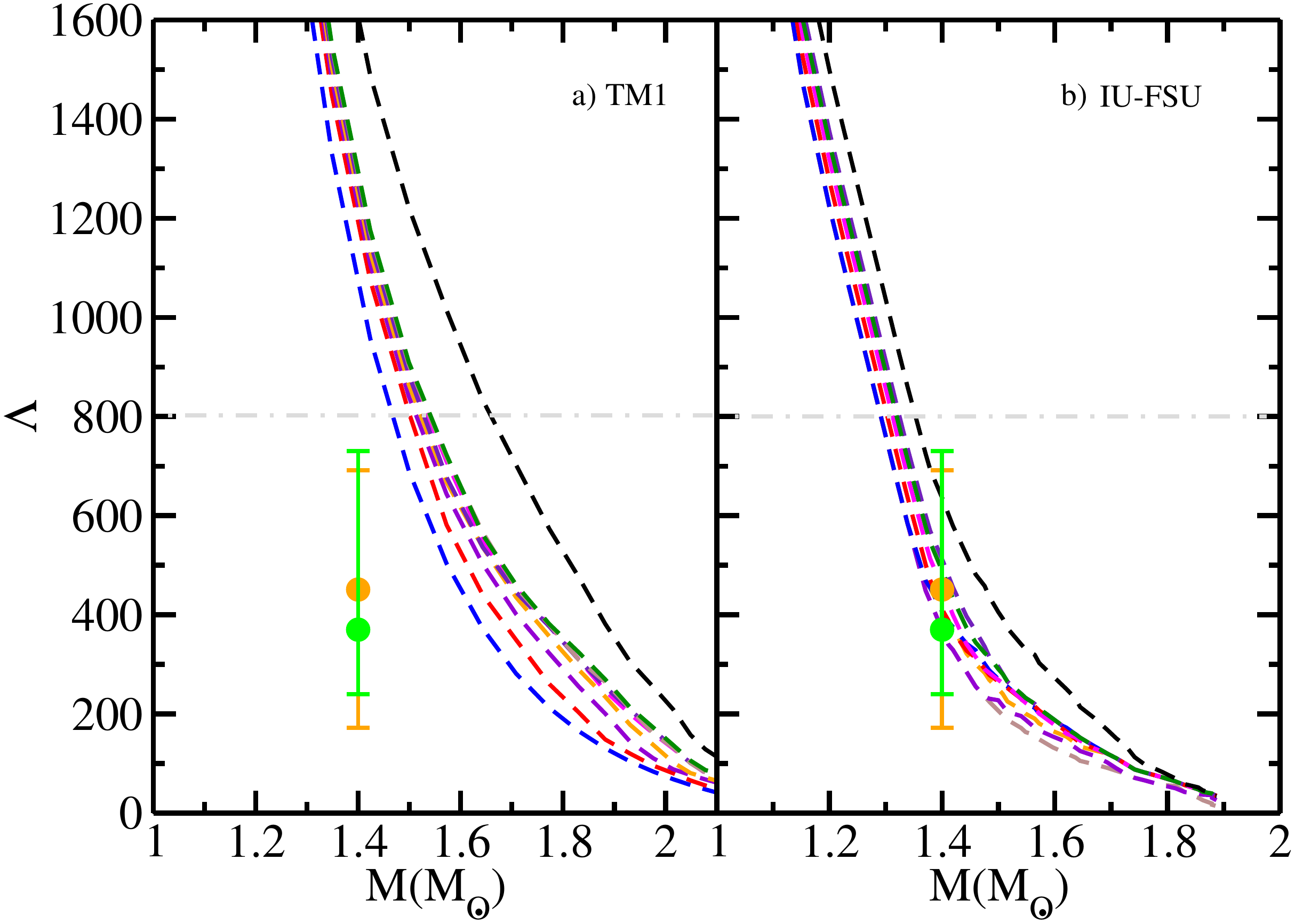}
	\caption{Same as Fig.~\ref{FIG:5}, but for a) TM1 and b) IU-FSU core EoS.}
	\label{FIG:6}
\end{figure}
\begin{figure}[hbt!]
	\centering
	\includegraphics[width=0.75\textwidth]{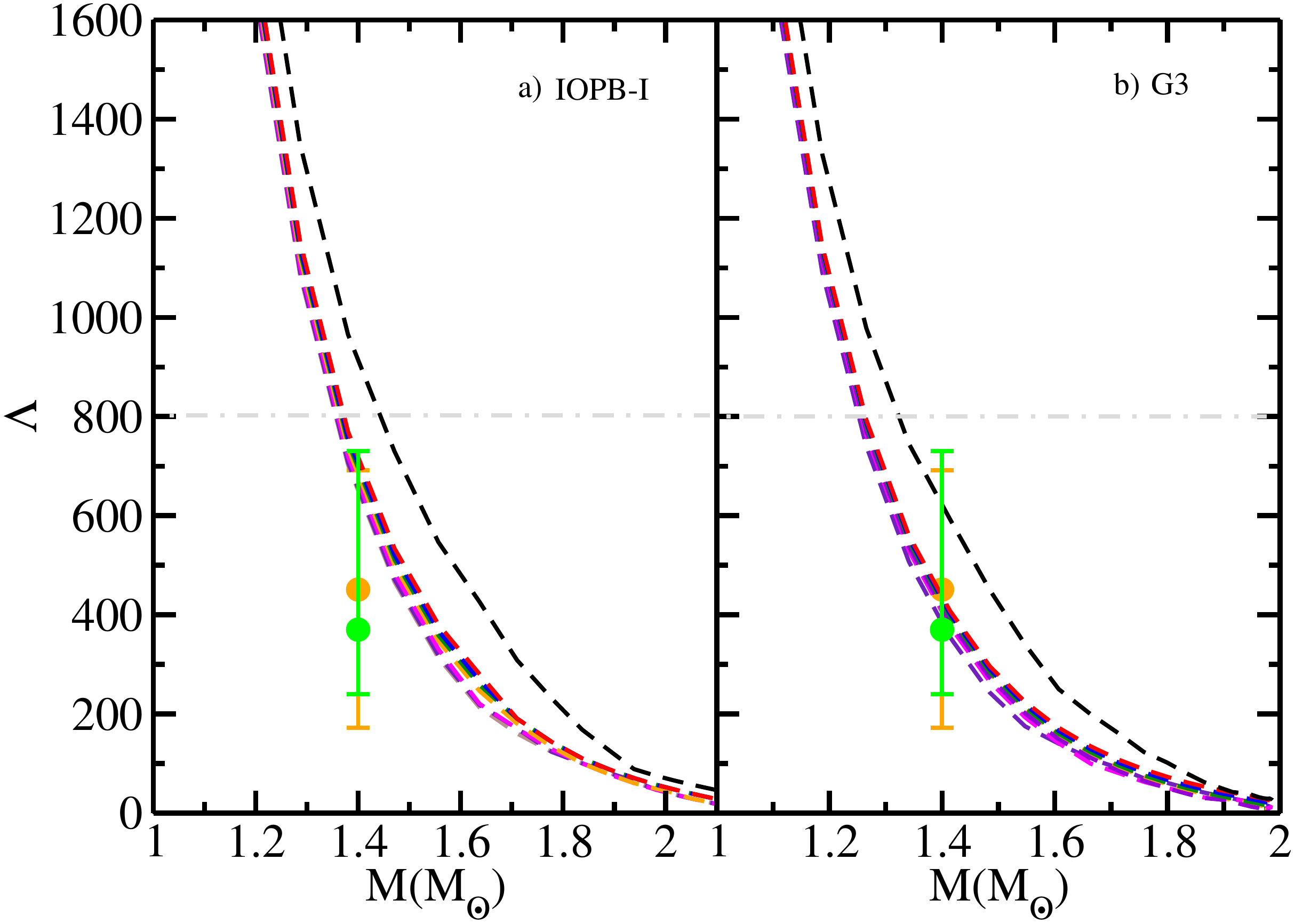}
	\caption{Same as Fig.~\ref{FIG:5}, but for a) IOPB-I and b) G3 core EoS.}
	\label{FIG:7}
\end{figure}

Fig.~\ref{FIG:6} shows the dimensionless tidal deformability for TM1 and IU-FSU core EoS with different crust EoSs. The unified TM1 EoS (TM1 crust + TM1 core) provides a low value for tidal deformability, $\Lambda_{1.4}$ = 1060. This value increases upto to $\Lambda_{1.4}$ =  1587 for TM1 NS without inner crust. For IU-FSU, the unified EoS gives $\Lambda_{1.4}$ = 368 and $\Lambda_{1.4}$ = 637 without inner crust EoS. For other non-unified EoSs, the variation in the tidal deformability at 1.4$M_{\odot}$ is very small for both TM1 and IU-FSU core EoSs.

Fig.~\ref{FIG:7} shows the dimensionless tidal deformability for IOPB-I and G3 core sets. For IOPB-I, the FSU crust EoS predicts the smallest value of $\Lambda_{1.4}$ = 637, while the other crust EoSs determine the value in the range 640-730. Similarly for  G3 set, the IU-FSU gives a low value for tidal deformability, $\Lambda_{1.4}$ = 349, while others provide a value in the range 393-450. The value increases with the increase in the value of the symmetry energy slope parameter. The NS without inner crust provides a value of $\Lambda_{1.4}$ = 914 and 620 for IOPB-I and G3 sets respectively.

In the simplest form, the moment of inertia $I$ is defined as the ratio of the angular momentum to the angular velocity of an NS, $I=J/\Omega$. In terms of the angular frequency $\omega$, the moment of inertia is defined  as \cite{LATTIMER2000121}
\begin{equation}
I\approx\frac{8\pi}{3}\int_{0}^{R} (\mathcal{E}+P)e^{-\phi(r)} \Big[1-\frac{2m(r)}{r}\Big]^{-1}\frac{\bar{\omega}}{\Omega}r^4 dr,
\end{equation}
where $\bar{\omega}$ is the dragging angular velocity of a rotating star, satisfying the boundary conditions
\begin{equation}
\bar{\omega}(r=R)=1-\frac{2I}{R^3}, \frac{d \bar{\omega}}{dr}|_{r=0}=0,
\end{equation}
In the present work, the NSs rotating at the Kepler frequency are studied, hence the numerical calculations for the moment of inertia are performed using the RNS code.
The MI of an NS has been calculated by various groups \cite{Stergioulas2003,Paschalidis2017}, but the variation in the value of MI with different inner crust EoS hasn't been calculated.
\begin{figure}[hbt!]
	\centering
	\includegraphics[width=0.75\textwidth]{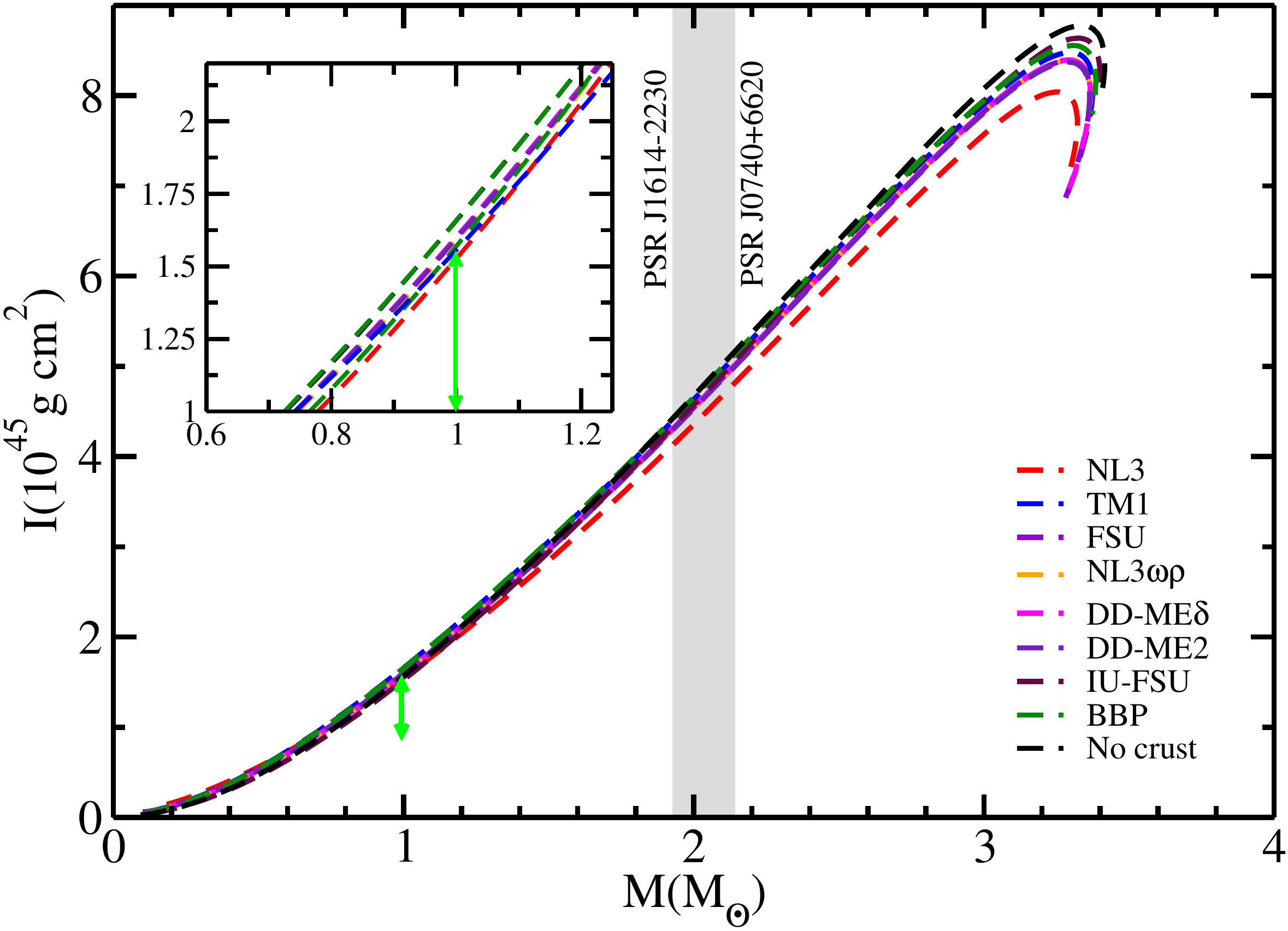}
	\caption{Variation of moment of inertia $I$ with the mass of an NS for NL3 core EoS with different crust EoS. The green arrow represents the constraints on the moment of inertia from PSR J0737-3039A \protect\cite{Landry_2018,PhysRevD.99.123026} obtained from the GW 170817 data analysis \protect\cite{PhysRevLett.119.161101,PhysRevLett.121.161101} . }
	\label{FIG:8}
\end{figure}
The variation in the moment of inertia for RNS with NL3 core EoS and different crust EoSs is shown in Fig.~\ref{FIG:8}. As clear from the figure, the change in the moment of inertia with different crust EoSs is very small. For unified NL3 EoS, the value of $I$ is found to be 1.53 $\times$ 10$^{45}$ g cm$^2$ well satisfying the constraint from PSR J0737-3039A  $I$ = 1.53$_{-0.24}^{+0.38}$ $\times$ 10$^{45}$ g cm$^2$. For the non-unified EoSs, the moment of inertia increases with the symmetry energy slope parameter $L_{sym}$ as they predict a large radius.

\begin{figure}[hbt!]
	\centering
	\includegraphics[width=0.75\textwidth]{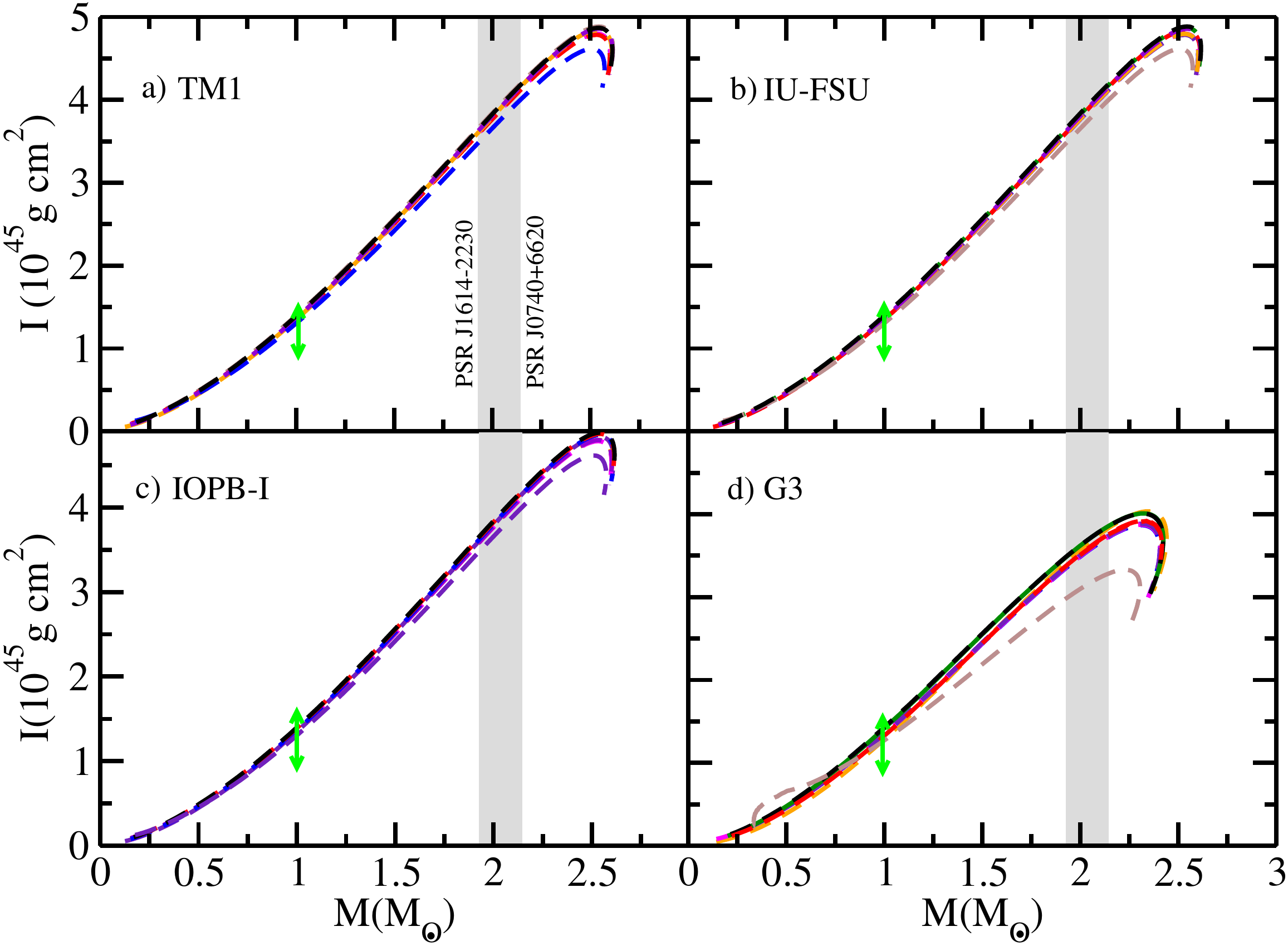}
	\caption{Same as Fig.~\ref{FIG:8}, but for a) TM1, b) IU-FSU, c) IOPB-I  and d) G3 core EoSs. }
	\label{FIG:9}
\end{figure}

For TM1, IU-FSU, IOPB-I, and G3 core EoSs, the moment of inertia variation with the NS mass is shown in Fig.~\ref{FIG:9}. For the TM1 and IU-FSU, the unified EoS provides a small moment of inertia $I$ = 1.31 \& 1.29 $\times$ 10$^{45}$ g cm$^2$, respectively. With no unified EoS available for IOPB-I and G3 sets, the low symmetry energy slope parameter crust EoS provides a lower value of the moment of inertia, $I$ = 1.27 \& 1.22 $\times$ 10$^{45}$ g cm$^2$, respectively. The moment of inertia is approximated by the relation $I\propto MR^2$ which shows that it increases almost linearly with the NS mass for all models. The NS radius starts to decrease as soon as the maximum mass is achieved which allows the moment of inertia to drop sharply. Since $I$ is proportional to the mass linearly and square of the radius, it is more sensitive to the density dependence of nuclear symmetry energy and its derivatives like slope parameter, which influence the NS radius \cite{Worley_2008}. By using different inner crusts for a given core EoS, the change in the radius at the canonical mass is observed, which affects the moment of inertia. Therefore the contribution to the moment of inertia due to the crust part of the star is much less than the core part.

\begin{figure}[hbt!]
	\centering
	\includegraphics[width=0.75\textwidth]{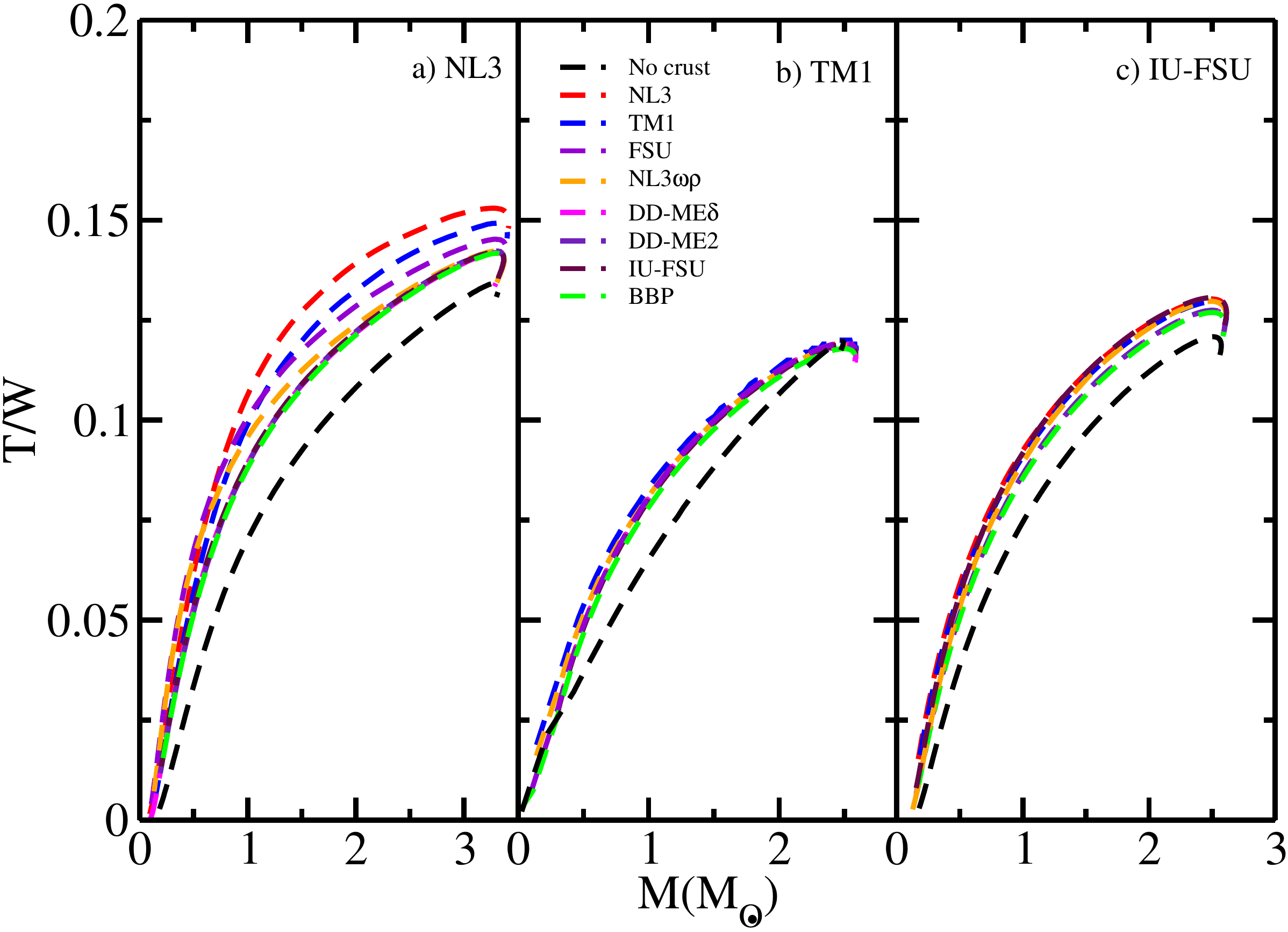}
	\caption{The variation of the Rotational kinetic energy to the gravitational potential energy ratio with the NS mass for a) NL3, b) TM1 and c) IU-FSU parameter sets with different crust EoSs. }
	\label{FIG:12}
\end{figure}

An important quantity that characterizes the rotation of a star is the T/W ratio, which is defined as the ratio of the rotational kinetic energy T to the gravitational potential energy W. For a given rotating star, if this ratio is greater than a critical value, the star becomes dynamically unstable. The critical value of the T/W ratio is not fixed. Some measurements predict a value of 0.27 as the critical limit \cite{1985ApJ...298..220T} and some show it to be in the range 0.14-0.27 \cite{2001ApJ...550L.193C}. Fig.~\ref{FIG:12} displays the variation in the T/W ratio of a RNS with the gravitational mass. With the increase in the central density, the angular velocity increases which in turn produces a star with a higher value of T/W ratio. The unified EoS for NL3, TM1, and IU-FSU parameter sets predict the highest value of the T/W ratio as compared to other non-unified equation of states. For NL3 set, the unified EoS ($L_{crust}$ = $L_{core}$ = 118.3 MeV) has T/W ratio of 0.15, while for TM1 and IU-FSU unified EoSs, the ratio is 0.12 and 0.13, respectively. The EoS without the inner crust part predicts a value of 0.13, 0.11, and 0.12 for NL3, TM1 and IU-FSU sets respectively.
\begin{table}[ht]
	\centering
	\caption{Variation (in percent) in the maximum mass, corresponding radius and the radius at 1.4$M_{\odot}$ of an NS without inner crust and the corresponding inner crust.} 
	{\begin{tabular}{p{1.0cm}p{1.2cm}p{1.2cm}p{1.2cm}p{1.2cm}p{1.2cm}cccc} 
			Model &  & BBP & IU-FSU & DD-ME2 & DD-ME$\delta$ & NL3$\omega \rho$ & FSU & TM1 & NL3 \\
			\hline			
			
			&  $\Delta M_{max}$ & 0.25&0.14&0.11&0.07&0.11&0.11&0.32&0.58\\
			{\small NL3} & $\Delta R_{max}$ & 1.90&2.62&0.87&1.22&0.93&0.79&0.73&1.19\\
			& $\Delta R_{1.4}$ & 3.28&3.12&3.22&3.30&3.54&4.74&6.51&7.57\\
			\hline
			
			&  $\Delta M_{max}$ & 0.60&1.65&1.51&1.70&1.38&1.56&0.64&1.65\\
			{\small TM1} & $\Delta R_{max}$ & 4.51&2.12&1.85&1.92&4.68&2.98&4.46&3.29\\
			& $\Delta R_{1.4}$ & 9.69&10.22&10.19&10.16&10.13&10.11&12.73&11.28\\
			\hline
			
			&  $\Delta M_{max}$ & 0.10&0.00&0.05&0.26&0.10&0.05&0.15&0.21\\
			{\small IU-FSU} & $\Delta R_{max}$ & 1.35&2.07&1.41&2.23&1.25&1.26&1.71&1.79\\
			& $\Delta R_{1.4}$ & 2.91&3.78&3.55&3.42&3.30&3.19&2.78&2.53\\
			\hline
			
			&  $\Delta M_{max}$ &0.04&0.02&0.06&0.04&0.04&0.07&0.04&0.04\\
			{\small IOPB-I} & $\Delta R_{max}$ &1.11&1.17&0.66&0.61&0.73&0.54&0.92&1.30 \\
			& $\Delta R_{1.4}$ & 1.84&2.57&2.34&2.29&2.26&2.89&1.85&1.79\\
			\hline
			
			&  $\Delta M_{max}$ & 0.05&0.00&0.15&0.20&0.15&0.30&0.04&0.50\\
			{\small G3} & $\Delta R_{max}$ & 3.94&5.74&4.02&3.79&4.24&3.28&0.92&3.50\\
			& $\Delta R_{1.4}$ & 11.50&13.91&11.82&11.66&9.45&11.36&9.56&10.17\\
			\hline
			\hline
		\end{tabular}\label{tbl2}}
\end{table}

Table \ref{tbl2} shows the deviation in the properties of a static NS like maximum mass, the corresponding radius and the radius at the canonical mass for the given parameter sets. The deviations are calculated by considering the NS with different inner crust EoS with respect to the NS without inner crust. The NS without inner crust predicts a very large radius at the canonical mass, while the unified EoS (same crust and core EoS) for any parameter set gives the low radius NS, which satisfies the radius constraints as explained before. For a given EoS, the deviation between an NS without inner crust and with crust is large, implying that the unified EoS gives a better estimate of the radius at the canonical mass as compared to other non-unified EoSs. It is clear that the variation in the maximum mass and the corresponding radius are very small for EoSs, but the radius at 1.4$M_{\odot}$ is highly impacted by the inner crust EoS. For NL3 core EoS, the variation in the radius at $R_{1.4}$ is maximum for NL3 inner crust $\approx$ 7.57$\%$ which is around 1.2 km. The maximum variation in the radius, $R_{1.4}$, for IOPB-I and G3 core EoSs is with the FSU and IU-FSU inner crust EoS, respectively. Such large deviations in the radius, $R_{1.4}$, show that a proper choice of inner crust EoS is important to calculate the mass and radius of an NS with small uncertainties in these values. 


The constraints on the inner crust EoS of an NS and the proper matching of inner crust with core EoS help consider the nuclear and astrophysical applications of the RMF model. A core EoS with a smaller symmetry energy slope parameter implies small symmetry energy at high densities \cite{Zhang2019}. For a model with higher symmetry energy at sub-saturation density, the inner crust properties of an NS are affected in addition to the pasta phases as shown in Refs. \cite{PhysRevC.90.045802,PhysRevC.89.045807}. Studies have shown that for a complete unified EoS, the inner crust part should either be from the same model or the symmetry energy slope parameter should match. Thus the crust-core transition allows the construction of a stellar EoS and a precise measurement of the NS properties for both static and rotating stars.

\section{Conclusion}
\label{sec:3}
The NS properties like mass and radius were investigated using the relativistic mean-field (RMF) model. To study the effect of symmetry energy and its slope parameter on an NS, the inner crust EoSs with different symmetry energy slope parameters have been used. For the outer crust, the BPS EoS is used for all sets as the outer crust part doesn't affect the NS maximum mass and radius. For the inner crust part, the NL3, TM1, FSU, NL3$\omega \rho$, DD-ME$\delta$, DD-ME2, and IU-FSU parameter sets have been used whose slope parameter varies from 118.3-47.2 MeV. For the core part, NL3, TM1, IU-FSU, IOPB-I, and G3 parameter sets are used. The unified EoSs are constructed by properly matching the inner crust EoS with outer crust and core EoS using the thermodynamic method. The EoSs constructed for the spherical and symmetrical NS under charge neutral and $\beta$-equilibrium conditions are taken as the input into the TOV equation to obtain NS properties. It is seen that although the NS maximum mass and the corresponding radius do not change by a large amount, the radius at the canonical mass, $R_{1.4}$, are largely impacted by using inner crust EoSs with different symmetry energy slope parameter. By varying the slope parameter from low to high values, the radius $R_{1.4}$ also increases. The effect of $L_{sym}$ on the NS maximum mass, radius and the radius at 1.4$M_{\odot}$ are calculated and the variation of about 2 km is found in the radius at the canonical mass. The properties like mass, radius, the moment of inertia, T/W ratio of maximally rotating stars are also calculated using the same combination of EoSs. It is seen that similar to SNS, the maximum mass and the corresponding radius for RNS does not vary much, but the radius at the canonical mass is affected by the slope parameter. The moment of inertia doesn't vary too much with a change in the symmetry energy slope parameter $L_{sym}$. The kinetic to potential energy ratio also varies with the change in the symmetry energy slope parameter of the crust. Such RNS properties are more related to the mass and the radius of the star and the variation in the radius at the canonical mass influences the other properties of the star. 

Several different aspects need to be further studied in the current work. A unified EoS for the parameter sets like IOPB-I and G3 with both crust and core part described by the same model with different slope parameters $L_{sym}$ will be a better investigation to see the behavior of radius and other NS properties at canonical mass.
\begin{savequote}[8cm]
\textlatin{The mysteries of universe are revealed to those who seek to know the truth of their own existence first.}

  \qauthor{---\textit{Anjali Chugh}}
\end{savequote}

\chapter{\label{ch:5-hadronquarkrns}Phase Transition in the context of GW190814}

\minitoc

\section{Introduction}
\label{intro}
The successful discovery of gravitational wave detection by LIGO and Virgo Collaborations (LVC) of a BNS merger GW170817 event \cite{PhysRevLett.119.161101,PhysRevLett.121.161101} has allowed us to study the dense matter properties at extreme conditions. The estimation of tidal deformability for NS provided a new constraint on the NS EoS. The total mass of the GW170817 BNS merger was found to be around 2.7$M_{\odot}$ with the heavier component of 1.16 - 1.60$M_{\odot}$ for low spin priors and the maximum mass approached 1.9$M_{\odot}$ for high spin priors \cite{PhysRevX.9.011001}. The variation of tidal deformability with the radius as $\Lambda \propto R^5$ provides a strong constraint on the nuclear EoS at high density. After GW170817, the second possible BNS event GW190425 occurred with a total mass of 3.3 - 3.7$M_{\odot}$. The mass of its components with high spin prior are around 1.12 - 2.52$M_{\odot}$ \cite{Abbott_2020}. Recently, a new gravitational wave event reported by LVC as GW190814 \cite{Abbott_2020a} was observed with a 22.2 - 24.3$M_{\odot}$ black hole and 2.50 - 2.67$M_{\odot}$ secondary component. The secondary component attracted a lot of attention as it has no measurable tidal deformability signatures and no electromagnetic counterpart. The mass of the secondary component of GW190814 lies in the lower region of the so-called mass-gap (2.5$M_{\odot}<M<5M_{\odot}$) which raises the question of whether it is a light black hole or a supermassive NS. To explain the secondary component of GW190814, many interesting works have been proposed recently regarding its nature as supermassive NS, lightest black hole, or fast pulsar \cite{PhysRevC.103.025808,10.1093/mnrasl/slaa168,2020PhRvL.125z1104T,Tsokaros_2020,2021ApJ...908L...1T}.

Different models with different parameterizations have been used in the literature to construct the NS EoS at supranuclear densities. 
The NM EoS at saturation density from many-body theories is well constrained. The extrapolation of these EoSs to higher densities $\approx 4-5 \rho_0$, where $\rho_0$ is the nuclear saturation density, describes the properties of NSs. However, only few EoSs like NL3 and recently proposed BigApple \cite{PhysRevC.102.065805,das2020bigapple} generate massive NSs with maximum mass of $\approx$ 2.6$M_{\odot}$.

The presence of exotic phases in the inner core of NSs has been studied over the past decade and the variation in the properties of NSs has constrained the EoS at high densities. \citet{Annala2020} has shown that the quarks are present in the NS core at several times the normal nuclear density. Hence the quark matter can exist inside the NSs in a deconfined phase \cite{PhysRevD.30.272,PhysRevD.30.2379} or as a mixed-phase of hadrons and quarks (hybrid star) \cite{PhysRevD.46.1274,zel_2010,refId0}. Depending upon the phase transition between outer hadronic matter and inner quark matter of the hybrid stars, the twin-star solution might appear as the mass-radius relation could exhibit two stable branches with the same maximum mass but different radius \cite{Gomes:2018bpw, PhysRevD.99.103009}. A steep first-order phase transition (large density jump) combined with an incredibly stiff quark equation of state can generate twins. It's seen that when twin-star solution appears, the tidal deformability also displays two distinct branches with the same maximum mass, which is different from the pure neutron and hybrid stars. However, a recent study shows that an NS with a maximum mass constraint of 2.5$M_{\odot}$ rules out the twin star solution \cite{PhysRevD.103.063042}.


In this work, we use a few recently obtained DD-RMF parameter sets which generate an NS with a maximum mass around 2.5$M_{\odot}$, thus implying the nature of the secondary component of GW190814 as a massive NS. Following a phase transition, the QM is studied using the Vector-Enhanced Bag (vBag) model \cite{Kl_hn_2015}. We use its mass to put additional constraints on the NS maximum mass and dense matter EoS.\par 

This chapter is organized as follows: in Sec.~\ref{sec:headings5}, the Vector-Enhanced Bag (vBag) model to discuss the quark matter is presented. The phase transition from hadron matter to quark matter is also discussed. The results Sec.~\ref{results5} is divided into two parts. In Sec.~\ref{snsresults}, the parameter sets DD-ME1, DD-ME2, DD-LZ1, DD-MEX, and DDV are used for spherically symmetric, static NS. The star matter properties such as mass, radii, and tidal deformability for the given parameter sets are calculated. The properties of the phase transition are also studied. In Sec.~\ref{rnsresults}, the star matter properties are calculated for a rotating neutron star with quark core for the parameter sets DDV, DDVT, DDVTD, DD-LZ1, and DD-MEX. The summary and concluding remarks are finally given in Sec.~\ref{summary}. The calculations presented in this chapter are based on the work from Refs. \cite{PhysRevC.103.055814,Rather_2021}.

\section{Formalism}
\label{sec:headings5}

\subsection{Vector-Enhanced Bag (vBag) model}
The commonly used effective models to explain the presence of quark matter in NS cores either mimic quark confinement while keeping the quark masses constant like the Bag model \cite{PhysRevD.9.3471,PhysRevD.17.1109,PhysRevD.30.2379} or do exhibit the Dynamic Chiral Symmetry Breaking (D$\chi$SB) without confinement like Nambu-Jona-Lasino (NJL) models \cite{PhysRev.122.345,PhysRev.124.246}. Both these types of models do not include the repulsive vector interactions, which is important in the study of NS properties as it allows the HSs to achieve 2$M_{\odot}$ limit which results from the recent constraints of PSR J1614-2230, PSR 0348+0432, and PSR J0740+6620. 

 We employ an extension of the bag model, Vector-Enhanced Bag (vBag) model \cite{Kl_hn_2015} which is an effective model accounting for D$\chi$SB and repulsive vector interactions. It also takes into consideration the phenomenological correction to the quark matter EoS that characterizes the deconfinement and is dependent on the hadron EoS used to build the phase transition. The repulsive vector interaction is important as it allows the HSs to attain 2$M_{\odot}$ limit on the maximum mass \cite{universe4020030}. Fits to the pressure of the chirally restored phase justifies the inclusion of flavor-dependent chiral bag constants. In addition, a deconfined bag constant is provided to minimize the energy per particle, favoring stable strange matter.

The expression for the energy density and the pressure in vBag model are given as \cite{vbageos} 
\begin{equation}
\mathcal{E}_Q = \sum_{f=u,d,s} \mathcal{E}_{vBag,f}-B_{dc},
\end{equation}
\begin{equation}
P_Q = \sum_{f=u,d,s} P_{vBag,f}+B_{dc},
\end{equation}
where $B_{dc}$ is the deconfined bag constant introduced to lower the energy per particle thereby favoring stable strange matter. $\mathcal{E}_{vBag,f}$
and $P_{vBag,f}$ are the energy density and pressure of a single quark flavor defined as:
\begin{equation}
\mathcal{E}_{vBag,f}(\mu_f) = \mathcal{E}_{FG,f}(\mu_f^*)+\frac{1}{2}K_{\nu}n_{FG,f}^2 (\mu_f^*)+B_{\chi,f},
\end{equation}   
\begin{equation}
P_{vBag,f}(\mu_f) = P_{FG,f}(\mu_f^*)+\frac{1}{2}K_{\nu}n_{FG,f}^2 (\mu_f^*)-B_{\chi,f},
\end{equation}
Here FG represents the ideal, zero temperature Fermi gas formula. $K_{\nu}$ parameter is a coupling constant resulting from the vector interactions and controls the stiffness of the quark matter EoS \cite{Wei_2019}. $B_{\chi,f}$ represents the bag constant for a single quark flavor. The chemical potential $\mu_f^*$ of the system is parameterized by the relation
\begin{equation}
\mu_f =\mu_f^* +K_{\nu}n_{FG,f}(\mu_F^*).
\end{equation}  
An effective bag constant is defined in the vBag model so that the phase transition to quark matter occurs at the same chemical potential
\begin{equation}
B_{eff}=\sum_{f=u,d,s}B_{\chi,f}-B_{dc}.
\end{equation}
This allows us to illustrate how $B_{eff}$ can be used in the case of two and three flavor quark matter in HSs.

For the quark matter, the charge neutrality and $\beta$-equillibrium conditions are given as:
\begin{equation}
\frac{2}{3}\rho_u -\frac{1}{2}(\rho_d+\rho_s)-\rho_e-\rho_u =0,
\end{equation}
\begin{equation}
\mu_s=\mu_d=\mu_u+\mu_e;
\mu_{\mu}=\mu_e.
\end{equation}
\subsection{Phase Transition}
In the present work, we employ both Gibbs and Maxwell methods to construct the phase transition between hadrons and quarks to determine how the change in the phase transition affects the NS properties such as mass, radius, and tidal deformability.

The equations governing the global charge-neutral conditions, Gibbs construction, are discussed in Sec.~\ref{phase}.

In Maxwell Construction, the local charge neutrality condition is defined as:
\begin{equation}
\rho_H(\mu_B,\mu_e) =0; \rho_Q(\mu_B,\mu_e)=0.
\end{equation}
The expressions for the pressure and chemical potential
are then given as:
\begin{equation}
P_H(\mu_B,\mu_e)=P_Q(\mu_B,\mu_e) = P_{MP},
\end{equation}
\begin{equation}
\mu_{B,H} =\mu_{B,Q}.
\end{equation}

\section{Results and Discussions}
\label{results5}
\subsection{Static Neutron stars}
\label{snsresults}
To study the hadron-quark phase transition and to determine the NS properties, several DD-RMF parameterizations are used.
The nucleon and meson masses and different coupling constants between nucleon and mesons along with the independent parameters $a,b,c,d$ for $\sigma$, $\omega$ and $\rho$ mesons for  DD-ME1, DD-ME2, DD-MEX, DD-LZ1, DDV, DDVT, and DDVTD  parameter sets are shown in Table \ref{tab2.3}.  The nuclear matter properties of the above-mentioned parameter sets are shown in Table \ref{tab2.4}. 
\begin{figure}[hbt!]
	\centering
	\includegraphics[width=0.75\textwidth]{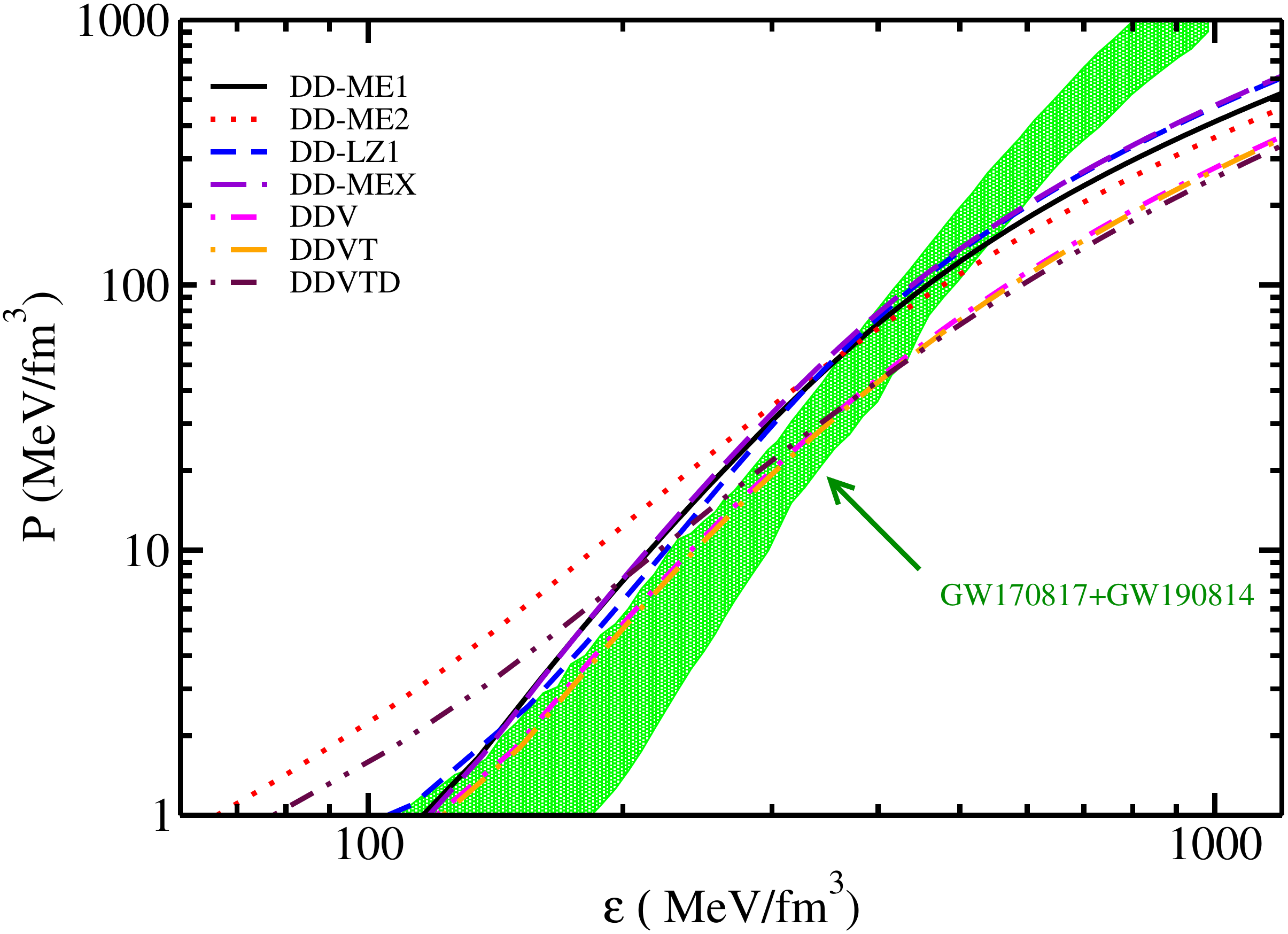}
	\caption{ Energy density vs pressure for the given DD-LZ1, DD-ME1, DD-ME2, DD-MEX, DDV, DDVT, and DDVTD parameter sets. The joint constraints from GW170817 and GW190814 shown are taken from \cite{Abbott_2020a}. }
	\label{fig1} 
\end{figure}

Fig.~\ref{fig1} shows the variation of pressure with energy density (EoS) for an NS in beta-equilibrium and charge-neutral conditions. The DD-ME2 parameter set produces stiff EoS at low densities while DD-LZ1 and DD-MEX produce stiff EoS in the high-density region. DDV set produces soft EoS at both low and high densities and hence defines an NS with low maximum mass as compared to other sets. The stiff EoS produced by the given parameter sets results in a high pressure due to strong vector potentials. The recent constraints on the EoS from GW170817 and GW190814 are also shown in the shaded area. The joint constraints were calculated by assuming a spectral distribution of EoS conditioned in GW170817 and re-weighted each EoS by the probability that its maximum mass is at least as large as the secondary component of GW190814. This was introduced by considering the GW190814 event as an NS-Black hole (NSBH) merger, with its secondary component assumed to be an NS. For this scenario, the maximum mass should be not less than the secondary component of GW190814, which constraints the distribution of EoSs compatible with astrophysical data \cite{Abbott_2020a}.  With energy density less than $\mathcal{E} \approx$ 600 MeV/fm$^3$, the DD-RMF EoSs satisfy the constraints from the gravitational waves. As the energy density increases, the pressure from the obtained EoSs starts to lower than the gravitational wave constraints. For the unified EoS, the Baym-Pethick-Sutherland (BPS) EoS \cite{Baym:1971pw} is used for the outer crust part. For the inner crust, the EoS in the non-uniform matter is generated by using the DD-ME2 parameter set in Thomas-Fermi approximation \cite{PhysRevC.94.015808}.  

To construct the hadron-quark phase and determine the phase transition properties of a three-flavor configuration, we consider the effective bag constant with a value of $B_{eff}^{1/4}$ = 130 MeV and $B_{eff}^{1/4}$ = 160 MeV. The value of $K_\nu$
parameter varies for two flavor and three flavor configurations. In the present study, we  will keep its value fixed at $K_{\nu} = 6$ GeV$^{-2}$.
\begin{figure}[hbt!]
	\centering
	\includegraphics[width=0.75\textwidth]{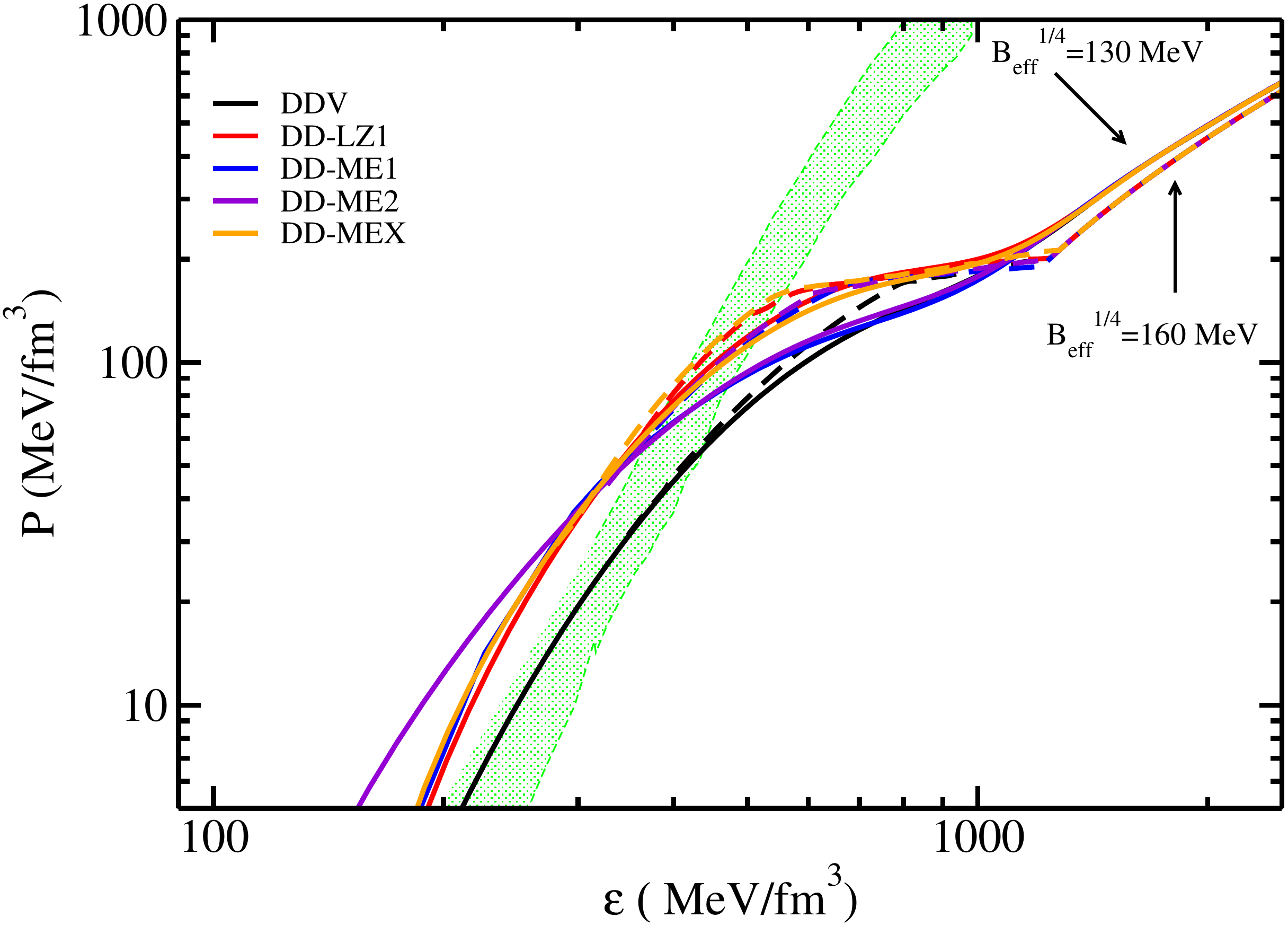}
	\caption{ Equation of state for the hadron-quark phase transition for different DD-RMF hadronic parameter sets and three flavor quark matter at $B_{eff}^{1/4}$ = 130 \& 160 MeV using Gibbs construction. The solid (dashed) lines represent the phase transition at $B_{eff}^{1/4}$ = 130 MeV ($B_{eff}^{1/4}$ = 160 MeV). }
	\label{fig2} 
\end{figure}

Fig.~\ref{fig2} shows the hadron-quark phase transition
with DD-RMF parameter sets for hadronic matter and vBag model for quark matter using the Gibbs method for constructing mixed-phase. The global charge neutrality condition ensures a smooth transition between the two phases. For effective bag constant $B_{eff}^{1/4}$ = 160 MeV, the phase transition takes place at higher density as compared to the bag value $B_{eff}^{1/4}$ = 130 MeV. In DDV EoS, the transition to quark matter at $B_{eff}^{1/4}$ = 130 MeV starts from the energy density $\mathcal{E} \approx$ 400MeV/fm$^3$ and ends at around $\mathcal{E} \approx$ 1200 MeV/fm$^3$ which corresponds to the density range $\rho_B =(2.47-4.03)\rho_0$. For $B_{eff}^{1/4}$ = 160 MeV, the phase transition region exists from $\rho_B = (3.69-5.31)\rho_0$. Similarly for DD-MEX EoS, the phase transition begins from 2.45 to 4.44$\rho_0$ and 3.09 to 5.57$\rho_0$ for bag values 130 and 160 MeV, respectively. It is clear that the phase transition for higher bag values occurs at higher densities and with large mixed-phase region.
\begin{figure}[hbt!]
	\centering
	\includegraphics[width=0.75\textwidth]{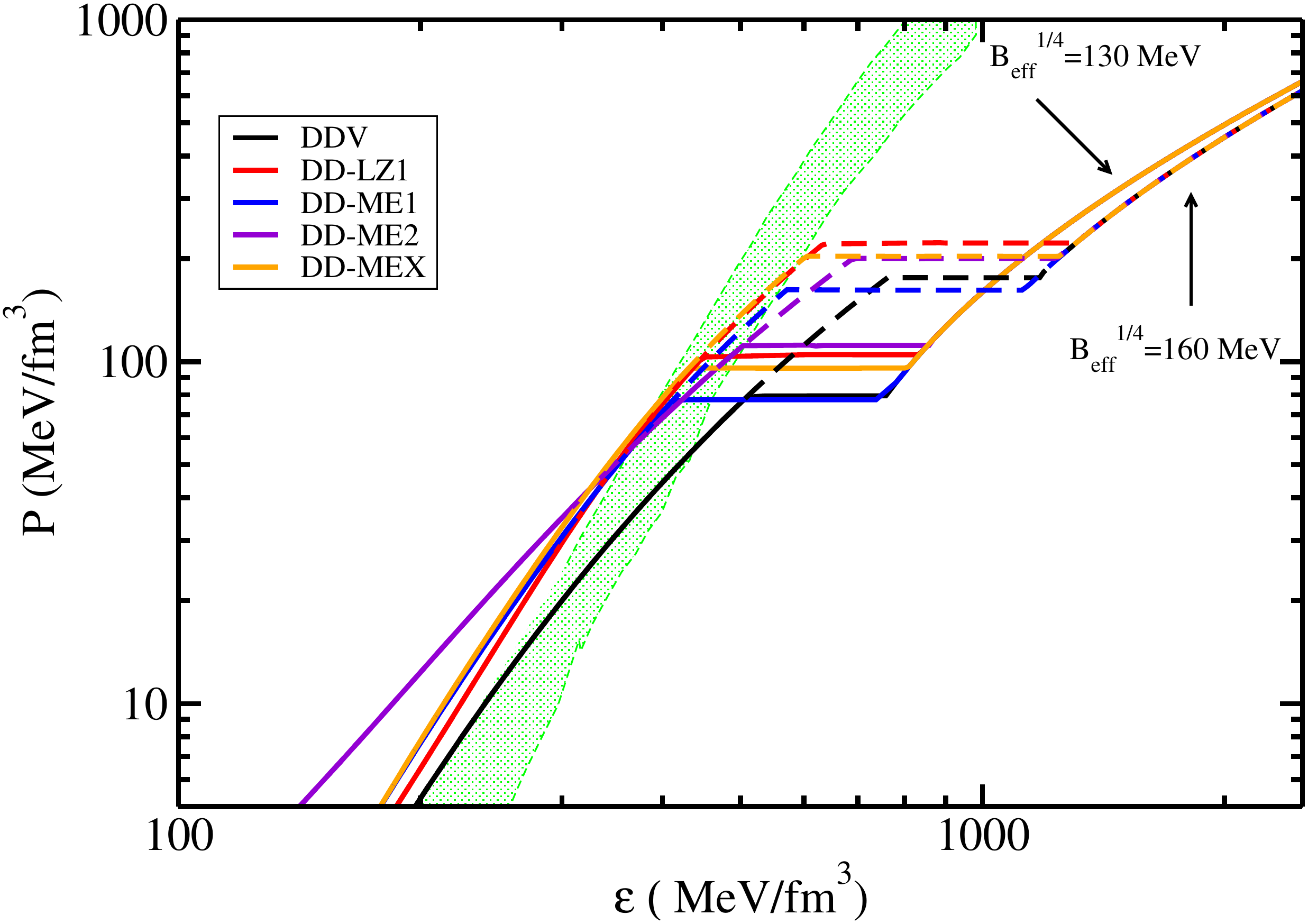}
	\caption{ Same as Fig.~\ref{fig2} but using Maxwell construction. }
	\label{fig3} 
\end{figure}

Fig.~\ref{fig3} represents the hadron-quark phase transition using the Maxwell construction method. The local charge neutrality condition allows the phase transition to take place at constant pressure which results in a sharp shift from hadron matter to quark matter. For DDV EoS, the phase transition occurs in the density region $(2.62-3.91)\rho_0$ for $B_{eff}^{1/4}$ = 130 MeV and $(4.08-5.23)\rho_0$ for bag constant $B_{eff}^{1/4}$ = 160 MeV. As clear from Figs. \ref{fig2} and \ref{fig3}, the EoS at low densities (hadronic) satisfies the joint GW constraints. The phase transition to quark matter satisfies this constraint at beginning of the mixed-phase for a low value of bag constant. As the pure quark phase begins, the softness of EoS shifts away from the joint GW170817 and GW190814 constraints. This implies that a too stiff EoS is required for a quark matter to satisfy these constraints. 
\begin{table}
	\centering 
	\caption{Phase transition density for hadron-quark matter at $B_{eff}^{1/4}$ = 130 \& 160 MeV using both Maxwell and Gibbs construction methods. $\rho_{MP}$ represents density of the mixed-phase region in terms of the saturation density $\rho_{0}$ which has the dimensions of fm$^{-3}$. }
	\begin{tabular}{p{1.8cm}|p{1.8cm}cp{1.8cm}c|p{1.8cm}cp{1.8cm}c}
		\hline
		&&&&\multicolumn{2}{c}{$\rho_{MP}(\rho_0)$}& \\
		\hline
		\multirow{3}{*}{EoS} & \multicolumn{4}{c|}{Gibbs Construction} & %
		\multicolumn{4}{c}{Maxwell Construction}\\
		\cline{2-9}
		& \multicolumn{2}{c}{130 MeV} & \multicolumn{2}{c|}{160 MeV} & \multicolumn{2}{c}{130 MeV} & \multicolumn{2}{c}{160 MeV} \\

		\hline
		DDV&2.47-4.03&&3.69-5.31&&2.62-3.91&&4.08-5.23& \\
		
		DD-LZ1&2.56-4.23 &&3.04-5.43 &&2.71-4.18 &&3.21-5.24& \\
		
		DD-ME1&2.70-4.18&&3.04-5.47&&2.89-4.11&&3.39-5.41& \\
		
		DD-ME2& 2.87-4.39&&3.42-5.53&&3.01-4.35&&3.75-5.44&\\
		
		DD-MEX&2.45-4.44&&3.09-5.57&&2.49-4.28&&3.47-5.49& \\
		\hline
	\end{tabular}
	\label{tab3}
\end{table}
Table \ref{tab3} shows the phase transition density region between hadron and quark matter at bag values $B_{eff}^{1/4}$ = 130 and 160 MeV using both Maxwell and Gibbs construction. The mixed-phase region exists between $(2-6)\rho_0$, where $\rho_0$ is the nuclear saturation density, which is important in obtaining NSs with a maximum mass larger than 2$M_{\odot}$ \cite{PhysRevC.66.025802}. The increase in the value of bag constant delays the phase transition and softens the pure quark phase \cite{PhysRevD.88.063001}. Also, the phase transition in the case of GC starts earlier than MC which is consistent with the work from Refs. \cite{PhysRevD.85.023003}. However, the width of the mixed-phase region in GC is much broader than in MC and it increases further for GC as the bag constant increases. These properties certainly affect the mass and radius.
\begin{figure}[htb!]
	\centering
	\includegraphics[width=0.75\textwidth]{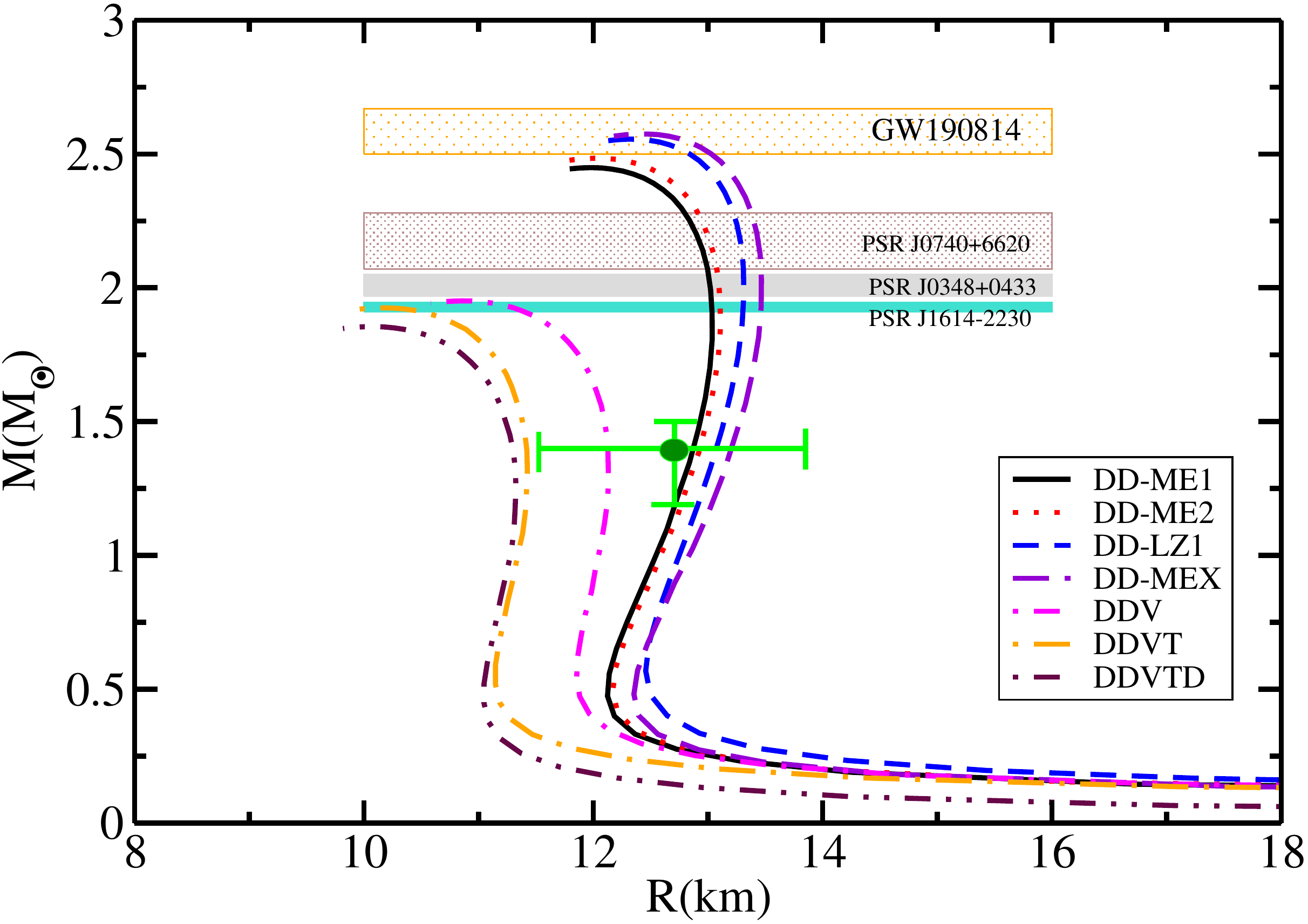}
	\caption{ Mass vs Radius profiles for pure DD-LZ1, DD-MEX, DDV, DDVT, and DDVTD parameters for a static NS. The recent constraints on mass from the various gravitational wave data and the pulsars (shaded region) \cite{Abbott_2020a,Demorest2010,Antoniadis1233232,Cromartie2020} and radii \cite{Miller_2019,Riley_2019} are also shown.}
	\label{fig5.10}
\end{figure}

Fig.~\ref{fig5.10} displays the hadronic mass vs radius curves for DD-LZ1, DD-MEX, DDV, DDVT, and DDVTD parameter sets. The DD-LZ1 set produces an NS with a maximum mass of 2.55$M_{\odot}$  and with a radius of 12.30 km. DD-MEX set produces a 2.57$M_{\odot}$ NS with a 12.46 km radius. Both these parameter sets satisfy the constraints from recent gravitational wave data GW190814 and recently measured mass and radius of PSR J0030+0451, $M=1.34_{-0.16}^{+0.15}$$M_{\odot}$ and $R=12.71_{-1.19}^{+1.14}$ km by NICER. The DDV, DDVT, and DDVTD predict a maximum mass of 1.95$M_{\odot}$, 1.93$M_{\odot}$ and 1.85$M_{\odot}$ for a static NS with 12.11, 11.40 and 11.33 km radius at canonical mass, $R_{1.4}$, respectively. DDV and DDVT satisfy the mass constraint from PSR J1614-2230 and radius constraint from PSR J0030+0451. The DDVTD parameter set produces an NS with a slightly lower maximum mass than PSR J1614-2230. The shaded regions display the constraints on the maximum mass of an NS from PSR J1614-2230 (1.928 $\pm$ 0.017$M_{\odot}$), PSR J0348+0432 (2.01 $\pm$ 0.04$M_{\odot}$), MSP J0740+6620 (2.14 $^{+0.10}_{-0.09}$$M_{\odot}$), and GW190814 (2.50-2.67$M_{\odot}$).

\begin{figure}[hbt!]
	\centering
	\includegraphics[width=0.80\textwidth]{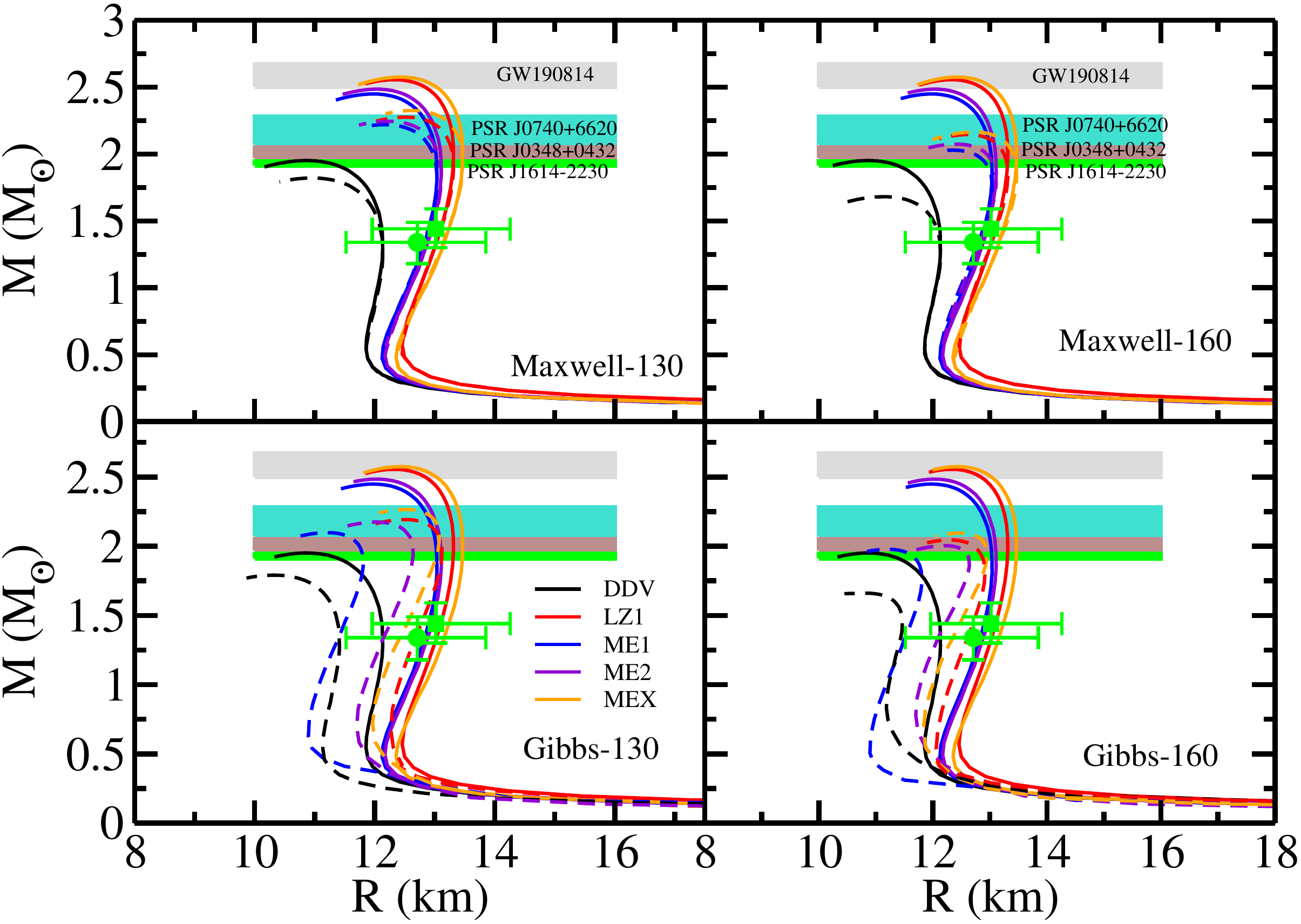}
	\caption{ Mass-Radius profile for pure hadronic DD-RMF parameters and HSs for different bag constants. The solid (dashed) lines represent MR plot for pure hadronic matter (hybrid NSs). The upper panels display the HS configuration with MC and the lower panels represent the same configuration with GC. The recent constraints on the mass \cite{Abbott_2020a,Demorest2010,Antoniadis1233232,Cromartie2020} and on the radii from NICER's observation \cite{Miller_2019,Riley_2019} are also shown.}
	\label{fig4}
\end{figure}
To understand the configuration of the star produced by a given EoS model, the mas-radius profile is analyzed. The HS models are divided according to the phase transition construction and will be assessed by the pure hadronic configurations. 

Fig.~\ref{fig4} shows the mass-radius curves for pure hadronic DD-RMF EoSs (solid lines) and hybrid NS EoS (dashed lines) at different bag constants. The upper two panels represent the HS configuration with Maxwell construction at two different bag values $B_{eff}^{1/4}$ = 130 \& 160 MeV, while the lower panels represent the HS configuration with Gibbs construction at the same bag values 130 \& 160 MeV. The pure hadronic EoSs produce an NS with a maximum mass of $\approx$ 2.5$M_{\odot}$ for DD-LZ1, DD-ME1, DD-ME2, and DD-MEX parameter sets, while DDV set produces an NS with a maximum mass of 1.95$M_{\odot}$. The vBag model parameters $K_{\nu}$ and $B_{eff}$ control the type of curves that result after the phase transition. While the $K_{\nu}$ parameter controls the stiffness of the curves, $B_{eff}$ triggers the location of the phase transition along the curve. In the upper two panels of Fig.~\ref{fig4}, the MR curves produce a sharp discontinuous transition from hadron matter to quark matter due to sharp phase transition in Maxwell construction. The NS maximum mass is reduced from 2.555$M_{\odot}$ to 2.275$M_{\odot}$ for DD-LZ1 hybrid EoS at $B_{eff}^{1/4}$ = 130 MeV and further reduced to 2.146$M_{\odot}$ at $B_{eff}^{1/4}$ = 160 MeV. Other hybrid EoSs follow a similar pattern.
 All the HS configurations satisfy the recently observed mass and radius constraints except that from DDV HS configuration whose maximum mass lies below the 1.9$M_{\odot}$ from PSR J1614-2230 at both 130 \& 160 MeV effective bag constant. This implies that a DDV EoS with phase transition to quark matter is too soft to satisfy the recent astrophysical constraints on the maximum mass and radius.

The HS models with Gibbs construction produce mass-radius curves with a smooth transition from hadron to quark matter because of the smoothly mixed-phase in GC. The maximum mass of HS DD-LZ1 decreases from 2.555$M_{\odot}$ to 2.192$M_{\odot}$  at $B_{eff}^{1/4}$ = 130 MeV and to 2.043$M_{\odot}$ at $B_{eff}^{1/4}$ = 160 MeV. We see that the maximum mass in MC is higher than that for the GC case. It is because of the delayed phase transition in MC than the GC that allows the star to stay longer in the hadronic phase. The radius at the maximum mass changes from 12.297 to 12.475 km and 12.355 km at bag constant 130 \& 160 MeV, respectively. For MC, the radius changes to 12.428 and 12.574 km respectively. This shows that the Maxwell construction produces an NS with a large maximum mass and radius as compared to the Gibbs construction. However, the radius at the canonical mass, $R_{1.4}$, is the same for MC as the pure hadronic star but smaller for GC as seen in Fig.~\ref{fig4}. Thus the maximum mass of the given NS configurations is lowered to satisfy the 2$M_{\odot}$ constraint. 
\begin{figure}[hbt!]
	\centering
	\includegraphics[width=0.80\textwidth]{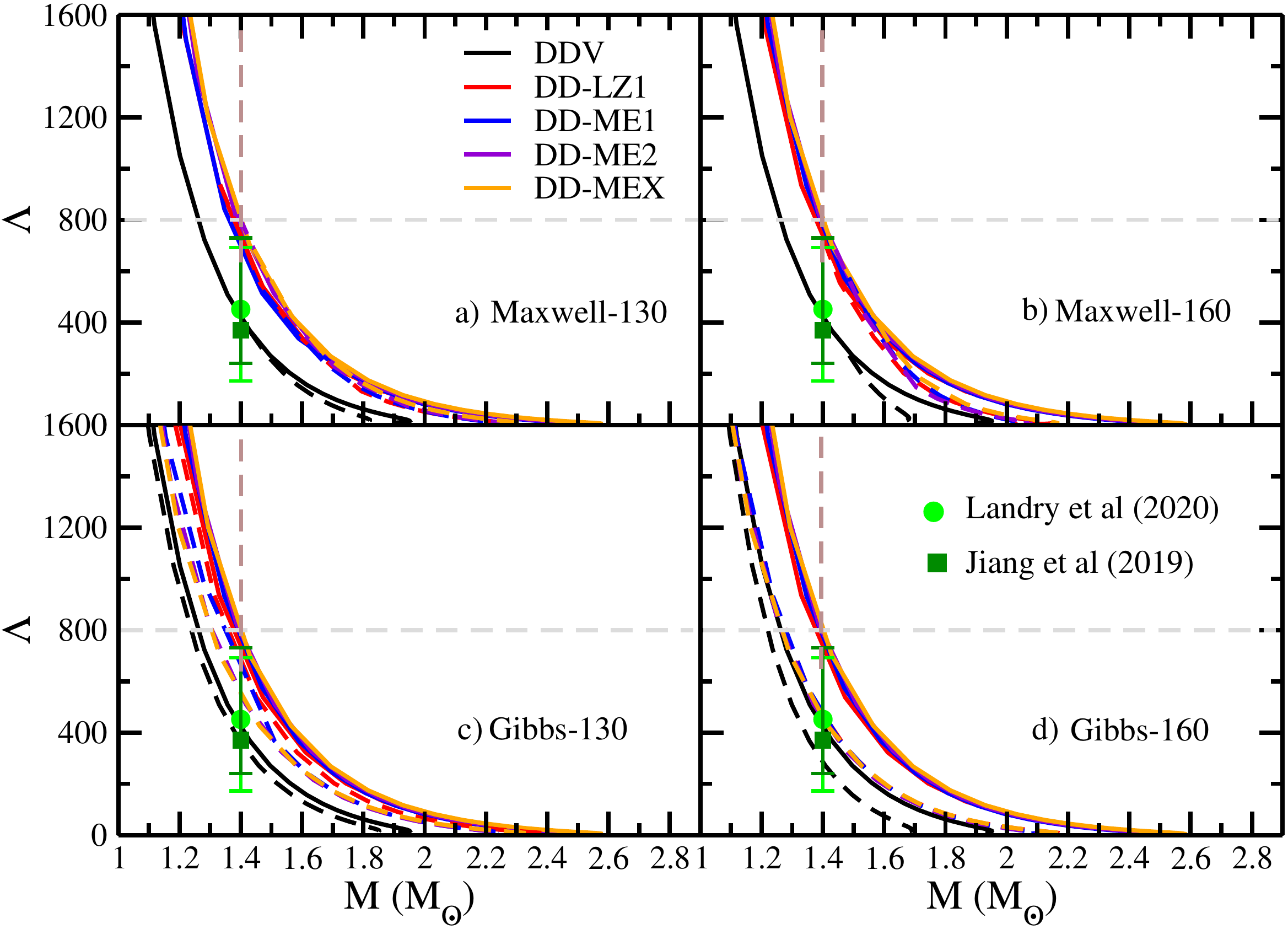}
	\caption{ The dimensionless tidal deformability ($\Lambda$) as a function of NS mass corresponding to DDV, DD-LZ1, DD-ME1, DD-ME2, and DD-MEX EoSs and their HS configurations. The solid (dashed) lines represent MR plot for pure hadronic matter (hybrid NSs). The upper panels display the HS configuration with MC and the lower panels represent the same configuration with GC. The brown dashed line represents the NS canonical mass. The grey dashed line shows the upper limit of $\Lambda_{1.4}$ value from GW170817 \cite{PhysRevLett.119.161101}. The non-parametric constraints on the tidal deformability of canonical NS mass are shown \cite{PhysRevD.101.123007}. The constraint from joint PSR J0030+0451, GW170817 and the nuclear data analysis \cite{Jiang_2020}.}
	\label{fig5}
\end{figure}

Fig.~\ref{fig5} displays the dimensionless tidal deformability as a function of NS mass for pure hadronic EoSs and HS configurations. As seen, the tidal deformability decreases with NS mass and becomes very small at the maximum mass. The softer EoS like pure hadronic DDV has tidal deformability at 1.4$M_{\odot}$,  $\Lambda_{1.4}$ = 392.052, while for other stiffer EoSs, the value lies in the range $\Lambda_{1.4}$ = 690-790, which is well constrained by the upper limit on tidal deformability from GW170817 data \cite{PhysRevLett.119.161101}. The non-parametric constraints on the tidal deformability given by $\Lambda$ = $451_{-279}^{+241}$ is shown \cite{PhysRevD.101.123007}. The constraint from the joint PSR J0030+0451, GW170817 and the nuclear data analysis at the canonical mass, $\Lambda_{1.4}$ = $370_{-130}^{+360}$ \cite{Jiang_2020} is also shown. For the phase transition with Maxwell construction, we see that the tidal deformability at the canonical mass remains the same as the pure hadronic one, while for the Gibbs phase transition, the tidal deformability decreases from 728.351 to 698.233 and 536.173 for effective bag constants 130 \& 160 MeV, respectively. The tidal deformability for the obtained HS configurations using the  Gibbs phase transition is favored over Maxwell construction by the tidal constraint from GW170817. We see that for the parameters used and phase transition construction, the maximum mass of each hybrid star curve is different from each other, which implies that no two curves produce a twin star that shares the same maximum mass and hence shows two branches of tidal deformability.

Table \ref{tab4} shows the NS properties like maximum mass, radius, the radius at the canonical mass, and the dimensionless tidal deformability for pure hadronic phase and HSs. All the NS matter properties have been calculated for HS configurations using both Gibbs and Maxwell construction to see how the global and local charge neutrality conditions shape up the mass radius of an NS. It is clear from the table that the radius at the canonical mass and the tidal deformability of an NS remains the same for Maxwell transition as that for pure hadronic matter, while both decrease for Gibbs transition. The obtained properties of HSs with Gibbs transition satisfy all the constraints from the recent observations of mass, radius and tidal deformability. 
\begin{center}
	\begin{table}[ht]
		\centering
		\caption{NS matter properties Maximum mass ($M_{max}$), corresponding radius ($R_{max}$),  canonical mass radius ($R_{1.4}$), and dimensionless tidal deformability ($\Lambda_{1.4}$) for pure hadron matter and HS configurations at effective bag constants $B_{eff}^{1/4}$ = 130 \& 160 MeV. The HS properties with both Gibbs as well as Maxwell construction are shown. }
		\begin{tabular}{|c|c|cccccccc|}
			\hline
			\multirow{3}{*}{\makecell{Star \\ properties}} &\multirow{3}{*}{\makecell{Pure \\ Hadronic}}& \multicolumn{4}{c|}{Gibbs Construction} & %
			\multicolumn{4}{c|}{Maxwell Construction}\\
			\cline{3-10}
			& &\multicolumn{2}{c}{130 MeV} & \multicolumn{2}{c}{160 MeV} & \multicolumn{2}{c}{130 MeV} & \multicolumn{2}{c|}{160 MeV} \\
			\cline{3-10}
			& &\multicolumn{8}{c|}{DDV EoS}  \\
			\hline
			$M_{max}(M_{\odot})$&1.951&1.793&&1.665&&1.821&&1.680& \\
			$R_{max}$(km)&10.851&10.298&&10.805&&10.943&&11.211& \\
			$R_{1.4}$(km)&12.132&11.459&&11.427&&12.132&&12.132& \\
			$\Lambda_{1.4}$&392.052&356.261&&297.815&&392.052&&392.052&\\
			\cline{3-10}
			& &\multicolumn{8}{c|}{DD-LZ1 EoS}  \\
			\hline
			$M_{max}(M_{\odot})$&2.555&2.192&&2.043&&2.275&&2.146& \\
			$R_{max}$(km)&12.297&12.475&&12.355&&12.428&&12.574& \\
			$R_{1.4}$(km)&13.069&12.752&&12.706&&13.069&&13.069& \\
			$\Lambda_{1.4}$&728.351&698.233&&536.173&&728.351&&728.351&\\
			\cline{3-10}
			& &\multicolumn{8}{c|}{DD-ME1 EoS}  \\
			\hline
			$M_{max}(M_{\odot})$&2.449&2.106&&1.974&&2.219&&2.027& \\
			$R_{max}$(km)&11.981&11.211&&10.128&&12.162&&12.349& \\
			$R_{1.4}$(km)&12.898&11.507&&11.543&&12.898&&12.898& \\
			$\Lambda_{1.4}$&689.342&658.047&&495.146&&689.342&&689.342&\\
			\cline{3-10}
			& &\multicolumn{8}{c|}{DD-ME2 EoS}  \\
			\hline
			$M_{max}(M_{\odot})$&2.483&2.174&&2.008&&2.246&&2.074& \\
			$R_{max}$(km)&12.017&12.187&&12.013&&12.204&&12.391& \\
			$R_{1.4}$(km)&12.973&12.224&&12.247&&12.973&&12.973& \\
			$\Lambda_{1.4}$&733.149&572.844&&475.367&&733.149&&733.149&\\
			\cline{3-10}
			& &\multicolumn{8}{c|}{DD-MEX EoS}  \\
			\hline
			$M_{max}(M_{\odot})$&2.575&2.246&&2.095&&2.325&&2.164& \\
			$R_{max}$(km)&12.465&12.547&&12.506&&12.659&&12.754& \\
			$R_{1.4}$(km)&13.168&12.497&&12.451&&13.168&&13.168& \\
			$\Lambda_{1.4}$&791.483&594.376&&462.753&&791.483&&791.483&\\
			\hline
		\end{tabular}
		\label{tab4}
	\end{table}
\end{center}
\subsection{Rotating Neutron stars}
\label{rnsresults}
To study the maximally rotating hybrid stars, the global charge neutrality between two different phases has been employed. The effective bag model with an effective bag constant $B^{1/4}$ is used to study the QM. The coupling constant parameter $K_{\nu}$ is fixed at 6 GeV$^{-2}$ for the three flavor configurations. Three different values of effective bag constant are used $B_{eff}^{1/4}$ = 130, 145 and 160 MeV. Parameter sets that produce stiff EoS such as DD-LZ1 and DD-MEX and soft EoS such as DDV, DDVT, and DDVTD sets have been used.
%

\begin{figure}[htb!]
	\centering
	\includegraphics[width=0.75\textwidth]{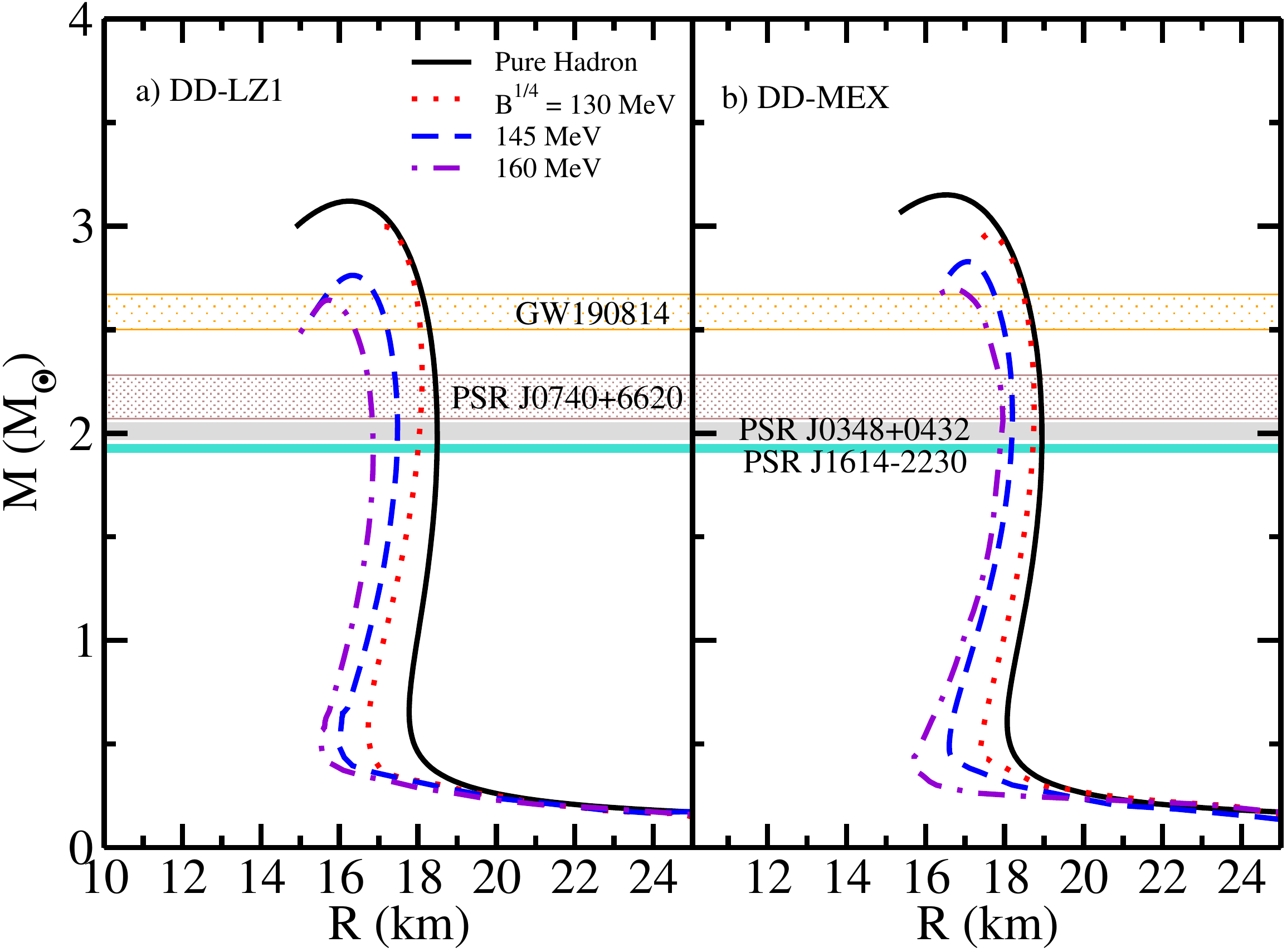}
	\caption{ Mass-Radius profile for pure hadronic and hybrid rotating NSs for (a) DD-LZ1 and (b) DD-MEX parameter sets at bag values $B_{eff}^{1/4}$ = 130, 145 \& 160 MeV. The shaded regions represent recent constraints on the mass from various measured astronomical observables.  }
	\label{fig5.11}
\end{figure}
The RNS mass-radius profile for DD-LZ1 and DD-MEX parameter sets are shown in Fig.~\ref{fig5.11}. The solid lines represent the pure hadronic star while the dashed lines represent the HS at different bag constants. The effective bag constant $B_{eff}^{1/4}$ is written as $B^{1/4}$ for convenience. The DD-LZ1 EoS produces a pure hadronic RNS with a maximum mass of 3.11$M_{\odot}$ with a radius of 18.23 km. With the phase transition from HM to QM, the maximum mass and the corresponding radius decrease with the increase in the bag constant. For the DD-LZ1 set, the maximum mass decreases from 3.11$M_{\odot}$ to 2.98$M_{\odot}$ for $B^{1/4}$ = 130 MeV and to 2.75$M_{\odot}$ and 2.64$M_{\odot}$ for $B^{1/4}$ = 145 and 160 MeV, respectively. The radius $R_{1.4}$ decreases from 18.32 km for pure HM to 16.64 km for hybrid star matter at 160 MeV bag value. Similarly, for the DD-MEX parameter set, the maximum mass for pure hadronic matter is 3.15$M_{\odot}$ at radius 16.53 km which reduces to 2.69$M_{\odot}$ at 16.63 km for bag constant of 160 MeV. Thus, while the pure hadronic RNS predicts a large maximum mass, the phase transition to QM lowers the maximum mass and the radius thereby satisfying the maximum mass constraint from GW190814. These results imply that the secondary component of GW190814 could be a possible fast-rotating hybrid star.
\begin{figure}[htb!]
	\centering
	\includegraphics[width=0.75\textwidth]{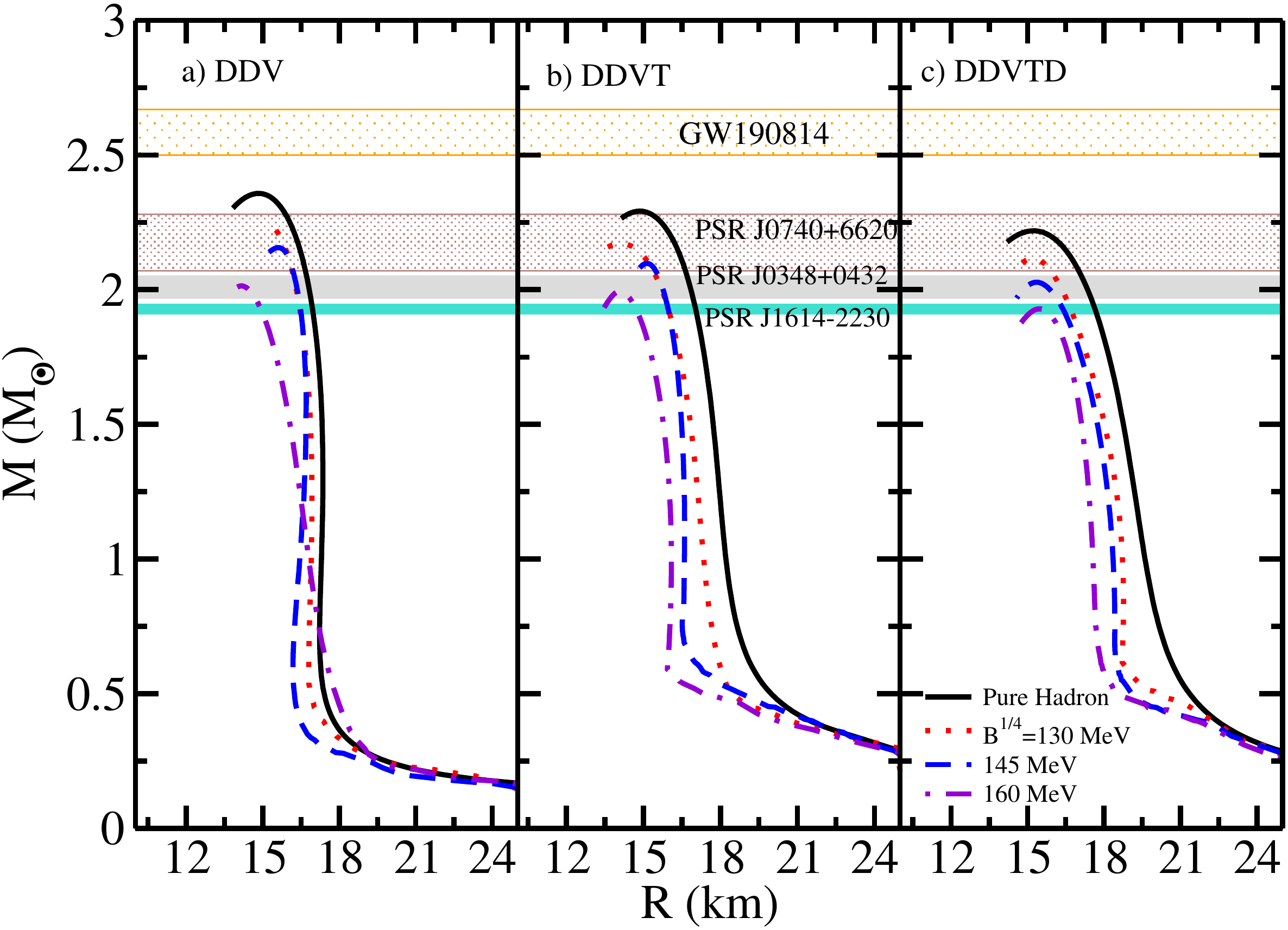}
	\caption{ Same as Fig.~\ref{fig5.11}, but for (a) DDV, (b) DDVT and (c) DDVTD EoSs. }
	\label{fig5.12}
\end{figure}

Fig.~\ref{fig5.12} displays the mass-radius relation for hadronic and hybrid rotating NS with DDV, DDVT, and DDVTD EoSs. The maximum mass for an RNS with DDV EoS is 2.37$M_{\odot}$ with a 17.41 km radius at the canonical mass. Both the maximum mass and the radius decrease to 2.23$M_{\odot}$, 2.13$M_{\odot}$, 2.01$M_{\odot}$ and 16.91, 16.68, 16.13 km for bag constants $B^{1/4}$ = 130, 145, and 160 MeV, respectively, thereby satisfying the 2$M_{\odot}$ constraint. For DDVT, the maximum mass reduces from 2.28$M_{\odot}$ to 1.99$M_{\odot}$. $R_{1.4}$ also decreases from 17.82 km to 16.01 km. Similarly for the DDVTD EoS, the RNS maximum mass reduces to 1.93$M_{\odot}$ from 2.21$M_{\odot}$ at $B^{1/4}$ = 160 MeV. For all the parameter sets, the phase transition to QM lowers the maximum mass which satisfies the 2$M_{\odot}$ limit.
 
\begin{figure}[htb!]
	\centering
	\includegraphics[width=0.75\textwidth]{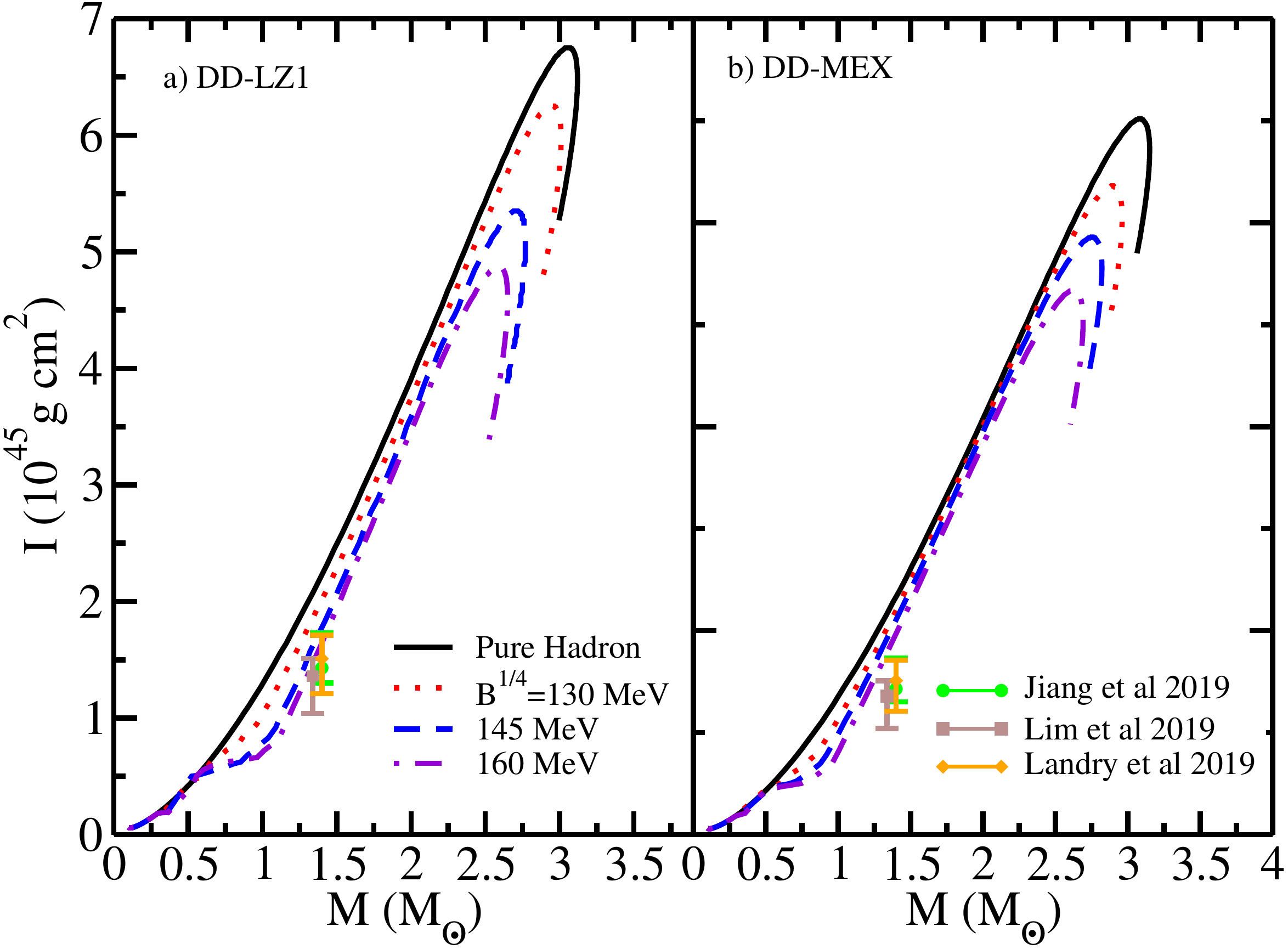}
	\caption{ Moment of inertia variation with the gravitational mass for (a) DD-LZ1 and (b) DD-MEX EoSs. The constraints on canonical moment of inertia are also shown \cite{PhysRevC.100.035802}. The constraint from joint PSR J0030+0451, GW170817 and the nuclear data analysis are shown by green bar \cite{Jiang_2020}. The predicted moment of inertia of pulsar J0737-3039A using Bayesian analysis of nuclear EoS is shown by brown bar \cite{PhysRevD.101.123007}. }
	\label{fig5.13}
\end{figure}

The measurement of the NS moment of inertia is important because it follows a universal relation with the tidal deformability and the compactness of an NS.
The moment of inertia as a function of gravitational mass for the RNS is displayed in Fig.~\ref{fig5.13}. The constraint on the moment of inertia obtained from the joint PSR J0030+0451, GW170817 and the nuclear data analysis predicting $I_{1.4}=1.43_{-0.13}^{+0.30}\times 10^{38}$ kg.m$^2$ is given in Ref. \cite{Jiang_2020}. The predicted moment of inertia of pulsar PSR J0737-3093A,  $I_{1.338}=1.36_{-0.32}^{+0.15}\times 10^{45}$ g.cm$^2$ is also given \cite{PhysRevD.101.123007}. For pure hadronic matter, DD-LZ1 and DD-MEX EoSs predict an NS with a moment of inertia 2.22 and 2.35 $\times$ 10$^{45}$ g.cm$^2$, respectively. The phase transition to the QM reduces the moment of inertia to a value 1.65 and 1.93 $\times$ 10$^{45}$ g.cm$^2$ for DD-LZ1 and DD-MEX parameter sets at bag constant $B^{1/4}$=160 MeV, which satisfies the constraint from Refs.~\cite{Jiang_2020,PhysRevD.101.123007,PhysRevC.100.035802}.

\begin{figure}[htb!]
	\centering
	\includegraphics[width=0.75\textwidth]{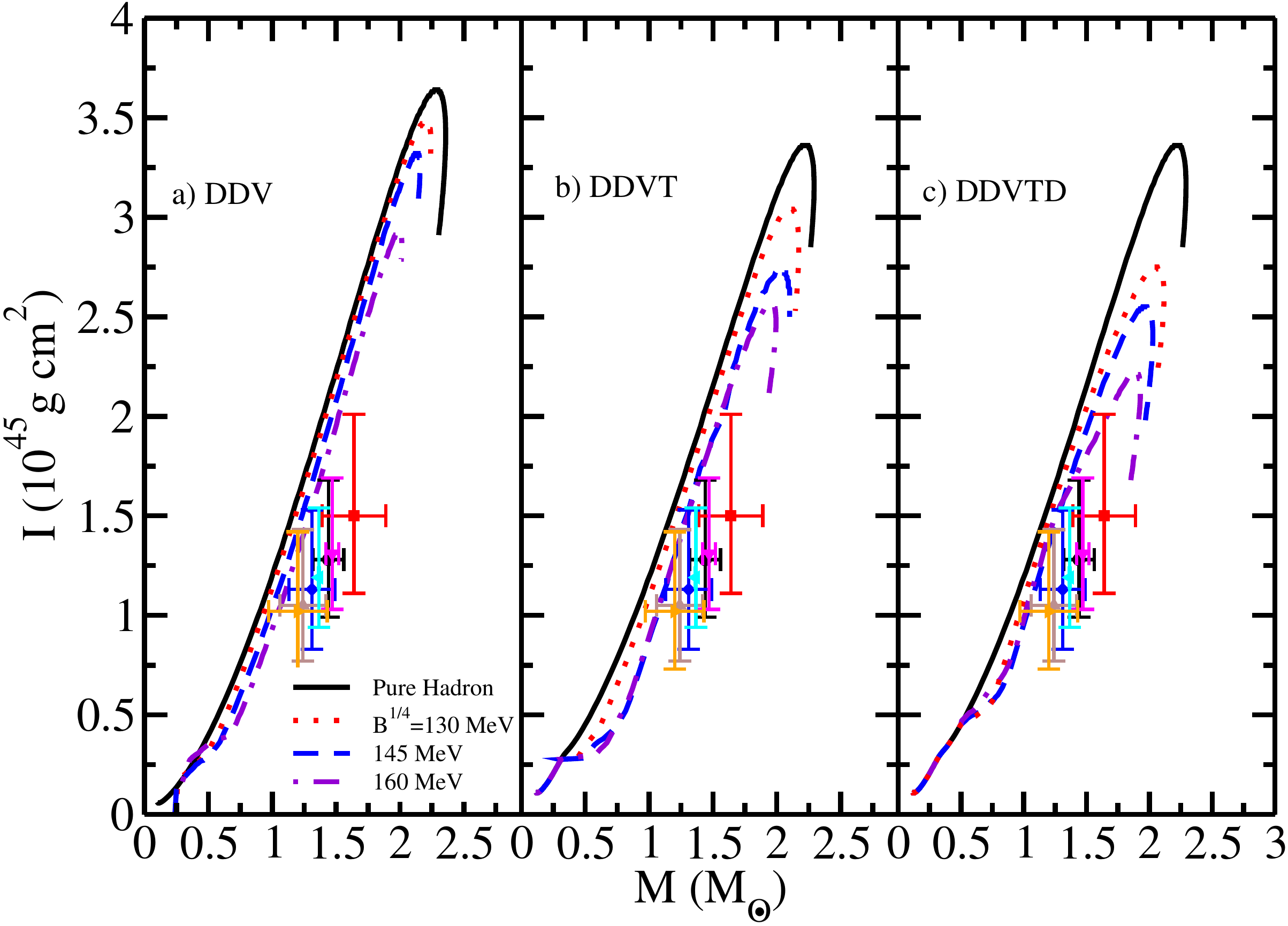}
	\caption{ Same as Fig.~\ref{fig5.13} but for (a) DDV, (b) DDVT and (c) DDVTD parameter sets. The constraints on the moment of inertia of MSPs obtained from universal relations with GW170817 are shown \cite{PhysRevD.99.123026}.}
	\label{fig5.14}
\end{figure}
Fig.~\ref{fig5.14} displays the moment of inertia variation with the gravitational mass for DDV, DDVT, and DDVTD parameter sets. The solid lines represent the pure hadronic matter, while the dashed lines represent the hadron-quark mixed-phase at bag constants $B^{1/4}$ = 130, 145 and 160 MeV. The constraints on the moment of inertia obtained from millisecond pulsars (MSPs) with GW170817 universal relations are shown in Ref.~\cite{PhysRevD.99.123026}. For the DDV EoS, the moment of inertia of a pure hadronic star is found to be 2.01 $\times$ 10$^{45}$ g.cm$^2$ while for the DDVT and DDVTD EoSs, the value is found to be 1.95 and 1.88 $\times$ 10$^{45}$ g.cm$^2$, respectively.  For the hybrid EoS, the moment of inertia is lowered to a value of 1.71$\times$ 10$^{45}$ g.cm$^2$ for the DDV set at bag constant 160 MeV. For DDVT and DDVTD sets, this value reduces to 1.68 and 1.64 $\times$ 10$^{45}$ g.cm$^2$ respectively for a 160 MeV bag constant. The phase transition to QM produces an NS with the moment of inertia that satisfies the constraints from various measurements.

A useful parameter to characterize the rotation of a star is the ratio of rotational kinetic energy $T$ to the gravitational potential energy $W$, $\beta=T/W$. For a RNS, if $\beta > \beta_d$, where $\beta_d$ is the critical value, the star will be dynamically unstable. The critical value $\beta_d$ for a rigidly rotating star is found to be 0.27 \cite{10.1093/gji/21.1.103-a,1985ApJ...298..220T}. However, for different angular-momentum distributions, the value lies in the range 0.14 to 0.27 \cite{1996ApJ...458..714P,1995ApJ...444..363I,2001ApJ...550L.193C}. 

\begin{figure}[htb!]
	\centering
	\includegraphics[width=0.75\textwidth]{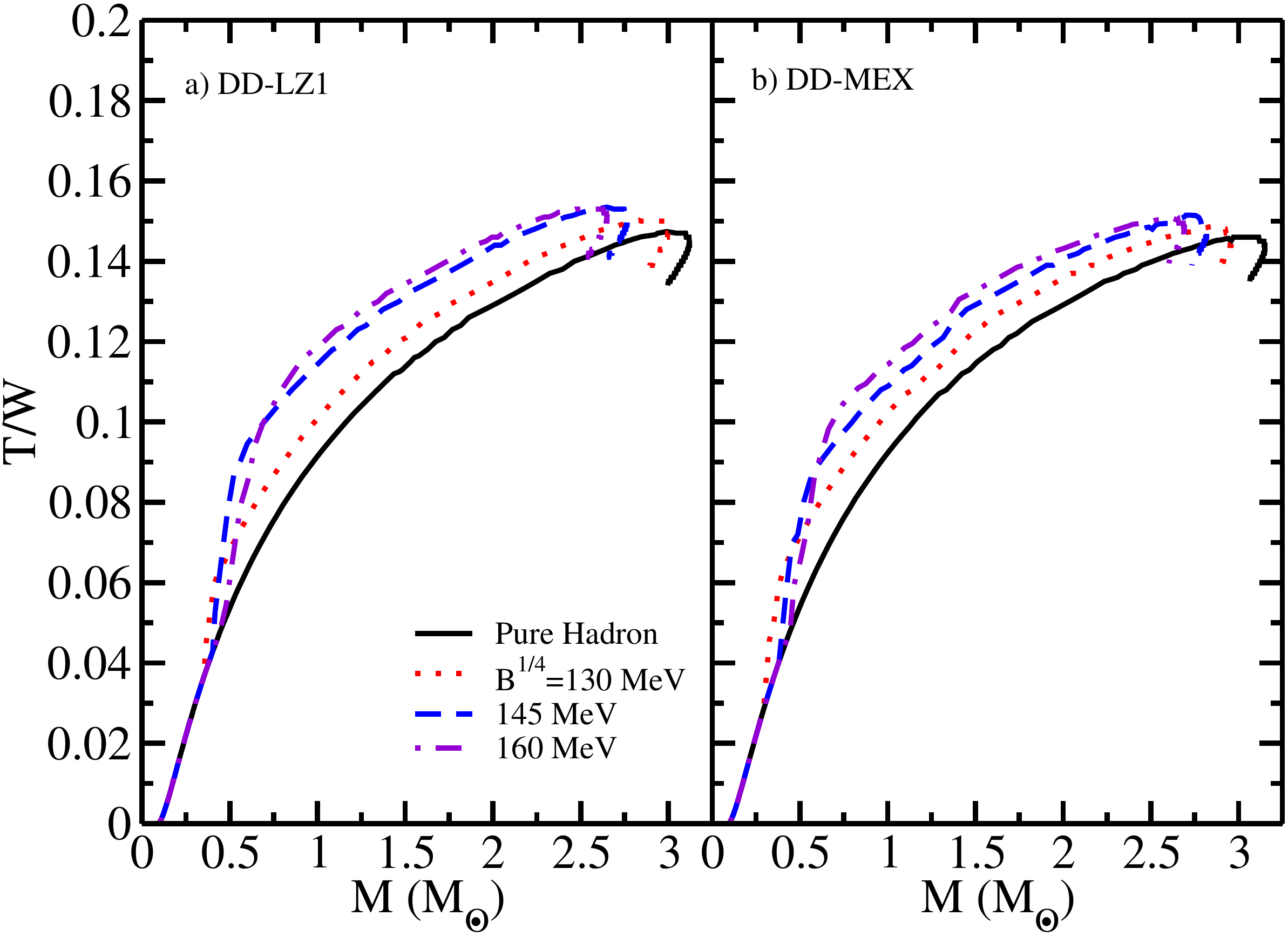}
	\caption{ Variation in the ratio of rotational kinetic energy to the gravitational potential energy $T/W$ with gravitational mass for (a) DD-LZ1 and (b) DD-MEX EoSs. Solid lines represent pure hadronic stars while the dashed lines represent hybrid stars at bag constants $B^{1/4}$ = 130, 145 \& 160 MeV.}
	\label{fig5.19}
\end{figure}

The variation in the $T/W$ ratio of the pure hadron and HS with the gravitational mass is shown in Fig.~\ref{fig5.19}. The $T/W$ ratio for pure hadronic stars is 0.147 and 0.145 for DD-LZ1 and DD-MEX parameter sets, respectively. The HSs have a large $T/W$ ratio and increase with bag constant. For DD-LZ1 set, the ratio increases from 0.150 at $B^{1/4}$ = 130 MeV to 0.153 at $B^{1/4}$ = 160 MeV. For the DD-MEX set, the ratio increases to 0.149 and 0.151 for bag values 130 and 160 MeV, respectively. The reason for large value of the $T/W$ ratio in HSs is because the quark stars are bound by the strong interaction, unlike hadron stars which are bound by gravity.
\begin{figure}[htb!]
	\centering
	\includegraphics[width=0.75\textwidth]{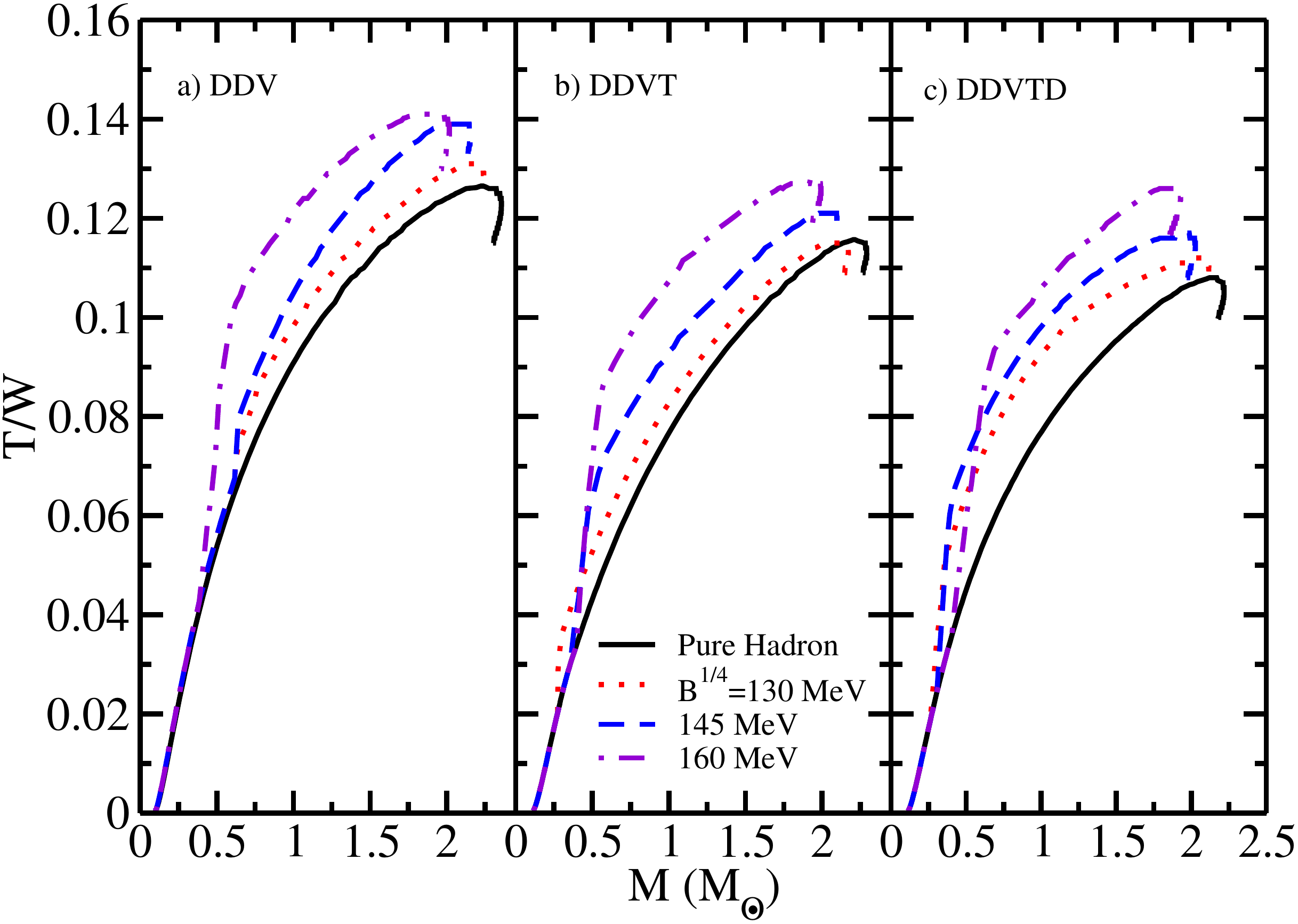}
	\caption{ Same as Fig.~\ref{fig5.19} but for (a) DDV, (b) DDVT and (c) DDVTD EoSs.}
	\label{fig5.20}
\end{figure}

Fig.~\ref{fig5.20} depicts the $T/W$ variation with the gravitational mass for DDV, DDVT, and DDVTD parameter sets. For the DDV EoS, the pure hadronic star predicts a $T/W$ ratio of 0.127, which lies below the critical value $\beta_d$. For hybrid stars, this ratio increases 0.142 for a bag constant of 160 MeV thereby satisfying the critical $\beta_d$ limit and hence becomes dynamically unstable and emits gravitational waves. Similarly, for DDVT and DDVTD EoS, the pure hadron star produces a ratio of 0.115 and 0.108 while the HS at $B^{1/4}$ = 160 MeV gives a value of 0.127 and 0.125, respectively.

Einstein's field equations provide Kerr space-time for so-called Kerr black holes which can be fully described by the angular momentum $J$ and gravitational mass $M$ of rotating black holes \cite{spin.parameter, PhysRevD.92.023007}. The condition $J\ge GM^2/c$ must be satisfied to define a stable Kerr black hole. The gravitational collapse of a massive RNS constrained to angular-momentum conservation creates a black hole with mass and angular momentum resembling that of an NS. Thus, it is an important quantity used in the study of black holes as well as RNSs. The Kerr parameter leads to the possible limits on the compactness of an NS and also can be an important criterion for determining the final fate of the collapse of a rotating compact star \cite{PhysRevC.101.015805,spin.parameter}. The Kerr parameter is described by the relation
\begin{equation}
\kappa=\frac{cJ}{GM^2}
\end{equation}
where $J$ is the angular momentum and $M$ is the gravitational mass of the rotating NS. The Kerr parameter for black holes is a fundamental quantity with a maximum value of 1, but it is important for other compact stars as well.  
\begin{figure}[htb!]
	\centering
	\includegraphics[width=0.75\textwidth]{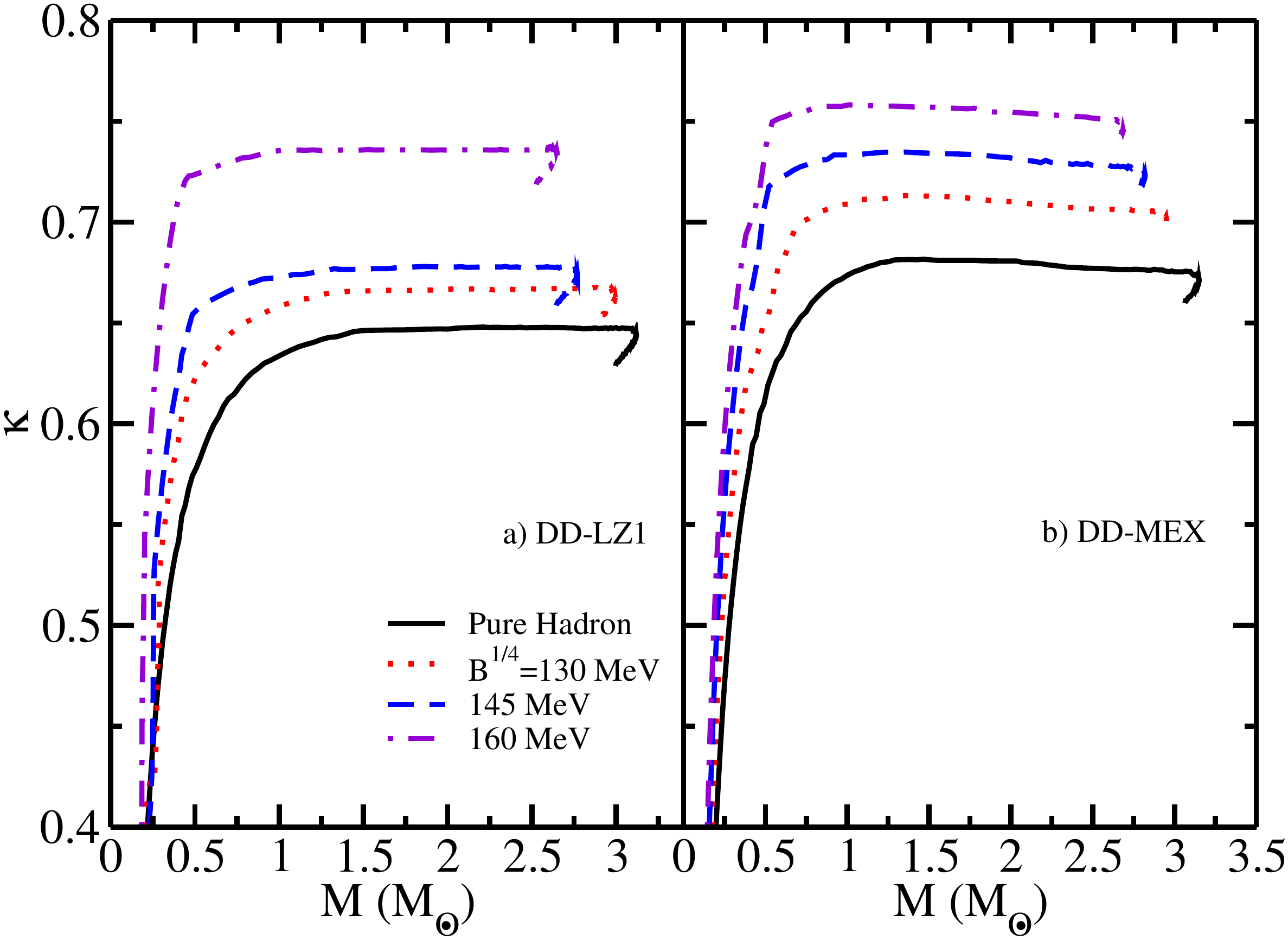}
	\caption{ Kerr parameter $\kappa$ as a function of gravitational mass for (a) DD-LZ1 and (b) DD-MEX EoSs. The plot shows both pure hadronic stars (solid lines) and hybrid stars (dashed lines) at different bag constants.}
	\label{fig5.21}
\end{figure}
To constrain the Kerr parameter for NSs, we studied the dependence of the Kerr parameter on the NS gravitational mass as displayed in Figs. \ref{fig5.21} and \ref{fig5.22} for the given parameter sets. From Fig.~\ref{fig5.21}, the Kerr parameter for pure hadronic DD-LZ1 and DD-MEX parameter sets is found to be 0.64 and 0.67 respectively. This parameter increases for the hybrid stars with a maximum value of 0.73 at $B^{1/4}$ = 160 MeV for the DD-LZ1 set. For the DD-MEX set, the maximum value of the Kerr parameter is 0.75 at 160 MeV bag constant. For the DD-LZ1 parameter sets, the Kerr parameter remains almost unchanged once the star reaches a mass of around 1.4$M_{\odot}$ for pure hadronic matter and around 1.2$M_{\odot}$ for hybrid configurations. For DDV, DDVT, and DDVTD parameter sets as shown in Fig.~\ref{fig5.22}, the Kerr parameter value for pure hadronic stars at the maximum mass is 0.64, 0.62, and 0.61 respectively. For hybrid star configurations, the value increases to 0.75 for all parameter sets at bag constant $B^{1/4}$ = 160 MeV. The Kerr parameter for HS configurations remains almost identical to the hadron star up to almost 0.4$M_{\odot}$. Therefore, by definition, the gravitational collapse of an RNS cannot form a Kerr black hole.
\begin{figure}[htb!]
	\centering
	\includegraphics[width=0.75\textwidth]{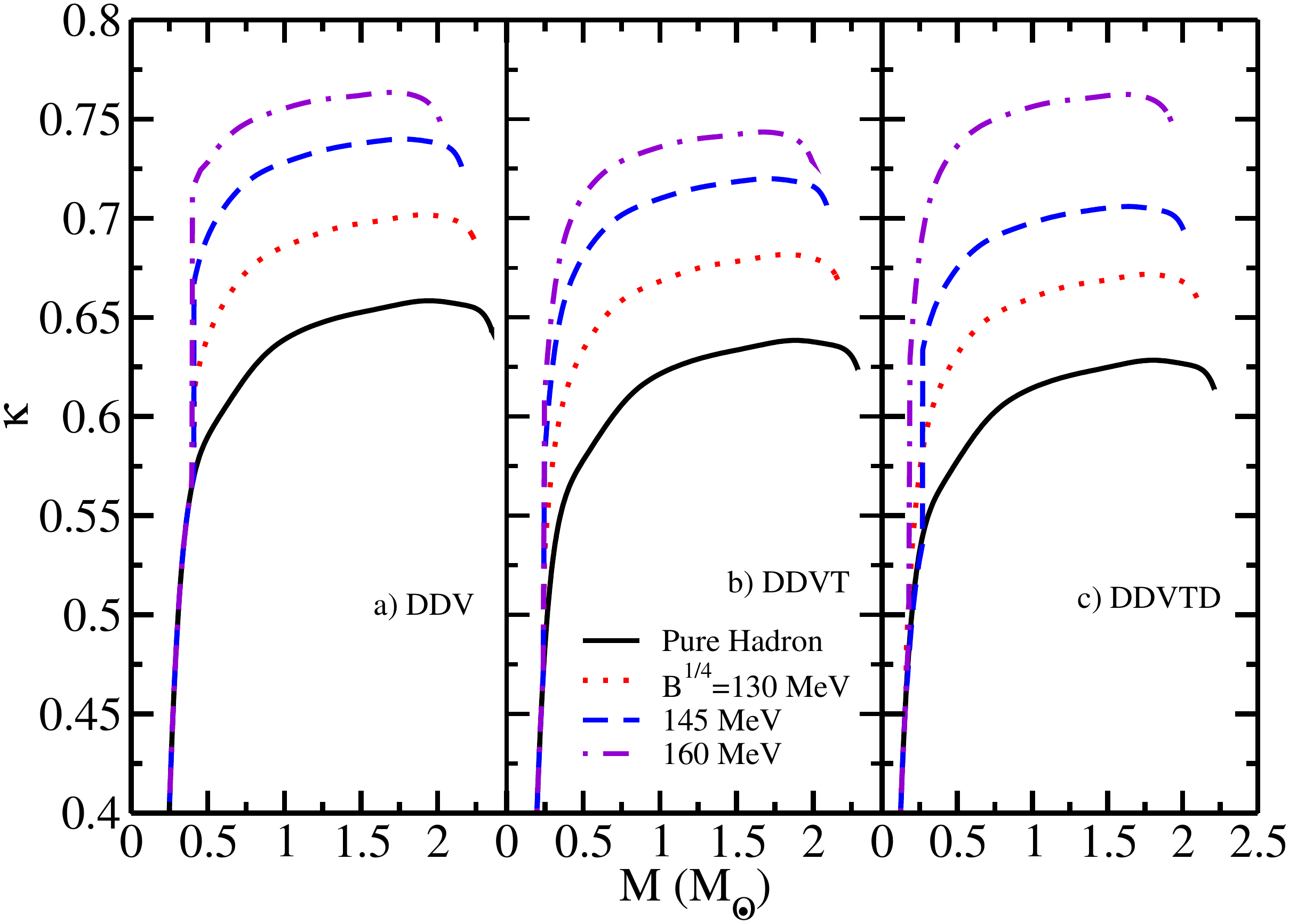}
	\caption{ Same as Fig.~\ref{fig5.21} but for (a) DDV, (b) DDVT, and (c) DDVTD EoSs.}
	\label{fig5.22}
\end{figure}

\begin{figure}[htb!]
	\centering
	\includegraphics[width=0.75\textwidth]{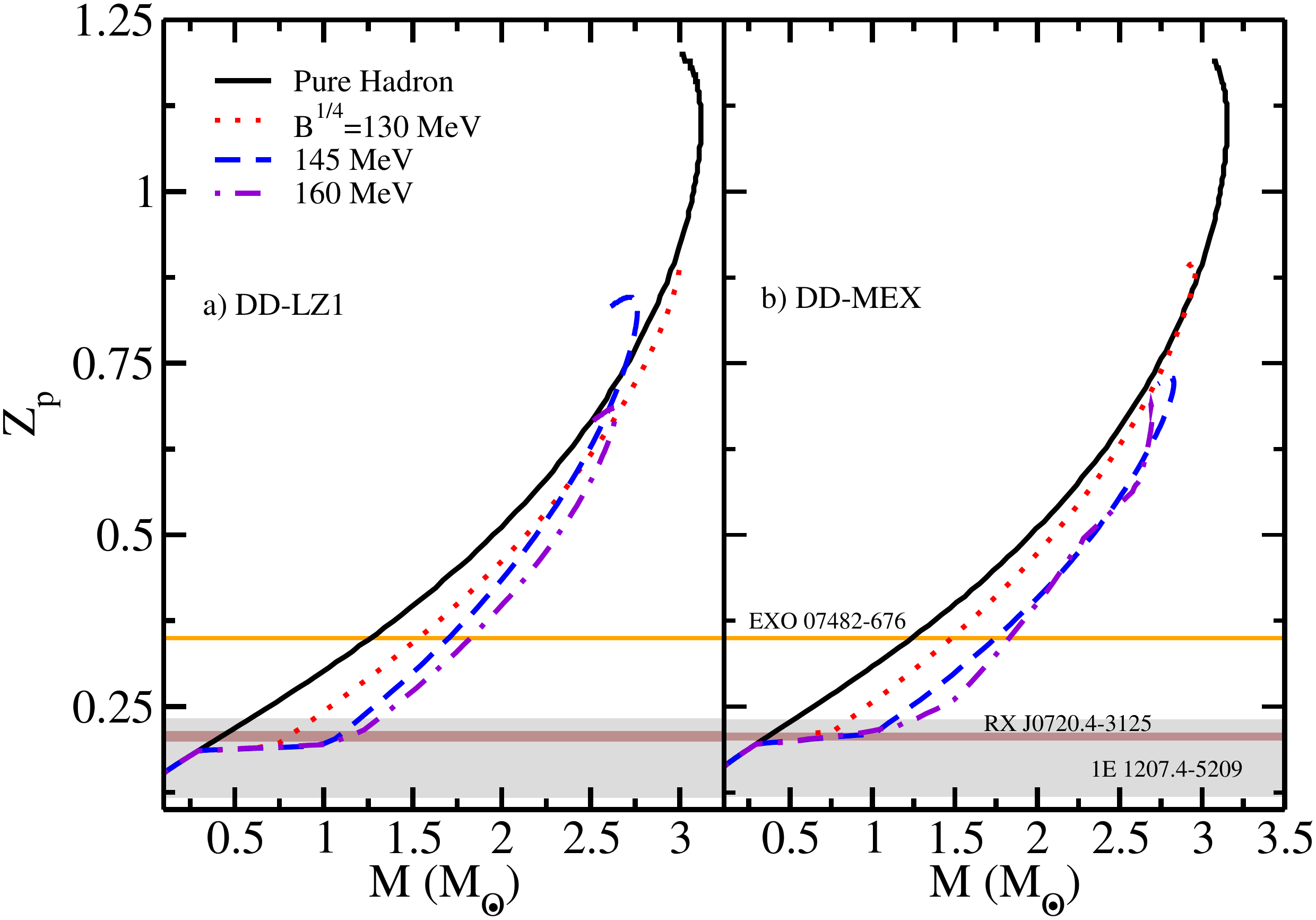}
	\caption{ Polar redshift vs gravitational mass for pure hadron stars and hybrid star configurations for (a) DD-LZ1 and (b) DD-MEX EoSs. The observational limits imposed on the polar redshift from 1E 1207.4-5209 (grey band) \cite{Sanwal_2002}, RX J0720.4-3125 (brown band) \cite{refId5}, and EXO 07482-676 (orange horizontal line) \cite{cottam} are shown. }
	\label{fig5.23}
\end{figure}

Another important quantity related to the NSs is the redshift which has been investigated deeply \cite{Xia_2009,1994ApJ...424..823C,1986ApJ...304..115F}. The measurement of redshift can impose constraints on the compactness, and in turn, on the NS EoS. For an RNS, if the detector is placed in the direction of the polar plane of the star, the polar redshift, also called gravitational redshift, can be measured. For a detector directed tangentially, the forward and backward redshifts can be measured. The expression for the polar redshift is given as
\begin{equation}
Z_P(\Omega)=e^{-2\nu(\Omega)}-1
\end{equation}
where $\nu$ is the metric function. The variation of the polar redshift with the gravitational mass is depicted in Fig.~\ref{fig5.23} for DD-LZ1 and DD-MEX EoSs. For pure hadronic stars, the polar redshift is found to be around 1.1 for both EoSs. With the QM present in the NSs, the polar redshift for DD-LZ1 decreases to a value 0.89, 0.84, and 0.64 for bag constants $B^{1/4}$ = 130, 145, and 160 MeV, respectively. Similarly, for the DD-MEX set, the redshift decreases up to 0.68 for the 160 MeV bag constant. The observational limits imposed on the redshift from 1E 1207.4-5209 ($Z_P$ = 0.12-0.23) \cite{Sanwal_2002} , RX J0720.4-3125 ($Z_P$ = 0.205$_{-0.003}^{+0.006}$) \cite{refId5}, and EXO 07482-676 ($Z_P$ = 0.35) \cite{cottam} are also shown. The redshift prediction of $Z_P$ = 0.35 for EXO 07482-676 was based on the narrow absorption lines in the X-ray bursts. However, it was later seen that the spectral lines from EXO 07482-676 may be narrower than predicted \cite{Baub_ck_2013}. Therefore the estimates of the redshift from EXO 07482-676 are uncertain. 
\begin{figure}[htb!]
	\centering
	\includegraphics[width=0.75\textwidth]{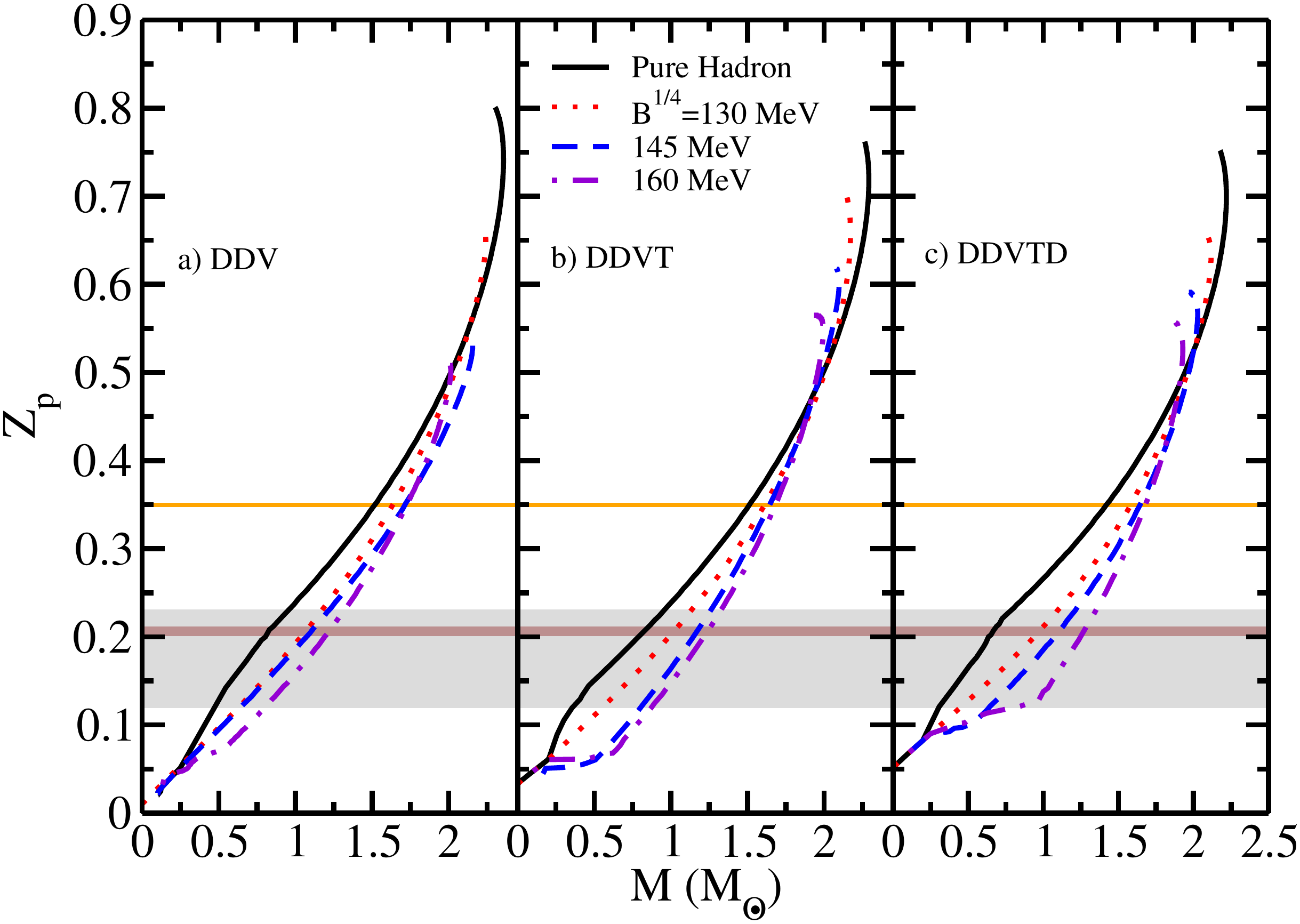}
	\caption{ Same as Fig.~\ref{fig5.23} but for (a) DDV, (b) DDVT, and (c) DDVTD EoSs. }
	\label{fig5.24}
\end{figure}

For the softer EoS group, the polar redshift variation with the gravitational mass is shown in Fig.~\ref{fig5.24} for both pure HM and HS configurations. For the DDV set, the polar redshift is found to be 0.75 for the maximum mass of a pure hadronic star and decreases to 0.50 for the hybrid star at a bag constant of 160 MeV. For DDVT and DDVTD EoSs, the redshift decreases from 0.72 and 0.70 for pure HM to 0.55 and 0.53 respectively for a hybrid star at $B^{1/4}$ = 160 MeV. The NS redshift provided by measuring the $\gamma$-ray burst annihilation lines has been interpreted as gravitationally redshifted 511 keV electron-positron pair annihilation from the NS surface \cite{Liang}. If this interpretation is correct, then it will support an NS with redshift in the range $0.2\le Z_P \le 0.5$ and thus will rule out almost every EoS studied in this work.\par 

\section{Conclusion}
\label{summary}
LVC recently detected GW190814, a black hole collision of mass 22.2 - 24.3$M_{\odot}$ with a compact object of mass 2.50 - 2.67$M_{\odot}$. Because there are no tidal signs or electromagnetic equivalents, the secondary object of GW190814 has drawn a lot of attention because of its nature as either the heaviest neutron star or the lightest black hole ever detected. We employed several latest DD-RMF parameter sets like DDV, DD-LZ1, DD-ME1, DD-ME2, and DD-MEX to study the star matter properties. To study the phase transition from hadron matter to quark matter, the vBag is employed for quark matter which accounts for the Dynamic Chiral Symmetry Breaking and repulsive vector interactions, explicitly. The hybrid star EoSs are generated by allowing a phase transition between hadron matter and quark matter using both Maxwell and Gibss methods. The free parameter in the vBag model, $K_{\nu}$, which controls the stiffness of the EoS curve is fixed at $K_{\nu}=6$ GeV$^{-2}$ for three flavor quark matter. The effective bag constant with values $B_{eff}^{1/4}$ = 130 \& 160 MeV are used. 

By solving the TOV equation for the obtained pure and hybrid EoSs under $\beta$-equilibrium and charge-neutral conditions, the NS properties such as mass, radius, and tidal deformability are calculated for all the configurations. The softer EoS DDV supports an NS with a maximum mass of 1.951$M_{\odot}$ at 10.851 km and tidal deformability at 1.4$M_{\odot}$, $\Lambda_{1.4}$ = 392.052. The presence of quarks lowers the maximum mass and the tidal deformability from 1.951 to 1.6$M_{\odot}$ and 392 to around 297 using both Gibbs and Maxwell construction. For the stiffer EoSs like DD-LZ1, DD-ME1, DD-ME2, and DD-MEX, the NS maximum mass generated lies in the range 2.44 - 2.57$M_{\odot}$ which satisfies the mass constraint from GW190814 data. However, the phase transition between hadron and quark matter reduces the maximum to around 2$M_{\odot}$
which satisfies the constraints from the GW170817 data. The tidal deformability is also lowered from 790 to around 500. 
 While the tidal deformability of the HSs remains the same as that of pure hadronic stars in maxwell construction, it decreases with an increase in the bag constant for Gibbs construction. 

The properties of rotating NSs are also studied with a hadron-quark phase transition using the Gibbs method.  In this case, the effective bag constant $B_{eff}^{1/4}$ is varied by taking the values 130, 145, and 160 MeV. The variation in the NS properties like maximum mass, radius, the moment of inertia, Kerr parameter, and polar redshift are studied.

For RNSs, the maximum mass is found to be > 3$M_{\odot}$ for the DD-LZ1 and DD-MEX EoSs which in presence of QM reduces to $\approx$ 2.6$M_{\odot}$ satisfying the recent GW190814 possible maximum mass constraint. For the softer EoS group, the RNS mass lies in the range 2.2$M_{\odot}$ - 2.3$M_{\odot}$ which then reduces with increasing bag constant to satisfy the 2$M_{\odot}$ limit. The radius also decreases with increasing bag constant. The moment of inertia for the stiffer group decreases from (2.2 - 2.3)$\times$10$^{45}$ g.cm$^2$ for pure hadron EoSs to 1.7 $\times$ 10$^{45}$ g.cm$^2$ for hybrid EoS satisfying the recent constraints. For the softer group of EoSs, the moment of inertia is lowered in the presence of QM to satisfy the constraints from GW170817 with universal relations. 

The ratio of rotational kinetic energy to the gravitational potential energy $\beta=T/W$ is studied to determine the dynamical stability of the RNS. For $\beta > \beta_d$ (= 0.14-0.27), the star is considered to be dynamically unstable and hence emits gravitational radiation. The $T/W$ ratio for rotating pure hadronic stars is found to be 0.147 and 0.145 for DD-LZ1 and DD-MEX EoSs. The QM phase transition tends to increase the $T/W$ ratio with decreasing mass. For a bag constant of 160 MeV, the ratio is found to be 0.153 and 0.151 for DD-LZ1 and DD-MEX EoSs, respectively. For a softer EoS group, this ratio lies below the critical limit for a pure hadronic star but increases to a value well within the critical limit. \par 
 For the given parametrization sets, the Kerr parameter is calculated whose value lies around 0.65 for the stiffer group and 0.6 for the softer group. Following the inverse relationship with the gravitational mass, the Kerr parameter increases in the presence of quarks. For both stiffer and softer EoS groups, the value attains a maximum value of 0.75, which remains almost unchanged as the mass increases beyond 1$M_{\odot}$. The dependence of polar redshift on the NS mass is also calculated. It is seen that the polar redshift decreases in presence of quarks. The redshift parameter measured for all hybrid star configurations lies well above the predicted value from EXO 07482-676, $Z_P$ = 0.35. 

Thus, it is clear that the presence of quarks inside the NS affects both static and rotating NS properties. Eliminating the uncertainties present in the values of these quantities will allow us to rule out very stiff and very soft EoSs. The measurement of tidal deformability for RNS will help us to constraint its properties and hence determine a proper EoS in the near future. Additional gravitational-wave observations of binary NS mergers and more accurate measurements of other NS properties like mass, radius, tidal deformability will allow the universal relation-based bounds on canonical deformability to be further refined. The theoretical study of a uniformly RNS, along with the accurate measurements, may offer new information about the equation of state in a high-density regime. Besides, NSs through their evolution may provide us with a criterion to determine the final fate of a rotating compact star.
\begin{savequote}[8cm]
\textlatin{No two things have been combined together better than knowledge and patience.}
  \qauthor{---\textit{Prophet Muhammad (SAW)}}
\end{savequote}

\chapter{\label{ch:6-magneticeos}Heavy Magnetic Neutron Stars}

\minitoc

\section{Introduction}

At densities about a few times normal nuclear density, the composition of matter inside NS is not known. With the increasing density, the appearance of exotic degrees of freedom, like quarks, inside NSs are possible and have been studied over the past decade \cite{Annala2020,PhysRevD.30.272,PhysRevD.30.2379,PhysRevD.46.1274,zel_2010,PhysRevD.88.085001}. The appearance of hyperonic matter under NS inner core conditions is energetically favored \cite{1960SvA.....4..187A,1985ApJ...293..470G}. The onset of hyperons reduces the pressure, leading to a softer EoS as they open new channels filling their Fermi sea, which decreases the maximum mass of NS by about 0.5$M_{\odot}$ \cite{GLENDENNING1982392,Chatterjee2016}. Several works has been performed recently considering hyperonic matter in NSs \cite{PhysRevC.100.015809,Li_2019,PhysRevD.102.041301,PhysRevC.95.065803}.
 
Heavy neutron stars are expected to contain exotic matter in their interior, even if they are rotating fast \cite{PhysRevC.97.035207}. Very massive and /or fast rotating stars could be the result of accretion, or even a previous stellar merger, both of which have been shown to enhance stellar magnetic fields \cite{Pons2019,PhysRevD.102.096025}.
With exceptionally high density, a magnetic field reaching $\approx$ 10$^9$ to 10$^{18}$ G \cite{Cardall_2001,10.1093/mnras/stu2706} is attainable in massive neutron star centers. Among various classes of compact stars available, Anomalous X-ray pulsars (AXPs) and the Soft gamma repeaters (SGRs), usually called magnetars \cite{2015SSRv..191..315M,doi:10.1146/annurev-astro-081915-023329,Mukherjee:2015sga}, are considered to have surface magnetic field within the order 10$^{14}$-10$^{15}$ G \cite{Harding_2006,Turolla_2015}. Fast radio bursts (FRBs) have also shown evidence of magnetars \cite{Margalit_2020,10.1093/mnras/staa1783}.
In the interior of the magnetars, the magnetic field cannot be measured directly and hence only estimated using theoretical models. The absolute largest value of magnetic field a star can possess in the interior (usually taken as an upper bound for the magnetic field) can be estimated using the relativistic version of the virial theorem \cite{1993A&A...278..421B}, for which the negative total energy implies that the magnitude of the gravitational potential energy must be greater than the magnetic field energy, which results in the magnetic field $\approx$ 10$^{18}$ G. As discussed in Refs.~\cite{1995A&A...301..757B,Cardall_2001}, assuming a poloidal configuration, the maximum allowed central magnetic ﬁeld that still fulfills Einstein's and Maxwell's equations is around a few times 10$^{18}$ G, the exact value being dependent on the equation of state. Similar limits exist in purely toroidal configurations \cite{Pili:2017yxd}. The magnitude of the magnetic field does not increase much beyond one order of magnitude within a star, regardless of the equation of state or magnetic field configuration, as pointed out in Refs.~\cite{Dexheimer:2016yqu,Pili:2017yxd}. This implies that, regardless of its feasibility, the star would have to present $\geq10^{16}$ G on the surface in order to reach a magnetic field $\geq10^{17}$ G or higher in the center. Most of the magnetar observations inferred magnetic fields of $\sim10^{15}$ G or below, the only exception being data from the source 4U 0142+61 for slow phase modulations in hard X-ray pulsations (interpreted as free precession) that suggests magnetic fields of the order of $10^{16}$ G \cite{Makishima:2014dua}. Ref.~\cite{DallOsso:2018dos} provides $10^{16}$ G as a low bound for the same source. To be strongly structurally changed by magnetic fields, either more sources have not yet been discovered because their magnetic fields do not fit in the magnetar model (used to infer magnetic fields through period-period derivative diagrams), their magnetic fields decay quickly due to misalignment of rotation and magnetic axes \cite{Lander:2018und}, or these stars are not stable, being formed for example in previous mergers \cite{Giacomazzo:2014qba,Most:2019kfe}. It should be mentioned that magnetic fields $\geq10^{15}$ G could be produced by three-dimensional simulations of supernova explosions with the main requirement to produce a strong field dynamo being sufficient angular momentum in the progenitor star \cite{Raynaud:2020ist}.

 Previous works have studied the effect of the magnetic field on the NS EoS and the stellar properties like mass and radius \cite{PhysRevLett.78.2898,PhysRevC.89.045805,PhysRevLett.79.2176,PhysRevD.90.063013,PhysRevD.91.023003}. The presence of a strong magnetic field deviates the neutron-star structure from spherical symmetry of the strongly and hence the spherically symmetric Tolman–Oppenheimer–Volkoff (TOV) equations can no longer be applied for studying their macroscopic structure \cite{10.1093/mnras/stu2706,Dexheimer:2016yqu}.
 
In the present work, we employ recently proposed density-dependent Relativistic Mean-Field (DD-RMF) parameter sets to study the properties of NSs. By reproducing different hyperon-hyperon optical potentials, the values of the hyperon couplings are obtained, using several different coupling schemes, which are then used to model hyperons in our calculations. 
 We further analyze the effect of magnetic fields on the nucleonic and hyperonic matter by employing a realistic chemical potential-dependent magnetic field \cite{Dexheimer:2016yqu}. The macroscopic stellar matter properties for the magnetic EoSs are obtained using the publicly available Language Objet pour la RElativit\'e Num\'eriquE (LORENE) library \cite{LORENE,1995A&A...301..757B,1993A&A...278..421B,PhysRevD.58.104020}, which solves the coupled Einstein-Maxwell field equations to determine stable magnetic star configurations. In this case, neutron stars become more massive, especially the ones containing hyperons in their interior.\par 
The main motivation behind our work is to demonstrate that a possible measurement of a neutron star mass $\sim$ 2.5$M_\odot$ does not necessarily rule out exotic degrees of freedom in its interior, as several works in the literature claim. In this regard, despite the fact that the GW190814 data has been available for about a year \cite{Abbott_2020a}, our conclusions differ significantly from all other works on the subject that have been published. 
We investigate the possibility of a star with a strong magnetic field inside. We investigate how different particle populations and nuclear interactions affect microscopic and macroscopic stellar properties and we illustrate how, if the secondary object in GW190814 is a massive neutron star, we can learn about the dense matter inside them. 

The chapter is organized as follows: in Sec.~\ref{sec:headings6}, the density-dependent RMF model is presented and the inclusion of the magnetic field for the beta-equilibrium EoS is discussed. The various DD-RMF parameter sets, nuclear matter properties and hyperon couplings are discussed in Sec.~\ref{para}. The nucleonic and hyperonic EoSs are described in Sec.~\ref{hyp} along with spherical solutions for star matter properties. Sec.~\ref{magneticfield} deals with the neutron and hyperon EoS with effects of magnetic fields, together with the discussion of their stellar properties. Sec.~\ref{sub3} presents additional results produced assuming different hyperon couplings. The results are summarized in Sec.~\ref{summary6}. The calculations presented in this chapter are based on the work from Ref.~\cite{rather2021heavy}.
\section{Formalism}
\label{sec:headings6}
The total Lagrangian density in the presence of a magnetic field is given as:
\begin{equation}
\mathcal{L}=\mathcal{L}_m +\mathcal{L}_f,
\end{equation}
where $\mathcal{L}_m$ represents the DD-RMF Lagrangian as given by Eq.~(\ref{ddeq}) along with the contribution from the baryon octet.
The Lagrangian density for the pure electromagnetic part is written as
\begin{equation}
\mathcal{L}_f=-\frac{1}{16\pi} F_{\mu \nu}F^{\mu \nu},
\end{equation}
where $F_{\mu \nu}$ is the electromagnetic field tensor, $F_{\mu \nu}$=$\partial_{\mu}A_{\nu}-\partial_{\nu}A_{\mu}$.  
where $B$ sums over the baryon octet ($n,p,\Lambda, \Sigma^+, \Sigma^0, \Sigma^-, \Xi^0, \Xi^- $)  and $l$ over $e^-$ and $\mu^-$. $\psi_B$ and  $\psi_l$ represent the baryonic and leptonic dirac fields, repsectively. Table \ref{baryonoctet} displays the baryon octet and lepton properties.
\begin{table}[hbtp!]
	\centering
	\caption{Baryon and lepton properties like mass $M$ in MeV, isospin projection in z-direction $I^3$, baryon charge $q_b$, electric charge $q_e$ and strangeness $s$. }
	\begin{tabular}{  p{1.8cm}|p{1.8cm}p{1.8cm}p{1.8cm}p{1.8cm}p{1.8cm} }
		\hline
		\hline
		Particles&$M (MeV)$&$I^3$& $q_b$&$q_e$&$s$ \\
		\hline
		$\Lambda$&1116&0&1&0&-1\\	
		$\Sigma^+$&1189&+1&1&+1&-1\\						
		$\Sigma^0$&1193&0&1&0&-1\\						
		$\Sigma^-$&1197&-1&1&-1&-1\\						
	    $\Xi^0$&1315&+1/2&1&0&-2\\											
		$\Xi^-$&1321&0&0&-1&0\\						
		$e^-$&0.511&0&0&-1&0\\
		$\mu^-$&105.7&0&0&-1&0\\									
		\hline
		\hline
	\end{tabular}
	\label{baryonoctet}
\end{table}
For the present case that includes all baryons from the octet, the NS chemical equilibrium condition between different particles are
\begin{align}\label{eq13}
	\mu_n &=\mu_{\Sigma^0} =  \mu_{\Xi^0}, \nonumber \\
	\mu_p &=\mu_{\Sigma^+}=\mu_n -\mu_e, \nonumber \\
	\mu_{\Sigma^-}& =\mu_{\Xi^-}=\mu_n +\mu_e,\nonumber \\
	\mu_{\mu}&=\mu_e.
\end{align}
The charge neutrality condition follows as
\begin{equation}
	\rho_p +\rho_{\Sigma^+}= \rho_e+\rho_{\mu^-}+\rho_{\Sigma^-}+\rho_{\Xi^-}.
\end{equation}
Inclduing the baryon octet, the expression for the energy density and pressure are
\begin{align} \label{eq6.7}
	\mathcal{E}_m &= \sum_{B} \frac{2}{(2\pi)^3}\int_{0}^{k_B} d^3 k E_B^* (k) + \frac{1}{2}m_{\sigma}^2 \sigma^2-\frac{1}{2}m_{\omega}^2 \omega^2-\frac{1}{2}m_{\rho}^2 \rho^2 \nonumber \\
	&+g_{\omega}(\rho_B)\omega \rho_B+\frac{g_{\rho}(\rho_B)}{2}\rho \rho_3,  \nonumber \\
	P_m&= \sum_{B} \frac{2}{3(2\pi)^3}\int_{0}^{k_B} d^3 k \frac{k^2}{E_B^* (k)} -\frac{1}{2}m_{\sigma}^2 \sigma^2+\frac{1}{2}m_{\omega}^2 \omega^2 \nonumber \\
	&+\frac{1}{2}m_{\rho}^2 \rho^2-\rho_B \sum_R (\rho_B),
\end{align}
where $E_{B}^*=\sqrt{k_{B}^2+M_B^{*2}}$. The rearrangment term $\sum\limits_{R}(\rho_B)$ contributes to the pressure only.

In the presence of magnetic field, the scalar and vector density for charged baryons $cb$, uncharged baryons $ub$ and leptons follow as \cite{Broderick_2000}
\begin{align}\label{eq6.8}
\rho_s^{cb}&=\frac{|q_{cb}|B M_{cb}^{*2}}{2\pi^2}\sum_{\nu=0}^{\nu_{max}}r_{\nu}  ln\Bigg(\frac{k_{F,\nu}^{cb}+E_F^{cb}}{\sqrt{M_{cb}^{*2}+2\nu|q_{cb}|B}}\Bigg), \nonumber \\
	\rho_s^{ub}&=\frac{M_{ub}^{*2}}{2\pi^2}\Bigg[E_F^{ub} k_F^{ub} -  M_{ub}^{*2} ln\Bigg(\frac{k_{F,\nu}^{ub}+E_F^{ub}}{M_{ub}}\Bigg)\Bigg],  \nonumber \\
	\rho^{cb}&= \frac{|q_{cb}|B}{2\pi^2}\sum_{\nu=0}^{\nu_{max}}r_{\nu} k_{F,\nu}^{cb}, \nonumber \\
	\rho^{ub}&=\frac{(k_F^{ub})^3}{3\pi^2},\nonumber \\
	\rho_l &=\frac{|q_l|B}{2\pi^2}\sum_{\nu=0}^{\nu_{max}}r_{\nu} k_{F,\nu}^l,
\end{align}
where $r_{\nu}$ is the Landau degeneracy of $\nu$ level.
The expressions for the baryon and lepton energy densities in the presence of magnetic field become
\begin{align}\label{eq6.9}
	\mathcal{E}_{cb}&= \frac{|q_{cb}|B}{4\pi^2}\sum_{\nu=0}^{\nu_{max}}r_{\nu} \nonumber \\
	&\times \Bigg[k_{F,\nu}^{cb}E_F^{cb}+(M_{cb}^{*2}+2\nu|q_{cb}|B) ln\Bigg(\frac{k_{F,\nu}^{cb}+E_F^{cb}}{\sqrt{M_{cb}^{*2}+2\nu|q_{cb}|B}}\Bigg)\Bigg], \nonumber \\
	\mathcal{E}_{ub} &= \frac{1}{8\pi^2}\Bigg[k_{F}^{ub}(E_F^{ub})^3+(k_F^{ub})^3E_F^{ub} -M_{ub}^{*4} ln \Bigg(\frac{k_{F}^{ub}+E_F^{ub}}{M_{ub}^*}\Bigg)\Bigg],\nonumber \\
	\mathcal{E}_{l}&= \frac{|q_l|B}{4\pi^2}\sum_{\nu=0}^{\nu_{max}}r_{\nu} \nonumber \\
	&\times \Bigg[k_{F,\nu}^{l}E_F^{l}+(m_{l}^{2}+2\nu|q_l|B) ln\Bigg(\frac{k_{F,\nu}^{l}+E_F^{l}}{\sqrt{m_{l}^2+2\nu|q_l|B}}\Bigg)\Bigg].
\end{align}
The expressions for the energy density and pressure in presence of a magnetic field can be obtained by solving the energy-momentum tensor relation:
\begin{equation}
	T^{\mu\nu} = T_m^{\mu\nu}+T_l^{\mu\nu}, 
\end{equation}
where \cite{PhysRevD.81.045015,PhysRevD.65.056001}
\begin{align}\label{eq11}
	T_m^{\mu\nu} &= \mathcal{E}_m u^{\mu}u^{\nu}-P(g^{\mu\nu}-u^{\mu}u^{\nu}) \nonumber \\
	&+\mathcal{M}B \Bigg(g^{\mu\nu}-u^{\mu}u^{\nu}+\frac{B^{\mu}B^{\nu}}{B^2}\Bigg), \nonumber \\
	T_l^{\mu\nu} &= \frac{B^2}{4\pi}\Bigg(u^{\mu}u^{\nu}-\frac{1}{2}g^{\mu\nu}\Bigg)-\frac{B^{\mu}B^{\nu}}{4\pi}.
\end{align}
Here $\mathcal{M}$ is the magnetization per unit volume and $B^{\mu}B_{\mu}=-B^2$.
For the nuclear matter in presence of a magnetic field, the single particle energy of all charged baryons and leptons is quantized in the direction perpendicular to the magnetic field. For a uniform magnetic field locally pointing in the $z$-direction, $B=B\hat{z}$, the total energy of a charged particle becomes
\begin{equation}
	E_{cb}=\sqrt{k_z^2 +M^{*2}+2\nu|q|B},
\end{equation}
The quantity $\nu=\Big(n+\frac{1}{2}-\frac{1}{2}\frac{q}{|q|} \sigma_z\Big)=0,1,2,...$ indicates the Landau levels of fermions with electric charge $q$. $n$ is the orbital angular momentum quantum number and $\sigma_z$ is the Pauli matrix. 

The fermi momentum of all baryons charged $k_{F,\nu}^{cb}$ and leptons $k_{F,\nu}^l$ with fermi energies $E_F^{cb}$  and $E_F^l$, respectively, are given as:
\begin{align}
	k_{F,\nu}^{cb}&=\sqrt{(E_F^{cb})^2-M_{cb}^{*2}-2\nu |q_{cb}|B},\nonumber \\
	k_{F,\nu}^{l}& =\sqrt{(E_F^l)^2-m_l^{2}-2\nu |q_l|B}.
\end{align}
The highest value of $\nu$ is obtained with sum over Landau levels under the condition that the Fermi momentum of each particle is positive:
\begin{align}
	\nu_{max}&=\Bigg[\frac{(E_F^{cb})^2 -M_{cb}^{*2}}{2|q_{cb}|B}\Bigg], \nonumber \\
	\nu_{max}&=\Bigg[\frac{(E_F^l)^2 -m_{l}^{2}}{2|q_{l}|B}\Bigg], 
\end{align}
for charged baryons $cb$ and leptons, respectively.

 Following the energy density expression given by Eq.~(\ref{eq6.9}) in presence of magnetic field, the total energy density is
\begin{equation}
	\mathcal{E}=\mathcal{E}_m +\frac{B^2}{8\pi}.
\end{equation}
The total pressure in the perpendicular and the parallel directions to the local magnetic field are
\begin{align}
	P_{\perp}&=P_m-\mathcal{M}B +\frac{B^2}{8\pi},\nonumber \\
	P_{\parallel}&=P_m-\frac{B^2}{8\pi},
\end{align}
where the magnetization is calculated as
\begin{equation}
	\mathcal{M}=\partial P_m/\partial B.
\end{equation}
\section{Parameter Sets}
\label{para}
For the present study, we employ two recently proposed density-dependent DD-MEX and DD-LZ1 parameter sets.  Additionally, widely used parameter sets DD-ME1 and DD-ME2 are also used.
Table \ref{tab2.3} displays the nucleon and meson masses and the coupling constants between nucleon and mesons for the given parameter sets. The  parameters $a,b,c,d$ for $\sigma$, $\omega$ and $\rho$ mesons are also shown. 


The density-dependent coupling constants of the hyperons to the vector mesons are determined from the SU(6) symmetry as \cite{PhysRevC.64.055805,PhysRevC.53.1416,Tolos_2016}
\begin{align}\label{eq20}
	\frac{1}{2}g_{\omega\Lambda}&=\frac{1}{2}g_{\omega\Sigma}=g_{\omega\Xi}=\frac{1}{3}g_{\omega N},\nonumber \\
	\frac{1}{2}g_{\rho\Sigma}&=g_{\rho\Xi}=g_{\rho N}, g_{\rho \Lambda}=0. 
\end{align}
These couplings are calculated by fitting the hyperon optical potential obtained from the experimental data. 

The hyperon coupling to the $\sigma$ field is obtained, to reproduce the hyperon potential in the symmetric nuclear matter (SNM) derived from the hypernuclear observables \cite{RevModPhys.64.649}:
\begin{equation}
	U_{\Lambda}^N(\rho)=g_{\omega \Lambda}\omega_0 +\sum_{R}-g_{\sigma \Lambda}\sigma_0. 
\end{equation}
For the present study, we reproduce the following optical potentials for the hyperons \cite{particles3040043}:
\begin{align}\label{pot1}
	U_{\Lambda}^N(\rho_0)&=-30 MeV,\nonumber \\
	U_{\Sigma}^N(\rho_0)&=+30 MeV, \nonumber \\
	U_{\Xi}^N(\rho_0)&=-14 MeV.
\end{align} 
These potentials correspond to the value of the density-dependent scalar couplings $g_{\sigma \Lambda}/g_{\sigma N} = 0.6105$, $g_{\sigma \Xi}/g_{\sigma N} = 0.3024$ and $g_{\sigma \Sigma}/g_{\sigma N} = 0.4426$ .
\section{Results and Discussions}
\label{results}

\subsection{Nucleonic and Hyperonic neutron stars}
\label{hyp}
\begin{figure}[hbt!]
	\centering 
	\includegraphics[width=0.75\textwidth]{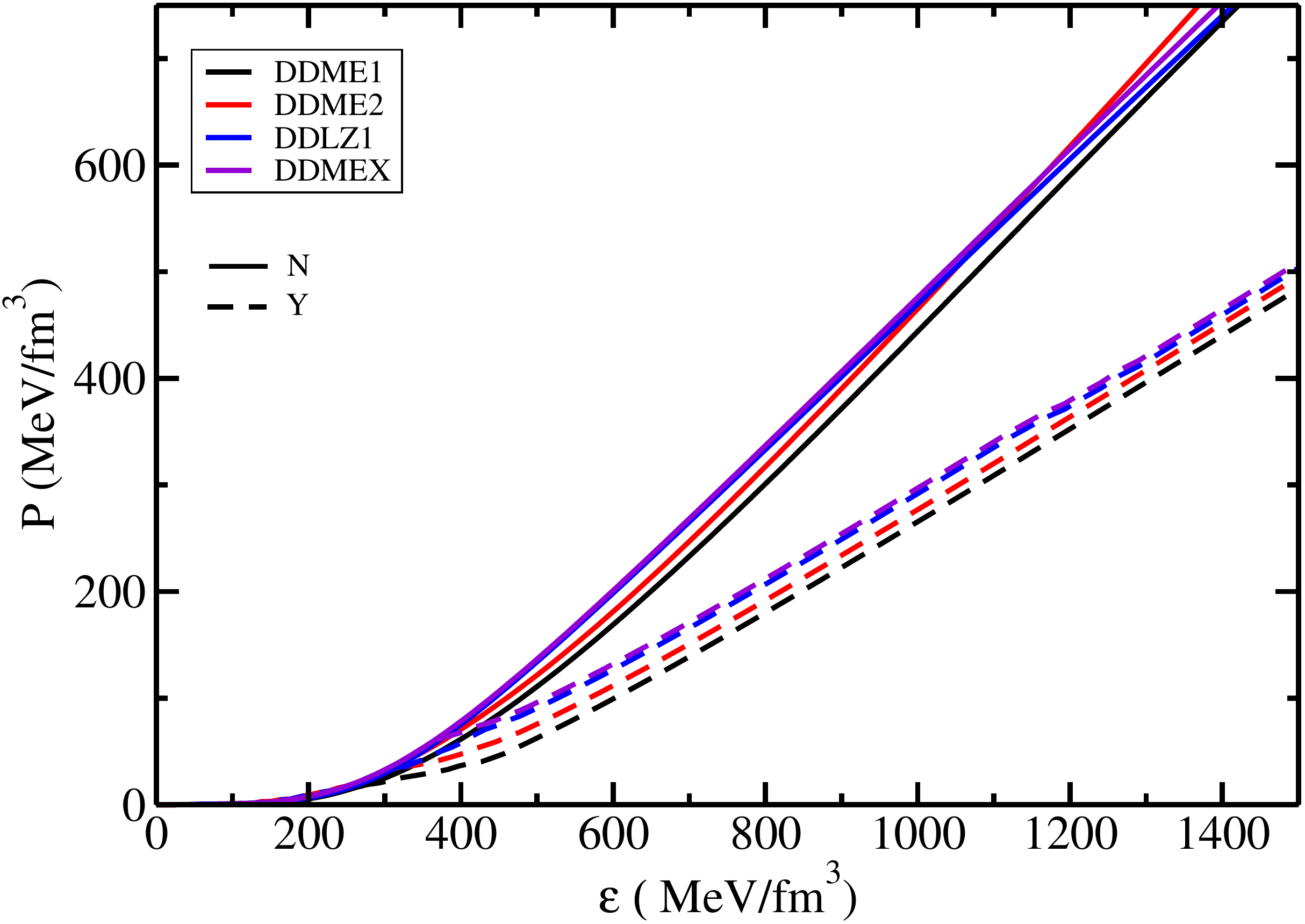}
	\caption{  Dense matter equation of state for DD-LZ1, DD-ME1, DD-ME2 and DD-MEX parameter sets. The solid lines represent pure nucleonic matter, while the dashed lines represent hyperonic matter including the entire baryon octet. }
	\label{fig6.1} 
\end{figure}

The EoS of pure nucleonic (solid lines) and hyperonic matter (dashed lines) under chemical equilibrium and charge neutrality conditions for several DD-RMF parameter sets are displayed in Fig.~\ref{fig6.1}. For pure nucleonic matter, the DD-MEX parameter set produces a stiff EoS at low-density region while as DD-ME2 set produces a stiff EoS in the high-density regime. The hyperons start to appear in the density range $\approx$ 300-400 MeV/fm$^3$ for all parameter sets. The onset of hyperonization softens the EoS (reduction in the pressure) due to the hyperons replacing the neutrons and opening new channels to distribute the Fermi energy. To build a unified EoS, the Baym-Pethick-Sutherland (BPS) EoS \cite{Baym:1971pw} has been used in the outer crust region. For the inner crust part, the EoS for non-uniform matter is generated by using the DD-ME2 parameter set in a Thomas-Fermi approximation \cite{PhysRevC.79.035804,PhysRevC.94.015808}. All the DD-RMF parameter sets considered are very similar in the outer crust density regime.
With the EoSs obtained for pure nucleonic and hyperonic matter, stellar matter properties like mass and radius are obtained by solving the TOV coupled differential equations \cite{PhysRev.55.364,PhysRev.55.374} for a static isotropic spherically symmetric stars (see Sec.~\ref{tovall}).
\begin{figure}[hbt!]
	\centering
	\includegraphics[width=0.75\textwidth]{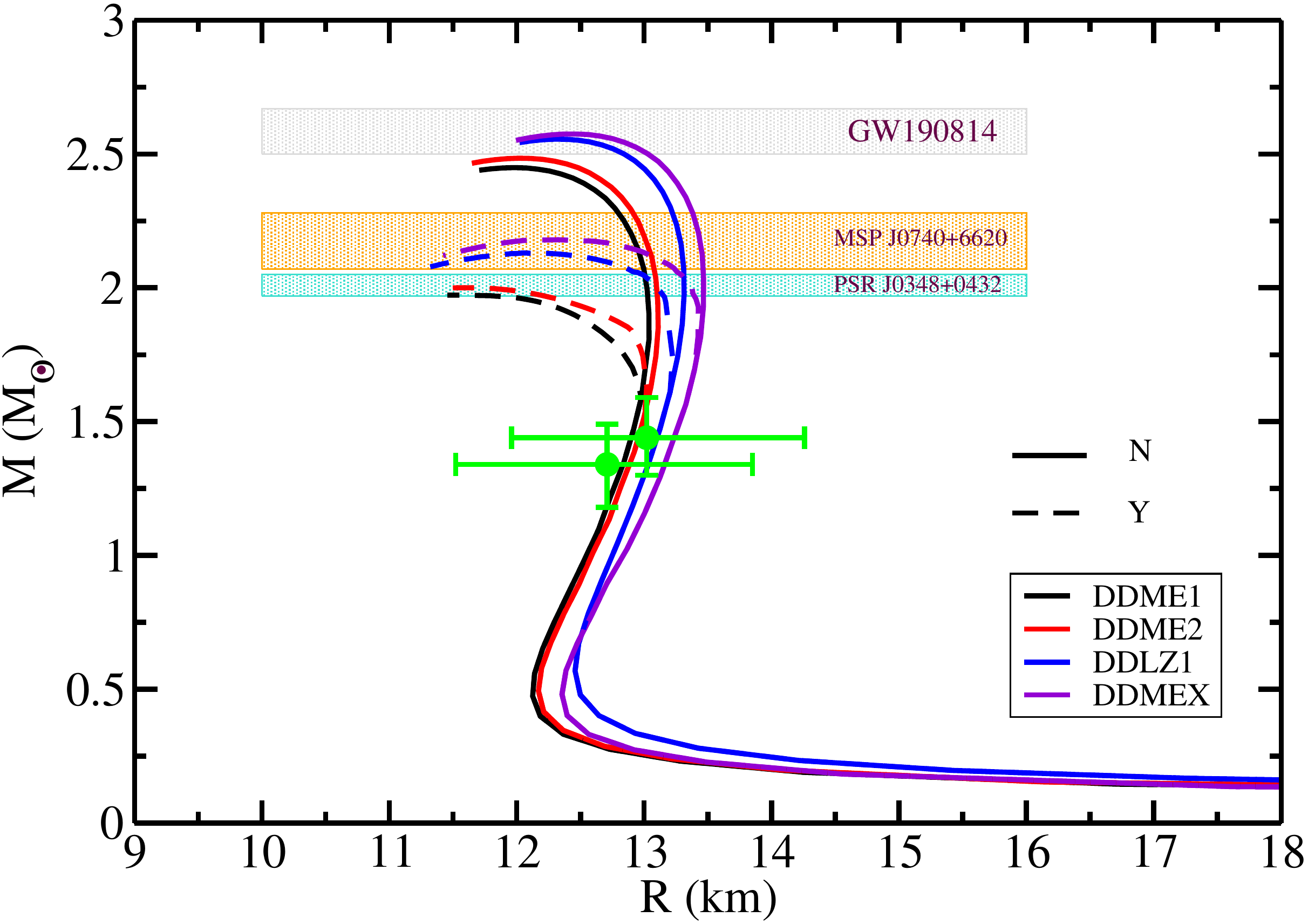}
	\caption{ Mass-radius relation for pure nucleonic (solid lines) and hyperonic (dashed lines) stars using several DD-RMF parameters. The colored areas show recent constraints inferred from GW190814, the massive pulsars MSP J0740+6620 and PSR J0348+0432 \cite{Abbott_2020a,Cromartie2020,Antoniadis1233232}. The constraints on the mass-radius limits inferred from  NICER observations \cite{Miller_2019,Riley_2019} are also shown.}
	\label{fig6.2} 
\end{figure}
\vspace{-0.5cm}
\begin{center}
	\begin{table}[hbt!]
		\centering
		\caption{Properties of pure nucleonic and hyperonic NS for different DD-RMF parameter sets, including maximum mass, respective radius and radius and dimensionless tidal deformability of a 1.4$M_{\odot}$ star.  }
		\begin{tabular}{ p{1.8cm}|p{1.2cm}p{1.2cm}p{1.4cm}
				p{1.2cm}|p{1.2cm}p{1.2cm}p{1.2cm}p{1.2cm} }
			\hline
			\hline
			&&\multicolumn{3}{p{2.5cm}|}{Neutron Star} & %
			&\multicolumn{3}{p{2.5cm}}{Hyperon Star}\\
			\cline{2-9}
			&\parbox[t]{0.2cm}{\centering $M_{max}$ \\ ($M_{\odot}$) }  &\parbox[t]{0.2cm}{\centering $R$ \\ (km) }&\parbox[t]{0.2cm}{\centering $R_{1.4}$  \\ (km) }&$\Lambda_{1.4}$&\parbox[t]{0.2cm}{\centering $M_{max}$ \\ ($M_{\odot}$) }  &\parbox[t]{0.2cm}{\centering $R$ \\ (km) }&\parbox[t]{0.2cm}{\centering $R_{1.4}$  \\ (km) }&$\Lambda_{1.4}$ \\
			\hline
			DD-ME1 &2.449&11.981&12.898&689.342&1.983&11.515&12.898&689.342\\
			DD-ME2 &2.483&12.017&12.973&733.149&2.013&11.674&12.973&733.149\\
			DD-LZ1 &2.555&12.297&13.069&728.351&2.130&12.067&13.069&728.351\\
			DD-MEX &2.575&12.465&13.168&791.483&2.183&12.238&13.168&791.483\\
			
			\hline
			\hline
		\end{tabular}
		\label{tab6.1}
	\end{table}
\end{center}
\vspace{-0.4cm}
Fig.~\ref{fig6.2} shows the mass-radius relation for pure nucleonic and hyperonic matter for the parameter sets DD-ME1, DD-ME2, DD-LZ1 and DD-MEX. The shaded areas represent the recent constraints on the NS maximum mass inferred from GW190814 ($M$ = 2.50-2.67$M_{\odot}$), the massive pulsars MSP J0740+6620 ($M$ = 2.14$_{-0.09}^{+0.10}$$M_{\odot}$) and PSR J0348+0432 ($M$ = 2.01$\pm 0.04$$M_{\odot}$). The constraints on the radius limits around the NS canonical mass inferred from PSR J0030+0451 by NICER experiment $R$ = 13.02$_{-1.06}^{+1.24}$ km at $M$ = 1.44$_{-0.14}^{+0.15}$$M_{\odot}$ \cite{Miller_2019} and $R$ = 12.71$_{-1.19}^{+1.14}$ km at $M$ = 1.34$_{-0.16}^{+0.15}$$M_{\odot}$ \cite{Riley_2019} are also shown. For pure nucleonic matter, the DD-LZ1 and DD-MEX parameter sets reach a maximum mass of 2.55 and 2.57$M_{\odot}$, with a radius of 12.30 and 12.46 km, respectively, indicating the possibility of GW190814 secondary component being a supermassive NS. The hyperonic counterparts of the given DD-RMF parameter sets, which soften the EoS, produce a maximum mass of 2.18 and 2.13$M_{\odot}$ with a radius of 12.24 and 12.07 km respectively. For DD-ME1 and DD-ME2 parameter sets, the NS maximum mass decreases from 2.45 and 2.48 to 1.98 and 2.01$M_{\odot}$ respectively, while the respective radius decreases by $\sim$ 0.5 km. The radius at the NS canonical mass, $R_{1.4}$, remains insensitive to the appearance of hyperons. The hyperonic configurations satisfy the maximum mass limit from the massive pulsar MSP J0740+6620 and PSR J0348+0432, but are inconsistent with the GW190814  potential constraint (see discussion in  \cite{PhysRevD.102.041301}). All the nucleonic and hyperonic configurations satisfy the mass-radius limits inferred from NICER experiment.
\begin{figure}[hbt!]
	\centering
	\includegraphics[width=0.75\textwidth]{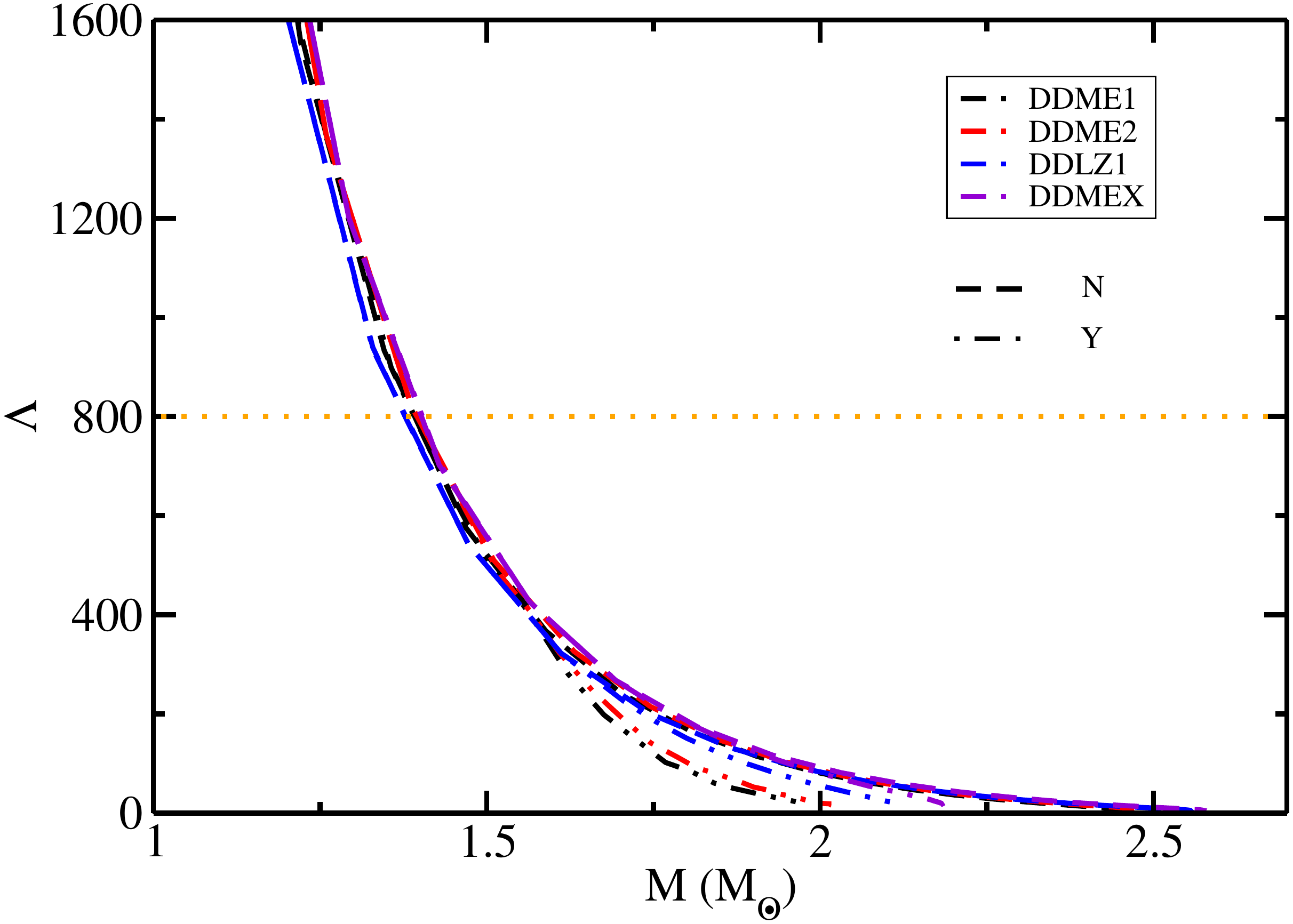}
	\caption{ Dimensionless tidal deformability variation with the NS mass for nucleonic (dashed lines) and hyperonic (dotted-dashed lines) stars using DD-LZ1, DD-ME1, DD-ME2 and DD-MEX parameter sets. The orange dotted line represents the upper limit on the dimensionless tidal deformabilty set by measurement from GW170817 \cite{PhysRevLett.119.161101}. }
	\label{fig6.3} 
\end{figure}

The variation of the dimensionless tidal deformability with the NS mass for the nucleonic and hyperonic stars based on the DD-RMF parametrizations are shown in Fig.~\ref{fig6.3}.  The dimensionless tidal deformability of pure nucleonic, as well as hyperonic matter for all the parameter sets lies well below the upper limit of $\Lambda_{1.4}$ = 800 obtained from the gravitational wave event GW170817 \cite{PhysRevLett.119.161101}. The shift in the tidal deformability for the hyperonic matter is seen as the mass increases. Table \ref{tab6.1} displays the different properties of neutron and hyperon stars obtained with different DD-RMF parameter sets. A subsequent analysis by the LVC suggest a much smaller upper limit of 580 on the tidal deformability, which is smaller than all the values displayed in Table \ref{tab6.1} \cite{PhysRevLett.121.161101}. However, this value  corresponds to the 50\% confidence region, the 90\% confidence region extracted from the same data includes the values we reproduce.

\subsection{Magnetic stars}
\label{magneticfield}
To study the effects of magnetic fields on our microscopic description of matter, we employ a chemical potential-dependent magnetic field which was derived from the solutions of the Einstein-Maxwell equations. The quadratic relation between the magnetic field and the chemical potential depends on the magnetic dipole moment and  is given by \cite{Dexheimer:2016yqu}
\begin{equation}
	B^*(\mu_B)=\frac{(a+b\mu_B +c\mu_B^2)}{B_c^2} \mu,
\end{equation}
with $\mu_B$ being the baryon chemical potential in MeV and $\mu$ the dipole magnetic moment in units of Am$^2$ to produce $B^*$ in units of the electron critical field  $B_c=4.414\times 10^{13}$ G. The coefficients $a$, $b$ and $c$  taken as $a=-0.0769$  G$^2$/(Am$^2$), $b=1.20\times 10^{-3}$  G$^2$/(Am$^2$ MeV) and $c=-3.46\times 10^{-7}$  G$^2$/(Am$^2$ MeV$^2$) are obtained from a fit for the magnetic field in the polar direction of a star with a baryon mass of 2.2$M_{\odot}$.
\begin{figure}[hbt!]
	\centering
	\includegraphics[width=0.75\textwidth]{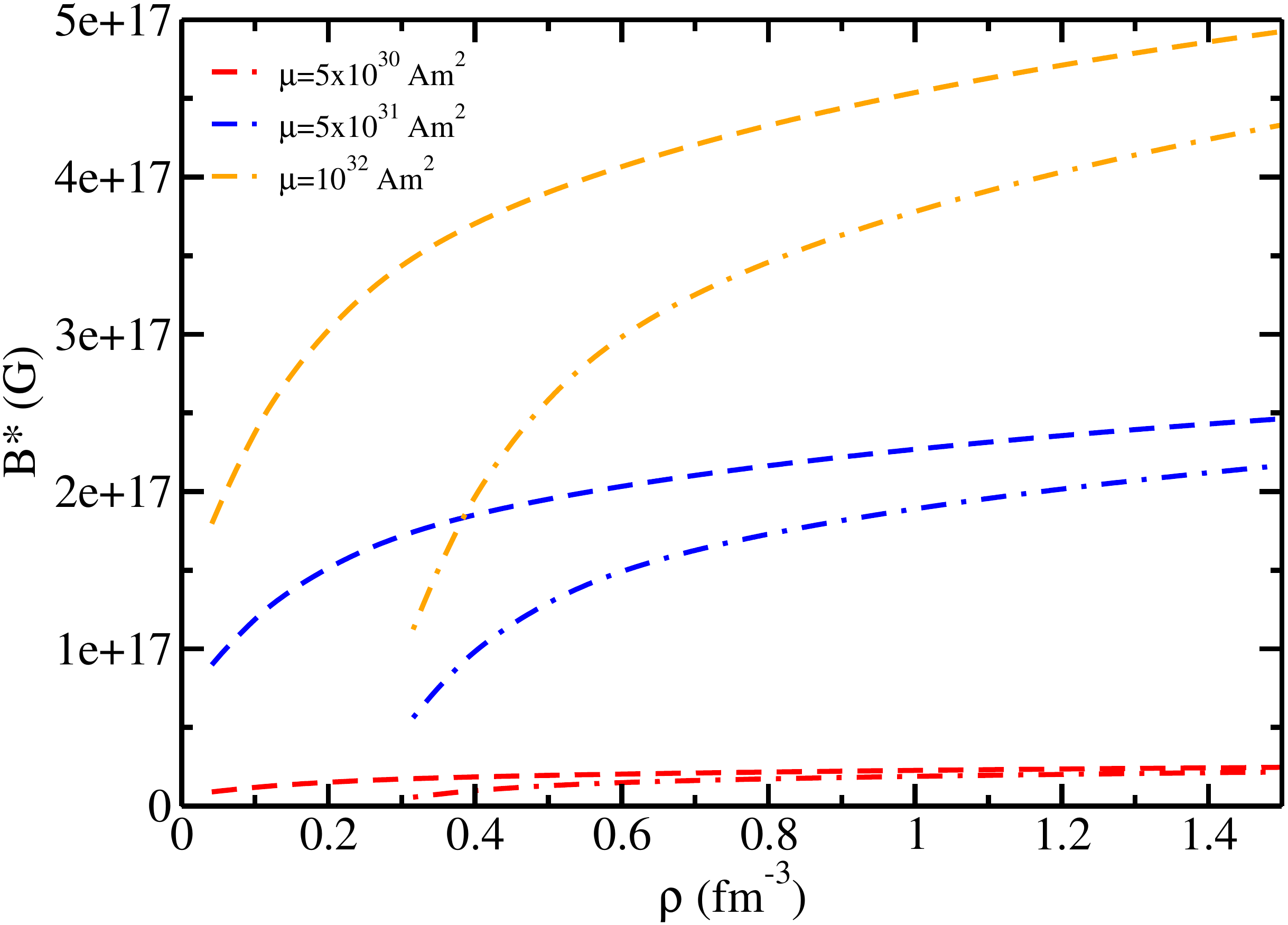}
	\caption{ Magnetic field profile as a function of baryon density for DD-MEX EoS with different values of magnetic dipole moment. The dashed lines represent the profile for NS without hyperons while the dotted-dashed lines represent hyperon stars.}
	\label{fig6.4} 
\end{figure}
Fig.  \ref{fig6.4}  displays the magnetic field profile as a function of baryon density for a 2.2$M_{\odot}$ baryonic mass star obtained for the DD-MEX EoS. The magnetic field effect on the DD-ME2 EoS has already been calculated \cite{particles3040043} and DD-ME1 EoS predicts similar behavior. However, they have used spherically symmetric TOV equations for central magnetic fields $\sim$ 10$^{18}$ G with a density-dependent and universal profile for the magnetic field. Since the DD-MEX EoS predicts a heavier NS than other parameter sets, we choose this parameter set to study magnetic effects and verify whether this model predicts the possibility of GW190814 secondary component to be a hyperonic magnetar. In Fig.~\ref{fig6.4}, the dashed curves represent a neutron star without hyperons and dotted-dashed curves represent an NS with hyperons. It is clear that the magnetic field produced by the NS without hyperons is larger. This illustrates that the transition from $\mu_B$ to $\rho_B$ is model and particle population dependent. For magnetic dipole moment greater than 10$^{31}$ Am$^2$, the magnetic field produced at large densities is larger than 10$^{17}$ G, which is strong enough to cause a large deformation in the NS structure. The values of the magnetic field produced at the surface and at large densities using different values of the magnetic dipole moment for NSs with and without hyperons are shown in Table \ref{tab6.2}. For a magnetic dipole moment 10$^{32}$ Am$^2$, the magnetic field produced at large densities is greater than 4$\times$ 10$^{17}$ for both cases. 

\begin{table}[hbt!]
	\centering
	\caption{Magnetic field at low densities B$_s$ (corresponding to the stellar surfaces) and at high values of densities B$_c$ calculated for DD-MEX EoS at 2.2$M_{\odot}$ baryonic mass for a neutron star and a hyperon star. }
	\begin{tabular}{  p{2.0cm}|p{2.2cm}p{2.2cm}|p{2.2cm}p{2.2cm} }
		\hline
		\hline
		&\multicolumn{2}{p{2.7cm}|}{Neutron Star} & %
		\multicolumn{2}{p{2.7cm}}{Hyperon Star}\\
		\hline
		$\mu$ (Am$^2$)& $B_{s}$ (G)&$B_{c}$ (G)&$B_{s}$ (G)&$B_{c}$ (G) \\
		\hline
		5$\times$10$^{30}$ &1.01$\times$10$^{15}$&2.59$\times$10$^{16}$&6.65$\times$10$^{15}$&1.96$\times$10$^{16}$\\
		5$\times$10$^{31}$&8.98$\times$10$^{16}$&2.28$\times$10$^{17}$&5.83$\times$10$^{16}$&1.89$\times$10$^{17}$\\
		10$^{32}$&1.79$\times$10$^{17}$&4.55$\times$10$^{17}$&1.12$\times$10$^{17}$&3.77$\times$10$^{17}$\\
		\hline
		\hline
	\end{tabular}
	\label{tab6.2}
\end{table}
\begin{figure}[hbt!]
	\centering
	\includegraphics[width=0.75\textwidth]{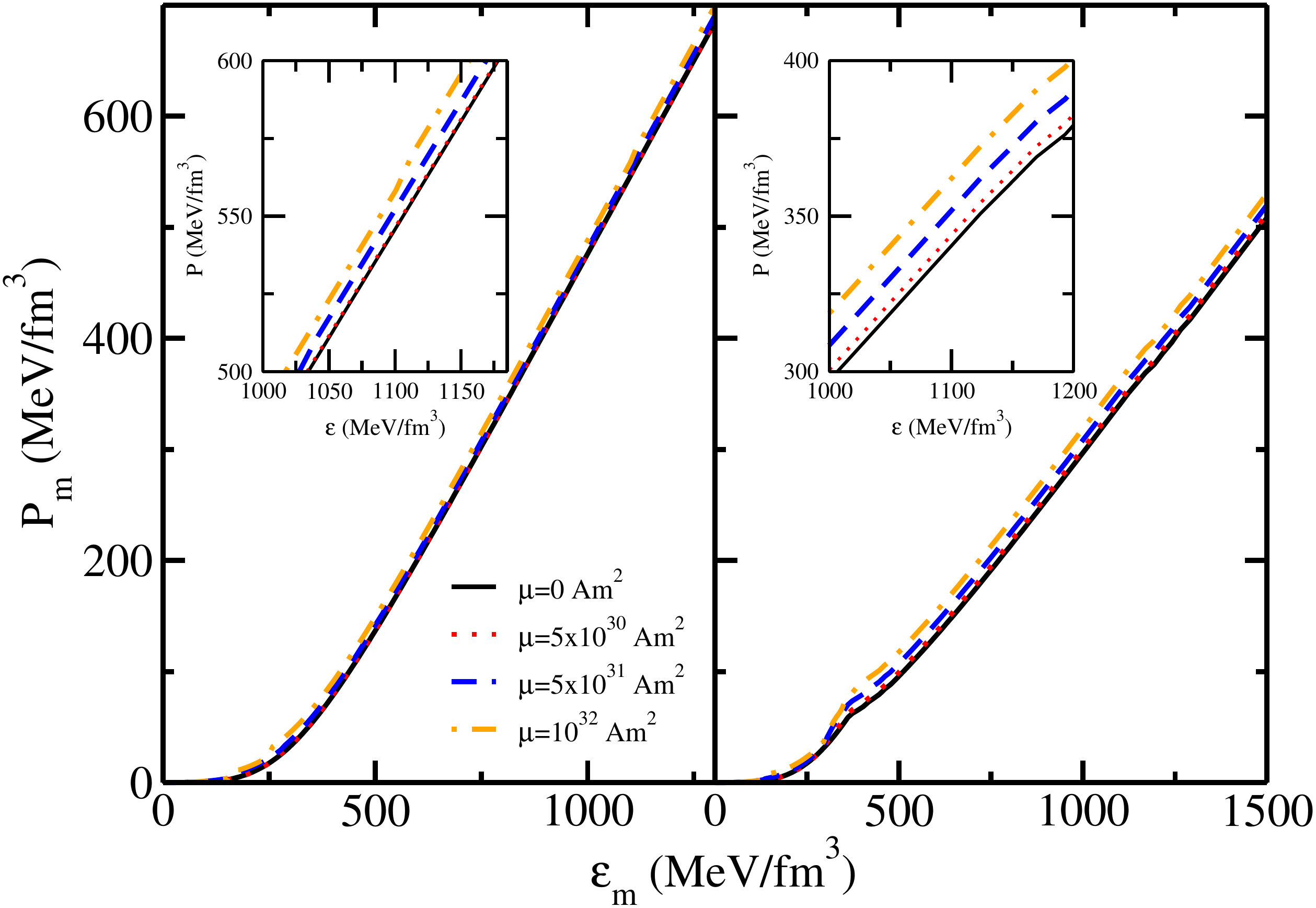}
	\caption{ Variation of matter pressure in the transverse direction (P$_{\perp}$) vs energy density for DD-MEX parameter set without and with magnetic field effects at different values of magnetic dipole moment. Left panel depicts the EoSs without hyperons and right panel depicts the EoS with hyperons. The insets in each panel show the variation in the pressure at a higher value of the energy density for different magnetic moments. }
	\label{fig6.5} 
\end{figure}
%
%
%

\begin{figure}[h]
	\centering
	\subfloat[Subfigure 1 list of figures text][ $\mu$ = 0 Am$^2$ ]{
		\includegraphics[width=0.485\textwidth]{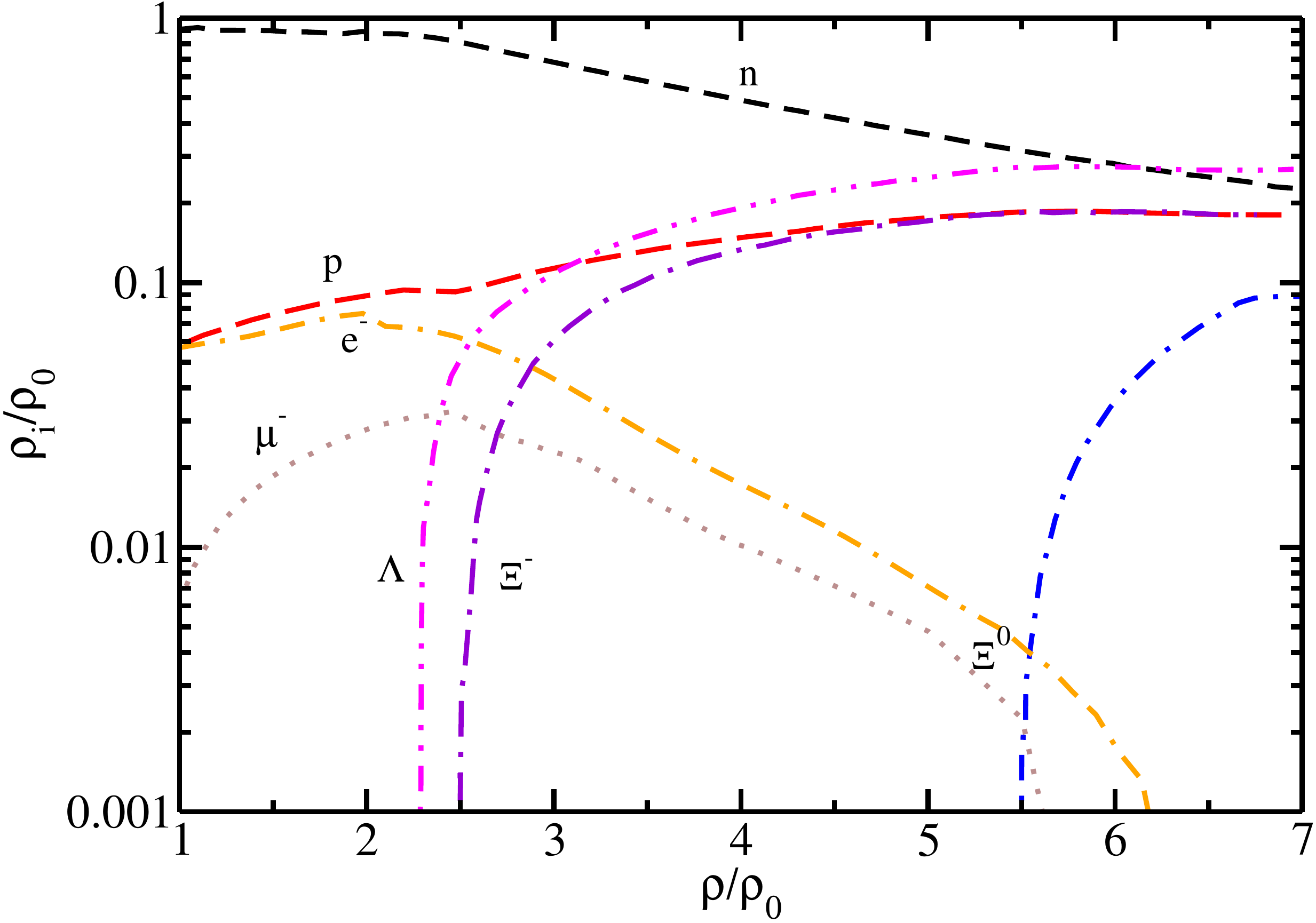}
		\label{fig6a}}
	\subfloat[Subfigure 2 list of figures text][ $\mu$ = 5$\times$10$^{30}$ Am$^2$]{
		\includegraphics[width=0.485\textwidth]{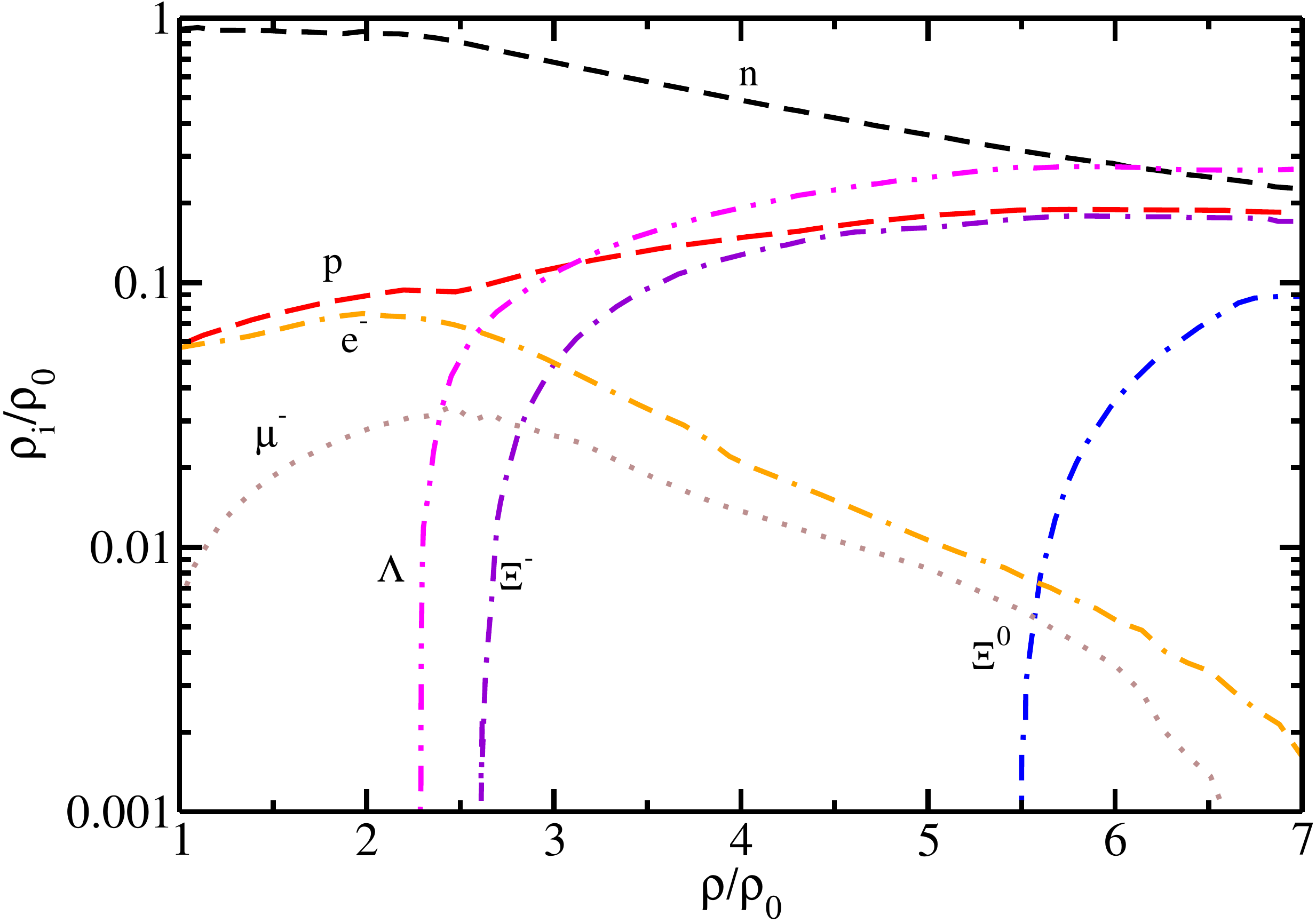}
		\label{fig6b}}
	\qquad
	\subfloat[Subfigure 3 list of figures text][ $\mu$ = 5$\times$10$^{31}$ Am$^2$]{
		\includegraphics[width=0.485\textwidth]{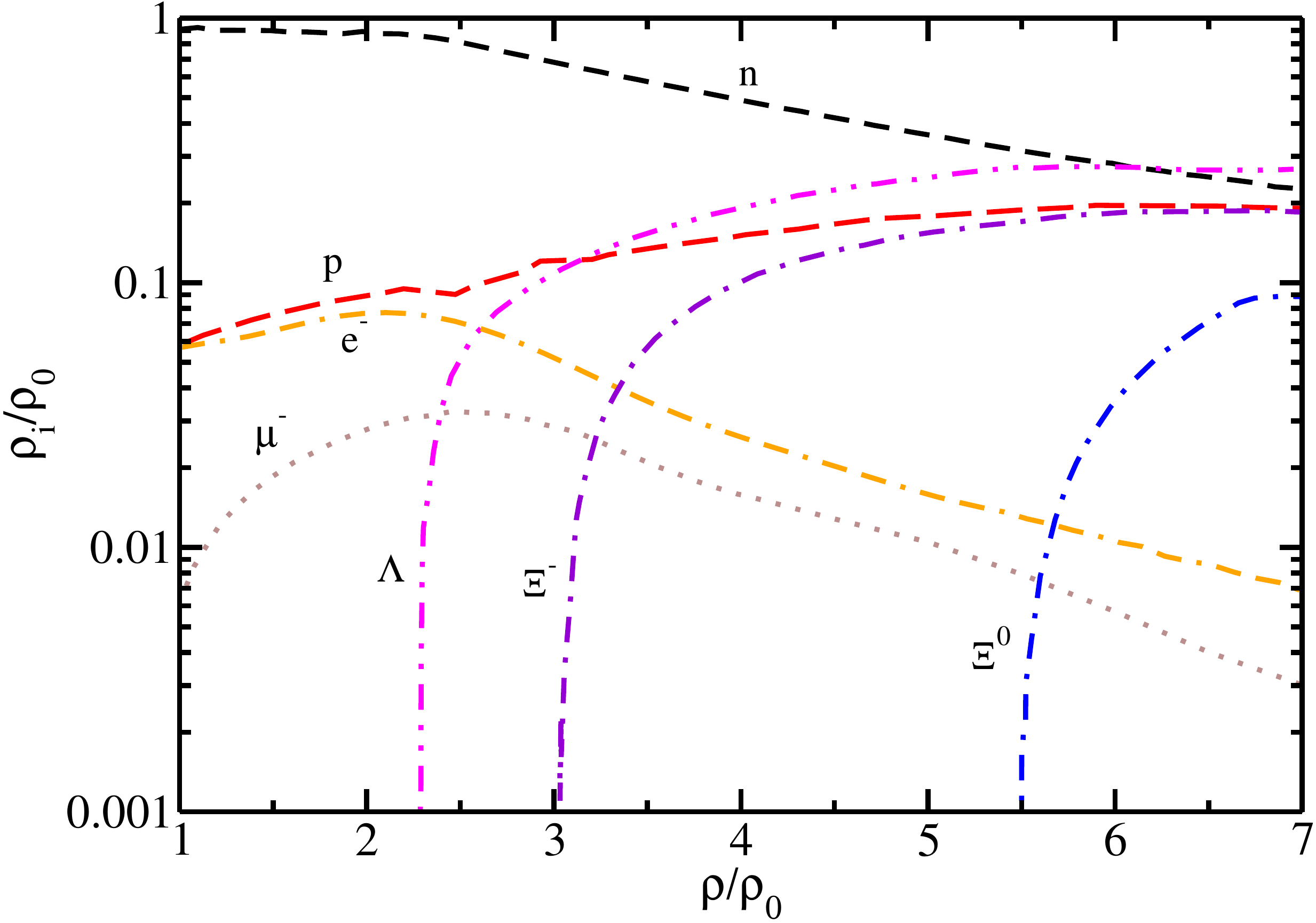}
		\label{fig6c}}
	\subfloat[Subfigure 4 list of figures text][ $\mu$ = 10$^{32}$ Am$^2$]{
		\includegraphics[width=0.485\textwidth]{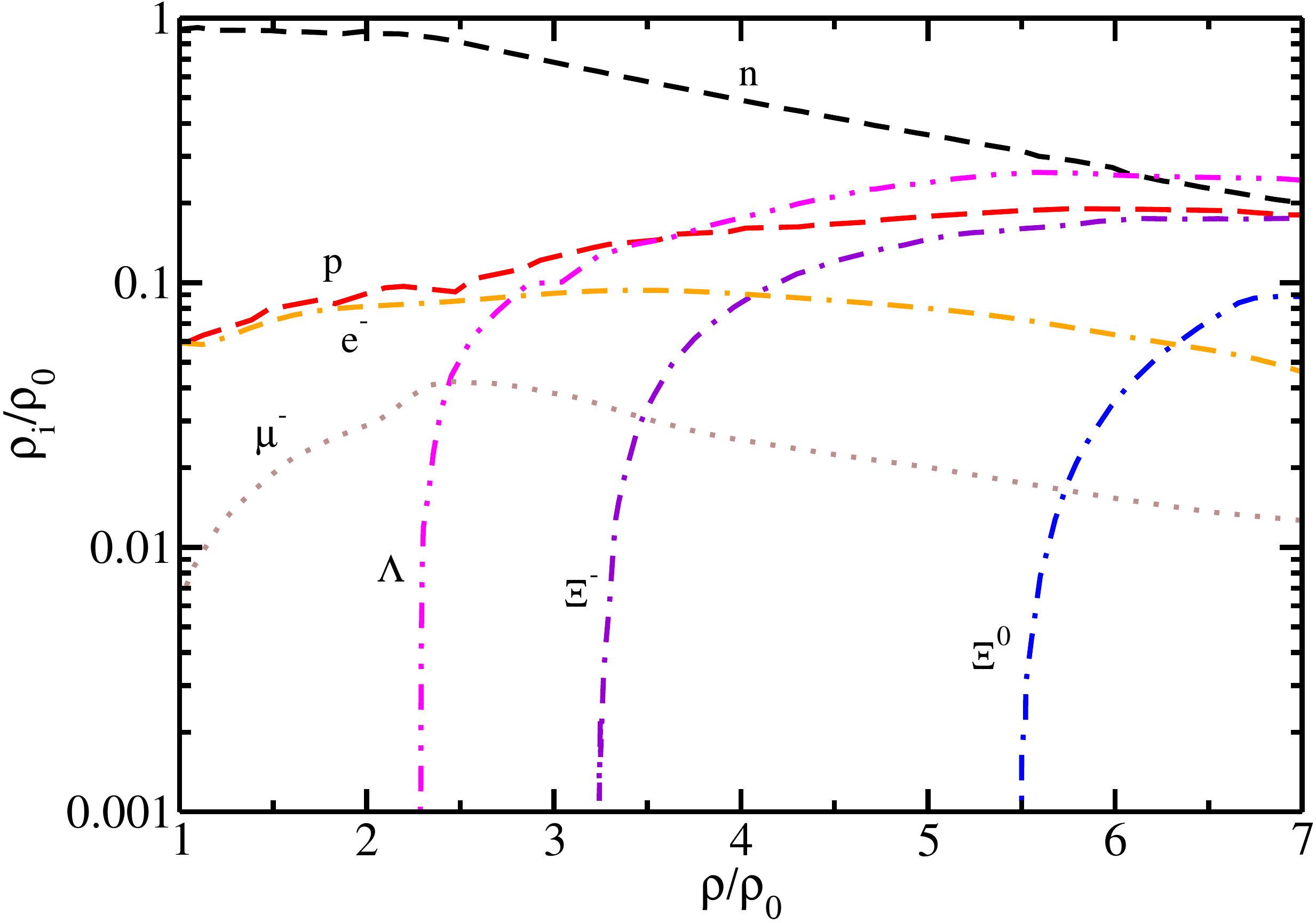}
		\label{fig6d}}
	\caption{Particle fraction of the baryons and leptons as a function of normalized baryon density for DD-MEX model without magnetic field (a) and with magnetic field with different magnetic dipole moments b) $\mu$ = 5$\times$10$^{30}$ Am$^2$, c) $\mu$ = 5$\times$10$^{31}$ Am$^2$ and d) $\mu$ = 1$\times$10$^{32}$ Am$^2$.}
	\label{fig6.6}
\end{figure}
Fig.~\ref{fig6.5} shows the variation of the transverse pressure vs the energy density for the DD-MEX parameter set with and without hyperons. The solid line represents the variation in the pressure without including magnetic field ($\mu$ = 0) while the other lines represent the variation obtained with the magnetic field at magnetic dipole moments $\mu$ = 5$\times$10$^{30}$ Am$^2$, 5$\times$10$^{31}$ Am$^2$ and 10$^{30}$ Am$^2$. The insets show the pressure at higher values of the energy density. It is clear that the change in the pressure at a given value of energy density is larger for an NS with hyperons as compared to the NS without hyperons, which implies that the EoS with hyperons becomes stiffer than without hyperons (when compared to the $B$ = 0 case) in the presence of a strong magnetic field. The reason for such behavior will be discussed in the following. The magnetic field produced at the magnetic dipole moment 5$\times$10$^{30}$ Am$^2$ is of the order of 10$^{16}$ G at the center, which is small enough to be indistinguishable from the zero magnetic case. For higher magnetic dipole moments, the magnetic field produced $\approx$ 4 $\times$ 10$^{17}$ G is strong enough to increase the matter pressure to higher values, thus producing a distinguishable effect.  

Fig.~\ref{fig6.6} shows the particle fractions as a function of baryon density for a beta stable neutron star matter obtained using the DD-MEX EoS. Fig.~\ref{fig6.6} panel \subref{fig6a} displays the fractions in the absence of magnetic field, while panels \subref{fig6b}, \subref{fig6c}, and \subref{fig6d} depict the particle fractions in the presence of magnetic fields with fixed magnetic dipole moments $\mu$ = 5$\times$10$^{30}$ Am$^2$, $\mu$ = 5$\times$10$^{31}$ Am$^2$, and $\mu$ = 5$\times$10$^{32}$ Am$^2$, respectively. Clearly, in all the cases, the $\Lambda$ particle is the dominant hyperonic component, which starts appearing in the density range 2 - 3 $\rho_0$ \cite{PhysRevC.81.035803,VIDANA2013367,rather2018role,doi:10.1142/S0218301318500970}. The neutral $\Xi$ hyperon appears at a density $\approx$ 5.5 $\rho_0$ for $B$ = 0, which remains unaltered with the inclusion of the magnetic field. As expected, the charged particles are more strongly affected by the magnetic field and an increase in their population is seen with the increase in the magnetic field-strength. For $B$=0 (and all other cases), the $e^-$ and $\mu^-$ population is large at low densities, which suppresses the appearance of $\Xi^-$ hyperons. With an increase in the magnetic dipole moment, the magnetic field  strength increases, which shifts the appearance of $\Xi^-$ hyperon from $\approx$ 2.5 to around $\approx$ 3.5 $\rho_0$. At this threshold, the density of the negatively charged leptons $e^-$ and $\mu^-$ starts to drop, as the charge neutrality condition from Eq.~(\ref{eq13}) allows the $\Xi^-$ hyperon to take over. Similarly, the appearance of the $\Lambda$ hyperon accelerates the disappearance of neutrons, as both are neutral particles. Overall, we see that the addition of a magnetic field increases the population of leptons (re-leptonizes) and correspondingly decreases the population of hyperons (de-hyperonizes), which renders the EoS stiffer \cite{Tolos_2016}.

Because of the repulsive nature of the $\Sigma$ potential, the formation of $\Sigma^0$ and $\Sigma^-$ is suppressed for the densities considered in the present work \cite{PhysRevC.64.025801,particles3040043}. The absence of $\Sigma$ hyperons is supported by the fact that no bound $\Sigma^-$ hypernuclei have been found yet, despite several searches \cite{HARADA2015312}. The inclusion of strong magnetic fields does not change this feature.

\begin{figure}[hbt!]
	\centering
	\includegraphics[width=0.75\textwidth]{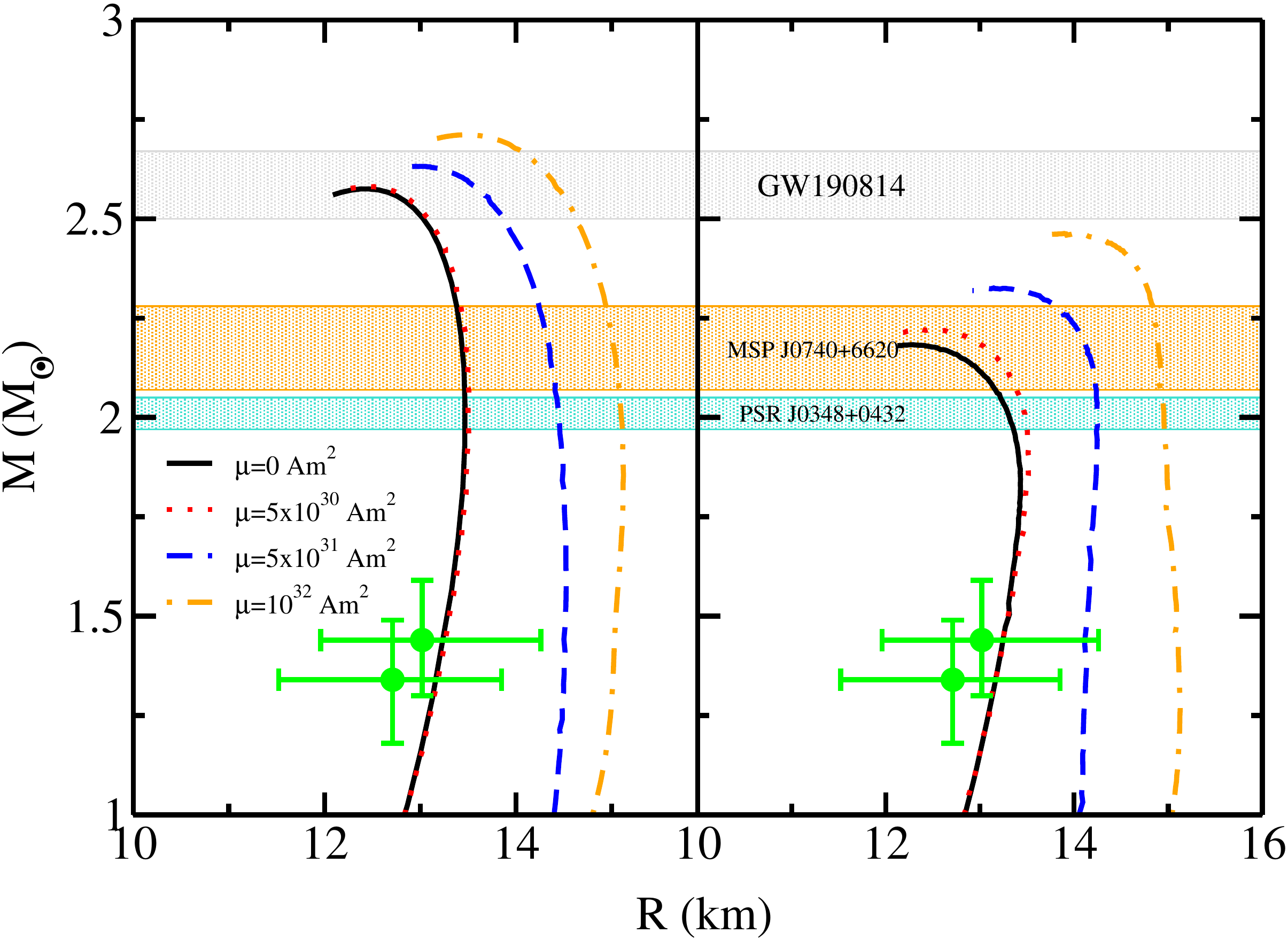}
	\caption{ Relation between mass and circumferential radius for an NS without magnetic field and with magnetic field effects considering different magnetic dipole moments without hyperons (left panel) and with hyperons (right panel) using the  DD-MEX parameter set. The colored areas show the recent constraints inferred from GW190814, MSP J0740+6620, and PSR J0348+0432 \cite{Abbott_2020a,Cromartie2020,Antoniadis1233232}. The constraints on the mass-radius limits inferred from  NICER  \cite{Miller_2019,Riley_2019} are also shown.}
	\label{fig6.7} 
\end{figure}
\begin{figure}[hbt!]
	\centering
	\includegraphics[width=0.75\textwidth]{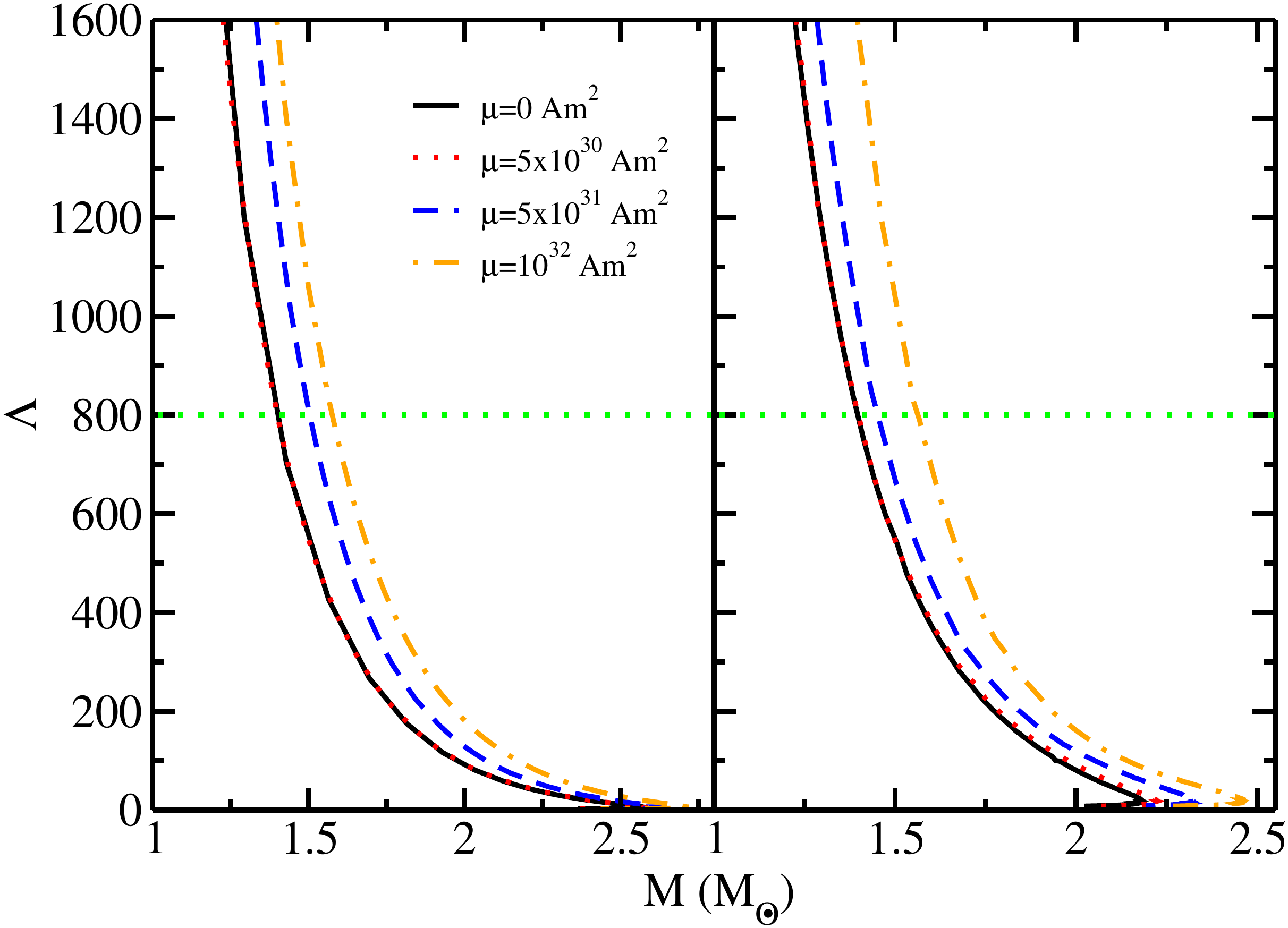}
	\caption{ Dimensionless tidal deformability ($\Lambda$) as a function of NS mass for EoS without magnetic field (solid line) and with magnetic field effects considering different magnetic dipole moments. The left panel shows the results obtained for NS without hyperons, while the right panel shows the results for NS with hyperons.  The orange dotted line represents the upper limit on the dimensionless tidal deformability set by measurement from GW170817 \cite{PhysRevLett.119.161101}.}
	\label{fig6.8} 
\end{figure}
The effect of magnetic field on the mass-radius relation of an NS with and without hyperons is displayed in Fig.~\ref{fig6.7}. These calculations are performed for DD-MEX EoS using the LORENE library \cite{LORENE}. Different values of magnetic dipole moment are used to obtain different values of the magnetic field at the stellar surface and the center. As can be seen, the NS maximum mass without hyperons increases from 2.575$M_{\odot}$ for $B$ = 0 to 2.711$M_{\odot}$ for $\mu$ = 10$^{32}$ Am$^2$. The corresponding radius changes from 12.465 to 13.474 km. The radius at 1.4$M_{\odot}$ increases by around 1 km. With the hyperons included, the mass increases from 2.183 to 2.463$M_{\odot}$ when magnetic field effects are included with $\mu$ = 10$^{32}$ Am$^2$. The variation obtained in the mass-radius is larger for hyperonic stars than for the pure nucleonic stars due to the additional effect of de-hyperonization that takes place due to the magnetic field. The de-hyperonization results in the enhancement of the matter pressure $P_m$ for a given $\mathcal{E}_m$. For higher magnetic fields produced at magnetic dipole moments $\mu$ = 5$\times$ 10$^{31}$ and $\mu$ = 10$^{32}$ Am$^2$, the effect of the magnetic field is seen to be very large at smaller stellar masses. For low magnetic fields, pure nucleonic stars still satisfy the possible maximum mass constraint from the GW190814 event, implying the possibility of its secondary component being a magnetar. The radius constraints inferred from NICER experiments are satisfied by both nucleonic as well as hyperonic stars with low magnetic fields. For central magnetic fields $\approx$ 7 $\times$ 10$^{16}$ G, the EoS obtained for pure nucleonic matter satisfies the radius constraints from NICER measurement. For hyperonic matter, a lower magnetic field $\approx$ 4 $\times$ 10$^{16}$ G produces an hyperon star with a radius that satisfies NICER constraints. This is due to the fact that for magnetic fields less than 10$^{17}$ G, the deformation produced in the stellar structure is negligible and hence the variation in radius is too small when compared to the non-magnetic case. 

Fig.~\ref{fig6.8} shows the variation in the dimensionless tidal deformability as a function of gravitational mass with magnetic field effects and considering different values of magnetic dipole moment. The results for EoS with and without hyperons are shown. For pure nucleonic stars, the tidal deformability increases to a value $\Lambda_{1.4} \approx$ 1500  for a central magnetic field of 4.55$\times$ 10$^{17}$ G produced fixing the magnetic dipole moment to $\mu$ = 10$^{32}$ Am$^2$, thus violating the constraint on the dimensionless tidal deformability from GW170817, which provides an upper limit of 800 on $\Lambda_{1.4}$ at 90\% confidence \cite{PhysRevLett.119.161101}. The properties of neutron stars and hyperon stars at different values of the magnetic dipole moment, which correspond to different magnetic field values at the surface and the center, are shown in Table \ref{tab6.4}. As clear, a small magnetic field produces an NS with tidal deformability larger than the upper limit from GW170817. This confirms that the BNS merger event GW170817 did not consist of magnetars. For hyperonic stars, the value at 1.4$M_{\odot}$ remains unchanged when considering a magnetic dipole moment of $\mu$ = 5$\times$ 10$^{30}$ Am$^2$. As the magnetic dipole moment increases, stronger magnetic fields increase the stellar radius, which allows the tidal deformability $\Lambda_{1.4}$ to reach a value around 1550. For an NS and an hyperon star with dimensionless tidal deformability well within the limit of GW170817 at 90\% confidence, a magnetic field with a maximum value of $\approx$ 2 $\times$ 10$^{16}$ G is required.
\begin{center}
	\begin{table}[hbt!]
		\centering
		\caption{Stellar properties: Maximum mass ($M_{max}$), corresponding radius ($R$),  canonical mass radius ($R_{1.4}$), and dimensionless tidal deformability ($\Lambda_{1.4}$) of pure nucleonic and hyperonic star for DD-MEX EoS without magnetic field and with magnetic field effects for different values of the magnetic dipole moment.  }
		\begin{tabular}{ p{1.8cm}|p{1.2cm}p{1.2cm}p{1.2cm}p{1.2cm}|
				p{1.2cm}p{1.2cm}p{1.2cm}p{1.2cm} }
			\hline
			\hline
			\multirow{3}{1.5cm}{$\mu$ (Am$^2$)}
			&&\multicolumn{3}{p{2.5cm}|}{Neutron Star} & %
			&\multicolumn{3}{p{2.5cm}}{Hyperon Star}\\
			\cline{2-9}
			&\parbox[t]{0.2cm}{\centering $M_{max}$ \\ ($M_{\odot}$) }  &\parbox[t]{0.2cm}{\centering $R$ \\ (km) }&\parbox[t]{0.2cm}{\centering $R_{1.4}$  \\ (km) }&$\Lambda_{1.4}$&\parbox[t]{0.2cm}{\centering $M_{max}$ \\ ($M_{\odot}$) }  &\parbox[t]{0.2cm}{\centering $R$ \\ (km) }&\parbox[t]{0.2cm}{\centering $R_{1.4}$  \\ (km) }&$\Lambda_{1.4}$ \\
			\hline
			0 &2.575&12.465&13.168&791.483&2.183&12.238&13.168&791.483\\
			5$\times$ 10$^{30}$  &2.580&12.536&13.235&802.801&2.224&12.506&13.168&791.483\\
			5$\times$ 10$^{31}$ &2.632&13.024&14.507&1175.35&2.325&13.269&14.112&998.882\\
			10$^{32}$ &2.711&13.474&15.027&1517.09&2.463&13.894&15.105&1559.194\\
			
			\hline
			\hline
		\end{tabular}
		\label{tab6.4}
	\end{table}
\end{center}
\begin{figure}[hbt!]
	\centering
	\includegraphics[width=0.75\textwidth]{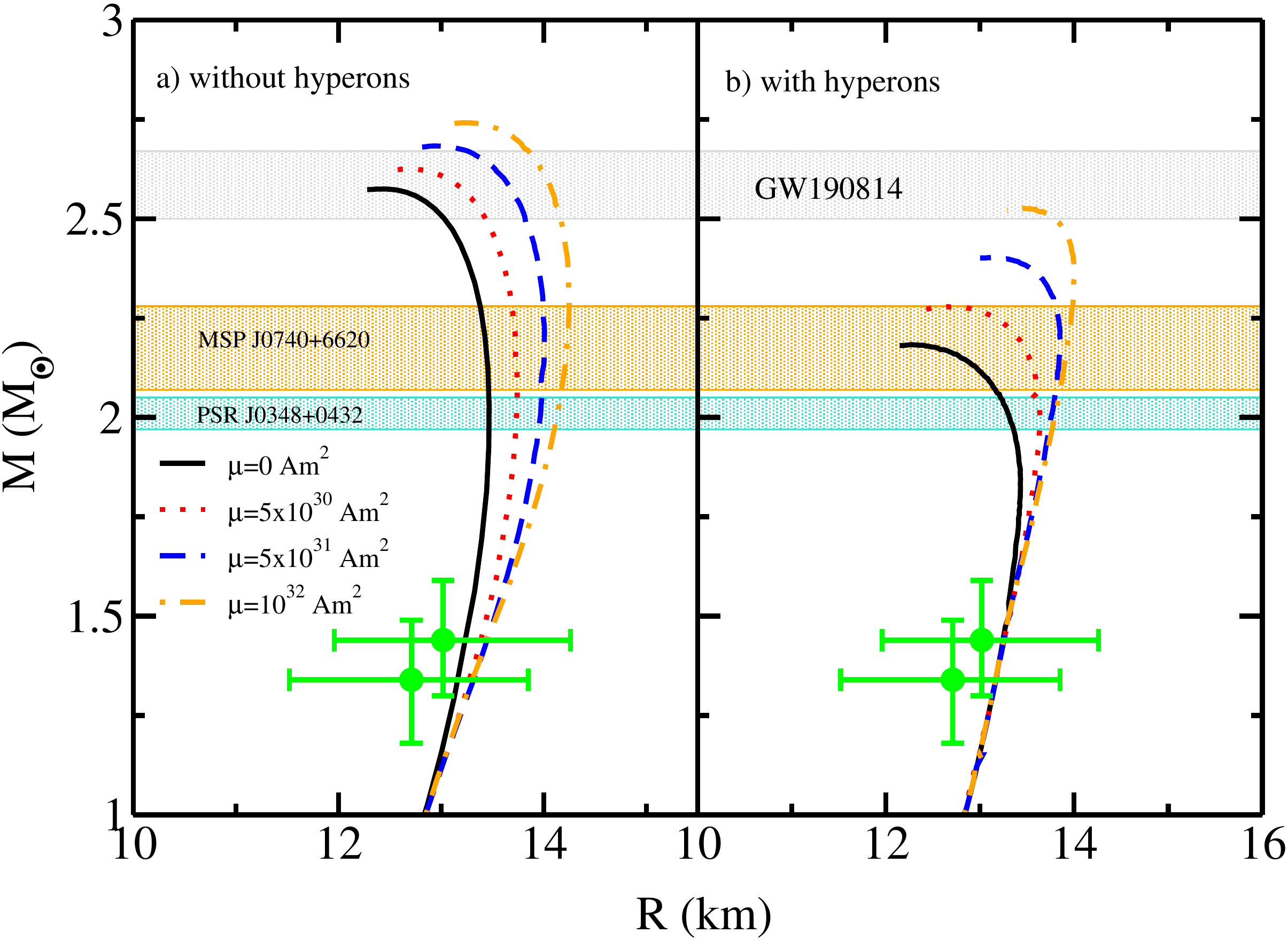}
	\caption{ Relation between mass and radius for an NS without magnetic field and with magnetic field effects considering different magnetic dipole moments without hyperons (left panel) and with hyperons (right panel) for DD-MEX parameter set by solving general relativity spherically symmetric solutions (TOV) for a static NS. The colored areas show the recent constraints inferred from GW190814, MSP J0740+6620, and PSR J0348+0432 \cite{Abbott_2020a,Cromartie2020,Antoniadis1233232}. The constraints on the mass-radius limits inferred from  NICER  \cite{Miller_2019,Riley_2019} are also shown.}
	\label{fig6.7a} 
\end{figure}
\vspace{-0.5cm}
Fig.~\ref{fig6.7a} displays the mass radius profile for DD-MEX parameter set without magnetic field and with magnetic field considering different magnetic dipole moments by solving general relativity spherically symmetric solutions (TOV) for a static NS. As the magnetic field increases by changing magnetic dipole moment, the maximum mass increases by about 0.1$M_{\odot}$ for NS and 0.2$M_{\odot}$ for hyperon star. The change in the radius at canonical mass is very small. Even with the strong magnetic field produced by magnetic dipole moment $\mu$ = 10$^{32}$ Am$^2$, the radius satisfies all the constraints. Thus, we see that neglecting the deformation effects by solving the spherically symmetric TOV equations, leads to an overestimation of the mass and an underestimation of the radius. This happens because the extra magnetic energy that would deform the star is being added to the mass due to the imposed spherical symmetry.  
\begin{figure}[hbt!]
	\centering
	\includegraphics[width=0.75\textwidth]{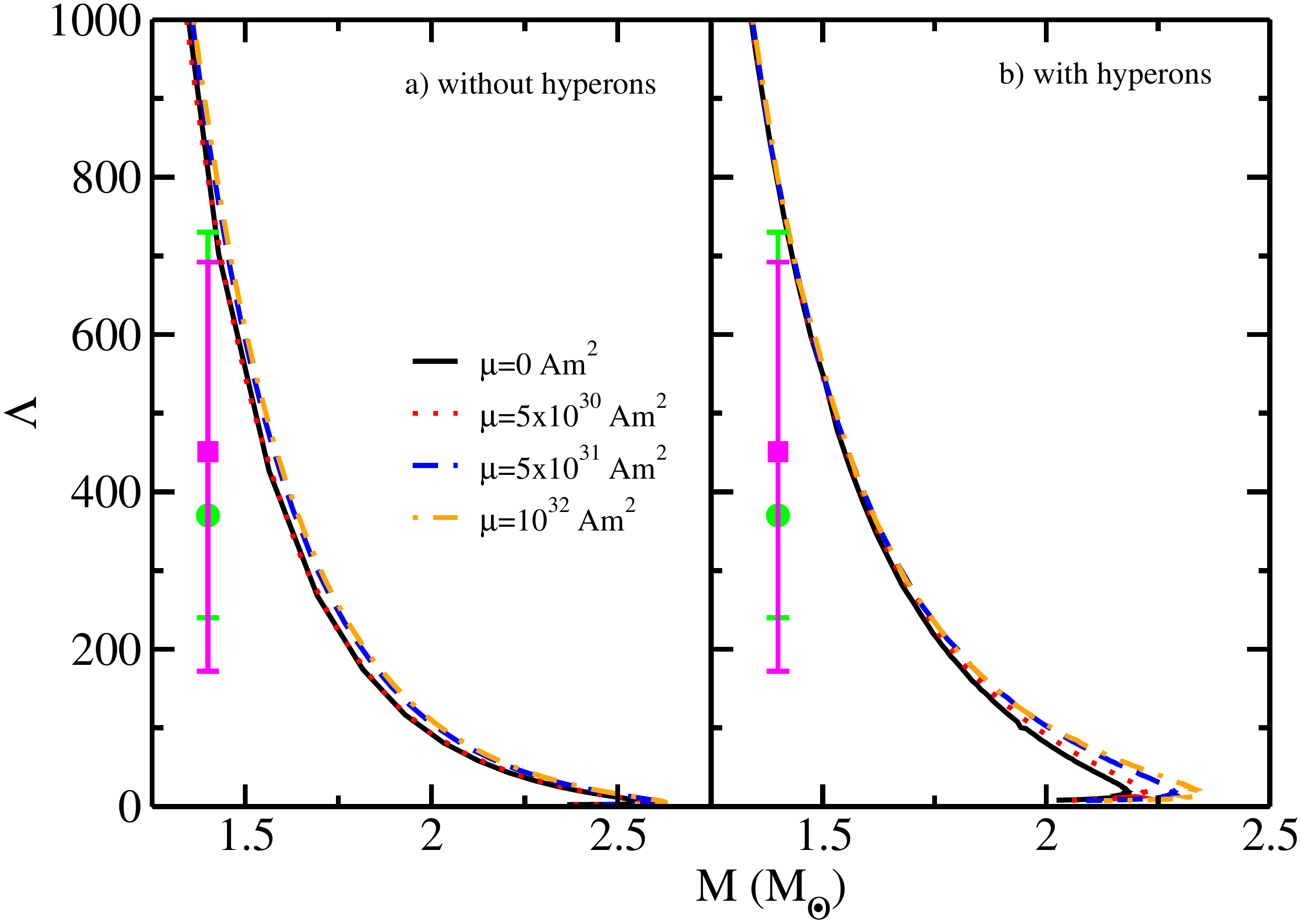}
	\caption{ Dimensionless tidal deformability ($\Lambda$) as a function of NS mass for EoS without magnetic field (solid line) and with magnetic field effects considering different magnetic dipole moments using results from relativity spherically symmetric solutions (TOV). The left panel shows the results obtained for NS without hyperons, while the right panel shows the results for NS with hyperons. The orange dotted line represents the upper limit on the dimensionless tidal deformability set by measurement from GW170817 \cite{PhysRevLett.119.161101}.}
	\label{fig6.8a} 
\end{figure}

For similar reasons, the dimensionless tidal deformability calculated by using the results from spherically symmetric TOV solutions is small as compared to the LORENE calculation (Fig.~\ref{fig6.8}). Fig.~\ref{fig6.8a} displays the tidal deformability variation with the NS mass for EoS without magnetic field (solid line) and with magnetic field effects considering different magnetic dipole moments using results from relativity spherically symmetric solutions (TOV). The results obtained from TOV equations are displayed in Table \ref{tab6.4a}.
\begin{center}
	\begin{table}[hbt!]
		\centering
		\caption{Stellar properties: Maximum mass ($M_{max}$), corresponding radius ($R$),  canonical mass radius ($R_{1.4}$), and dimensionless tidal deformability ($\Lambda_{1.4}$) of pure nucleonic and hyperonic star for DD-MEX EoS without magnetic field and with magnetic field effects for different values of the magnetic dipole moment by solving general relativity spherically symmetric solutions (TOV) for a static NS.   }
		\begin{tabular}{ p{1.8cm}|p{1.2cm}p{1.2cm}p{1.2cm}p{1.2cm}|
				p{1.2cm}p{1.2cm}p{1.2cm}p{1.2cm} }
			\hline
			\hline
			\multirow{3}{1.5cm}{$\mu$ (Am$^2$)}
			&&\multicolumn{3}{p{2.5cm}|}{Neutron Star} & %
			&\multicolumn{3}{p{2.5cm}}{Hyperon Star}\\
			\cline{2-9}
			&\parbox[t]{0.2cm}{\centering $M_{max}$ \\ ($M_{\odot}$) }  &\parbox[t]{0.2cm}{\centering $R$ \\ (km) }&\parbox[t]{0.2cm}{\centering $R_{1.4}$  \\ (km) }&$\Lambda_{1.4}$&\parbox[t]{0.2cm}{\centering $M_{max}$ \\ ($M_{\odot}$) }  &\parbox[t]{0.2cm}{\centering $R$ \\ (km) }&\parbox[t]{0.2cm}{\centering $R_{1.4}$  \\ (km) }&$\Lambda_{1.4}$ \\
			\hline
			0 &2.575&12.465&13.168&791.483&2.183&12.238&13.168&791.483\\
			5$\times$ 10$^{30}$  &2.625&12.694&13.357&844.82&2.279&12.465&13.172&799.21\\
			5$\times$ 10$^{31}$ &2.683&12.951&13.403&884.103&2.402&13.222&13.195&828.415\\
			10$^{32}$ &2.741&13.245&13.411&956.86&2.526&13.475&13.209&884.03\\
			
			\hline
			\hline
		\end{tabular}
		\label{tab6.4a}
	\end{table}
\end{center}
\subsection{Additional hyperon couplings}
\label{sub3}
\begin{figure}[hbt!]
	\centering
	\includegraphics[width=0.75\textwidth]{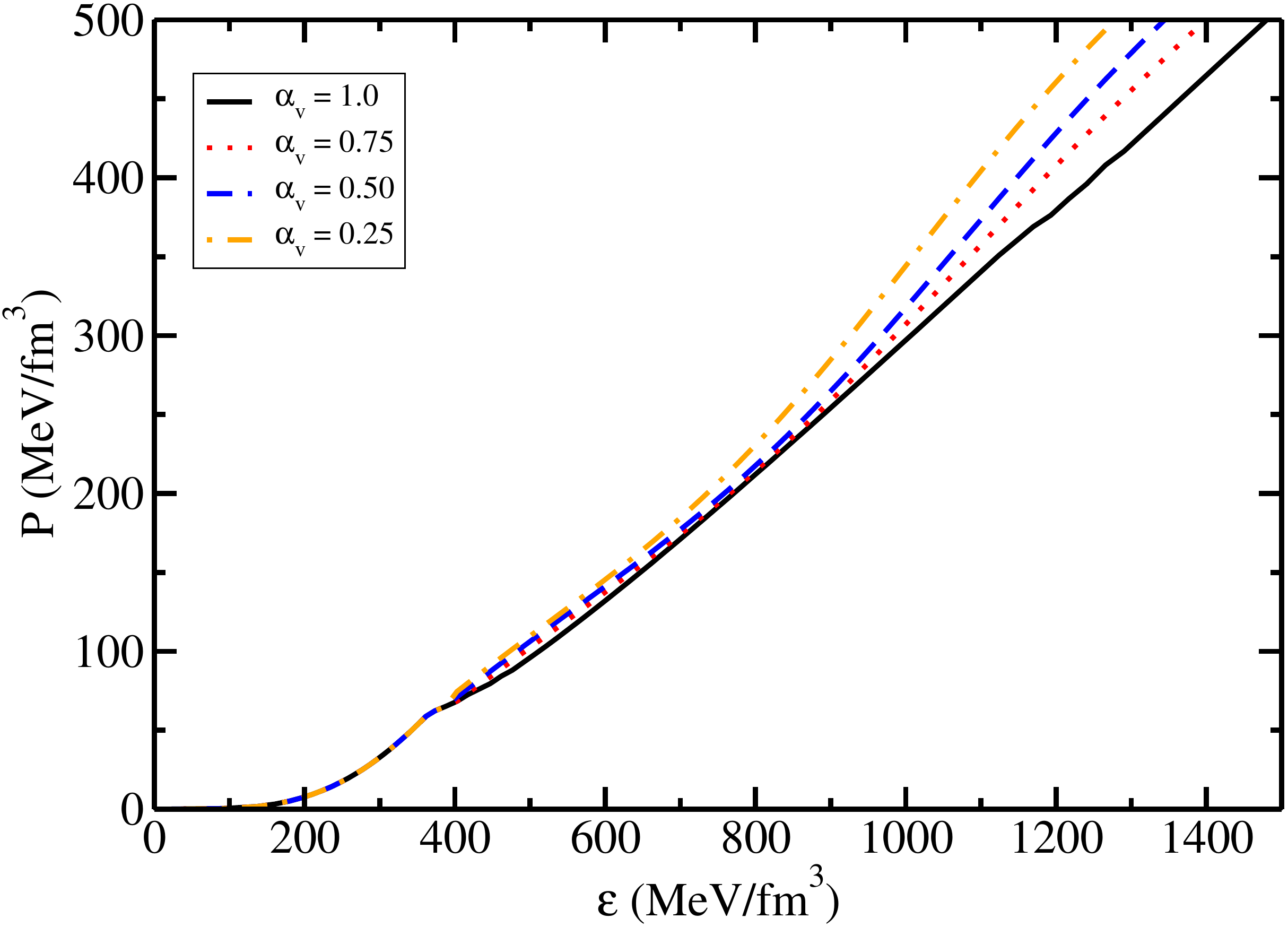}
	\caption{ EoS for DD-MEX parameter set at different values of the parameter $\alpha_v$.  }
	\label{fig6.9} 
\end{figure}

To investigate how different hyperon couplings and hyperon potentials affect the results we presented so far, we use a more general symmetry group SU(3) to determine the coupling constants of all baryons and mesons \cite{PhysRevC.89.025805,LOPES2021122171}. For the $\omega$ meson, we have 
\begin{equation}
	\frac{g_{\omega \Lambda}}{g_{\omega N}} = \frac{4+2\alpha_v}{5+4\alpha_v},
	\frac{g_{\omega \Sigma}}{g_{\omega N}} = \frac{8-2\alpha_v}{5+4\alpha_v},
	\frac{g_{\omega \Xi}}{g_{\omega N}} = \frac{5-2\alpha_v}{5+4\alpha_v}.
\end{equation}
For the $\rho$ meson
\begin{equation}
	\frac{g_{\rho \Sigma}}{g_{\rho N}}=2\alpha_v,
	\frac{g_{\rho \Xi}}{g_{\rho N}}=2\alpha_v-1,
	\frac{g_{\rho \Lambda}}{g_{\rho N}}=0.
\end{equation}
The hyperon-scalar meson coupling constants are fixed so as to reproduce the following optical potentials \cite{PhysRevLett.67.2414,DOVER1984171,SCHAFFNER199435}:
\begin{equation}
	\begin{gathered}
		U_{\Lambda}=-28 MeV,\\
		U_{\Sigma}=+30 MeV,\\
		U_{\Xi}=-18 MeV.
	\end{gathered}
\end{equation}
The coupling constants at different values of the parameter $\alpha_v$ are displayed in Table \ref{newcoup}. For $\alpha_v$ = 1.0, the hybrid SU(6) group is recovered. 
\begin{table}
	\centering
	\caption{Hyperon meson coupling constants for different values of the hyperon coupling parameter $\alpha_v$. }
	\begin{tabular}{  p{1.8cm}|p{1.8cm}p{1.8cm}p{1.8cm}p{1.8cm} }
		\hline
		\hline
		&$\alpha_v$ = 1.0&$\alpha_v$ = 0.75&$\alpha_v$ = 0.50&$\alpha_v$ = 0.25 \\
		\hline
		$g_{\omega \Lambda}/g_{\omega N}$&0.667&0.687&0.714&0.75\\
		$g_{\omega \Sigma}/g_{\omega N}$&0.667&0.812&1.00&1.25\\
		$g_{\omega \Xi}/g_{\omega N}$&0.333&0.437&0.571&0.75\\
		$g_{\rho \Sigma}/g_{\rho N}$&2.0&1.5&1.0&0.5\\
		$g_{\rho \Xi}/g_{\rho N}$&1.0&0.5&0.0&-0.5\\
		$g_{\sigma \Lambda}/g_{\sigma N}$&0.610&0.626&0.653&0.729\\
		$g_{\sigma \Sigma}/g_{\sigma N}$&0.403&0.514&0.658&0.850\\			
		$g_{\sigma \Xi}/g_{\sigma N}$&0.318&0.398&0.500&0.638\\						
		\hline
		\hline
	\end{tabular}
	\label{newcoup}
\end{table}
The particle population of hyperons and other particles depend upon the hyperon meson coupling constants, which vary with the parameter $\alpha_v$. Changing the value of $\alpha_v$ from 1.0 to 0.25 suppresses strange particles. The suppression of $\Xi^-$  increases the lepton fraction at lower value of $\alpha_v$. The neutrons and protons are the most populated particles at $\alpha_v$ = 0.25. 
Fig.~\ref{fig6.9} displays the EoS obtained using different values of the parameter $\alpha_v$. As the $\alpha_v$ value decreases from 1.0 to 0.25, the stiffness of the EoS increases due to the increase in the value of $\omega$ meson couplings.

\begin{figure}[!hbtp]
	\centering
	\includegraphics[width=0.75\textwidth]{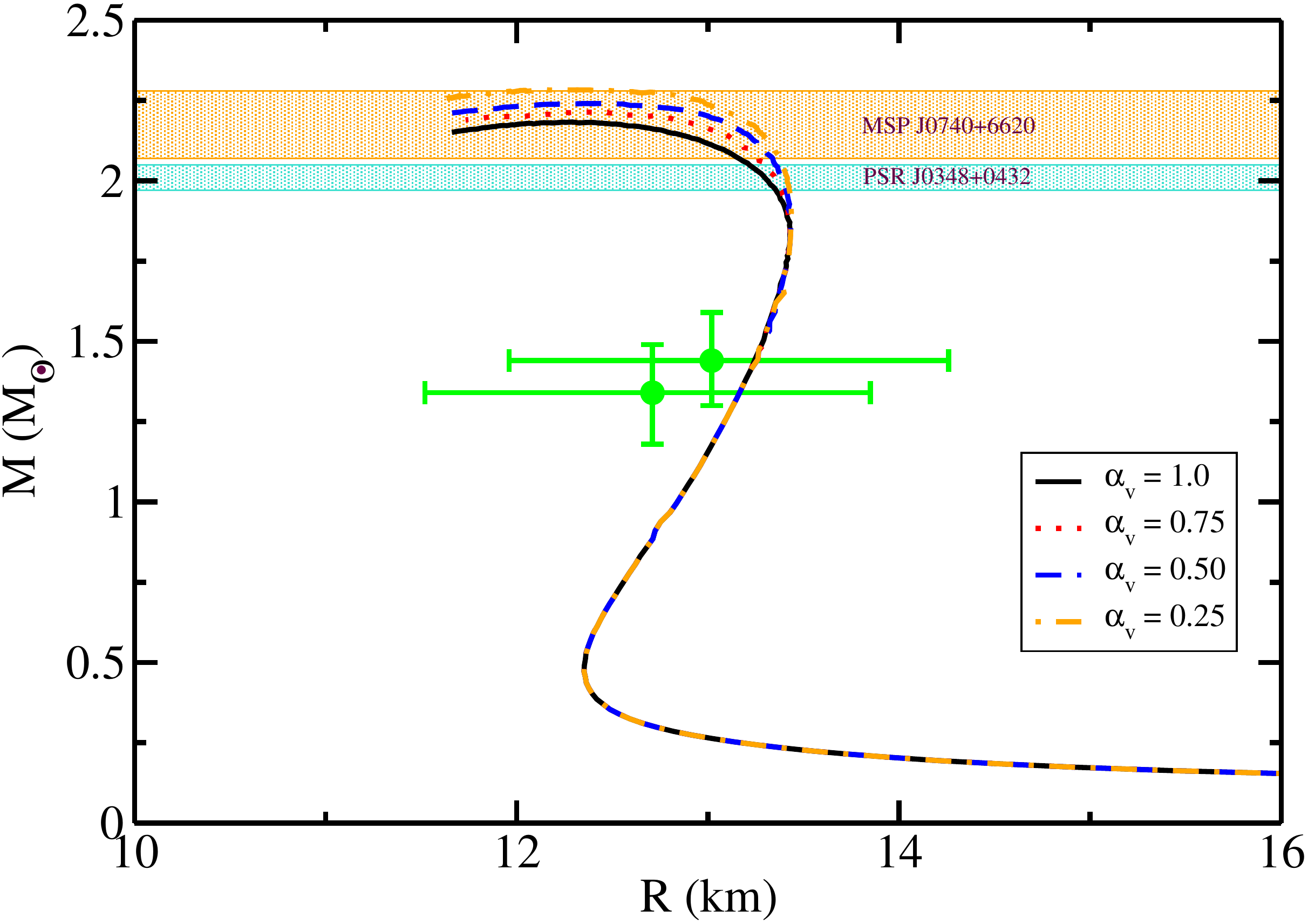}
	\caption{ Mass-radius for an NS with hyperons at different values of the parameter $\alpha_v$ using the  DD-MEX parameter set. The shaded regions show the recent constraints inferred from GW190814, MSP J0740+6620, and PSR J0348+0432 \cite{Cromartie2020,Antoniadis1233232}. The green overlaid arrows show constraints on the mass-radius limits inferred from  NICER  \cite{Miller_2019,Riley_2019}.}
	\label{fig6.10} 
\end{figure}

\begin{figure}[hbt!]
	\centering
	\includegraphics[width=0.75\textwidth]{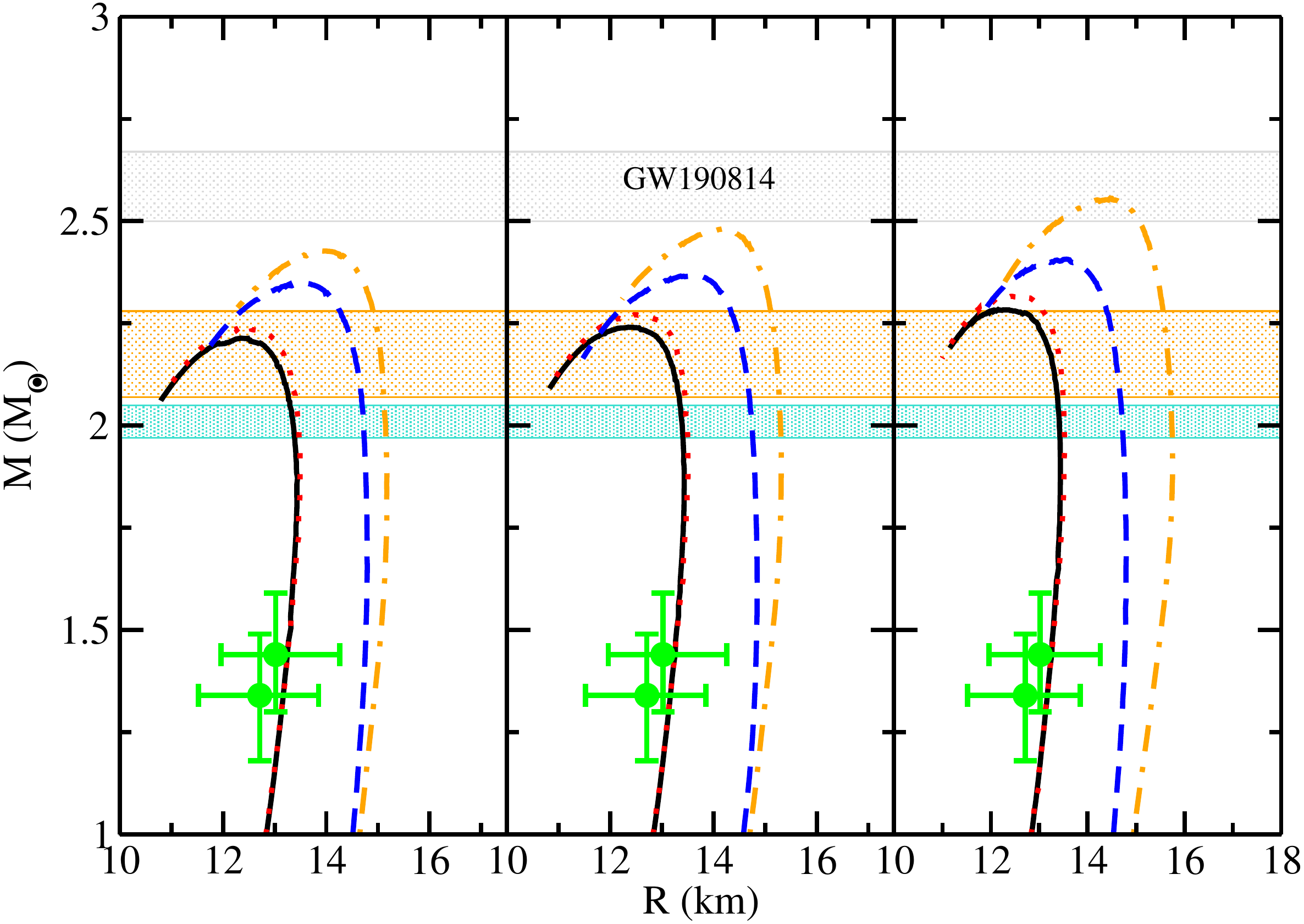}
	\caption{ Relation between mass and circumferential radius for a hyperonic star with magnetic field effects considering different magnetic dipole moments at different hyperon couplings  $\alpha_v$=0.75 (left panel),  $\alpha_v$=0.50 (middle panel) and  $\alpha_v$=0.25 (right panel) using the  DD-MEX parameter set. The colored areas show the recent constraints inferred from GW190814, MSP J0740+6620, and PSR J0348+0432 \cite{Abbott_2020a,Cromartie2020,Antoniadis1233232}. The constraints on the mass-radius limits inferred from  NICER  \cite{Miller_2019,Riley_2019} are also shown.}
	\label{fig6.11} 
\end{figure}

Fig.~\ref{fig6.10} displays the mass-radius relation for hyperon stars without magnetic field effects at different values of $\alpha_v$. Lowering the value of $\alpha_v$  stiffens the EoS, which increases the maximum mass to 2.283$M_{\odot}$ for $\alpha_v$ = 0.25 (from 2.183  for $\alpha_v$ = 1.0).  It is to be mentioned that the stellar properties of hyperon stars obtained at $\alpha_v$ = 1.0  almost resemble that obtained from the hyperon potentials given by Eq.~(\ref{pot1}). The change in the hyperon potentials alters the values of sigma meson couplings by a small fraction and, hence, the changes obtained in the particle population and stellar properties are negligible.
With the addition of magnetic field effects, the results of different hyperon couplings on hyperonic EoSs is determined. Fig.~\ref{fig6.11} displays the relation between mass and radius for a hyperon star at different $\alpha_v$ values without magnetic field and with magnetic field effects considered at different values of magnetic dipole moment. For $\alpha_v$ = 0.75, the maximum mass increases to a value 2.437$M_{\odot}$ for a magnetic dipole moment $\mu$ = 10$^{32}$ Am$^2$, which corresponds to a central magnetic field of 3.77 $\times$ 10$^{17}$ G. Similarly, for $\alpha_v$ = 0.50 and 0.25, the maximum mass reaches a value 2.480 and 2.556$M_{\odot}$, respectively, at $\mu$ = 10$^{32}$ Am$^2$. This implies that the secondary component of GW190814 could be a hyperonic magnetar. For the present, hyperon couplings with a magnetic dipole moment $\mu$ = 5 $\times$ 10$^{31}$ Am$^2$, which corresponds to a central magnetic field greater than 10$^{17}$ G, the deviation in the hyperon star radius at canonical mass is very large, around 1.5 km, as compared to that obtained from previous couplings. But with even stronger magnetic field, the deviation obtained is less in the present case. This implies that a change in hyperon couplings affects the stellar properties, especially the radius at canonical mass at a strong central magnetic field.
\begin{figure}[hbt!]
	\centering
	\includegraphics[width=0.75\textwidth]{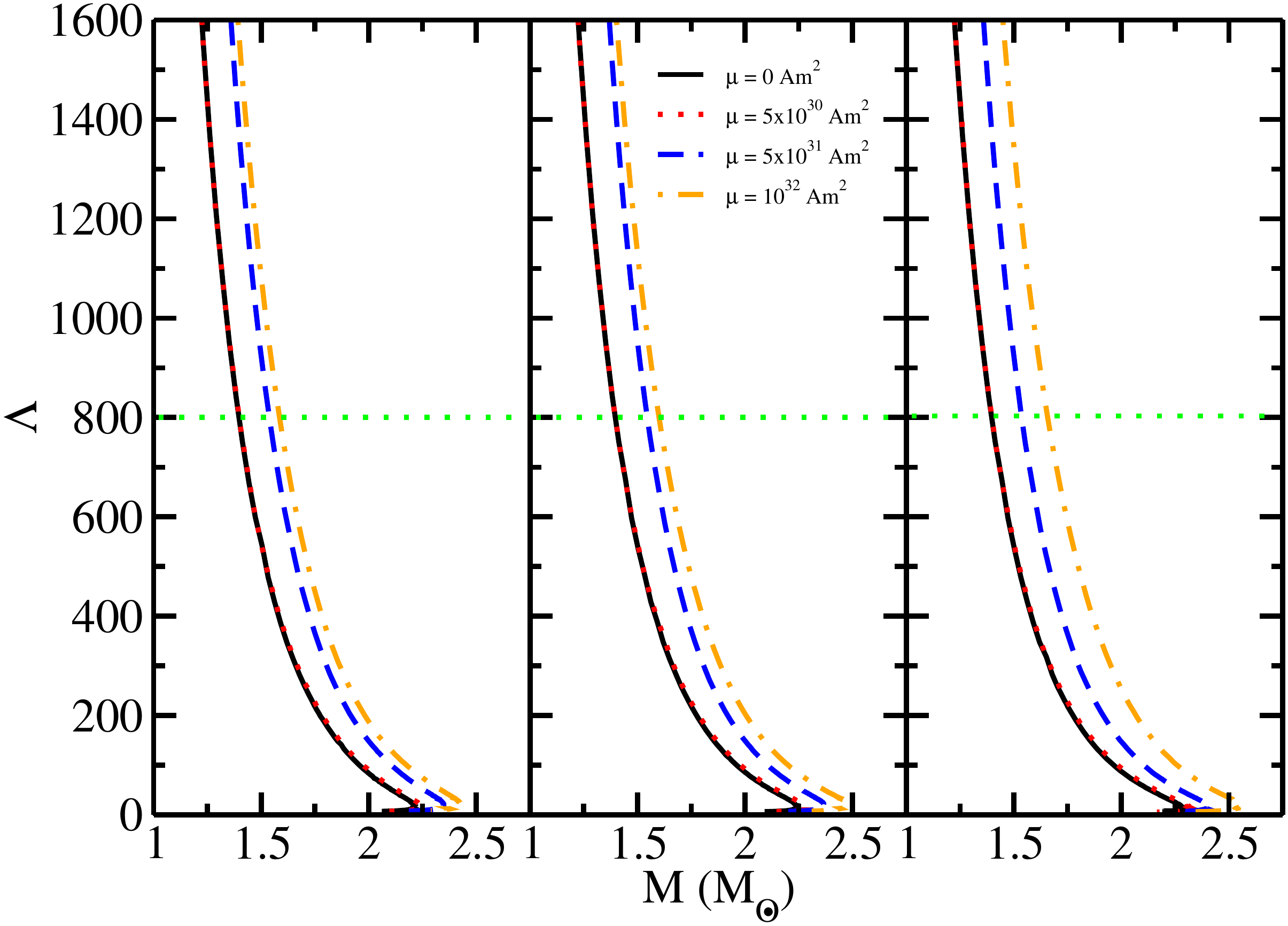}
	\caption{ Dimensionless tidal deformability  as a function of NS mass for EoS without magnetic field (solid line) and with magnetic field effects considering different magnetic dipole moments. Different panels show the results obtained for hyperonic EoSs at different values of the parameter $\alpha_v$. The orange dotted line represents the upper limit on the dimensionless tidal deformabilty set by measurement from GW170817 \cite{PhysRevLett.119.161101}.}
	\label{fig6.12} 
\end{figure}
\begin{figure}[hbt!]
	\centering
	\includegraphics[width=0.75\textwidth]{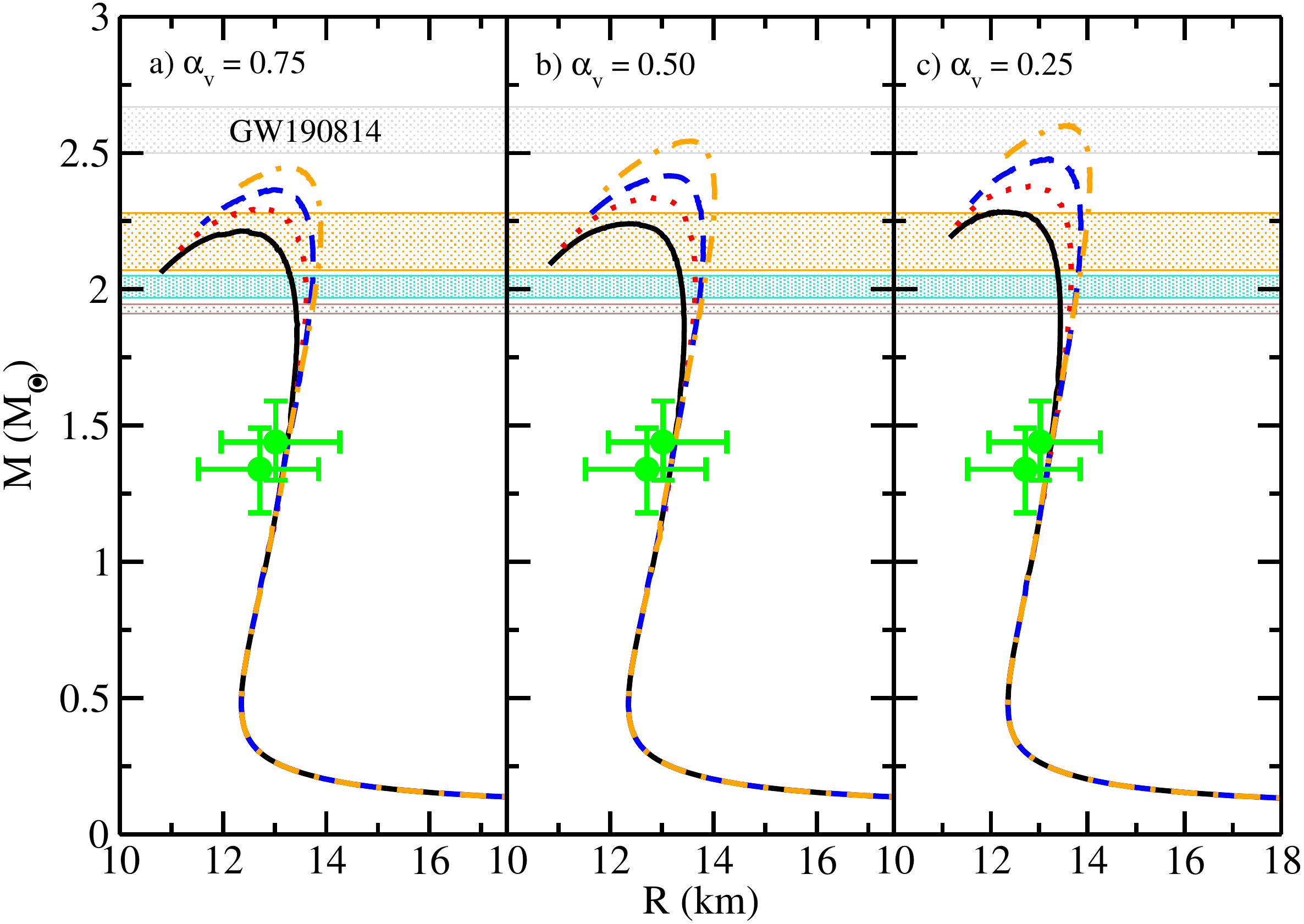}
	\caption{ Same as Fig.~\ref{fig6.11}, but using solutions from spherically symmetric TOV equations.}
	\label{fig6.11a} 
\end{figure}
\begin{figure}[hbt!]
	\centering
	\includegraphics[width=0.75\textwidth]{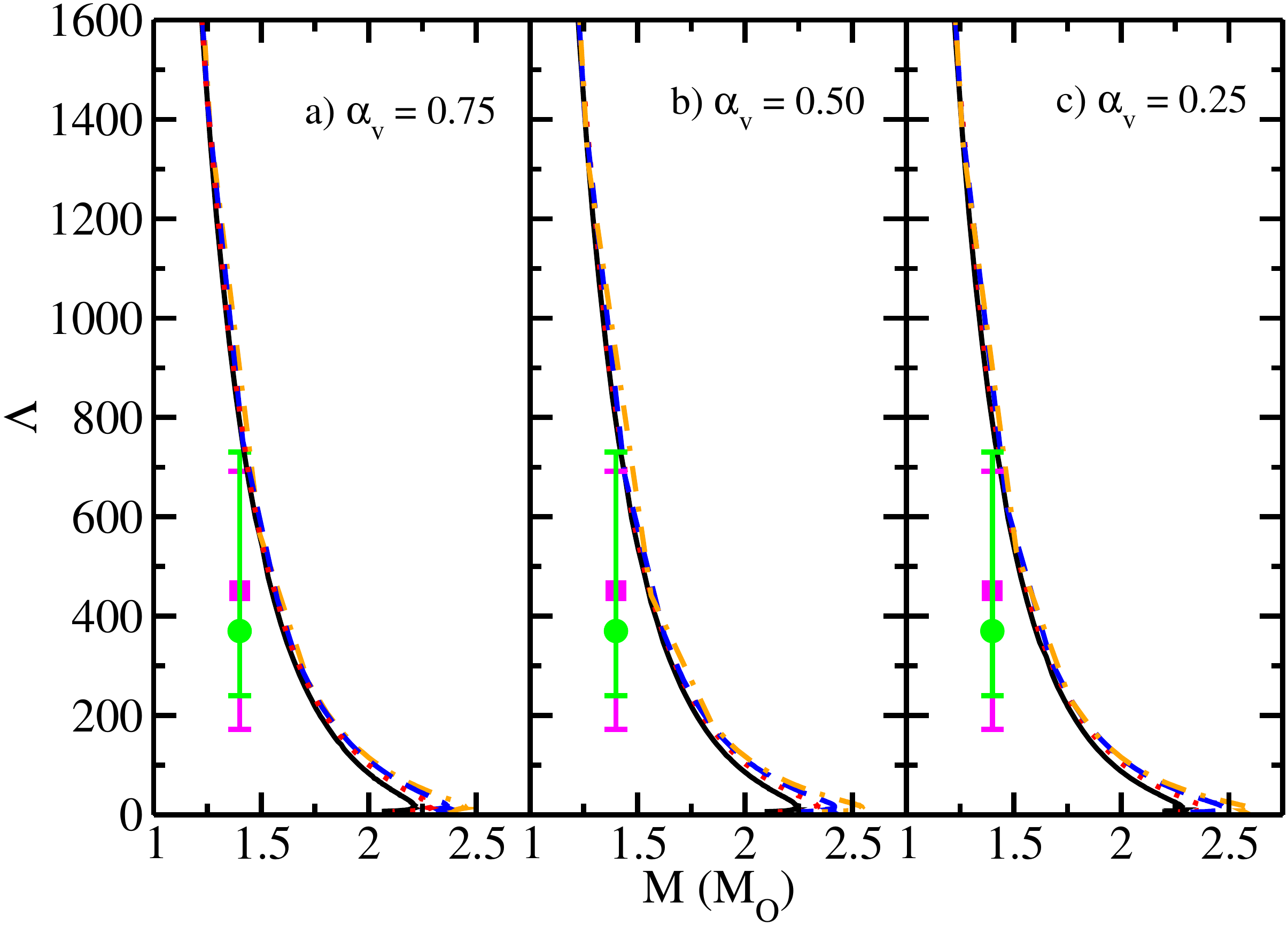}
	\caption{ Same as Fig.~\ref{fig6.12}, but using solutions from spherically symmetric TOV equations.}
	\label{fig6.12a} 
\end{figure}

Fig.~\ref{fig6.12} displays the variation in the dimensionless tidal deformability of a hyperonic star with magnetic field effects considered at different values of magnetic dipole moment. For a small value of the magnetic dipole moment at different values of $\alpha_v$, the $\Lambda$-M curve follows a similar pattern as for the previous hyperon couplings. Since the variation in radius at canonical mass is large for the present couplings, the tidal deformability, $\Lambda_{1.4}$ obtained is also large. For $\alpha_v$ = 0.25, the dimensionless tidal deformability reaches a value of $\approx$ 1900 for magnetic dipole moment 10$^{32}$ Am$^2$. The properties of hyperon star using different hyperon coupling values with and without magnetic field effects considered at different magnetic dipole moments are shown in Table \ref{tab6.5}.

Figs.~\ref{fig6.11a} and \ref{fig6.12a} display the mass-radius and tidal deformability for hyperonic stars at different values of the parameter $\alpha_v$ with and without magnetic field effects considering different magnetic dipole moments, using solutions from spherically symmetric TOV equations. Similar to Figs.~\ref{fig6.7a} and \ref{fig6.8a}, the standard TOV solutions lead to an overestimation of the mass and underestimation of the radius. Table \ref{tab6.6} displays the star matter properties obtained for hyperonic stars at different values of the parameter $\alpha_v$ using TOV solutions.
\begin{sidewaystable}[!htbp]

		\caption{Stellar properties: Maximum mass ($M_{max}$), corresponding radius ($R$),  canonical mass radius ($R_{1.4}$), and dimensionless tidal deformability ($\Lambda_{1.4}$) of hyperonic star for DD-MEX EoS at different values of the parameter $\alpha_v$ without magnetic field and with magnetic field with different values of the magnetic dipole moment.  }
		\begin{tabular}{ p{1.5cm}|p{1.4cm}p{1.5cm}p{1.5cm}p{1.5cm}|
				p{1.4cm}p{1.5cm}p{1.5cm}p{1.5cm}|				
		p{1.4cm}p{1.5cm}p{1.5cm}p{1.5cm} }
			\hline
			\hline
			\multirow{3}{0.2cm}{$\mu$ (Am$^2$)}&&\multicolumn{3}{p{2.5cm}|}{$\alpha_v$ = 0.75} & %
			&\multicolumn{3}{p{2.5cm}}{$\alpha_v$ = 0.50}
			&\multicolumn{3}{p{2.5cm}}{$\alpha_v$ = 0.25}\\
			\cline{2-13}
			&\parbox[t]{0.2cm}{\centering $M_{max}$ \\ ($M_{\odot}$) }  &\parbox[t]{0.2cm}{\centering $R$ \\ (km) }&\parbox[t]{0.2cm}{\centering $R_{1.4}$  \\ (km) }&$\Lambda_{1.4}$&\parbox[t]{0.2cm}{\centering $M_{max}$ \\ ($M_{\odot}$) }  &\parbox[t]{0.2cm}{\centering $R$ \\ (km) }&\parbox[t]{0.2cm}{\centering $R_{1.4}$  \\ (km) }&$\Lambda_{1.4}$&\parbox[t]{0.2cm}{\centering $M_{max}$ \\ ($M_{\odot}$) }  &\parbox[t]{0.2cm}{\centering $R$ \\ (km) }&\parbox[t]{0.2cm}{\centering $R_{1.4}$  \\ (km) }&$\Lambda_{1.4}$\\
			\hline
			
			0  &2.217&12.335&13.168&791.483&2.239&12.335&13.168&791.483&2.283&12.453&13.168&791.483\\
			5$\times$ 10$^{30}$ 
			&2.240&12.543&13.213&815.398&2.271&12.632&13.224&869.281&2.316&12.652&13.237&914.012\\
			5$\times$ 10$^{31}$  &2.349&13.522&14.741&1338.714&2.366&13.708&14.821&1397.389&2.407&13.563&14.938&1425.185\\
			10$^{32}$  &2.437&13.995&14.992&1510.783&2.480&14.224&15.069&1615.288&2.556&14.469&15.403&1879.880\\
			
			\hline
			\hline
		\end{tabular}
		\label{tab6.5}
\end{sidewaystable}
\begin{sidewaystable}[!htbp]
	
	\caption{Stellar properties: Maximum mass ($M_{max}$), corresponding radius ($R$),  canonical mass radius ($R_{1.4}$), and dimensionless tidal deformability ($\Lambda_{1.4}$) of hyperonic star for DD-MEX EoS at different values of the parameter $\alpha_v$ without magnetic field and with magnetic field with different values of the magnetic dipole moment by solving general relativity spherically symmetric solutions (TOV).  }
	\begin{tabular}{ p{1.5cm}|p{1.4cm}p{1.5cm}p{1.5cm}p{1.5cm}|
			p{1.4cm}p{1.5cm}p{1.5cm}p{1.5cm}|				
			p{1.4cm}p{1.5cm}p{1.5cm}p{1.5cm} }
		\hline
		\hline
		\multirow{3}{0.2cm}{$\mu$ (Am$^2$)}&&\multicolumn{3}{p{2.5cm}|}{$\alpha_v$ = 0.75} & %
		&\multicolumn{3}{p{2.5cm}}{$\alpha_v$ = 0.50}
		&\multicolumn{3}{p{2.5cm}}{$\alpha_v$ = 0.25}\\
		\cline{2-13}
		&\parbox[t]{0.2cm}{\centering $M_{max}$ \\ ($M_{\odot}$) }  &\parbox[t]{0.2cm}{\centering $R$ \\ (km) }&\parbox[t]{0.2cm}{\centering $R_{1.4}$  \\ (km) }&$\Lambda_{1.4}$&\parbox[t]{0.2cm}{\centering $M_{max}$ \\ ($M_{\odot}$) }  &\parbox[t]{0.2cm}{\centering $R$ \\ (km) }&\parbox[t]{0.2cm}{\centering $R_{1.4}$  \\ (km) }&$\Lambda_{1.4}$&\parbox[t]{0.2cm}{\centering $M_{max}$ \\ ($M_{\odot}$) }  &\parbox[t]{0.2cm}{\centering $R$ \\ (km) }&\parbox[t]{0.2cm}{\centering $R_{1.4}$  \\ (km) }&$\Lambda_{1.4}$\\
		\hline
		
		0  &2.217&12.335&13.168&791.483&2.239&12.335&13.168&791.483&2.283&12.453&13.168&791.483\\
		5$\times$ 10$^{30}$ 
		&2.293&12.873&13.208&811.905&2.345&12.952&13.212&814.006&2.377&13.001&13.217&825.210\\
		5$\times$ 10$^{31}$  &2.361&13.067&13.217&816.106&2.414&13.251&13.220&847.619&2.482&13.268&13.234&875.63\\
		10$^{32}$  &2.445&13.206&13.234&904.342&2.545&13.639&13.237&912.05&2.603&13.663&13.245&926.045\\
		
		\hline
		\hline
	\end{tabular}
	\label{tab6.6}
\end{sidewaystable}

\section{Conclusion}
\label{summary6}
We model massive nucleonic and hyperonic stars that fulfill the constraints set by the observation of the possibly most massive neutron star (NS) ever detected (in the secondary object of the gravitational wave event GW190814) using a density-dependent relativistic mean-field model (DD-RMF). This simple, yet powerful formalism provides the freedom necessary to fulfill simultaneously nuclear and astrophysical constraints.\par 
The results obtained from the TOV equations show that the hyperons soften the EoS, lowering the maximum mass of NS to around 2.2$M_{\odot}$, thereby satisfying more conservative massive constraints from the astrophysical observations. The radius of the NS canonical mass is seen to be insensitive to the appearance of hyperons. Both the nucleonic and the hyperonic stars satisfy the constraints from mass-radius limits inferred from NICER observations and tidal deformability constraints from the LIGO and VIRGO collaborations. \par 
We also study the effects of strong magnetic fields on DD-RMF nucleonic and hypeornic EoSs. We investigated the EoS and particle populations using a realistic chemical potential-dependent magnetic field that was developed by solving Einstein-Maxwell equations. For very strong magnetic fields, spherically symmetric solutions obtained by solving the TOV equations lead to an overestimation of the mass and the radius and, hence, cannot be used for determining stellar properties. For this reason, we used the LORENE library to determine the stellar properties of magnetic NSs. For low values of the magnetic dipole moment, implying lower strengths of magnetic fields, the EoS resembles the non-magnetic one. For higher magnetic dipole moments, the EoS stiffens at higher energy density. The amount of stiffness is larger in the case of hyperonic EoSs than in the case of pure nucleonic ones. As the magnetic field strength is increased, the particle fractions of leptons ($e^-$ and $\mu^-$) increase at higher densities and the appearance of charged hyperons ($\Xi^-$) is delayed. \par 
The stiffening in the EoS caused by the changes in population described above increases the maximum mass of magnetic stars. For a small dipole moment of 5$\times$ 10$^{30}$ Am$^2$, the nucleonic mass-radius profiles resemble the non-magnetic case due to the lower magnetic field produced. For higher magnetic dipole moments, although the variation in the maximum mass is small, still satisfying the GW190814 mass constraint, a variation of about 1 km is seen in the radius of the NS canonical mass. For hyperonic stars, the maximum mass increases by $\approx$ 0.3$M_{\odot}$. 

The change in the dimensionless tidal deformability for NS masses is also studied. It is seen that, for a higher value of the magnetic dipole moment, the tidal deformability of the canonical mass surpasses the upper limit of 800 set by the measurement from GW1701817 at 90\% confidence, which is consistent with the acknowledgment of such objects possessing low values of magnetic fields. No such measurement of tidal deformability is available for GW190814 as it has no tidal signatures. But if in the future, measurements of tidal deformability of massive neutron stars such as the one measured for the secondary object in GW190814 are $\le$ 800, it would imply that object could not be a hyperonic magnetar.\par
In particular, when different coupling schemes are considered, the maximum mass reproduced satisfies the GW190814 mass limit implying that its secondary component can be a possible hyperonic magnetar. It is seen that for a central magnetic field approaching 10$^{17}$ G, the radius at canonical mass increases by about 1.5 km as compared to the previous couplings where the radius changes by around 1 km. For an even stronger magnetic field, the different coupling scheme for hyperons increases the radius by 0.2 km in comparison to the previous one, in which case it increases by 1 km. Thus, we see that different hyperon couplings and different hyperon potentials populate the star matter differently and, hence, change significantly the stellar properties.\par 
%
%
%

\begin{savequote}[8cm]
\textlatin{Let the mind be enlarged ... to the grandeur of the mysteries and not the mysteries contracted to the narrowness of the mind.}

  \qauthor{---\textit{Francis Bacon}}
\end{savequote}

\chapter{\label{ch:7-conclusion}Summary and Conclusions}

With the nuclear many-body issue being addressed primarily through phenomenological models, it is critical to test/validate the models and underlying assumptions over a wide range of phenomena involving nuclear interactions. Several contemporary nuclear models seek for this broader application by providing a consistent account of finite nuclei throughout the nuclear chart, infinite matter and neutron stars. The relativistic mean-field (RMF) models modified to incorporate higher-order interactions are among the most successful in this area. The primary goal of this thesis work is to investigate how these expanded versions of RMF models, which are effective in the limited nuclear domain, may explain neutron star features (NS). The repercussions of accepting the presence of exotic matter, like quarks and hyperons, at the core of NS are heavily emphasized. 

The two methods of extending RMF models by introducing: (i) higher-order couplings and (ii) density-dependent couplings have proven useful in explaining dense infinite matter and NS, with the same parameters that have been successfully tested by reproducing several properties of finite nuclei throughout the nuclear chart. The theoretical basis for the various models under consideration is provided methodically with as much information as feasible. The field equations, pressure and energy density expressions, and field equations were deduced for all of the models included in this thesis work. The computer codes produced in this thesis study were used to perform all calculations.

This thesis touched on various ongoing research areas, including nuclear matter, neutron stars, and gravitational waves. We used the effective field theory motivated RMF (E-RMF) and density-dependent RMF (DD-RMF) formalism for studying the hadron matter, which has been very successful in the past, to analyze finite nuclei throughout the nuclear chart, infinite nuclear matter and neutron stars. Exotic phases such as quarks have been explored using the MIT Bag model and its extensions such as the vBag model at various bag constants.

After providing a detailed description of the theoretical models (both non-relativistic and relativistic) used in the study of nuclear matter and neutron stars in Chapter \ref{cha-lit}, a brief introduction to several contemporary occurrences for infinite nuclear matter are described in Chapter \ref{ch:1-intro}. Chapter \ref{ch:2-litreview} focuses on the mathematical derivations that have been included throughout our work. We started with the extended RMF Lagrangian density with $\delta$ meson and $\sigma$-$\omega$ cross-coupling, which contain large number of terms with all sorts of self-and cross-coupling. The density-dependent coupling RMF (DD-RMF) model is also used to study the hadron matter that allows a consistent calculation of NSs and yield results that are comparable with other models. It incorporates the properties of the Dirac-Brueckner model using microscopic interactions at various densities as input. The extrapolation to higher densities is more constrained than for the phenomenological RMF calculations that use only information from the limited density range of finite nuclei for the determination of their parameters. The RMF and DD-RMF parameter sets along with their nuclear matter properties used in this thesis work are also discussed. In addition, using the energy-momentum tensor, the equation of state is calculated and numerous formulas for the properties of symmetric nuclear matter, pure neutron matter, and infinite nuclear matter are produced in the RMF approximation.

In Chapter \ref{ch:3-hybridtstar}, we investigate the hybrid EoS created by combining hadron and quark matter under Gibbs circumstances. The E-RMF model for hadron matter with recently reported parameter sets is used, as is the MIT bag model for quark matter with varied bag constant. For hybrid EoS, nuclear matter properties such as symmetry energy ($J$), slope parameter ($L$), and incompressibility ($K$) are determined. It is discovered that the values of symmetry energy $J$ and other variables are quite high for a hybrid EoS and that they increase with the bag constant, while the incompressibility decreases with $B$.
For various bag constants, star matter properties such as mass and radius are determined. It is observed that a bag constant value in the range $B^{1/4}$ = 130-160 MeV is acceptable for explaining quark matter in neutron stars. The results obtained with bag values less than 130 MeV and larger than 160 MeV disagree with the recently measured gravitational wave observables.

In Chapter \ref{ch:4-innercrust}, the RMF model is used to examine the NS properties such as mass, radius, tidal deformability. The inner crust EoSs with varied symmetry energy slope parameters were utilized to investigate the effect of symmetry energy and its slope parameter on an NS. The BPS EoS is used for all sets for the outer crust because the outer crust component does not affect the NS maximum mass and radius. For the inner crust part, several parameter sets such as  NL3, TM1, FSU, NL3$\omega \rho$, DD-ME$\delta$, DD-ME2, and IU-FSU are used. The parameter sets NL3, TM1, IU-FSU, IOPB-I, and G3 are used for the core portion. Using the thermodynamic technique, the unified EoSs are built by appropriately matching the inner crust EoS with the outer crust and core EoS.
The radius $R_{1.4}$ rises as the slope parameter is changed from low to high values. The influence of $L_{sym}$ on the NS maximum mass, radius, and radius at 1.4$M_{\odot}$ is computed and a difference of around 2 km in the radius at the canonical mass is discovered. The parameters of a maximally rotating star such as mass, radius, the moment of inertia, and T/W ratio, are likewise determined using the same combination of EoSs. The highest mass and associated radius for RNS, like SNS, do not change substantially. However, the radius at the canonical mass is impacted by the slope parameter.

Chapter \ref{ch:5-hadronquarkrns} studies the hadron-quark phase transition in the context of gravitational wave event GW190814. To investigate the properties of stellar matter, we used many recent DD-RMF parameter sets, including DDV, DDVT, DDVTD, DD-LZ1, DD-ME1, DD-ME2, and DD-MEX. To investigate the phase transition from hadron matter to quark matter, the Vector-Enhanced Bag (vBag) model for quark matter is used, which explicitly accounts for Dynamic Chiral Symmetry Breaking (D$\chi$SB) and repulsive vector interactions.
 Both the Maxwell and Gibss methods are used to construct the mixed-phase between hadrons and quarks. The free parameter in the vBag model, $K_{\nu}$, which controls the stiffness of the EoS curve is fixed at $K_{\nu}$ = 6 GeV$^{-2}$ for three flavor quark matter. The effective bag constant with values $B_{eff}^{1/4}$ = 130 \& 160 MeV are used.
 We see that the hadronic EoSs generated using the latest DD-RMF parameterization fulfill the mass constraint from the GW190814 data, enabling us to investigate the idea of the GW190814 secondary component being a supermassive NS. The phase transition from hadron matter to quark matter reduces NS properties such as mass, radius, and tidal deformability to fulfill the requirements from the GW170817 data, imposing further limits on the NS maximum mass and hence on the dense matter EoS. Apart from the static NS properties, the maximally rotating NS properties such as mass, stellar radius, the moment of inertia, Kerr parameter, etc., are also studied. It is seen that with the given parameter sets, the secondary component of GW190814 is a possible supermassive rotating NS with a strange quark core. 
 
 In Chapter \ref{ch:6-magneticeos}, Using a DD-RMF model, we simulate enormous nucleonic and hyperonic stars that satisfy the limitations imposed by the discovery of perhaps the most massive neutron star (NS) ever recorded (in the secondary object of the gravitational wave event GW190814). The effects of strong magnetic fields on nucleonic and hyperonic EoSs are investigated. We studied the EoS and particle populations using a realistic chemical potential-dependent magnetic field created by solving Einstein-Maxwell equations. The spherical symmetric solutions obtained by solving the TOV equations result in an overestimation of the mass and radius under very strong magnetic fields and hence cannot be used to determine stellar parameters. As a result, we used the LORENE library to determine the stellar properties of magnetic NSs. At low values of the magnetic dipole moment, the EoS resembles the non-magnetic one, implying lower magnetic field intensities.

 In the future, I hope to investigate the NS properties in the presence of exotic matter, such as quarks, utilizing effective field theory-driven RMF (E-RMF) and perturbative QCD (pQCD) with recent extensions, while keeping the observational restrictions in mind. I will investigate the effect of isospin on the nuclear matter—quark matter phase transition at high density and finite temperature by incorporating the nonperturbative interactions between quarks that induce color superconductivity in the 2SC phase. This will aid in the discussion of the conditions under which this state of matter might be formed in the laboratory and its possible observable signs in the context of the quest for the nuclear matter—quark matter mixed phase in heavy-ion studies. However, depending on the temperature, isospin asymmetry, and superconducting gap value, the phase transition could involve conventional quark matter or color superconducting matter. Therefore the effect of isospin dependency on the phase transition will be investigated using normal quark matter as well as color superconducting phase, with varying density and temperature, to examine the properties of the mixed-phase. The Dark Matter (DM) admixtured NSs with color-superconducting phases will be used to study the post-merger analysis of gravitational waves and the oscillation modes in NSs and compare the results obtained without considering DM.
 
 I would also investigate the influence of introducing more exotic particles in the E-RMF and pQCD modeling of magnetic NSs. Kaons, $\Delta$-resonances, and deconfined QM are examples of these (with and without mixtures of phases). To identify the magnetic field intensities that are relevant for modifying the macroscopic properties of NSs and demonstrate how the largest effects created by increasing magnetic field are seen on the NS properties using general relativistic spherically symmetric solutions (TOV equations) as well as the LORENE library, emphasizing the importance of such calculations.




\setlength{\baselineskip}{0pt} 


\nocite{*}
\label{Bibliography}

\lhead{\emph{Bibliography}} 

%
%

\end{document}